\documentclass[11pt]{book}
\author{Fabrizio Messina}
\usepackage[pdftex]{color}
\usepackage{overcite}
\usepackage[italian,english]{babel}
\usepackage{amsmath,amssymb,amsfonts,fancyhdr,graphicx,pdfpages}
\usepackage[margin=15pt, font=footnotesize, labelfont=bf]{caption}
\makeatletter
\renewcommand\@biblabel[1]{#1.}
\makeatother
\fancypagestyle{plain}{%
\fancyhead{}

} \normalbaselineskip=14pt \normalbaselines
\include{Definizioni}
\usepackage[
bookmarks=true,%
bookmarksopen=true,%
bookmarksopenlevel=0,%
bookmarksnumbered=true,%
hypertexnames=false,%
breaklinks=true,%
colorlinks,%
linkcolor=red,
linktocpage,%
pdfkeywords={amorphous silica, point defects, diffusion, hydrogen, E'%
center, silicon dangling bond, UV laser radiation, in situ optical absorption},%
pdfsubject={Experimental PHD Thesis on the effects of laser
radiation on amorphous silicon dioxide}, pdftitle={Role of hydrogen
on the generation and decay of point defects in amorphous silica
exposed to UV laser radiation},pdfauthor={Fabrizio Messina}
]{hyperref}
\begin{document}
\frontmatter
\pdfbookmark[0]{Title}{}
\includepdf[pages={1,2}, scale=1, offset=0 -2.8cm]{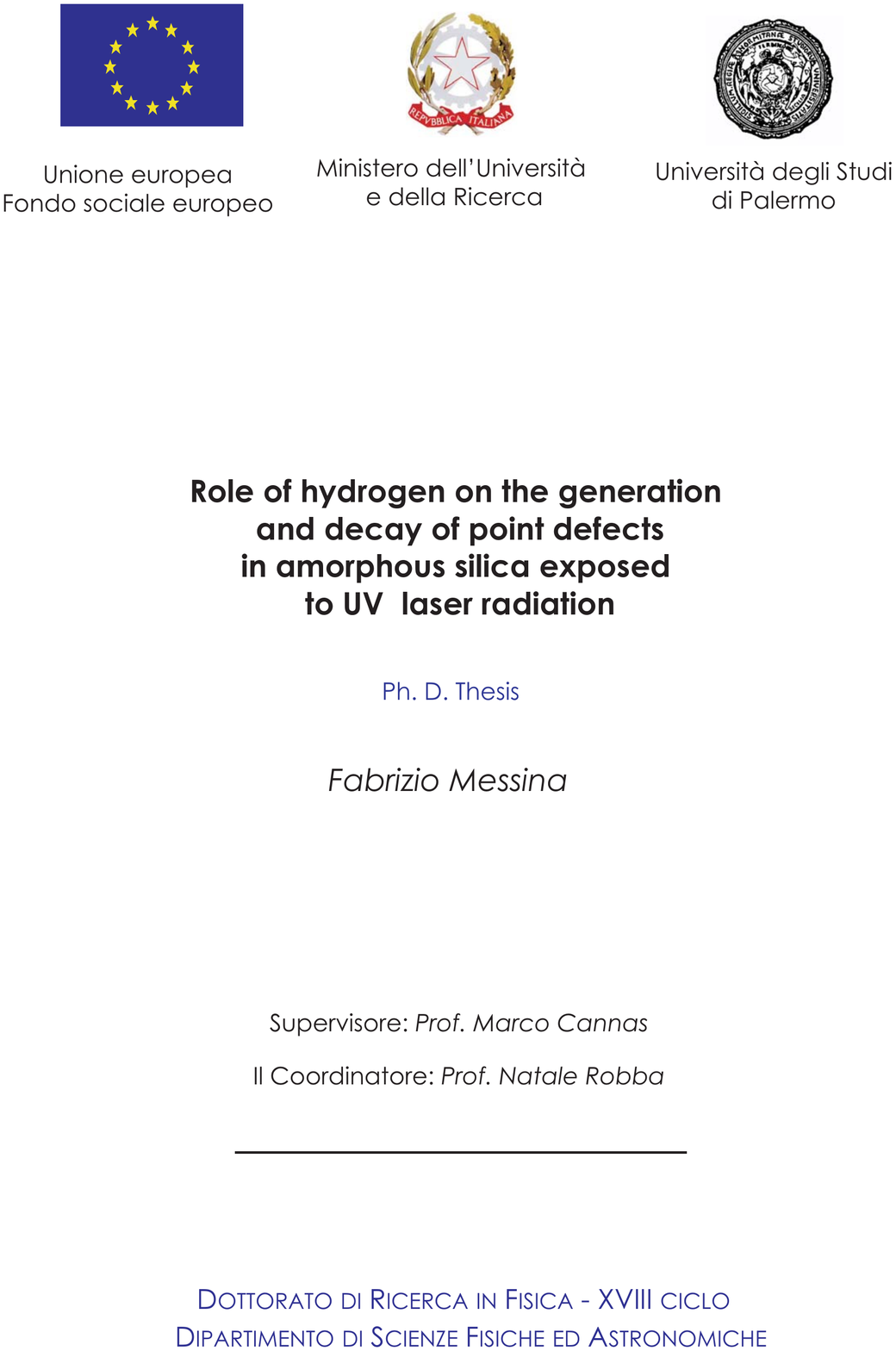}
 \pagestyle{empty} \hrule \begin{Huge}
\center \emph{Acknowledgements}\label{preface}
 \vspace{20pt} \hrule
\end{Huge} \vspace{20pt} \begin{small}

\textsl{The time spent in writing a doctoral Thesis provides a
rather unique opportunity to look back and gain a new perspective on
several years of work. In doing so, I feel the need to thank the
many people that have contributed in different ways to the
realization of this work.}

\textsl{I wish to express my sincerest gratitude to Prof. Marco
Cannas, for generously sharing with me his scientific experience,
which has permitted me to continuously improve my skills while
allowing the progressive development of my autonomy. But even more,
I thank him for his persistent support, and for being gifted with a
firm enthusiasm and a solid optimism which really help to transform
even the hard work in an enjoyable experience. It is also a real
pleasure for me to thank Prof. R. Boscaino. Aside from his
outstanding ability to suggest, direct and organize future research
work, I am grateful to him for having closely followed my activity,
thus allowing me to benefit from his very large experience and
knowledge. I thank Dr. S. Agnello, who has always been exceedingly
generous in sharing his time in plenty of stimulating scientific
discussions, and whose experimental support has been always at hand.
Moreover, I genuinely acknowledge the many important suggestions
that during these years have repeatedly come from Prof. F. M.
Gelardi and Prof. M. Leone, whether during formal or informal
discussions. With all these people I am really indebted for having
given an invaluable contribution to this work and, even more, to my
formation as a young researcher.}

\textsl{I wish to spend also a few words on the colleagues who have
shared with me the PhD experience. The peculiar mix of friendship
and professional collaboration which I have been establishing with
them, particularly with the members of my research group, is a very
important part of the human experience I have gone through during
the last years. It has been a real pleasure to work among these
people.}

\textsl{In these years I have been given several occasions to get
advantage from valuable discussions with scientists from all around
the world. Apart from improving my formation as a physicist, these
meetings have provided important steps forward in the understanding
of the experimental results which I was working on. Among these
people, I express a particular gratitude to Prof. V. A. Radzig.,
Prof. A. Shluger, Prof. A. Thrukhin, and especially to Dr. L. Skuja,
who offered me very detailed and useful comments, suggestions and
corrections on the Thesis. I thank also Prof F. Persico, for
providing me with a wider standpoint on my research, as well as for
carefully proofreading my English. Finally, I would like to express
my appreciation for the work of Prof. N. Robba, the coordinator of
the PhD activities in my department, and I gratefully acknowledge
the technical assistance received by Mr. G. Lapis, Mr. G. Napoli,
and Mr. G. Tricomi.}

\textsl{I wish this to be an occasion to thank my family. Among the
many reasons that I could choose to motivate my gratitude, I would
like to mention their everlasting support and their commitment in
setting up throughout the years the necessary boundary conditions
which have permitted me to arrive here. Also my friends in my
private life truly deserve my most sincere gratefulness for being
close to me during these years and for helping me in times of
trouble.}

\textsl{Last but not least, I thank Elisa. You have been at my side
and believed in me during these years, and your enduring love is a
privilege that I really hope to deserve. This Thesis is dedicated to
you because I regard every achievement in my life as a success of
the two of us.}

\end{small}

\pdfbookmark[0]{Acknowledgements}{Acknowledgements}
\newpage
\clearpage{\pagestyle{empty}\cleardoublepage} \pagestyle{fancy}

\includepdf[pages={1,2}, scale=1, offset=0 -2.8cm]{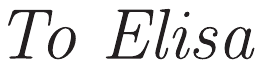}
\renewcommand{\contentsname}{Contents}
\tableofcontents
\mainmatter
\chapter*{Introduction\markboth{Introduction}{}}
\addcontentsline{toc}{chapter}{Introduction} \hspace{0.8cm}Amorphous
silicon dioxide, or silica (\sil), is a material of major scientific
and technological interest, which has been for several decades a
very active subject of investigation frequently crossing the border
between physics and materials science. Research has been motivated
by the circumstance that silica may be regarded as a simple
archetypal system helpful to understand the general properties of
amorphous insulators, as well as by the several applications of the
material, which at the moment remains one of the most important in
optical and microelectronic technologies. Notwithstanding many years
of research and the wide availability of the basic experimental
techniques necessary for the investigation, many of the properties
of silica still remain poorly understood and lively debated, so that
the subject continues to attract an intense research activity aiming
to completely unravel the complex puzzle of its physical properties
\cite{Erice,Nalwa,Devinebook}.

Many investigations dealing with \sil{} focus their attention on the
properties of point defects, usually generated by exposure of the
material to laser or ionizing radiation. On the one hand, the
presence of defects may significantly affect the properties of
silica that are exploited by applications. On the other hand, some
basic physical properties of the defects embedded in an amorphous
solid are not thoroughly understood at the moment, and it took
considerable ingenuity to elucidate the microscopic structure of
even the most common defects in \sil{}\cite{Erice,Nalwa,Devinebook}.
The generation of defects upon irradiation features quite a complex
phenomenology. In particular, the type of centers produced by
radiation and their concentration strongly depend on several
factors, such as the presence of even small concentration of
impurities, or the characteristics of the radiation being used. A
common means of inducing point defects in silica is exposure of the
material to laser radiation. Not surprisingly, apart from being a
useful tool to generate defects to be studied, the interaction of
laser light with the material has become in the years a research
field in itself, also in this case motivated by strong technological
demands due to the wide use of lasers in applications. A specific
defect may usually be generated by several mechanisms depending on
the experimental conditions, but the current understanding of
several of these processes is still qualitative, and many important
questions are unanswered at the moment. For these reasons, the
interest in performing further investigations on these topics is
still alive\cite{Erice,Nalwa,Devinebook}.

This Thesis reports an experimental research work on the effects of
laser irradiation on amorphous silica. The investigation is mainly
focused on the kinetics of generation and decay of point defects
induced by laser radiation. As we are going to show, these processes
are strongly conditioned by the presence of hydrogen, which is able
to diffuse spontaneously in silica even at room temperature. In the
specialized literature on \sil{}, most of the current knowledge on
defect-related issues is based on stationary studies, namely
investigations that look at the defects mainly after the end of any
possible time-dependent effect altering their concentration. In
contrast, we argue that kinetic investigations often provide
additional information worthy to be collected and discussed. This
idea underlies all this work. Our experimental approach is based on
the combined use of several spectroscopic techniques, as mandatory
to perform comprehensive studies in this field. In particular, some
of the most valuable information come from the use of \emph{in situ}
optical absorption to investigate the transient kinetics of point
defects generated by laser radiation.

The Thesis is organized in a Part I, comprising chapters
\ref{backI}-\ref{mm1}, and a Part II, comprising chapters
\ref{mm2}-\ref{chaconc}. The purpose of Part I is to provide a
background on the specialized literature dealing with the effects of
laser irradiation on \sil{}. The approach is by no means
comprehensive, as the attention is mainly focused on a selection of
relevant and interconnected topics that our work aims to clarify.
The last chapter of Part I includes a brief theoretical background
on the experimental methodologies used in our investigation. Part II
reports the experiments and their main results. After the
description of the experimental setups and of the instruments
(chapter \ref{mm2}), the chapters from \ref{overview} to \ref{PRL}
present the results of our investigation; the order is dictated by
the sake of clarity rather than by relevance reasons. Finally, in
chapter \ref{chaconc} we draw the main conclusions and briefly
sketch some proposals for further work. Most of the results
presented in this work have been published as papers on scientific
Journals. Bibliographic references to these papers and to a few
others on closely related topics are available in a
"\hyperlink{LRP}{List of related papers}" included at the end of
Part II.

\part{Background}
\chapter{Effects of laser irradiation on amorphous silica}\label{backI}
\hspace{0.8cm}In this chapter we open the background part of the
Thesis with a discussion of some general properties of amorphous
silica and its defects, after which we propose a review of the
current knowledge on some of the most important effects induced by
laser irradiation in \sil{}.
\section{Structure and basic point defects in silica} \label{strdef}
\begin{figure}[htb!]
\begin{center}
%\makebox[\textwidth]{\framebox[5cm]{\rule{0pt}{5cm}}}
\includegraphics[width=.6\textwidth]{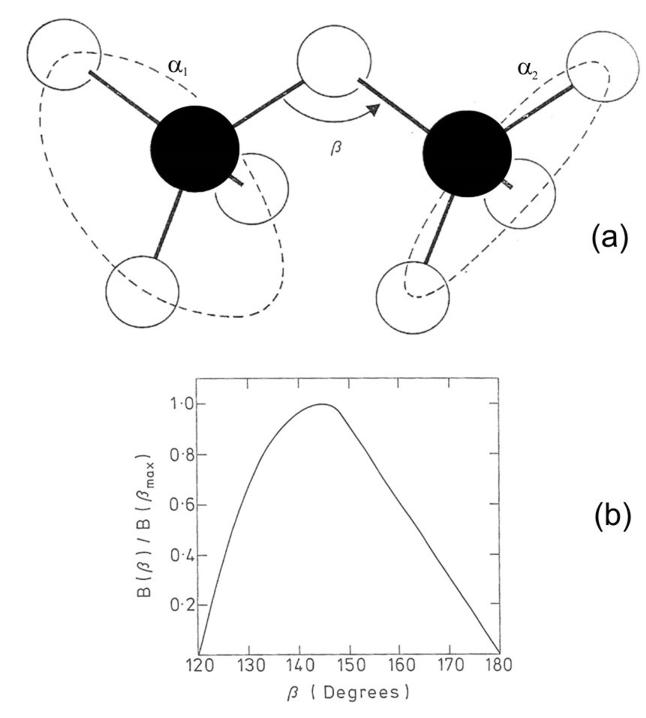}
\end{center}
\caption{Panel (a): structure of \sil, with Si atoms in black and O
atoms in white. The three angles $\alpha_{1}$, $\alpha_2$ and
$\beta$ define the spatial configuration of two connected
tethraedra. Panel (b): statistical distribution of $\beta$, as
derived from X-ray scattering measures. Figure adapted from Mozzi
\emph{et al.}\cite{warren} and Bell \emph{et al.}\cite{Bell72} Other
more recent investigations of the distribution of $\beta$ based on
different experimental techniques have suggested a significantly
narrower distribution than panel (b)\cite{Erice}. }
\label{silicastructure}\end{figure}

\hspace{0.8cm}The most widely accepted description of the
microscopic structure of amorphous silica is known as the Continuous
Random Network (CRN) model, and is mainly based upon the evidences
coming from X-ray and neutron scattering studies of the solid
\cite{Erice,Devinebook,zachariasen,warren,Bell72}. Within the CRN
model, the basic structural unit of silica is a SiO$_4$ tethraedron
with each silicon atom forming four bonds with oxygen first
neighbors (\figurename~\ref{silicastructure}), while the solid
consists in an infinite repetition of basic units connected by
sharing an oxygen atom. For what concerns the basic unit (with an
O\,--\,Si\,--\,O angle of 109.5$^{\circ}$), the structure of silica
closely resembles that of the crystalline solid with the same
chemical composition, i.e. quartz, at least in its most common
crystalline form, $\alpha-quartz$. The main structural difference is
that the angles defining the relative spatial orientation of each
pair of connected tethraedra are statistically distributed in
silica, differently from quartz where they assume fixed values. Due
to the stochastic nature of the spatial configuration, the structure
of silica is amorphous in that it lacks translational invariance,
and thus long-range order. Although the term is sometimes used to
indicate both the crystalline and the amorphous form of silicon
dioxide, in this work only the latter will be referred to as
\emph{silica}.

From the standpoint of solid state physics, silica is a wide bandgap
insulator. Due to its large bandgap ($\sim$9\,eV) the material, when
pure, is optically transparent from infrared (IR) up to ultraviolet
(UV). In addition, it is characterized by excellent insulating
properties, a high radiation resistance and good mechanical and
thermal stability. For these reasons, silica is one of the key
materials at the basis of the current telecommunication and
computing technologies. Indeed, it is the material of choice for
many optical applications, ranging from photolithography to optical
fibers, particularly when usability in the far UV is required, while
in electronics it is found as a thin insulating layer covering
silicon, in metal-oxide-semiconductor (MOS) transistors and other
components
\cite{Erice,Nalwa,Devinebook,GriscomReview91,SkujaSPIE01}. Finally,
silica is being used to manufacture novel experimental devices, such
as photonic crystal fibers and nanowires \cite{mazur,russell}.
Although in recent years other insulators, such as ZrO$_2$ and
HfO$_2$, have been proposed to substitute silica in selected
applications, particularly in microelectronics,\cite{hafnia} at the
moment \sil{} still remains one of the most important technological
materials, so that investigation is in progress with the purpose of
further improving our ability to control its macroscopic properties.

One of the basic issues in research on silica is the study of point
defects and of their generation and conversion processes. A point
defect in the intrinsically disordered structure of silica can be
defined as any deviation from the 'perfect' structure, as described
by definition by the CRN model, provided that it is localized in a
region whose dimensions are comparable to the interatomic
distance\cite{Erice}. The experimental investigation of point
defects is founded on the use of several spectroscopic techniques.
Each of them may be sensitive only to some types of defects,
depending on their properties, and is characterized by its own
advantages and drawbacks. The most common are Electron Spin
Resonance (ESR), Optical Absorption (OA) in the visible, in the UV,
or in the infrared, Raman and photoluminescence (PL) spectroscopies.
The minimum defect concentration detectable by
typical\footnote{\label{detlimit}Here and in the following, the
detection limits refer to standard commercial instruments, and to
sample sizes of the order of 1\,mm.} ESR, OA, PL measurements
usually ranges from $\sim$10$^{13}$cm$^{-3}$ to
$\sim$10$^{16}$cm$^{-3}$ depending on the specific defect, although
the sensitivity can be much higher in particular experimental
conditions\footnote{In particular, PL measurements with high power
laser sources and very low-noise detectors, may allow in some cases
to detect even a single luminescent center\cite{Erice}. }. Often,
the combined use of different techniques allows to infer information
not available by examining separately the results of the single
observations \cite{Erice}. Among these different techniques, the
ESR, albeit applicable only to paramagnetic centers, possesses an
unsurpassed ability to yield direct structural information on the
defects, thereby allowing the development of their microscopic
models. Indeed, also the structural models for diamagnetic defects
are often founded on the observed conversion into a paramagnetic
defect whose structure is known by ESR.

From a historical point of view, it was initially thought that the
knowledge accumulated on crystalline point defects could be
straightforwardly extended to defects embedded in an amorphous
structure. Nonetheless, the study of simple model systems such as
amorphous silica and silicon, showed that the defects found in a
disordered solid feature new and different characteristics peculiar
of the amorphous state, with a quite unforeseen complexity that
still prevents a thorough understanding comparable to that available
for the crystalline case\cite{Erice,SkujaReview98}. In particular,
some types of defects that exist in silica are not found in quartz,
meaning that the amorphous structure provides new degrees of freedom
to incorporate or stabilize the defects. An example of an
amorphous-specific defect class is that of \emph{isolated dangling
bonds}, namely under-coordinated atoms surrounded by a defect-free
network in which all other atoms are regularly
coordinated\cite{SkujaReview98}. Besides, the properties of
'amorphous defects' are conditioned by the characteristic
inhomogeneous site-to-site distribution of structural parameters,
such as bond length and angles, with relevant consequences on their
spectroscopic features. At the moment, the current understanding of
these issues is mostly qualitative \cite{SkujaSPIE01,SkujaReview98}.

The study of point defects is also connected to the issue of
investigating the effects of irradiation. As anticipated, exposure
of the material to several forms of radiation, from lasers to
$\gamma$, X, or $\beta$, results in the generation of defects;
actually, this often occurs by transformation, or \emph{conversion},
of other defects preexisting in the \emph{as-grown} material before
irradiation, and known as \emph{precursors}. Generally the defects
found in the as-grown material are prevalently diamagnetic, whereas
paramagnetic centers are detected in measurable concentrations only
in irradiated specimens. In several cases, the growth of point
defects alters the macroscopic properties of the material with
subsequent degradation of the technological performance, an
important example being the loss of optical transparency due to the
absorption bands associated to most centers. Still, in other cases
the generation or conversion of point defects can be exploited to
induce \emph{ad hoc} new and useful properties
\cite{Erice,Nalwa,Ikushima00,Devinebook,SkujaSPIE01,GriscomReview91}.
In this context, most investigation efforts in recent times focus on
the effects of \emph{laser} irradiation \cite{SkujaSPIE01}. Indeed,
exposure of silica to laser light is common in many applications,
such as photolithography, a technique used to produce
micro-electronics circuits by irradiation of a sensitive material
with laser light through a mask \cite{Nalwa}. Here, the progressive
reduction of the typical circuit dimensions has led to an increase
of the photon energy $hc/\lambda$ used in the process, resulting in
an increasing tendency of laser photons to damage the optical
components used in the lithographic system; hence, the need to
understand and control the basic laser-induced damage processes in
\sil. On the other hand, lasers have been found to be particularly
effective in inducing \emph{controlled} variations of the properties
of silica, in particular when the material is conveniently doped
with suitable impurities\cite{Nalwa,Ikushima00,Erice}. As many of
these processes are related to point defects, it is easily
understandable why the investigation of laser-induced generation and
conversion of defects in \sil{} is currently a very active field.

\begin{figure}[t!]
\begin{center}
%\makebox[\textwidth]{\framebox[5cm]{\rule{0pt}{5cm}}}
\includegraphics[width=.7\textwidth]{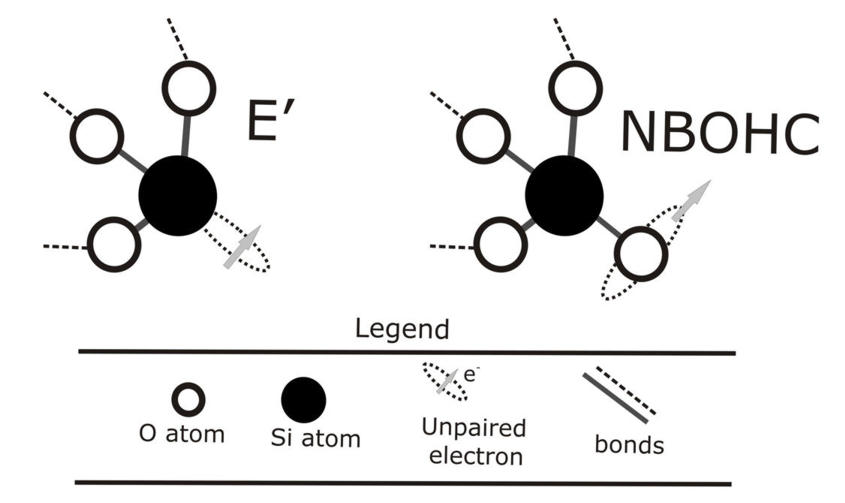}
\end{center}
\caption{Schematic representation of the structure of the $E'$ and
NBOHC centers in \sil{} } \label{modellicentri}\end{figure}

Two among the simplest defects that may be expected \emph{a priori}
to exist in the structure of amorphous silica are the \emph{silicon
dangling bond} ($\equiv$Si$^\bullet$) and the \emph{oxygen dangling
bond}, ($\equiv$Si\,--\,O$^\bullet$), where each symbol "-"
indicates a bond with an oxygen atom, and the symbol "$\bullet$"
represents an unpaired electron. Both centers are expected to be
paramagnetic due to the presence of an unpaired electron in their
structure. Indeed, research has demonstrated that these two
structures (represented in \figurename~\ref{modellicentri}) are
actually very common in \sil{} after irradiation. In the specialized
literature, they are referred to as the \textbf{E$'$}
center\cite{Weeks56,WeeksSonder,Erice,SkujaSPIE01,GriscomReview91},
and the Non Bridging Oxygen Hole Center (\textbf{NBOHC})
\cite{clarifynbohc,Erice,SkujaSPIE01,GriscomReview91,SkujaJNCS94,SilinSPCM78},
respectively.

The NBOHC center is detectable in silica by its characteristic ESR
signal, as well as by its optical activity, consisting in three
absorption bands, at 2.0\,eV, 4.8\,eV and 6.4--6.8\,eV, which excite
a photoluminescence emission peaked at 1.9\,eV
\cite{clarifynbohc,Erice,SkujaJNCS94,SilinSPCM78,CannasPRB04,HosonoSSC}.
The structure of the defect, featuring an unpaired $p$ electron on
an O atom bonded to a 3-fold coordinated Si
($\equiv$Si\,--\,O$^\bullet$), was consistently inferred by ESR
investigations, radio-chemical arguments, and by the detailed study
of its photoluminescence activity
\cite{clarifynbohc,Erice,SkujaJNCS94,SilinSPCM78}.

The $E'$ center is an almost ubiquitous defect in irradiated silica:
it is found virtually in every specimen exposed to radiation. It is
accompanied by a characteristic absorption band peaked at 5.8\,eV,
which is going to play a central role in our experimental
investigation
\cite{Erice,SkujaSPIE01,Nalwa,BoscainoNIM96,NishikawaJNCS94}. In
discussing the properties of the $E'$ center, it is necessary to go
beyond the scheme in \figurename~\ref{modellicentri}. Indeed, to be
precise the expression "$E'$-center" is used in literature to refer
to a \emph{variety} of paramagnetic defects in amorphous silica or
crystalline quartz that share a common property: their unpaired
electron spin is localized on a $sp^3$ orbital of a 3-fold
coordinated Si atom
($\equiv$Si$^{\bullet}$)\cite{Erice,SkujaSPIE01,GriscomReview91}.
However, several possible varieties of the $E'$ center have been
proposed to exist in silica
glass,\cite{GriscomNIM84,GriscomReview91} their structure still
representing a quite debated problem in the current literature: all
of them \emph{comprise} the $\equiv$Si$^{\bullet}$ basic moiety,
whose structure was unambiguously clarified by the careful analysis
of the spectroscopic properties of the ESR signal
\cite{Weeks56,GriscomSSC74,GriscomPRB79,FeiglSSC74,Erice,StirlingJPCM05,GriscomReview91},
but they differ for the environment of the primary structure, and
are distinguishable, at least in principle, either by their
spectroscopic properties or on the basis of their generation
mechanism.

Hystorically, the first variety of $E'$ to be identified was the
$\equiv$Si$^{\bullet}$ $^+$Si$\equiv$ structure in crystalline
quartz, which is generated by hole trapping on a pre-existing oxygen
vacancy ($\equiv$Si\,--\,Si$\equiv$):
\begin{equation} \label{genEdasisipre}
\equiv\text{Si\,--\,Si}\equiv\text{ + h$^+$ } \longrightarrow \text{
}\equiv\text{Si}^{\bullet}\text{
 $^{+}$}\text{Si}\equiv\end{equation}
The Si\,--\,Si bond in the vacancy initially involves two electrons.
After hole trapping, the remaining electron localizes asymmetrically
on only one of the two Si atoms, thus giving the structure at the
right side of (\ref{genEdasisipre})
\cite{Erice,SkujaSPIE01,GriscomReview91,FeiglSSC74,GriscomSSC74}.
This type of $E'$ is widely believed to exist also in amorphous
silica, generated either by laser or ionizing
radiation\cite{ImaiJNCS94,SkujaSPIE01,Erice,Devinebook,BuscarinoPRB06,GriscomNIM84,SkujaReview98}.
In this work, it will be referred to as
\emph{vacancy}-$\mathbf{E'}$.\footnote{The traditional
classification of the sub-types of the $E'$ center in \sil{} is
mainly based on the differences among their ESR signals, rather than
on their structural
properties\cite{Erice,GriscomNIM84,GriscomReview91}. In literature,
the \emph{vacancy}-$E'$ is often referred to as the $E'$$_{\gamma}$
center;\cite{SkujaReview98,Erice,GriscomNIM84,GriscomReview91} we
prefer here to use an alternative denomination in order to better
emphasize the structural difference with repect to the isolated
dangling bond. } However, differently from the case of quartz, in
principle the $\equiv$Si$^{\bullet}$ defect can exist in silica also
as an \textbf{isolated dangling bond} \cite{SkujaSPIE01}. It is
unclear at the moment which of the two is prevalent in the amorphous
material. A further type of $E'$ which will be referred to in the
following, is the so-called $\mathbf{E'_{\beta}}$
center,\cite{GriscomNIM84,GriscomReview91} whose currently accepted
model is $\equiv$Si\,--\,H $\equiv$Si$^\bullet$; this defect is
supposed to arise from trapping of a H atom on an oxygen
vacancy\footnote{namely the reaction: $\equiv$Si\,--\,Si$\equiv$ + H
$\Longrightarrow$ $\equiv$Si\,--\,H $\equiv$Si$^{\bullet}$},
followed by a structural relaxation at the end of which the unpaired
electron points away from the former vacancy
\cite{Erice,SkujaReview98,GriscomReview91}. Such a local
rearrangement must be supposed in order to account for the
unobserved hyperfine interaction (see chapter \ref{mm1}) between the
electron and the proton. For completeness, we recall that at least
two more types of $E'$ have been proposed by the several studies on
point defects in silicon
dioxide\cite{Erice,SkujaReview98,GriscomNIM84,GriscomReview91}: the
$E'$$_{\delta}$ and the $E'$$_{\alpha}$. Recent investigations have
provided strong evidence that the structure of the former consists
in an unpaired electron delocalized over four $sp^3$ orbitals of
nearly equivalent Si atoms\footnote{Accepting this structural model,
the inclusion of the $E'$$_{\delta}$ within the category of $E'$
centers relies on a broader interpretation of the above definition
of $E'$.},\cite{BuscarinoPRL05,BuscarinoPRB06,BuscarinoReview} while
the $E'$$_{\alpha}$ has been proposed to be a variant of the
\emph{vacancy}-$E'$ in which the unpaired electron points away from
the vacancy and interacts with an extra oxygen in the \sil{}
network\cite{BuscarinoPRL06}.

Apart from the detailed structure of the several sub-types of $E'$,
it is still discussed at the moment if all of them contribute to the
"usual" 5.8\,eV absorption band \cite{SkujaReview98}. This problem
is complicated by the circumstance that the electronic transition
responsible for this absorption has not been clarified yet. In
particular, theoretical calculations have ascribed the optical
transition to a "charge transfer" of the unpaired electron from the
dangling bond to the positively charged $^+$Si$\equiv$ group which
is expected according to
(\ref{genEdasisipre})\cite{PacchioniPRB98b,PacchioniPRL98}; on this
basis, the 5.8\,eV band should be a property of the
\emph{vacancy}-$E'$ defect, but certainly not of the isolated
dangling bond. However, other theoretical works have questioned this
finding \cite{OReillyPRB83,SkujaReview98}, and it has been also
predicted a weaker transition falling in the same energy region but
entirely comprised within the silicon dangling bond
\cite{PacchioniPRB98b,PacchioniPRL98}. For these reasons, the
character of the 5.8\,eV absorption band of $E'$ has not been
conclusively estabilished \cite{SkujaSPIE01,SkujaReview98}.

As common in literature, in this work the term "$E'$ center", as
well as the expression "silicon dangling bond", will be used in the
general sense of a defect comprising the $\equiv$Si$^{\bullet}$
fragment \emph{and} absorbing at 5.8\,eV, as these are the two most
basic features of this defect in \sil{}. More specifically, the
expression "isolated" dangling bond will be used for the $E'$ center
that consists \emph{only} in the $\equiv$Si$^{\bullet}$ structure.

\begin{figure}[htb!]
\begin{center}
%\makebox[\textwidth]{\framebox[5cm]{\rule{0pt}{5cm}}}
\includegraphics[width=.7\textwidth]{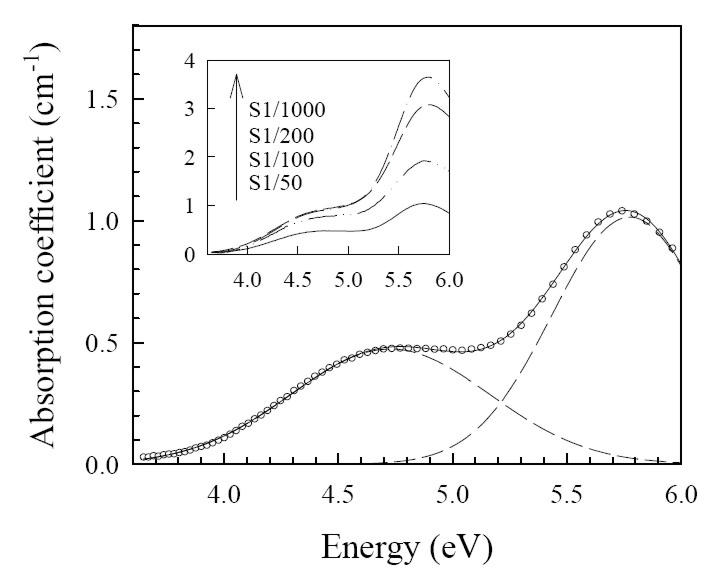}
\end{center}
\caption{Optical absorption induced in a synthetic silica sample by
$\gamma$ irradiation, showing the 5.8\,eV and 4.8\,eV bands related
to $E'$ and NBOHC centers respectively. Figure taken from Cannas
\emph{et al.}\cite{nostroboh} } \label{inducedoa}\end{figure}

Aside from being the two most "basic" defects in \sil{}, $E'$ and
NBOHC are very important also from the technological point of view.
In fact, the generation of these two centers is the main cause of
degradation of the UV transparency of silica upon irradiation, due
to their wide absorption bands peaked at 5.8\,eV and 4.8\,eV
(\figurename~\ref{inducedoa}) \cite{Erice,SkujaSPIE01,Nalwa}.
Several works have investigated the generation processes of the two
defects under laser irradiation, showing a complex landscape in
which many formation channels are possible, depending on the
specific laser wavelength and intensity being used, as well as on
the manufacturing procedure of the material. However, the
understanding of the generation mechanisms often remains at a
qualitative level, thus calling for more investigations to clarify
the several open
issues.\cite{SkujaSPIE01,SkujaReview98,Erice,Nalwa,nostroboh,Devinebook}.

Many other defects, such as oxygen-deficient centers, peroxy
radicals, interstitial oxygen, and so on have been identified, some
of them still being actively studied, and the picture becomes even
more complicated if we include another important class of defects in
silica: impurity-related centers. In this context we recall that
point defects are usually classified in extrinsic or intrinsic
depending on the presence or absence of an impurity in their
structure, respectively (in the specific case of \sil{}, the term
"impurity" indicates any chemical element which is neither Si nor
O). $E'$ and NBOHC are examples of intrinsic defects; extrinsic
defects due to the presence of impurities are always present in
variable concentrations in the material. On the other hand, selected
impurities can be deliberately added by doping to induce many useful
properties
\cite{SkujaSPIE01,SkujaReview98,Erice,Nalwa,GriscomReview91,Devinebook}.
Germanium and Hydrogen have assumed a particularly important role in
the last two decades. In fact, Ge-doped silica has been shown to
feature interesting optical properties, such as photosensitivity and
nonlinearity, not observed in pure \sil{} and partially related to
point defects \cite{Erice,Nalwa,Ikushima00}. Hydrogen is by far the
most common impurity in \sil{}, whose presence is basically
unavoidable also in high purity samples. Hydrogen is able to diffuse
in silica, even at low temperatures, and conditions the response of
the material to irradiation by reacting with point defects.
Furthermore, diffusion and reaction of mobile species like hydrogen
in \sil{} are quite interesting also from a fundamental point of
view, as their features are a fingerprint of the amorphous nature of
the solid
\cite{Shelby77,Shelby94,Schmidt98,GriscomJNCS84,GriscomJAP85,KajiharaPRL02}.

The experiments presented in this Thesis are relevant to the
understanding of the generation mechanisms and the properties of
laser-induced $E'$ centers in silica. In addition, some of our
findings concern conversion processes of Ge-related defects. Hence,
to provide a background for the presentation of the results, in the
following sections of this chapter we are going to review in more
detail the current understanding of laser-induced effects in \sil{},
particularly with regard to the generation of $E'$ and the
conversion of Ge-related defects.

\section{Exposure of \texorpdfstring{\sil}{silica} to laser radiation}
\subsection{Overview} \hspace{0.8cm}The complex interaction processes
between laser radiation and matter are a very timely research
subject. In general, laser light interacts with solids by coupling
with electron or nuclear degrees of freedom, resulting in
articulated photothermal and photochemical effects of which the
generation of point defects is just one of the end products. The
availability of progressively increasing laser intensities in a wide
range of wavelengths has provided research and industry with an
invaluable set of tools for material processing. In fact, lasers are
currently used for deposition, etching, ablation, and controlled
amorphization, they are capable to generate hot plasmas, to induce
controllable modifications in the optical properties, and so on
\cite{UVlasersbook,SkujaSPIE01}. In this sense two categories of
lasers have been found to be particularly effective, thus being the
subject of strong technological and scientific interest: pulsed
high-power UV lasers and femtosecond pulsed lasers. The former emit
pulses with a few ns duration, and typical energy density from tens
to hundreds mJcm$^{-2}$ per pulse, this corresponding to peak
intensitites\footnote{The intensities, however, can be highly
increased above these values by focusing the laser beam.} around
10$^{10}$\,Wcm$^{-2}$. The most commonly used are excimer lasers
(using KrF, ArF, XeCl, F$_2$ as the active medium) or frequency
multiplied Nd:YAG lasers. Femtosecond lasers (the most common being
the Ti:Sapphire) emit in the infrared (IR) spectral range and have a
much higher peak intensity due to the very short pulse duration.
Also, they can be converted to shorter wavelengths by using
frequency-multiplying nonlinear devices \cite{UVlasersbook}.

In regard to silica, the effects of laser radiation began to be
intensively studied during the eighties, following the first
observation in 1984 of the $E'$ center induced by laser radiation
\cite{StathisPRB84}. Nowadays, laser irradiation has been
demonstrated to induce a variety of macroscopic effects in pure or
doped silica. The most common and relevant for applications is
transparency loss, but more complex modifications of the material
have been demonstrated, such as variations of the refractive
index\cite{MiuraNIMB98,Qiu98,KawamuraAPL01,KawamuraAPL02},
laser-induced birefringence,\cite{BorrelliAPL02} or optical
nonlinearity\cite{CorbariAPL02},
ablation\cite{KawamuraAPL01,LiAPL02},
densification\cite{FioriPRB86,SewardJNCS97,ShimboJAP00,KawamuraAPL01},
crystallization,\cite{Devinebook} and more. From the microscopic
standpoint, many of these effects are at least partially ascribable
to generation and conversion of point defects, although laser
irradiation may also cause extended (non-local) structural
rearrangements of the \sil{} matrix. Finally, some of these
processes are strongly enhanced when the material is enriched in
some impurities like Germanium, as we are going to see in more
detail in the following.\cite{Erice, Ikushima00, Devinebook,Nalwa}

After two decades of research on laser-induced processes in \sil{}
it is now known that every defect induced by ionizing radiation can
be observed, under suitable conditions, in laser-irradiated
silica\cite{SkujaSPIE01}. From a purely structural point of view,
laser-induced defects are virtually identical to those growing upon
much more energetic $\gamma$ or $\beta$ exposure; on the other hand,
processes caused by laser irradiation are characterized by some
peculiarities, such as \emph{selectivity}, i.e. laser light at a
given wavelength often acts only on a specific precursor, thereby
exciting only a particular defect generation process among the many
which can be induced simultaneously by ionizing radiation
\cite{SkujaSPIE01}. In this sense, laser irradiation may be a very
useful tool if the aim is to study selectively the properties and
the generation mechanisms of a specific point defect.

\subsection{Generation of $E'$ center}\label{geneprimosection} \hspace{0.8cm} Stathis and
Kastner were the first to report the generation of point defects in
\sil{} upon exposure to laser light \cite{StathisPRB84}: irradiation
of a high purity silica sample with KrF and ArF pulsed excimer
lasers (photon energies 5.0\,eV and 6.4\,eV respectively) or with a
F$_2$ pulsed laser (7.9\,eV) was observed to generate the $E'$
center, detected by electron spin resonance (ESR) spectroscopy. The
effect was found to be strongly dependent on laser photon energy
\cite{StathisPRB84}.

\begin{figure}[htb!]
\begin{center}
%\makebox[\textwidth]{\framebox[5cm]{\rule{0pt}{5cm}}}
\includegraphics[width=.7\textwidth]{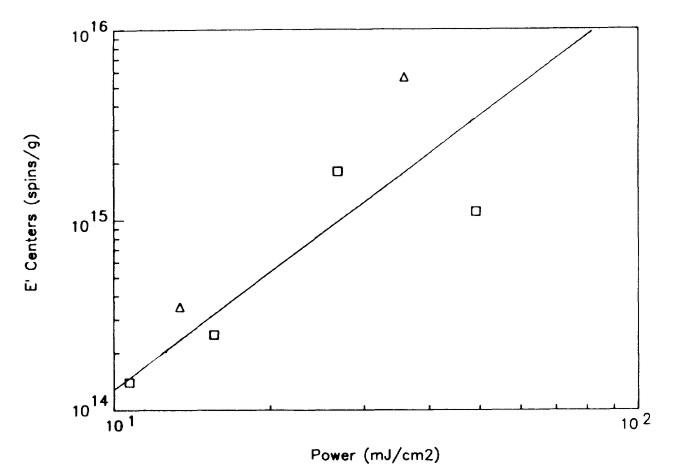}
\end{center}
\caption{Concentration $\rho$ of $E'$ centers induced in a synthetic
silica sample by 6.4eV ArF pulsed laser radiation, as a function of
energy density per pulse $I$. The line with slope
$\alpha$=1.9$\pm$0.7 is a least-square fit of the experimental data
with the function $\rho$=k$I$$^{\alpha}$. The value of $\alpha$,
close to 2, evidences the involvement of a two photon process.
Squares and triangles represent two different irradiations. Figure
taken from Tsai \emph{et al}.\cite{TsaiPRL88} }
\label{twophotonTsai}\end{figure}

High purity silica samples like those used by Stathis and Kastner
are basically transparent to excimer laser photons (whose energies
are lower than the silica bandgap $\sim$9\,eV) due to the absence of
midgap electronic levels that may be introduced by the presence of
impurities or pre-existing intrinsic defects\footnote{This is not
completely true for the F$_2$ laser, which features a 7.9\,eV photon
energy close to the bandgap. Indeed, as a consequence of its
amorphous nature, \sil{} features localized electronic states just
below the conduction band and above the valence band\cite{Erice},
which are responsible for an interband absorption "tail" extending
to energies smaller than the bandgap}. Hence, at that time, it was
somewhat unexpectedly found that photons with energies lower than
the bandgap, which excite the material in an optical transparency
region, are able to produce defects efficiently. Indeed, prior to
this work, point defect generation in \sil{} had been observed only
under high energy radiation, such as $\gamma$, $\beta$, electrons or
neutrons. Therefore, many authors tried to investigate in more
detail the defects generation mechanism, suggesting in particular
that it could involve multi-photon processes. Convincing
experimental proofs followed a few years after, when several
independent experiments evidenced that the concentration of induced
$E'$ centers depends quadratically on incident KrF or ArF laser
power (\figurename~\ref{twophotonTsai}), strongly suggesting two
photon absorption to be involved in $E'$ generation
\cite{TsaiPRL88,AraiAPL88,NishikawaPRB93}.

Nevertheless, the specific mechanism at the basis of laser-induced
damage remained quite debated. Broadly speaking, the immediate
consequence of a two photon absorption by the \sil{} matrix is the
generation of free electron-hole pairs or excitons
\cite{Trukhinreviewexc, Ashcroft, Dekker}, depending on the
excitation energy. If free charges are made available, they can move
and get trapped at a suitable precursor site, when available,
forming paramagnetic defects. Alternatively, they can recombine in
an exciton. The successive dynamics of the excitons is strongly
conditioned by the electron-phonon coupling. In fact in silica,
similarly to other solids, elementary excitations such as excitons
can spontaneously get trapped at a regular lattice site and become
immobile at low temperature
\cite{Devinebook,ShlugerPRB90,BeigiPRL05,SaetaPRL93,Trukhinreviewexc}.
This occurs because the electron-hole pair is able to create a
localized distortion in the lattice, which decreases the total
energy, resulting in the formation of a potential well in which the
quasi-particle is then trapped. In both silica and quartz this
process is spontaneous, and is accompanied by a transition of the
exciton to a triplet electronic state. The result of the trapping is
referred to as the \emph{self-trapped-exciton} (\textbf{STE}). The
STE has been for years the subject of intensive experimental and
theoretical work, aimed to elucidate accurately its transient
structure, dynamics and formation mechanism. Once formed, the STE
decays radiatively (with a measurable luminescence band peaked
between 2.0\,eV\,--\,2.3\,eV and a $\sim$\,1\,ms radiative lifetime)
or non-radiatively. The luminescence decay time of the exciton
results from a combination of the parallel radiative and
non-radiative channels; as a result, it decreases with temperature
due to the progressive thermal activation of the non-radiative decay
mechanisms.\cite{Devinebook,ShlugerPRB90,BeigiPRL05,SaetaPRL93,Trukhinreviewexc,
PetiteNIMB94,FisherPRL90,ShlugerJPC88,GuizardJPCM96,ShlugerPRB00,TanimuraPRL83,TanimuraJPC88,ItohPRB90}

The relevance of the STE in the present context arises from the fact
that it can serve as a transient product of irradiation whose energy
may be eventually spent in the formation of a point defect. The
non-radiative decay of a STE on an (initially) defect-free lattice
has been recognized in silica, as well as in other solids, as one of
the fundamental mechanisms for defect formation, which allows to
break the original Si\,--\,O\,--\,Si matrix locally, so as to form a
permanent
defect\cite{Erice,Devinebook,NishikawaPRB90,ImaiPRB93,DevinePRB90,
TsaiPRL91,FisherPRL90,ShlugerJPC88,ShlugerPRB90,ShlugerPRB00,
SaetaPRL93,BeigiPRL05,Trukhinreviewexc,GuizardJPCM96,PetiteNIMB94,TsaiPRL88,FukataAPL03}.
Due to the very nature of the process, in which the electron-phonon
coupling is involved, this defect generation mechanism is
characterized by an efficiency that is highly temperature-dependent.
The main experimental evidences supporting this process, which
permits $E'$ generation in \sil{} under laser irradiation, are
reviewed in the next subsection A. After that, in the following two
subsections B and C we discuss other two classes of generation
mechanisms of $E'$ that have been proposed in literature; both of
them, differently from the decay of STE on the defect-free silica
matrix, require the presence of pre-existing precursor defects.
\begin{figure}[t!]
\begin{center}
%\makebox[\textwidth]{\framebox[5cm]{\rule{0pt}{5cm}}}
\includegraphics[width=.5\textwidth]{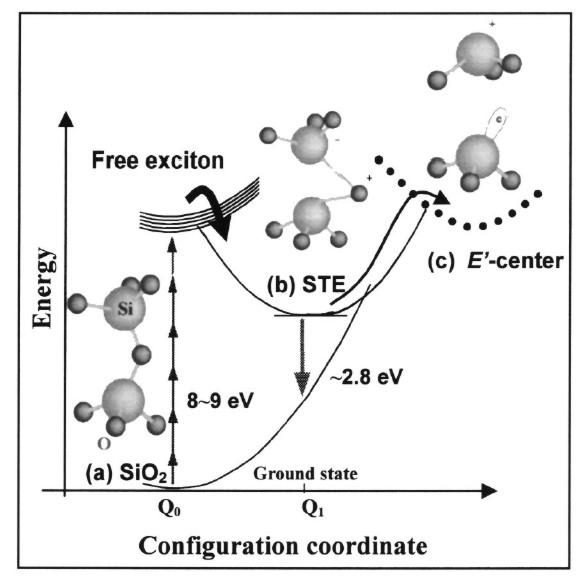}
\end{center}
\caption{Configuration coordinate diagram representing the
generation and self trapping processes of an exciton in
(crystalline) SiO$_2$, triggered by absorption of $\sim$9\,eV
radiation. After self trapping, the exciton can either decay
radiatively (2.8\,eV emission) or non radiatively; in the second
case, a stable $E'$ center is supposedly generated. Figure taken
from Fukata \emph{et al}.\cite{FukataAPL03} }
\label{STEfigure}\end{figure}

\subsubsection{A. Excitonic mechanism and generation from a regular
lattice site} \hspace{0.8cm}It was reported that XeCl laser
radiation at 4.0\,eV, lower than KrF and ArF lasers energy and less
than one \emph{half} of the silica bandgap, is unable to generate
$E'$ at comparable power levels \cite{AraiAPL88}. This is consistent
with the idea that electron-hole pairs formation is the first step
of the $E'$ generation process, since with 4.0\,eV radiation
\emph{three} photons are required to bridge the silica bandgap,
resulting in a much lower efficiency. A further step was made by
Tsai \emph{et al.} \cite{TsaiPRL88,TsaiPRL91}, which analyzed the
effect of highly focused ArF laser radiation on synthetic silica:
they were able to demonstrate experimentally that actually no
\emph{free} electrons in conduction band are generated by absorption
of two 6.4\,eV photons, in spite of the very high laser intensity
($>$100\,J\,cm$^2$ per pulse) used in their experiment. Hence, it
was argued that most of the energy deposited by laser radiation
results in the formation of excitons. Moreover, they measured a very
high $E'$ concentration up to 10$^{18}$\,cm$^{-3}$, higher than any
possible precursor, and reached without any tendency to saturation
with increasing irradiation dose. The experimental evidences led the
authors to propose the following model: laser-induced generation of
$E'$ occurs by non-radiative decay of STE on the defect-free silica
matrix, leading to the formation of a persistent defect.

\begin{figure}[t!]
\begin{center}
%\makebox[\textwidth]{\framebox[5cm]{\rule{0pt}{5cm}}}
\includegraphics[width=.7\textwidth]{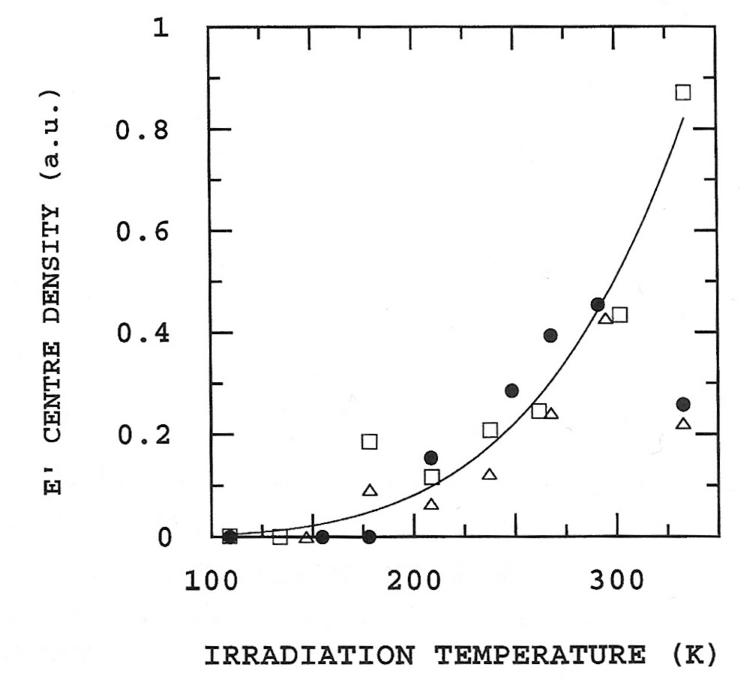}
\end{center}
\caption{Defect density as a function of irradiation temperature in
bulk \sil{} samples exposed to 5.0\,eV irradiation with
300\,mJcm$^{-2}$ energy density per pulse. Circles and triangles are
the concentrations of induced $E'$, while the squares represent the
concentrations of peroxy radicals produced by a post-irradiation
thermal treatment at 600\,K for one hour. Figure taken from Devine
\emph{et al}.\cite{Devinebook} } \label{devinetemp}\end{figure}

The involvement of STE in defect generation was also confirmed by
other two experimental findings. First, studying the concentration
of laser-induced $E'$ as a function of temperature, it was found
that the generation of defects is quenched below $\sim$150\,K under
radiation at 5.0\,eV (\figurename~\ref{devinetemp})
\cite{DevineMRSSP86,DevineNIMB94}, or grows with temperature up to
400\,K under radiation at 6.4\,eV \cite{RothschildAPL89}. This
behavior resulted to be anticorrelated with the luminescence decay
time of the exciton (as measured in quartz), as expected from the
idea that the generation of $E'$ is the end-product of the
non-radiative decay of STE.
Second, another work focused on the comparison between laser- and
$\gamma$-induced concentrations of $E'$ \cite{DevinePRL89}. It is
known that the primary effect of $\gamma$ exposure is creation of
electron-hole pairs by the Compton effect, which are expected to
spontaneously combine in STEs; hence, it was found that the growth
curves of $E'$ under $\gamma$ and laser are identical, if the
irradiation dose is expressed in both cases as the number of
generated electron-hole pairs. This finding strongly suggests the
existence of a common generation process for $E'$, independent of
the radiation source, and occurring via the excitonic mechanism.

Up to this point, we have described the $E'$ generation process as
due to the decay of a STE at a regular lattice site, without
mentioning the specific reaction by which the decay creates the
defect. In a first stage, this was supposed to be
\cite{TsaiPRL91,DevinePRB90}:
\begin{equation}
\label{genEdasiosiconO} \equiv\text{Si\,--\,O\,--\,Si}\equiv \text{
+ (STE) }\longrightarrow\text{ }\equiv\text{Si}^{\bullet}\text{
 $^{+}$}\text{Si}\equiv\text{ + O$^0$ + }e^-\end{equation}
where O$^{0}$ is an interstitial oxygen atom, $e^-$ is an electron,
and the first term at the right side is the $E'$ center.
Reaction~(\ref{genEdasiosiconO}) can be read saying that the self
trapping and the successive non-radiative decay of the exciton
causes the displacement of an oxygen ion to an interstitial
position; of course, one must assume that the transient structure is
eventually stabilized by migration of the oxygen, preventing
recombination with the $E'$. One of the most convincing proofs of
this generation scheme was the demonstration\footnote{The presence
of radiolytic O$_2$ was indirectly inferred by the observed decay of
$E'$ correlated to the formation of \emph{peroxy radicals}
\textbf{PRs} ($\equiv$Si\,--\,O\,--\,O$^{\bullet}$), after a high
temperature treatment, supposedly as a consequence of the reaction
of O$_2$ with $E'$ \cite{TsaiPRL91,DevineNIMB94}. } of the presence
in the lattice of interstitial oxygen O$_2$ of radiolytic origin
(presumably formed by diffusion and dimerization of O atoms), after
high power ArF irradiation of samples where O$_2$ was initially
absent\cite{TsaiPRL91,DevineNIMB94}. Process (\ref{genEdasiosiconO})
is in some sense the analogous under laser radiation of the
so-called \emph{knock on} process induced by $\gamma$ rays, namely
the displacement of an oxygen atom due to the impact on a lattice
site of a sufficently energetic Compton
electron\cite{Devinebook,TsaiPRL88}.

Another group of experiments suggested a slightly different picture
of defect generation upon non-radiative decay of STE. In fact,
several works reported a correlated growth of $E'$ and NBOHC centers
either under $\gamma$ or laser radiation
\cite{DevinePRB90,DevinePRB89,RothschildAPL89,ImaiPRB93,ImaiJNCS94,AraiPRB92},
both defects being usually monitored with ESR spectroscopy. This
finding can be explained as a consequence of another possible
rupture mechanism of the Si\,--\,O\,--\,Si bond:
\begin{equation}
\label{genEdasiosiconnbohc} \equiv\text{Si\,--\,O\,--\,Si}\equiv
\text{ + (STE) }\longrightarrow\text{
}\equiv\text{Si}^{\bullet}\text{
 + }^{\bullet}\text{O\,--\,Si}\equiv\end{equation}where the two terms
 at the right side are the $E'$ and the NBOHC
respectively. To accept reaction~(\ref{genEdasiosiconnbohc}), it is
necessary to explain the absence of broadening in the measured ESR
signal of the $E'$ and NBOHC centers, which would be expected due to
dipole-dipole interaction. To this aim, it was supposed that the two
defects after generation undergo a separation process so as to be
far enough (tens of angstroms) that this effect is negligible. Only
the defects that separate enough can survive recombination and
become permanent centers \cite{AraiPRB92,ImaiPRB93,DonadioPRB05}.

More recently, time-resolved pump-and-probe experiments on the
femtosecond scale have been carried out to study the self trapping
process of the exciton in \sil{}, demonstrating that it is actually
well represented by reaction~(\ref{genEdasiosiconnbohc}), rather
than (\ref{genEdasiosiconO}). In other words, the exciton self
trapping leads to the generation of a transient $E'$--NBOHC pair,
occurring in roughly 250\,fs \cite{SaetaPRL93}. Similar results have
been obtained by theoretical calculations
\cite{ShlugerJPC88,BeigiPRL05}. These studies, however, cannot
investigate the long-term stabilization of the defect pair leading
to the formation of a pair of \emph{permanent} centers; for this
reason, reaction~(\ref{genEdasiosiconO}) still cannot be excluded,
because after reaction (\ref{genEdasiosiconnbohc}), oxygen may be
eventually formed by decomposition of NBOHC, leading to the same net
effect as reaction (\ref{genEdasiosiconO}).

Finally, in some works it has been proposed that the occurrence of
process (\ref{genEdasiosiconO}) or (\ref{genEdasiosiconnbohc}) upon
STE decay could be \emph{site-selective}. In particular, process
(\ref{genEdasiosiconnbohc}) may occur preferably on \emph{strained
bonds}, namely sites in the inhomogenous structure of \sil{} where
the Si\,--\,O\,--Si bonds are weaker and can be more easy cleaved
\cite{KajiharaJNCS03,DonadioPRB05,HosonoPRL01}. Strained bonds can
be detected by vacuum ultraviolet (6\,eV\,--\,9\,eV) OA or Raman
measurements \cite{HosonoPRL01}; they are expected to be
particularly abundant in optical fibers \cite{KarlitschekOC98}, as a
consequence of the manufacturing process itself, or in silica
densified with suitable techniques. Consistently, correlated $E'$
and NBOHC generation has been reported in optical fibers
\cite{KarlitschekOC95}, and its efficiency in bulk silica was
observed to grow after densification \cite{DevinePRB90}. More
recently, process (\ref{genEdasiosiconnbohc}) has been proposed to
occur under F$_2$ laser radiation (7.9\,eV) \cite{KajiharaJNCS03},
by selective breaking of the strained bonds; with such a high photon
energy, the process can occur even by single photon absorption,
since the first electronic transition of the strained bond falls at
energies slightly lower than the silica bandgap.

\subsubsection{B. Generation from Oxygen Deficient
Centers}\label{ODCsection} 
\begin{figure}[htb!]
\begin{center}
%\makebox[\textwidth]{\framebox[5cm]{\rule{0pt}{5cm}}}
\includegraphics[width=.6\textwidth]{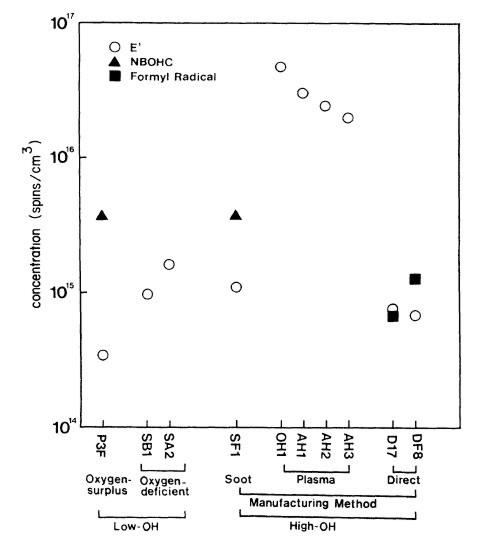}
\end{center}
\caption{Concentrations of several centers induced by 6.4\,eV ArF
laser radiation in silica samples prepared with different
manufacturing techniques. Figure taken from Nishikawa \emph{et
al}.\cite{NishikawaPRB90} } \label{varietamateriali}\end{figure}

\hspace{0.8cm}Up to this point, we have
described the experimental evidences regarding generation of $E'$
from an initially unperturbed silica matrix. Other works have
pointed out that processes starting from preexisting precursor
defects are often important and can prevail on the simpler
Si\,--\,O\,--\,Si bond
breaking\cite{NishikawaPRB90,AraiAPL88,ImaiPRB88,ImaiPRB91,NishikawaPRB93,NishikawaPRB93B}.
In particular,\cite{NishikawaPRB90} some studies have shown quite a
complex phenomenology not easily reducible to the picture proposed
so far: by comparing samples prepared with a variety of
manufacturing techniques it was pointed out a very strong dependence
of the induced defect species and of their concentrations on the
type of silica, as shown for example in
\figurename~\ref{varietamateriali}.

Such a variability strongly suggests that defect generation from
precursors, rather than transformation of the pure silica matrix, is
much more important than initially supposed. Nowadays it is
generally believed that generation from the defect-free matrix is
the prevailing generation mechanism of point defects and yields high
defect concentrations ($>$10$^{16}$\,cm$^{-3}$) only under very high
power focused excimer laser or femtosecond laser radiation, as it
occurs in a minority of
experiments\cite{TsaiPRL88,TsaiPRL91,SaetaPRL93,FukataAPL03,ZoubirPRB06}.
After recognizing the important role of precursors, the comparison
of samples prepared with different techniques has become nowadays a
standard approach to investigate the generation processes of point
defects.

In general, the transformation of a precursor into another point
defect under laser radiation may occur by several processes:
\textbf{(i)} direct one-, two-, or many- photon absorption at the
defect site (depending on the level scheme of the defect and on the
laser photon energy), which may result in a permanent transformation
of the structure, usually as a consequence of ionization or of the
breaking of a chemical bond. \textbf{(ii)} Trapping at the precursor
site of free electrons or holes made available by ionization of
another precursor or by two photon absorption by the \sil{} matrix.
\textbf{(iii)} Non-radiative decay at the precursor site of STEs
created in a first stage by two photon absorption by the \sil{}
matrix\cite{NishikawaPRB90,AraiAPL88,ImaiPRB88,ImaiPRB91,Erice}.
\textbf{(iv)} Processes mediated by the diffusion to the precursor
site of mobile ions, atoms or molecules made available by
irradiation. This class of mechanisms will be discussed in more
detail in the next chapter.

Some of the first experimental investigations dealing with
generation of $E'$ from precursors focused on oxygen-deficient
silica samples
\cite{AraiAPL88,ImaiPRB88,ImaiPRB91,NishikawaPRB93,NishikawaPRB93B},
where a specific class of generation mechanisms for $E'$ was pointed
out, namely conversion of pre-existing oxygen-deficient-centers
(ODCs). The ODCs in silica are found in two varieties: the
\textbf{ODC(I)}, absorbing at 7.6\,eV, and the \textbf{ODC(II)},
which absorbs at 5.0\,eV and emits photoluminescence at 2.7\,eV and
4.4\,eV \cite{SkujaSPIE01,SkujaReview98,SkujaJNCS92,SkujaSSC84}.
Both are usually found in significant concentrations in
oxygen-deficient silica due to the substoichiometry of the specimens
and have been supposed to be precursors for generation of $E'$ under
laser radiation.

The detailed understanding of the generation processes of $E'$ from
ODCs is a problem linked with that of elucidating the exact
structure of the two varieties of the oxygen vacancy. In detail, the
commonly accepted structural model for the ODC(I), based upon
several experimental evidences, is an oxygen vacancy, with a
$\equiv$\,Si\,--\,Si\,$\equiv$ structure
\cite{SkujaReview98,HosonoPRB91,ImaiPRB88}. Under ArF or F$_2$ laser
radiation, it was observed a reduction (\emph{bleaching}) of the
ODC(I) 7.6\,eV band concurrent to the growth of $E'$
\cite{NishikawaPRB93}. This finding suggests that the ODC(I) can be
a precursor for $E'$ by the following reaction:
\begin{equation} \label{genEdasisi}
\equiv\text{Si\,--\,Si}\equiv\text{ } \longrightarrow \text{
}\equiv\text{Si}^{\bullet}\text{
 $^{+}$}\text{Si}\equiv\end{equation} where the ionization of the vacancy (under
laser radiation at energies lower than 7.6\,eV) may occur by direct
two photon ionization or by hole trapping\footnote{in the latter
case, it coincides with reaction (\ref{genEdasisipre})}. As already
mentioned, for reaction~(\ref{genEdasisi}) to occur we must assume
that after ionization the remaining electron spontaneously localizes
on only one of the two adjacent Si atoms; in regard to oxygen
vacancies in quartz, this scheme has been supported by theoretical
calculations \cite{FeiglSSC74}. It is worth noting that the
laser-induced reduction of ODC(I) concentration is usually much
larger than the concentration of generated paramagnetic defects
\cite{NishikawaPRB93,SkujaSPIE01,SkujaReview98}, this meaning that
other unknown channels besides reaction~(\ref{genEdasisi})
contribute to the bleaching of the 7.6\,eV absorption band. The
generation of $E'$ by hole trapping on the oxygen vacancy is
generally regarded as one of the main mechanisms triggered by
$\gamma$-irradiation of amorphous silica
\cite{Erice,ImaiJNCS94,Devinebook,BuscarinoPRL06,BuscarinoPRB06}.

In contrast, the understanding of the role of ODC(II) is less
straightforward. Indeed, also in this case it has been observed a
reduction of ODC(II) absorption and luminescence signals accompanied
by a growth of $E'$, usually in concentration of a few
10$^{15}$\,cm$^{-3}$ \cite{ImaiPRB88,ImaiPRB91}. From this result,
and from the analysis of the power dependence, it was inferred that
ODC(II) can serve as a precursor for generation of the $E'$ by a two
photon process. Moreover, in samples that contain a high native
concentration of ODC(II) centers, a high concentration of induced
$E'$ is usually observed upon irradiation \cite{SkujaReview98}.
Starting from these observations, in literature it was proposed for
the ODC(II) a model of \emph{unrelaxed} oxygen vacancy, i.e. a
vacancy similar to the ODC(I) but with a different value of the
Si\,--\,Si bond length corresponding to another potential energy
minimum \cite{ImaiPRB88,KuzuuPRB95}.

On the other hand, another structural scheme has been proposed for
the same defect based on completely different observations: the
\emph{twofold coordinated Si} model, in which ODC(II) corresponds to
the structure =\,Si$^{\bullet\bullet}$, a silicon bonded with two
oxygen atoms and hosting an electron lone pair
\cite{SkujaJNCS92,SkujaSSC84}. This model has its own quite
convincing proofs, among which the observed conversion by hydrogen
trapping to \textbf{H(I)} center (=\,Si\,--H$^{\bullet}$), whose
structure has been established unambiguously by ESR spectroscopy and
calculations
\cite{SkujaJNCS92,SkujaSSC84,Vitko,RadtzigKK79,RadtzigPSS86,TsaiJNCS87,EdwardsNIMB88}.
Starting from the twofold coordinated Si model, it is not
immediately understandable how ODC(II) may be transformed in $E'$
after ionization; even so, theoretical calculations have suggested
this to be possible via a spontaneous local structural
rearrangement\footnote{To further complicate the whole picture,
several experimental evidences and theoretical calculations have
also suggested that excitation of the ODC(I) in the 7.6\,eV
absorption band can result in its conversion to a transient ODC(II)
center (in the excited singlet electronic state). Such unstable
defect manifests itself by emitting a luminescence emission signal
very similar to that associated to the standard ODC(II)
\cite{SkujaReview98,DonadioPRL01,TrukhinJNCS92,BoscainoPRB96,UchinoPRL01,AgnelloPRB03}
 }. \cite{DonadioPRL01} In conclusion, the issue of the ODCs structural models,
as well as the possible generation of $E'$ from these precursors are
two connected problems not completely clarified and both still open
at the moment.

\subsubsection{C. The Si\,--\,H precursor and instability of $E'$}
\hspace{0.8cm}Apart from the oxygen deficient centers, it was argued
that another class of possible precursors exists for the $E'$
center: they are extrinsic defects in the form Si\,--\,X, where X is
an impurity. One of the most important in literature and in relation
to the results of this Thesis is the Si\,--\,H group.
\begin{figure}[h!]
\begin{center}
\includegraphics[width=.6\textwidth]{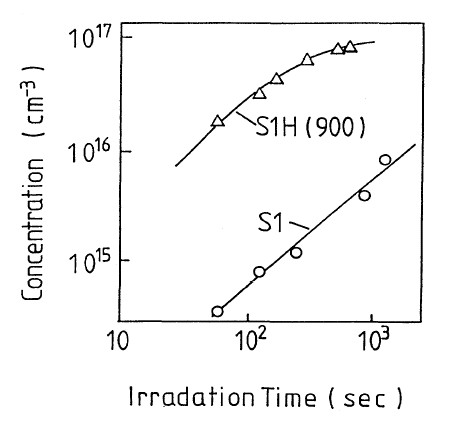}
\end{center}
\caption{Concentrations of $E'$ (measured by ESR) as a function of
ArF laser irradiation time, in a oxygen-deficient glass before (S1)
and after (S1H(900)) thermal treatment in hydrogen. Figure taken
from Imai \emph{et al}.\cite{ImaiPRB91} }
\label{comparisondaimai}\end{figure}

One of the strongest evidences that Si\,--\,H can be a precursor for
$E'$ came from an experiment in which an oxygen-deficient sample was
treated at high temperatures (T$>$600\,C$^{\circ}$) in a H$_2$
atmosphere. Upon thermal treatment, it was observed a reduction of
the 7.6\,eV absorption band due to the ODC(I) vacancies, accompanied
by an increase of the concentration of Si\,--\,H groups, as
evidenced by the growth of their typical Raman spectroscopic signal
at 2250\,cm$^{-1}$ \cite{ImaiPRB91}. This observation suggests that
the vacancies can be converted to Si\,--\,H groups by the following
reaction:\begin{equation} \label{genSi-HdaSiSI}
\equiv\text{Si\,--\,Si}\equiv\text{ + H$_2$ } \longrightarrow \text{
}\equiv\text{Si\,--H + H\,--Si}\equiv\end{equation} Due to
reaction~(\ref{genSi-HdaSiSI}), Si\,--\,H becomes the dominant
defect species in the H$_2$-treated substoichiometric material.
Then, it was observed that glasses prepared by (\ref{genSi-HdaSiSI})
are characterized by a much higher efficiency for $E'$ generation
(under 6.4\,eV laser irradiation) than before H$_2$ treatment
(\figurename~\ref{comparisondaimai}) \cite{ImaiPRB91}. This
experimental evidence in itself strongly suggests that Si\,--\,H
groups can be efficient precursors for the $E'$. Two\footnote{It is
worth noting that the two reactions (\ref{genEdaSi-HA}) and
(\ref{genEdaSi-HB}) give rise in principle to two different
varieties of $E'$. In particular, the former may be very similar to
the so-called $E'$$_{\beta}$ center, introduced in section
\ref{strdef}, while the latter is the usual \emph{vacancy}-$E'$
center. } possible photo-induced reactions were put forward
\cite{ImaiPRB91}:

\begin{equation} \label{genEdaSi-HA}
\equiv\text{Si\,--H + H\,--Si}\equiv\text{ }\longrightarrow \text{
}\equiv\text{Si}^{\bullet}\text{ H\,--Si}\equiv\text{ +
$\frac{1}{2}$ H$_2$}\end{equation}

\begin{equation} \label{genEdaSi-HB}
\equiv\text{Si\,--H + H\,--Si}\equiv\text{ }\longrightarrow \text{
}\equiv\text{Si}^{\bullet}\text{ $^+$Si}\equiv\text{ + H$_2$ +
}e^-\end{equation} supposedly occurring by a two-photon process, on
the basis of theoretical calculations that had predicted the lowest
$\sigma$-$\sigma^*$ electronic transition of the Si\,--\,H group to
be at about 8\,eV.\cite{Robertson88}

\begin{figure}[htb!]
\begin{center}
%\makebox[\textwidth]{\framebox[5cm]{\rule{0pt}{5cm}}}
\includegraphics[width=.7\textwidth]{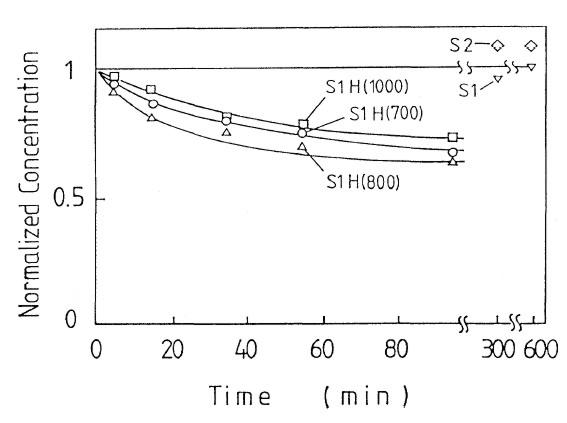}
\end{center}
\caption{Decay of $E'$ concentration after the end of laser
irradiation in the sample containing ODC [S1] and a sample
containing Si\,--\,H [S1H(900)], obtained by the high temperature
treatment of [S1] in a H$_2$ atmosphere. Figure taken from Imai
\emph{et al}.\cite{ImaiPRB91} } \label{decaydaImai}\end{figure}

The issue of $E'$ generation from Si\,--\,H is closely connected to
the problem of \emph{stability} vs \emph{instability} of the
paramagnetic center. In fact, as apparent from the above reactions,
the rupture of Si\,--\,H precursors gives rise both to $E'$ centers
and hydrogen (atoms or molecules). Now, it is known that H$_2$ in
silica is able to diffuse, even at room temperature, and
spontaneously reacts with some paramagnetic point defects
\cite{GriscomJNCS84,GriscomJAP85}. Hence, if $E'$ centers are
generated from Si\,--\,H, we should expect that after the end of
irradiation molecular hydrogen diffuses and recombinates back with
the defect restoring the Si\,--H bond by the \emph{inverse} of
reactions (\ref{genEdaSi-HA}) and (\ref{genEdaSi-HB}). Imai \emph{et
al.} confirmed this prediction by observing a partial
post-irradiation decay of the induced $E'$ taking place in a few
hours at room temperature (\figurename~\ref{decaydaImai}), as
apparent from ESR measurements performed at several delays from the
end of exposure. On the one hand, the observation of the anticipated
decay further supports the proposal of Si\,--\,H groups as possible
precursors of the $E'$ center under laser radiation; on the other
hand, however, we note that if one assumed $E'$ to be generated
\emph{only} by the above reactions, the concentration of hydrogen
produced together with $E'$ should be sufficient to
\emph{completely} cancel the paramagnetic defects in the
post-irradiation stage. Hence, the \emph{partial} decay in
\figurename~\ref{decaydaImai} indicates that, in the experimental
conditions explored by Imai \emph{et al.}, either a portion of
hydrogen was involved in a reaction with another unknown defect, or
that a second generation channel of $E'$ was active besides
Si\,--\,H breaking. Instability of point defects in \sil{} due to
reaction with mobile species is a much more general issue that
extends beyond this particular case. Since the study of the reaction
between $E'$ and H$_2$ is a central topic for this work, these
problems will be discussed in more detail in the next chapter.

The possibility that Si\,--\,H can be a precursor of $E'$ (both
under laser and ionizing radiation), at least in materials where its
concentration is increased artificially by a thermal treatment in
H$_2$, has been suggested in several other works
\cite{AfanasevJPCM00,NishikawaPRB90,NishikawaPRB93B,ImaiPRB93,ImaiJNCS94}.
It is worth noting that, while in the H$_2$-treated samples used by
Imai \emph{et al} Si\,--\,H should be present mainly in the form
predicted by eq.~(\ref{genSi-HdaSiSIrep}), i.e. a hydrogen-decorated
vacancy, in principle it is conceivable that Si\,--\,H may also
exist simply as an isolated defect, depending from the manufacturing
procedure of the material. Besides, Si\,--\,H can be found in couple
with Si\,--\,OH if the two are formed together by reaction of H$_2$
with Si\,--\,O\,--\,Si during the high temperature manufacturing
process of the silica specimen being used \cite{Shelby94,Doremus}.
Since these varieties of Si\,--\,H are not straightforwardly
distinguishable by their IR or Raman spectroscopic features, it is
not easy to known at the moment what is the prevalent arrangement of
the Si\,--\,H groups in a given silica sample.

Neglecting for the moment this rather subtle point, we return now to
the properties of the $E'$ center created from Si\,--\,H, whose
basic feature is that it is \emph{unstable} and decays, due to
recombination in the post-irradiation stage with free hydrogen
coming from the same generation process. In this respect, the $E'$
coming from Si\,--\,H is different from the $E'$ generated by the
previously described mechanisms, which is not reported to be
unstable at room temperature.

Since the $E'$ created by Si\,--H rupture is transient due to the
very nature of its generation process, ordinary (\emph{ex situ})
measurements, usually performed starting from a few minutes after
the end of irradiation, are inappropriate to thoroughly investigate
its properties and dynamics. In contrast, some papers proposed a
more deep investigation of the \emph{transient} $E'$ induced by
laser irradiation, by employing \emph{in situ} optical absorption
techniques \cite{LeclercAPL91,LeclercJNCS92,MarshallJNCS97,
HitzlerJNCS92,ShimboJJAP95,ShimboJJAP01,SmithAO00}. \emph{In situ}
OA techniques are measurements of the induced OA spectrum or
absorption coefficient at some fixed wavelength (for instance
215\,nm, corresponding to the peak of the 5.8\,eV absorption band of
$E'$ center) performed while the specimen is being exposed to laser
light. Measures usually continue also after the end of laser
irradiation, to investigate the post-irradiation kinetics of the
defect. As opposed to \emph{ex situ} measurements, the \emph{in
situ} techniques allow to investigate also the growth kinetics of
the defects and to access the first stages of the decay, so that
this approach is mandatory to obtain a satisfactory understanding of
the generation of the transient $E'$ coming from
Si\,--\,H.\footnote{Or in any other experimental situation in which
the generated defects are unstable and decay in the post-irradiation
stage.} The diffusion of \emph{in situ} techniques is relatively
recent in literature: in particular, PL \emph{in situ} measurements
have been applied to study the laser-induced generation and decay of
NBOHC by monitoring its emission at 1.9\,eV
\cite{KajiharaAPL01,KajiharaPRL02,Anedda}, and OA \emph{in situ} has
been recently employed to study the kinetics of defect generation
under ionizing radiation \cite{Simoneinsitu}.

\begin{figure}[htb!]
\begin{center}
%\makebox[\textwidth]{\framebox[5cm]{\rule{0pt}{5cm}}}
\includegraphics[width=.9\textwidth]{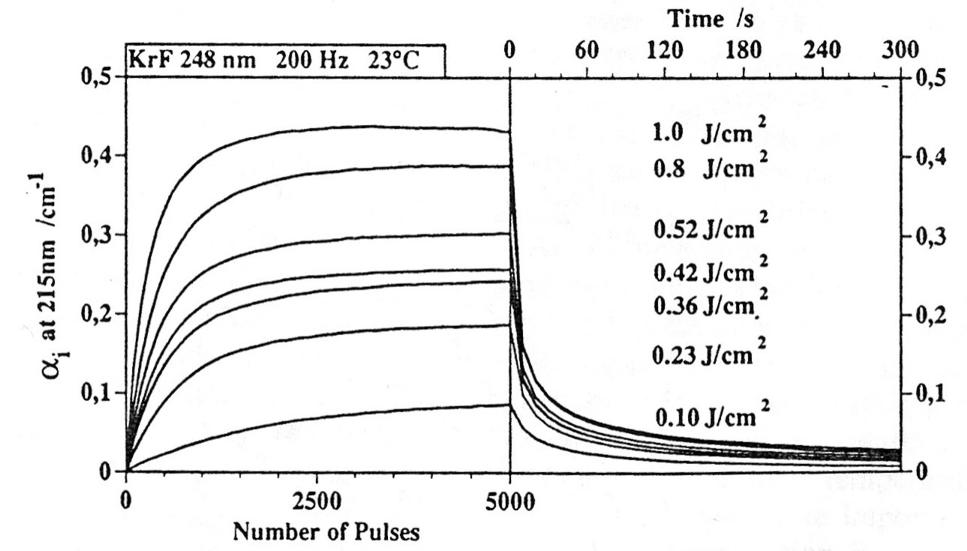}
\end{center}
\caption{Induced absorption coefficient at 5.8eV in high OH fused
silica during and after irradiation with KrF laser with several
different energy densities per pulse. Figure taken from Leclerc
\emph{et al}.\cite{LeclercJNCS92} }
\label{leclerckinetics}\end{figure}

One of the first \emph{in situ} studies was performed by Leclerc
\emph{et al.} \cite{LeclercAPL91,LeclercJNCS92}, first in bulk
\sil{} and later on optical fibers\cite{HitzlerJNCS92}. The authors
examined the kinetics of $E'$ under KrF 5.0\,eV laser radiation in
samples of high purity silica with high Si\,--\,OH content by
\emph{in situ} optical absorption. They found that the defects are
almost completely transient and rapidly disappear after the end of
irradiation (\figurename~\ref{leclerckinetics}). Such a relaxation
was not observed in samples with a low Si\,--\,OH content.

On the basis of the results, the authors put forward a model in
which the $E'$ center, which may be initially created from any
precursor, is converted by the relaxation process to some unknown
state $\overline{E'}$, spectroscopically invisible. Then, the
$\overline{E'}$ can be reconverted very efficiently to $E'$ by a
successive irradiation. Based on the observed temperature
dependence, they proposed that the relaxation is due to reaction
with H$_2$, and that the $\overline{E'}$ can actually be identified
with the Si\,--\,H. From the study of the dependence of process on
laser power, they concluded that generation of $E'$ occurs by a
single-photon process. It is worth noting that this finding is
apparently at variance with the above reported theoretical
predictions on the Si\,--\,H electronic transition; this issue is
even more controversial because no experimental data are available
at the moment on the UV absorption spectrum of Si\,--\,H. Finally,
Leclerc \emph{et al.} suggested on a qualitative basis that the
saturation of $E'$ concentration reached during laser exposure is
due to an equilibrium between generation and decay.

Other studies by Shimbo \emph{et al.}, dealing with transient $E'$
generated by KrF and ArF laser radiation,
\cite{ShimboJJAP95,ShimboJJAP01} examined in more detail the
dependence on the Si\,--\,OH content, finding that the maximum
induced absorption coefficient is inversely correlated to the native
concentration of Si\,--\,OH. In contrast to Leclerc \emph{et al.},
the power dependence of the transient $E'$ generation rate was found
to be sublinear.

Finally, Smith \emph{et al.} distinguished a fast and a slow decay
of $E'$, respectively due to recombination with atomic and molecular
hydrogen.\cite{SmithAO00} Furthermore, their data showed that the
decay of $E'$ is observed only in samples that already contain
dissolved hydrogen in \emph{free} form (H$_2$) prior to irradiation.
This conclusion contrasts with the original results by Imai \emph{et
al.} where the decay was observed even if hydrogen was initially
stored entirely as Si\,--\,H or Si\,--\,OH bonds. Therefore, it is
not clear whether in this case the $E'$ centers were actually
generated from Si\,--\,H. Finally, the results by Smith \emph{et
al.} suggested a linear single-photon $E'$ generation process.

In summary, although the application of \emph{in situ} measurement
techniques to the laser-induced $E'$ center has permitted several
step forwards in the understanding of its transient dynamics, many
issues are still debated due to the report of contradictory results.
In particular, given that Si\,--\,H has been regarded as one of the
main generation mechanism of transient $E'$ centers, an important
unclarified point is whether Si\,--\,H rupture occurs by a one- or a
two-photon process. In addition, some questions have never been
investigated, such as the detailed mathematical modeling of the
growth kinetics\footnote{For completeness we also recall that
Si\,--\,Cl groups, sometimes abundant in synthetic silica depending
on the manufacturing procedure, have been argued to be possible
precursors of the defect by a radiation-induced breaking process
similar to what occurs in the case of Si\,--\,H. In contrast,
breaking of Si\,--\,F, another relatively common impurity in \sil{},
is generally considered much less likely under laser radiation due
to the greater strength of the silicon-fluorine
bond.\cite{AfanasevJPCM00,NishikawaPRB90,NishikawaPRB93,ImaiPRB93,ImaiJNCS94}
Both Si\,--\,Cl and Si\,--\,F impurities in \sil{} can be detected
by Raman spectroscopy, sensitive to their stretching vibrational
modes. }.

\subsection{Ge-related defects}\label{Gedefsection} \hspace{0.8cm}From the technological
point of view, Ge-doped \sil{} has a very important application in
the fabrication of optical fibers, where it is commonly used as the
\emph{core} material, surrounded by a pure silica \emph{cladding}.
In fact, Ge-doping increases the refractive index of silica thus
permitting the guidance of light in this geometry by total
refraction\cite{Nalwa}. The scientific interest for Ge-doped silica
increased abruptly after the experimental demonstration by Hill
\emph{et al.} of the strong \emph{photosensitivity} of this material
\cite{Hill78}. In this context, the term photosensitivity indicates
the phenomenon by which exposure to laser irradiation induces a
modification of the refractive index $n$.
\begin{figure}[htb!]
\begin{center}
%\makebox[\textwidth]{\framebox[5cm]{\rule{0pt}{5cm}}}
\includegraphics[width=.8\textwidth]{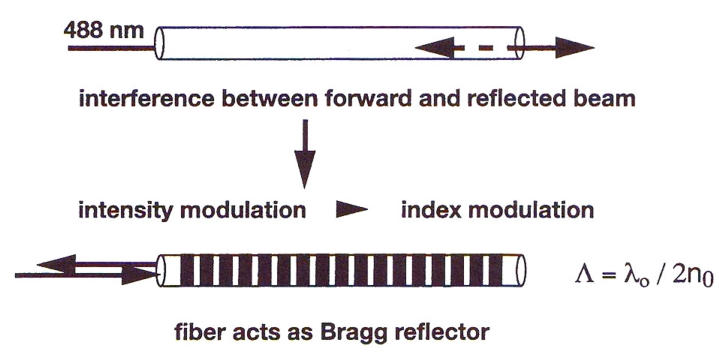}
\end{center}
\caption{Inscription of a Fiber Bragg Grating in a Ge-doped core
optical fiber. Figure adapted from Devine \emph{et
al}.\cite{Devinebook} } \label{fbg}\end{figure}

In the original experimental scheme by Hill \emph{et al.}
(\figurename~\ref{fbg}), two interfering coherent high-intensity
laser beams of the same wavelength $\lambda_0$ propagate within the
fiber in opposite directions, creating a standing wave pattern. The
exposure of the core material to laser radiation induces a variation
in the refractive index, which is modulated according to the
intensity pattern. The modulation of the refractive index may be
regarded as a grating-like structure, so that the fiber then behaves
like a filter; as a result, it reflects selectively at the entrance
light signals with wavelength near $\lambda_0$ \cite{Hill78}. Such a
device is referred to as a \emph{fiber bragg grating} (FBG) and
results to be a very versatile tool, both in telecommunications,
where it is used in multiplexing devices, and as a temperature or
stress sensor \cite{Erice,Bilodeau,Nalwa,Devinebook}.

More recent studies have demonstrated photosensitivity also in
glasses with different chemical compositions, including pure silica
under femtosecond laser radiation, but Ge-doped \sil{} remains one
of the most apt to technological purposes
\cite{Qiu98,MiuraNIMB98,ChiodiniPRBST}. It is thought that
photosensitivity may be exploited to build low-cost advanced optical
devices, as for example optical waveguides inscribed in the bulk of
glass by focused laser irradiation \cite{MiuraNIMB98}, or the
recently demonstrated three-dimensional optical memories, namely
devices where the "bit" is recorded as a localized modification of
$n$, with achieved storage densities of
10$^{12}$\,--\,10$^{13}$\,bits\,cm$^{-3}$ \cite{Qiu98,Watanabe00}.

Another very interesting optical property that can be induced in
Ge-doped \sil{} by particular procedures is optical nonlinearity. It
was first observed by Osterberg and Margulis that prolonged exposure
of an optical fiber to 1064\,nm light from a Nd:YAG laser results in
the generation of second harmonic light at 532\,nm, with
progressively increasing efficiency \cite{Osterberg86, Osterberg87}.
This finding came as a great surprise, because it is widely known
that glass, being a centrosymmetric material, has a zero
second-order nonlinear coefficient \cite{Harper,Ottica}: obviously
the induced nonlinearity must be due to the breaking of the native
symmetry property of the material. In the following years, special
techniques have been proposed, which are able to confer to Ge-doped
silica an artificial nonlinearity comparable to that of crystalline
materials commonly used in applications, e.g. LiNbO$_3$. One of most
effective is UV poling, consisting in laser UV irradiation under a
strong electric
field.\cite{Ikushima00,Fujiwara,IkushimaB,TakahashiAPL97} From the
technological point of view, the usability of glass as a nonlinear
active medium would bear many advantages, such as the low cost and
the straightforward integrability with other optical components
\cite{Kameyama01,Ikushima00,Nalwa}.

Hence, the demonstration of these properties of Ge-doped \sil{} led
to a strong interest in understanding the microscopic mechanisms at
the basis of photosensitivity and induced optical nonlinearity, with
the aim to control and enhance these properties of the material. It
was soon realized that photosensitivity is partly due to
laser-induced conversion processes of Ge-related point defects,
resulting in variations of the OA spectrum, which cause the
modification of $n$ through the Kramers-Kronig relationships .
Beyond point defect conversions also other mechanisms, such as
laser-induced densification, are supposed to contribute to
photosensitivity\cite{Nalwa,Erice,Dong,Essid,Ashcroft,AtkinsEL93}.
Similarly, also optical nonlinearity is linked to Ge-related point
defects. In particular, it has been proposed that one of the action
mechanisms of UV poling is ionization of Ge-related precursors
followed by trapping of the produced electrons in a non-homogeneous
spatial distribution, due to the presence of the electric poling
field. This charge redistribution implies the breaking of the
symmetry of the material so as to allow for second order
nonlinearity \cite{Ikushima00,TakahashiAPL97}.

\subsubsection{A. Defect models} \hspace{0.8cm}On these grounds, many
studies have tried to shed light on the problem of Ge-related
defects in silica. Germanium may be arranged within \sil{} in many
different configurations, each of which constitutes a specific point
defect. Since Ge and Si are isoelectronic elements, it is
qualitatively expected that many Ge-related point defects are
structurally identical to Si-related centers apart from the
substitution of Si with Ge. Actually, some exceptions are known to
this simple scheme. Research aims to clarify two main issues: to
associate the observed spectroscopic signals to specific structural
models, and to elucidate the photochemical processes induced by
irradiation, which usually convert diamagnetic precursors into
paramagnetic defects. The two problems are clearly connected,
because important clues to identify the structure of diamagnetic
defects often come from the observation of their conversion
processes \cite{Erice,Nalwa,NeustrevJPCM94}. As outlined hereafter,
the picture that has emerged from these studies is quite complicated
and still leaves many open questions.

\begin{figure}[htb!]
\begin{center}
%\makebox[\textwidth]{\framebox[5cm]{\rule{0pt}{5cm}}}
\includegraphics[width=.9\textwidth]{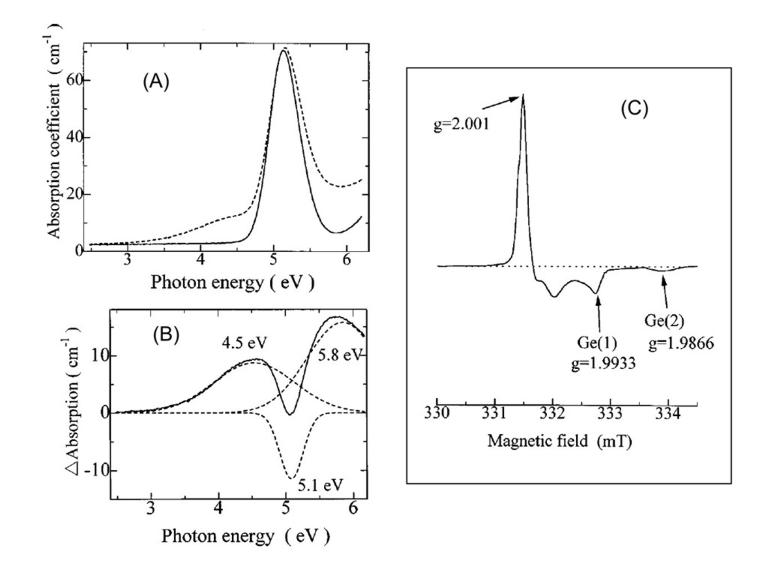}
\end{center}
\caption{Panel (A): native OA spectrum of a Ge-doped silica sample
(full line) and spectrum after irradiation (dashed line). Panel (B):
difference OA, showing the bleaching of the 5.1\,eV band and the
growth of two components at 4.5\,eV and 5.8\,eV. Panel (C): ESR
signal of the irradiated sample, resulting from the superposition of
components due to three Ge-related defects. Figure adapted from
Fujimaki \emph{et al}.\cite{FujimakiPRB98} }
\label{esempige}\end{figure}

Ge-containing silica samples usually present a native absorption
band peaked at $\sim$5.1\,eV (often referred to as B$_{2\beta}$ band
\cite{Tohmon}) prior to any irradiation. This signal has been
connected with Germanium oxygen deficient centers (GeODCs) and it
has been suggested to play a key role in the point defect conversion
processes at the basis of photosensitivity and nonlinear effects. In
fact, in experiments with excimer laser or UV lamp radiation, an
intensity reduction (\emph{bleaching}) of the native $\sim$5.1\,eV
OA band has been repeatedly observed, associated to the growth of
several paramagnetic signals and of new absorption signals (see an
example in \figurename~\ref{esempige}). This observation clearly
suggests that the defect(s) responsible for the B$_{2\beta}$ band
are converted by UV radiation to other
centers\cite{HosonoPRB92,FujimakiPRB98,NishiiPRB95,
NishiiAPL94,TakahashiAO03,NeustrevJPCM94,SakohOE03}.

The most common Ge-related paramagnetic defects that are detected by
ESR in irradiated Ge-doped \sil{} are the \textbf{Ge\,--E$'$}
center, the Germanium Electron Centers (GECs) \textbf{Ge(1)} and
\textbf{Ge(2)} and the \textbf{H(II)} center. The microscopic
structures of three of them, Ge\,--$E'$, Ge(1) and H(II), have been
unambiguously identified by ESR studies, further supported by
theoretical calculations, and are now widely accepted. The
Ge\,--$E'$, which is observed also in pure GeO$_2$, is structurally
identical to the $E'$ apart from substitution of Si with Ge
($\equiv$Ge$^\bullet$)
\cite{TsaiJAP87,PurcellPCG69,FriebeleJAP74,TamuraPRB04,PacchioniPRB00}.
An absorption band at 6.2\,eV\,--\,6.4\,eV has been attributed to
this
center\cite{AnoikinSLC91,HosonoJJAP96cn,NeustrevJPCM94,ChiodiniPRB99},
which is considered the most important in causing the refractive
index variation in FBGs\cite{AwazuJNCS97,TsaiOL97}. The Ge(1)
consists in an electron trapped at the site of a substitutional
4-fold coordinated Ge precursor (GeO$_4$$^{\bullet}$)$^{-}$
\cite{KawazoeJNCS85,NeustrevJPCM94,FriebeleinDIG86,TsaiDDD,PacchioniPRB00}.
A wide absorbtion band peaked at 4.4\,eV\,--\,4.6\,eV has been
correlated with this defect
\cite{FriebeleinDIG86,AnoikinSLC91,NeustrevJPCM94}. Finally, the
H(II) center consists in a Ge atom bonded to two oxygens and one
hydrogen, and hosting an unpaired electron (=Ge$^{\bullet}$\,--\,H)
\cite{Vitko,RadtzigPSS86,SkujaJNCS92,PacchioniPRB98}. No detailed
information is available on the absorption properties of the H(II)
centers, although they have been proposed to feature an emission at
1.83\,eV, excited by energy transfer from a nearby GeODC
center\cite{AwazuJAP93}.

In regard to the Ge(2), this is the paramagnetic Ge-related center
whose structural model is most controversial at the moment. Indeed,
it was initially considered as a variant of the Ge(1), i.e. an
electron trapped at the site of a GeO$_4$ unit, but differing from
Ge(1) for the number of Ge nearest neighbors ions; besides, it was
put in relationship with an absorption at 5.8\,eV
\cite{KawazoeJNCS85,NishiiPRB95,NishiiPRB99,FriebeleinDIG86,TsaiDDD}.
However, subsequent studies suggested that Ge(2) can be annihilated
by capturing an electron released by a donor ion,
\cite{AnoikinSLC91,NeustrevJPCM94} or that it can be created by
ionization of preexisting twofold coordinated Ge centers
\cite{FujimakiPRB98,YamaguchiPRB02}. On this basis, the alternative
model of a ionized twofold coordinated Ge (=Ge$^\bullet$) was
proposed. Also the attribution of the OA has been criticized, as
some authors have proposed the 5.8\,eV (in addition to the above
mentioned 4.4\,eV) band to be related to Ge(1) rather than to Ge(2)
\cite{ChiodiniPRB99,FujimakiPRB98}, and recent theoretical and
experimental investigations, some of which based upon Optically
Detected Magnetic Resonance (ODMR) spectroscopy, have suggested that
the absorption at 5.8\,eV may be due to a Ge-related diamagnetic
center, thus being unrelated to the Ge(2) \cite{PouliosJPCM00,
TakahashiJAP02, UchinoPRL00, UchinoPRB02}.

\begin{figure}[htb!]
\begin{center}
%\makebox[\textwidth]{\framebox[5cm]{\rule{0pt}{5cm}}}
\includegraphics[width=.6\textwidth]{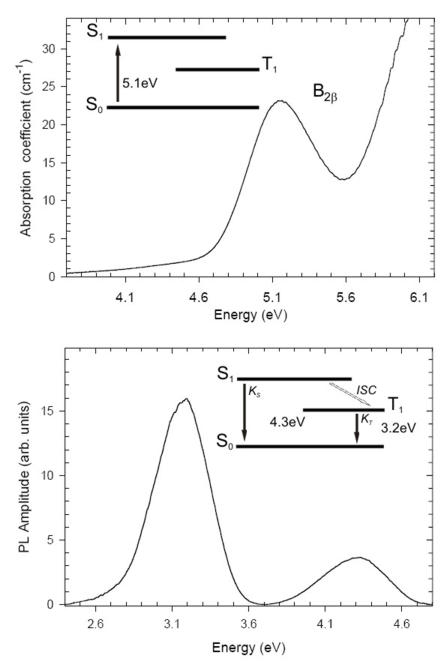}
\end{center}
\caption{Panel (A): B$_{2\beta}$ OA band as measured in a Ge-doped
silica sample. Panel (B): emission bands at 3.2\,eV (triplet) and
4.3\,eV (singlet) detected under excitation at 5.0\,eV. In both
panels: graphical representation of the electronic levels and
transitions responsible for the optical activity. Figure adapted
from Cannizzo.\cite{CannizzoPHD} } \label{schemab}\end{figure}

Also the structural model of the diamagnetic oxygen-deficient
precursor responsible of the $\sim$5.1\,eV band very common in
as-grown Ge-doped silica has been debated, similarly to what we have
seen for the Si-related ODCs, ODC(I) and ODC(II), introduced in
subsection \ref{ODCsection}-B. In detail, two defects have been
proposed to contribute to the Ge-related $\sim$5.1\,eV band: the
Germanium Lone Pair Center (\textbf{GLPC}) and the Neutral Oxygen
Vacancy (\textbf{NOV}).\paragraph{GLPC:}Exciting in the 5.1\,eV band
two luminescence bands, peaked at 3.1\,--\,3.2\,eV and
4.2\,--\,4.3\,eV, are detected
(\figurename~\ref{schemab})\cite{Skuja84,SkujaJNCS92,LeonePRB99,CannasPHD}.
The linear correlation between the three optical bands found in a
large number of samples, strongly suggests to ascribe them to a
single defect. The overall optical activity was attributed to a
twofold coordinated Ge structure (=Ge$^{\bullet\bullet}$), also
known as GLPC or GeODC(II)
\cite{Skuja84,SkujaJNCS92,LeonePRB99,CannasPHD}, which has the same
chemical structure of the intrinsic ODC(II), apart from substitution
of the silicon atom with the isoelectronic Ge.
 The
GLPC model is based on two main evidences: \textbf{(i)} the symmetry
properties of the center, deduced by PL polarization data, suggest
the defect to be in a twofold coordinated configuration,
\cite{Skuja84,SkujaJNCS92} and \textbf{(ii)} the observed conversion
(eq.~\ref{BinHII}) by hydrogen\footnote{A diffusing hydrogen
\emph{atom} is often indicated in the specialized literature by
H$^0$, so as to emphasize its charge neutrality and distinguish it
from the H$^+$ (proton) and H$^-$ ions. Since this distinction is
not necessary here, throughout this Thesis we are going to use
simply the symbol "H". } trapping into H(II)
\cite{RadtzigPSS86,AgnelloPRB00,SkujaReview98}:
\begin{equation} \label{BinHII}\text{=Ge}^{\bullet\bullet}\text{ + H}\text{ }\longrightarrow \text{
}\text{Ge}^{\bullet}\text{\,--\,H}
\end{equation} which strongly suggests the GLPC model thanks to the unambiguous knowledge of the structure of the H(II)
center obtained by ESR. This is an example of how the structural
information available on a paramagnetic defect (H(II)) can be an
indirect, though powerful, means of inferring the structure of its
diamagnetic precursor (GLPC). Finally, the twofold coordinated model
has been also confirmed by computational works
\cite{ZhangPRB97,PacchioniPRB98}.

From the detailed experimental and theoretical study of the optical
properties of the GLPC, it has been also clarified the scheme of its
electronic levels and transitions (see
\figurename~\ref{schemab}),\cite{SkujaJNCS92,PacchioniPRB98,CannasPHD,CannizzoPHD,Nalwa}
which we briefly describe here as the spectroscopic signals of the
GLPC will be reported among the results. The optical absorption band
at 5.1\,eV is associated to the promotion of the defect from the
ground singlet state S$_0$ to an excited singlet state S$_1$. From
the spectroscopic point of view, the OA is reported to peak at
5.13\,--\,5.16\,eV with a 0.45\,--\,0.48\,eV
FWHM.\cite{SkujaJNCS92,Nalwa,CannasPHD,CannizzoPHD,HosonoPRB92} From
S$_1$ the center can either go back to S$_0$ by the radiative decay
channel $k_S$, giving rise to the 4.2\,--\,4.3\,eV (singlet) band,
or undergo a non radiative transition (known as intersystem
crossing, or ISC) to the excited triplet state T$_1$. From T$_1$,
the defect decays radiatively to the ground state (channel $k_T$),
emitting in the 3.1\,--\,3.2\,eV (triplet) band\footnote{The level
scheme of the center includes another excited singlet state S$_2$
(not included in \figurename~\ref{schemab}), which can be excited at
$\sim$7.4\,eV, and from which a similar decay pattern is
observed.\cite{CannizzoPHD,CannasPHD} }. In this context, it is
worth noting that the intrinsic Si-related ODC(II) defect features a
very similar level scheme, mainly differing for the values of the
transition energies and of the decay
rates.\cite{SkujaJNCS92,PacchioniPRB98,CannasPHD,CannizzoPHD,Nalwa}

Although the optical activity pattern of the GLPC has been confirmed
by several experimental investigations, it is not universally
accepted at the moment. In fact, some authors have recently pointed
out a lack of correlation between the 5.1\,eV absorption and the two
luminescence bands,\cite{PoumellecJNCS03} or between the two
luminescence bands,\cite{Poulios03} under irradiation, thereby
concluding that the three signals cannot be attributed to a single
center.
\begin{figure}[h!]
\begin{center}
%\makebox[\textwidth]{\framebox[5cm]{\rule{0pt}{5cm}}}
\includegraphics[width=.6\textwidth]{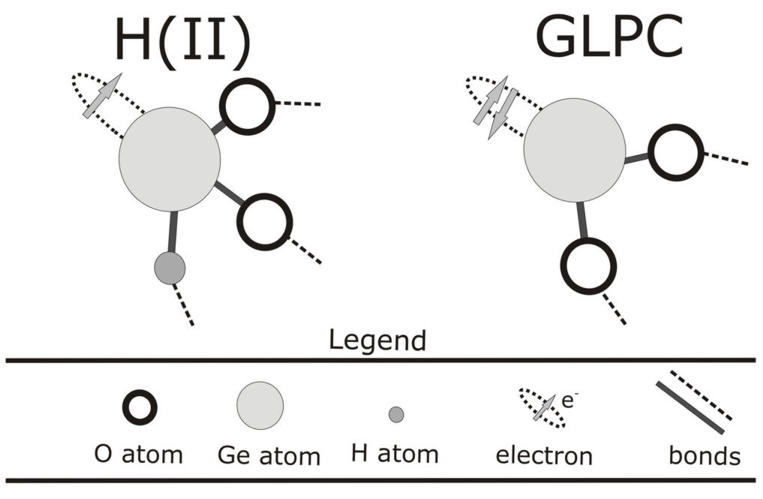}
\end{center}
\caption{Schematic representation of the structure of the GLPC and
H(II) centers in \sil{}. We stress that in the H(II) center, a (not
represented) significant portion of the wave function of the
unpaired electron is localized on the hydrogen atom, thus giving
rise to an hyperfine interaction between the electron and proton
spins. } \label{modellinige}\end{figure}

\paragraph{NOV:}On the other hand, it has been proposed that
another variety of oxygen-deficient Ge-related defect besides GLPC
contributes to the absorption at $\sim$5.1\,eV, but \emph{not} to
the two luminescence signals: the Neutral Ge-related Oxygen Vacancy
(NOV) (=Ge\,--\,Ge= or =Ge\,--\,Si=) \cite{HosonoPRB92}. The main
evidence of this model comes from the observation under exposure to
UV \emph{lamp} of the bleaching of an OA band peaked at 5.06\,eV
with 0.38\,eV FWHM (compare with the above reported parameters of
GLPC absorption), not accompanied by any reduction of the singlet
and triplet luminescence bands, but correlated to the growth of an
ESR signal due to the Ge\,--$E'$ center. This observation suggests
that the defect responsible for this 5.06\,eV component can be
ionized to give the Ge\,--$E'$ center. Based on the structure of the
Ge\,--\,$E'$, the simplest hypotesis is that such defect consists in
a Ge-related oxygen vacancy, the ionization of which generates the
Ge\,--$E'$ just as the ionization of an ordinary vacancy can
generate $E'$ (process \ref{genEdasisi}). Ionization of the
Ge-related vacancy is believed to occur by a single-photon process,
because an UV lamp is incapable of efficiently inducing multi-photon
processes due to its low intensity \cite{HosonoPRB92}. The NOV model
has been recently confirmed by theoretical calculations
\cite{TamuraPRB04}; nonetheless, adding more complexity to the
problem, other simulations have suggested that, aside from the NOV,
also the GLPC can be converted to Ge\,--$E'$ by a photochemical
laser-induced reaction \cite{UchinoPRL00}.

Although the issue of the GeODC remains open at the moment, based
upon the available experimental evidences it is likely that both
GLPC and NOV exists and contribute to the native $\sim$5.1\,eV
absorption of native Ge-doped \sil{}, while only the GLPC is
responsible for the luminescence activity. Finally, apart from
twofold (GLPC) and threefold (NOV) coordinated configurations, it is
generally accepted that Ge can be incorporated in \sil{} also as
substitutional four-fold coordinated Ge or as Ge
nanoclusters.\cite{HosonoPRB92,Grandi,ShigemuraJAP99} Given the
Ge-doping level, the prevalent arrangement of Ge in the as-grown
sample is expected to determine the effects of irradiation, and in
particular the paramagnetic defects appearing as a consequence of
laser exposure. Nevertheless, the pattern by which the population of
Ge impurities in the as-grown material distributes itself among the
several possible configurations, seems to strongly depend on the
details of the manufacturing procedure, in a way which is not fully
clarified\cite{HosonoPRB92,Grandi,ShigemuraJAP99}. As we are going
to see in the following, among the several possible Ge-related
defects in \sil{}, GLPC and H(II) play a particularly important role
in this work: in fact, GLPC happens to be the prevalent arrangement
of Ge in our samples prior to any treatment, while the generation of
H(II) after irradiation by reaction (\ref{BinHII}) will indicate the
presence of diffusing hydrogen in the material. The structure of
these two defects is represented in \figurename~\ref{modellinige}.

\subsubsection{B. Conversion processes}\hspace{0.8cm}\label{convgeiilabel}These controversies
about the structural models of some Ge-related defects correspond to
a parallel debate about the defect conversions triggered by UV lamp
or by excimer laser radiation.

We have already discussed the ionization of the NOV towards a
Ge\,--$E'$ center occurring by absorption of (single) $\sim$5eV
photons from an UV lamp\cite{HosonoPRB92}. In successive studies the
action mechanism of laser radiation was investigated
\cite{NishiiPRB95,HosonoPRB96,NishiiPRB99,FujimakiPRB96,FujimakiPRB98,FujimakiPRB99},
resulting in the proposal of two main models of laser-induced
processes in Ge-doped silica.

A first group of works,\cite{NishiiPRB95,HosonoPRB96,NishiiPRB99}
suggested that the primary process triggered by laser radiation on
Ge-related defects is the following: two-photon band to band
excitation results in pair generation of Self-Trapped \emph{Holes},
(STH),\footnote{The STH in \sil{} is a hole that gets spontaneously
trapped at the site of a bridging oxygen (=O$^{\bullet\bullet}$). It
is visible by ESR spectroscopy \cite{NishiiPRB95,Erice} } whose
structure is (=O$^{\bullet}$)$^{+}$, and GECs (i.e. Ge(1) and Ge(2)
centers), here modeled as electrons captured by a GeO$_4$ unit
(GeO$_4$$^{\bullet}$)$^{-}$:\begin{equation}
\label{modelnishii}\text{ GeO$_4$ + =O$^{\bullet\bullet}$
}\longrightarrow \text{
 (GeO$_4$$^{\bullet}$)$^{-}$ + (=O$^{\bullet}$)$^{+}$}
\end{equation} The efficiency of the band to
band excitation process was reported to be much higher than in pure
silica, partly because Germanium doping decreases the bandgap.
Reaction (\ref{modelnishii}) was proposed on the basis of the
correlated growth of the ESR signals of STH and GECs. However, this
model leaves aside the GLPC center, being unable to explain the
reduction of its PL signal which is also observed upon laser
irradiation. Hence, in other
studies,\cite{FujimakiPRB96,FujimakiPRB98,FujimakiPRB99} a different
model able to explain the role of the GLPC was suggested, based on
the picture in which the Ge(2) center corresponds to an ionized
GLPC. Within this alternative scheme, Ge(1)
(GeO$_4$$^{\bullet}$)$^{-}$ and Ge(2) (=Ge$^{\bullet}$) are
generated together by two-photon ionization of pre-existing twofold
coordinated Ge and subsequent electron trapping at GeO$_4$ sites
(eq.~\ref{modelfuji}):\begin{equation}\label{modelfuji}\text{
GeO$_4$ + =Ge$^{\bullet\bullet}$ }\longrightarrow \text{
(GeO$_4$$^{\bullet}$)$^{-}$ + =Ge$^{\bullet}$}\end{equation} It is
worth noting, however, that reactions (\ref{modelfuji}) and
(\ref{modelnishii}) are not incompatible in principle, provided that
we identify the (GeO$_4$$^{\bullet}$)$^{-}$ at the right side of
(\ref{modelnishii}) with the Ge(1) center and the =Ge$^{\bullet}$ at
the right side of (\ref{modelfuji}) with the Ge(2) center.

Finally, it has been pointed out that the presence of
\emph{hydrogen} can strongly influence Ge-related processes
\cite{AwazuJNCS97,TsaiOL97,MedjahdiOE03,MioIEEE05,FujimakiPRB99,LemaireEL93,Erice,NeustrevJPCM94,MioJNCS06CONKADER}.
In particular, loading of Ge-doped core optical fibers with H$_2$ is
known to increase photosensitivity of the material, so that it has
become a standard technique used to produce fibers suitable to
manufacture FBGs \cite{LemaireEL93,Erice,NeustrevJPCM94}. From the
microscopic point of view, it was found that the presence of H$_2$
strongly enhances the generation of Ge\,--$E'$
\cite{AwazuJNCS97,TsaiOL97,MedjahdiOE03,MioIEEE05,FujimakiPRB99},
which could explain the increasing photosensitivity, assuming that
the variations of the refractive index are due to the growth of the
Ge\,--$E'$ absorption band. According to a model proposed by Awazu
\emph{et al.} \cite{AwazuJNCS97}, the enhancement of Ge\,--$E'$
creation in the presence of H$_2$ may be due to the formation of
Ge\,--\,H bonds, which then act as efficient precursors for the
paramagnetic center. Nonetheless, the question is still open since
it is unclear how Ge\,--\,H can be efficiently formed upon H$_2$
treatment at \emph{room temperature},\cite{LemaireEL93,MedjahdiJCP}
and it appears strange that the Ge\,--$E'$ center features
interaction properties with hydrogen so different from the usual
Si-related $E'$, which is known to be \emph{passivated}, and not
enhanced, by the mobile species (as discussed in detail in the next
chapter). Aside from the generation of Ge\,--$E'$, the presence of
hydrogen is reported to influence also other Ge-related processes:
in particular, it has been reported that in H$_2$-loaded fibers or
bulk samples, no Ge(2) centers are observed after laser exposure,
whereas H(II) is detected in much higher concentration than in the
unloaded materials \cite{FujimakiPRB99,MedjahdiOE03}. This
observation suggests that Ge(2) can be passivated by reaction with
hydrogen, forming a hypothetical defect (=Ge--H)$^{+}$ that can be
easily converted into H(II) by electron trapping
\cite{FujimakiPRB99}. At the moment, further studies are needed to
thoroughly elucidate the interaction between hydrogen and
laser-induced conversion processes of Ge-related defects.

\chapter{Diffusion and reaction of mobile species in \texorpdfstring{\sil}{silica}}\label{backII}
The results presented in this Thesis deal with the generation and
decay kinetics of point defects induced in silica by UV laser
irradiation. As anticipated, we are going to show that these effects
are strongly conditioned by diffusion and reaction of mobile
hydrogen in the material. Hence, in this chapter we provide a
general background on diffusion- and hydrogen-related effects in
amorphous silica, in order to better contextualize our work with
respect to the available literature on related subjects.
\section{Post-irradiation kinetics and \\ annealing of radiation-induced
defects} 
\begin{figure}[htb!]
\begin{center}
\includegraphics[width=.8\textwidth]{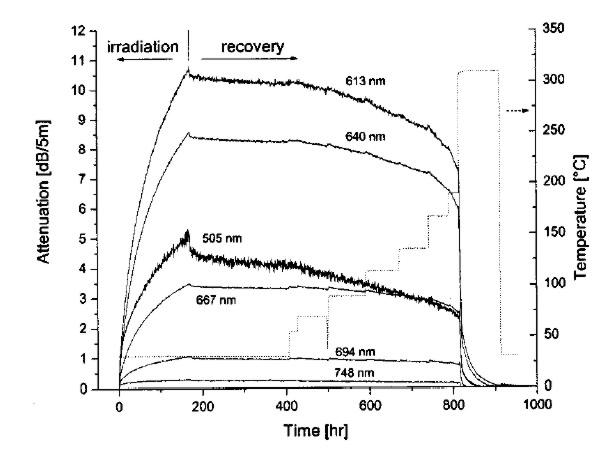}
\end{center}
\caption{Growth and recovery of the attenuation at several
wavelengths, as observed during and after $\gamma$ irradiation at
room temperature of an optical fiber. After a few days, the specimen
is thermal treated up to $\sim$600\,K (following the sequence
represented by the dotted line) to completely anneal the residual
induced absorption. Figure taken from Borgermans \emph{et
al}.\cite{BorgermansIEEE02} } \label{pikfibers}\end{figure}

\hspace{0.8cm} In some experiments it is found that the
point defects induced by irradiation at a temperature $T_i$ are
unstable after the end of exposure, and their concentration
continues to vary with time even after removal of the sample from
(or switching off) the irradiation source. These processes are
referred to as \emph{post-irradiation kinetics}: their most usual
consequence is a spontaneous decay of the generated defects after
that the irradiation is interrupted. Depending on the experimental
conditions, the decay can be partial or complete and its typical
time scale may vary from seconds to years, after which the defects
reach stationary
concentrations\cite{ImaiPRB91,LeclercJNCS92,SmithAO00,Erice,GriscomReview91,GriscomReview85,Devinebook,GriscomPRB01,KajiharaPRB06,GriscomPRL93,GriscomJAP85,BorgermansIEEE02,EvansIEEE88}.
More in general, the concentration of a specific defect can
\emph{grow}, rather than decay, in the post-irradiation stage, and
the observed effects can involve also centers that pre-existed in
the material before irradiation. We have already seen some examples
of post-irradiation kinetics in \figurename~\ref{decaydaImai} and
\figurename~\ref{leclerckinetics}. Similar effects have also been
observed and studied in irradiated optical fibers, where they result
in a partial recovery after the end of exposure of the transparency
lost due to the formation of point defects.
(\figurename~\ref{pikfibers}).
\cite{GriscomPRL93,BorgermansIEEE02,EvansIEEE88}.

\begin{figure}[htb!]
\begin{center}
\includegraphics[width=.7\textwidth]{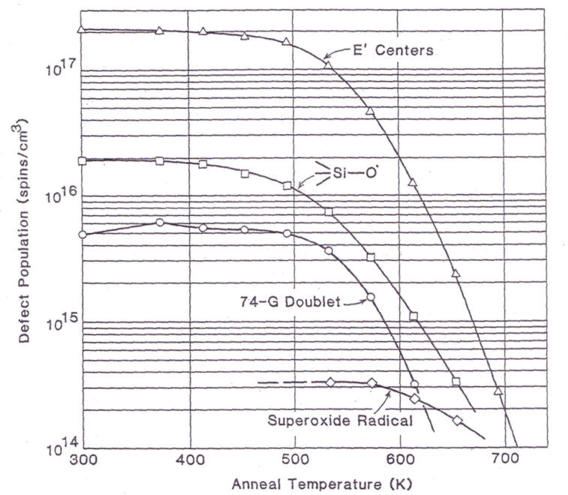}
\end{center}
\caption{Concentration variations of $E'$, NBOHC
($\equiv$Si\,--\,O$^{\bullet}$), H(I) centers (referred to as 74\,G
doublet) and peroxy radicals ($\equiv$Si\,--\,O\,--\,O$^{\bullet}$,
referred to as Superoxide Radical) observed during an isochronous
thermal treatment performed up to 700\,K. In this case, the defects
had been induced in the sample by $\gamma$ irradiation at room
temperature, and their decay after heating the specimen is due to
reactions with diffusing water of radiolytic origin. Figure taken
from Griscom.\cite{GriscomFromBook} }
\label{annealinggriscom}\end{figure}

After that the defects generated by irradiation at the temperature
$T_i$ have reached their stationary concentrations, it is commonly
observed that they become again unstable if the specimen is heated
at temperatures higher than a given threshold $T_d$$>$$T_i$. Hence,
upon such a \emph{thermal treatment} one observes some of the
radiation-induced defects to disappear by conversion in other
centers, the conversion processes strongly depending on the
temperature range, on the sample and on the irradiation
conditions.\footnote{Also by a thermal treatment, in general the
defects can either grow or decrease in concentration, and the
observed effects can involve also centers that pre-existed in the
material before irradiation.} Nevertheless, a general rule of thumb
is that the \emph{paramagnetic} centers appeared as a consequence of
irradiation disappear, or are \emph{annealed}, when a sufficiently
high temperature is reached, as shown for example in
\figurename~\ref{annealinggriscom}. A thermal treatment can be a
useful method to "cancel" the effects of irradiation
\cite{Erice,GriscomReview91,GriscomReview85,Devinebook}. While a
complete erasure of the induced defects is always possible, a
thermal treatment sometimes allows a selective cancelation of
radiation damage by a proper choice of a (not too high) temperature
$T_d$. From another point of view, thermal treatments are a common
means to investigate the stability of the induced point defects as a
function of temperature so as to indirectly infer information on
their properties.\footnote{Thermal treatments can be distinguished
in two categories. Given a base temperature $T_b$ (typically
coinciding with $T_i$), and a time interval $\Delta$t, an
\emph{isochronous} treatment consists in a sequence of cycles in
each of which the sample is kept at some $T_d$>$T_b$ for a
$\Delta$t, and then brought back to $T_b$ to perform measurements.
After that, the procedure is repeated at a higher T but using the
same $\Delta$t, so that the process can be followed as a function of
temperature. The other possibility is an \emph{isothermal}
treatment, in which both $T_d$ and $\Delta$t are the same for all
the cycles, so as to study the process as a function of time.}

Although the occurrence of post-irradiation kinetics and annealing
effects has been known for a long time, its influence on the
concentration of defects induced by irradiation has been in some
sense underestimated in literature. Indeed, most works on point
defect generation in \sil{} have investigated only the stationary
concentrations of induced centers, measured after the conclusion of
the post-irradiation kinetics, or have simply not discussed the
possible influence of the post-irradiation effects. In contrast, it
is worth pointing out that only the observation of the complete
growth and decay kinetics of the defects can give detailed
information on their dynamics and properties. In particular, the
stationary concentrations generally represent only the net result of
generation and annealing at the temperature $T_i$.

Several physical processes may in principle contribute to the
annealing of defects. Historically, the basic mechanism was first
believed to be the recombination of electron-hole pairs produced by
irradiation. Nonetheless, successive studies led to recognize that
the prevalent process is often the diffusion and reaction in \sil{}
of small chemical species like hydrogen, oxygen, or water, which are
able to react with point defects thus resulting in their conversion
in other centers. Nowadays, the idea that several molecules are able
to diffuse in \sil{} in a wide temperature range and react with
defects is widely accepted and supported by a large set of
experimental data
\cite{Erice,GriscomReview91,GriscomReview85,GriscomJNCS84,GriscomFromBook,Devinebook}.
Consistently with this interpretation, it is usually found that the
annealing temperature of a given point defect strongly depends on
the history of the specific sample being used, which is thought to
determine the nature and the concentration of the diffusing agents
that cause the disappearance of the center. Often, the reaction with
a diffusing chemical species converts an optically active defect to
another center that is "invisible" for what concerns optical
absorption in a range of technological interest: in this case, the
reaction is usually referred to as the \emph{passivation} of the
defect.

Since diffusion in a solid is a thermally activated process (see
next section), each mobile species is characterized by a typical
minimum temperature that must be reached to activate its migration
through the silica matrix. Hence, the temperature determines the
possible chemical species that, if present in sufficient
concentration, can influence the effects of irradiation by reacting
with point defects. In general, the diffusing species may be either
already present in the \sil{} matrix before irradiation, depending
on the manufacturing procedure of the material, or they can be made
available in the sample as a consequence of the irradiation itself,
typically by a radiolysis process. Another important general rule is
that paramagnetic defects, due to the presence of an unpaired
electron, are much more reactive than diamagnetic defects, so that
they are more easily passivated by reaction with hydrogen, oxygen or
water.
\cite{Erice,GriscomReview91,GriscomReview85,GriscomJNCS84,GriscomFromBook,Devinebook}.

Diffusion-related effects in \sil{} are a very lively research
topic; indeed, also in recent times many experimental and
computational works continue to investigate the diffusion mechanisms
of several mobile species in silica and their interactions with the
matrix and with point defects
\cite{KajiharaPRL02,KajiharaPRL04,RashkeevPRL01,GodetPRL06,BakosPRB04,BongiornoPRL01,AgnelloPRB06}.
In this context, \emph{hydrogen} assumes a particular importance
because it is the most reactive among the chemical species that
efficiently diffuse in silica already at room temperature or below
\cite{GriscomJNCS84,KajiharaPRL02,KajiharaPRB06,Shelby94}.
Furthermore, hydrogen is the most common impurity in silica,
virtually present in every variety of the material. In the next
section we give a general background of diffusion processes in
silica and diffusion-limited reactions, after which we proceed to
describe the main experimental results concerning the reactivity of
point defects in \sil{} with hydrogen.

\section{Diffusion in silica} \hspace{0.8cm}By definition,
diffusion is the motion of a chemical species driven by a gradient
of chemical potential, which in the simplest case arises from a
concentration gradient \cite{Atkinsbook,Doremus}. It is often a good
approximation to suppose the flux density $\overrightarrow{J}$ of
the chemical species to be linearly proportional to the spatial
gradient of its concentration
$\rho$:\begin{equation}\label{Fick}\overrightarrow{J}=-D\nabla
\rho\end{equation}Eq.~(\ref{Fick}) is called the first Fick law, and
defines a proportionality constant $D$ called the \emph{diffusion
constant}. Diffusion can be regarded as the continuum limit of the
well-known problem of random walk in a lattice. The mathematical
problem of diffusion can be treated by adding to the Fick law the
continuity equation and setting appropriate boundary conditions.
Without entering in the detail of the solutions, it is useful for
the present purposes to introduce another quantity of interest: the
\emph{mean diffusion length} $L_d$, which represents the mean
distance covered by a diffusing molecule in a given time interval
$t$. In a three-dimensional geometry, $L_d$ is given by
$L_d=\sqrt{6Dt}$ \cite{Atkinsbook,Doremus,Avraham}.

Diffusion in solids is a complex problem discussed in many
experimental and theoretical works. Both the detailed understanding
of the microscopic mechanisms at the basis of diffusion and the
measure of the diffusion coefficients are still open problems in
some systems, among which the amorphous solids. In general, the
principal and simplest microscopic diffusion mechanisms in a solid
are two: \emph{vacancy} diffusion and \emph{interstitial} diffusion.
In vacancy diffusion, the elementary process at the basis of the
macroscopic transport of matter is the jump of an atom to a
neighboring vacancy of the lattice\footnote{Vacancies are always
present even in a "perfect" lattice, as inferred by simple
thermodynamic considerations \cite{Ashcroft} }. This mechanism is
very important for substitutional impurities in metals and alloys
and is active also in ionic solids \cite{Doremus,Shewmon}. In the
alternative scheme, \emph{interstitial} diffusion, small molecules
of a given chemical species are dissolved in the solid by being
incorporated in interstitial positions, and diffuse by jumping to
one of the nearest-neighbor solubility sites \cite{Doremus,Shewmon}.
This mechanism is dominant in many covalent solids, such as \sil{},
which have relatively open structures, so as to allow for an easy
dissolution of sufficiently small molecules in interstitial
positions, and where the energy to break bonds is large, so that
vacancy formation is unfavorable \cite{Doremus,Shewmon}.

\begin{figure}[htb!]
\begin{center}
\includegraphics[width=.5\textwidth]{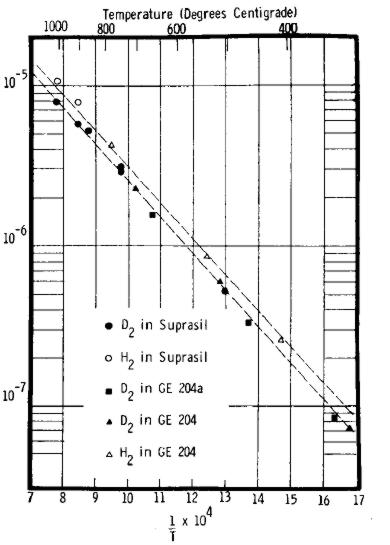}
\end{center}
\caption{Arrhenius plot of the diffusion coefficient of molecular
hydrogen and deuterium measured in several types of silica samples.
The straight lines correspond to Arrhenius temperature dependences,
eq.~(\ref{arrh}). Figure adapted from Lee.\cite{LeeJCP63} }
\label{arrheniuslee}\end{figure}

In the study of diffusion, apart from the interest connected to the
present work, silica glass is particularly important as it is the
only solid for which extensive measurements for many atoms and
molecules in a wide temperature range have been performed
\cite{Doremus,Shewmon}. From the microscopic point of view, the
motion of a molecule from an interstitial site to the neighboring
one in the solid requires to overcome a potential barrier due to the
necessary rearrangement of the closest atoms; hence, diffusion is an
\emph{activated} process, and the macroscopic diffusion coefficient
usually depends on temperature according to an \emph{Arrhenius
equation}:
\begin{equation}\label{arrh}D=D_0\exp{\left(-\frac{E_a}{k_B T}\right)}
\end{equation}where $D_0$ is called the pre-exponential factor, $k_B$ is
the Boltzmann constant, $T$ is the absolute temperature and $E_a$ is
the activation energy for the process. From an experimental point of
view, the parameters $E_a$ and $D_0$ can be determined by a simple
linear fit after reporting the diffusion coefficient measured at
several temperatures in a so-called \emph{Arrhenius plot} (log$D$ vs
1/T). Indeed, in such a graph eq.~(\ref{arrh}) corresponds to a
straight line (see for example \figurename~\ref{arrheniuslee}).

Physically, the value of $E_a$ (typically a fraction of 1\,eV) may
be interpreted as an estimate of the potential barrier that
separates neighboring interstitial sites. It has been pointed out
that for a given solid, $E_a$ is closely connected to the
dimensions\footnote{In the simplest treatment of the problem of
interstitial diffusion in a solid, the mobile atom or molecule is
regarded for simplicity as a not deformable object characterized by
a size $d$. $d$ is assumed to be unaffected by the insertion of the
molecule in the solid; therefore, it is usually estimated from the
kinetic theory of gases starting from measurements of the viscosity
of the mobile specie in gaseous form outside the
solid.\cite{Doremus} It was found that the functional dependence
between $E_a$ and $d$ in \sil{} is remarkably well reproduced by the
equation: $E_a=\alpha(d-d_0)^2$, where $\alpha$ and $d_0$=0.38{\AA}
are constants. This finding can be interpreted by the following very
simple argument: one imagines that the mobile species dissolves
within the solid in interstices connected by doorways of average
radius $d_0$: then, $E_a$ is proportional to the elastic energy
required to dilate such doorways from $d_0$ to $d$, so as to allow
the jump of the diffuser from one interstice to a neighboring one.
\cite{Doremus} } of the diffusing species, being lowest for atoms or
small molecules like H$_2$. For what concerns silica, its
microscopic structure is relatively porous, meaning that the
fraction of space occupied by the Si and O ions and by their
electronic clouds is lower than in many other solids. For this
reason, the measured values of $E_a$ are usually lower than in other
insulators \cite{Doremus}. The pre-exponential factor $D_0$ can be
connected to the entropy of activation for diffusion and the
vibrational frequency of the molecule inside each interstitial
position.\cite{Doremus,Shewmon} Its typical order of magnitude in
\sil{} is 10$^{-4}$\,cm$^2$s$^{-1}$. Up to this point, we have
described diffusion in \sil{} as the thermally activated motion of a
chemical species described by the simple eqs.~(\ref{Fick}) and
(\ref{arrh}). Actually, there are at least two additional
complications that must be taken into account for a thorough
description of this process.

\paragraph{Reaction with the matrix:} When the temperature is increased above a certain limit, which
may depend on the experimental conditions, the solid cannot be more
regarded as a passive lattice within which the migration of the
mobile species takes place. For example, at $T$$<$700\,K H$_2$
diffuses with essentially no chemical interaction with the silica
matrix, but at increasing temperatures the picture gets more
complicated because of the occurrence of chemical reactions of H$_2$
with the glass, the most basic being the reaction of H$_2$ with
Si\,--\,O\,--\,Si groups.\footnote{Namely the reaction
Si\,--\,O\,--\,Si + H$_2$ $\Longrightarrow$ Si\,--\,OH + H\,--\,Si,
that produces Si\,--\,OH and Si\,--\,H (immobile) impurities in the
network. } At even higher temperatures, one must take into account
also the spontaneous decomposition of preexisting Si\,--\,OH groups,
which makes available additional
hydrogen\cite{LeeJCP62,LeeJCP63,Shelby77,ShelbyJAP80,GriscomJNCS84,Doremus,LouJNCS03}.
When measuring the diffusion coefficient $D$, these processes
influence the estimated values, so that one has to introduce an
\emph{effective} diffusion coefficient $D_e$. The actual $D$ can
still be estimated from $D_e$ by more or less complex treatments, as
thoroughly discussed in specialized texts.\cite{Doremus}. Such
diffusion-reaction kinetics are particularly important for water,
which significantly reacts with the lattice in most of the
temperature range in which it is mobile\cite{Nalwa,Doremus}. Due to
these effects, the values of $D$ for H$_2$O in \sil{} reported in
literature appear to be very variable according to the experimental
conditions\cite{Doremus,MoulsonTBCS60}. For what concerns the
present work, the main mobile species we are interested with is
H$_2$. From this point of view, we stress that the temperature of
our experiments ($<$500\,K) was in any case low enough to neglect
the influence of any reaction with the matrix on the diffusion
process.

\paragraph{Non-Arrhenius behavior:}The diffusion constant is usually measured by experiments founded
on the controlled permeation of a gas through a membrane at several
temperatures\footnote{A notable exception to this approach is atomic
hydrogen H, whose value of $D$ was estimated by experiments dealing
with its reactions with point defects
\cite{GriscomJAP85,BrowerAPL82,ShkrobPRB96} }. When the measurements
are carried out on a sufficiently extensive temperature range, often
it is evidenced that the experimental data present small deviations
from a precise Arrhenius behavior, eq.~(\ref{arrh}). In particular,
it is often observed that the slope in an Arrhenius plot tends to
increase with temperature. In silica, this has been experimentally
evidenced in a clear way at least for helium and neon
\cite{Doremus,GriscomJNCS84,ShelbyKeeton,SwetsJPC61,Shelby72}. More
in general, slightly different values of the diffusion parameters of
a given species are often obtained by fitting with eq.~(\ref{arrh})
data collected in different temperature ranges. For example, for He
it was reported by Swets \emph{et al.} that
$D_0$=3.0$\times$10$^{-4}$\,cm$^{2}$\,s$^{-1}$ and $E_a$=0.24\,eV
for 300\,K$<$$T$$<$600\,K while
$D_0$=7.4$\times$10$^{-4}$\,cm$^{2}$\,s$^{-1}$ and $E_a$=0.29\,eV
for 600\,K$<$$T$$<$1300\,K \cite{SwetsJPC61}. The observed
deviations from a purely Arrhenius behavior are usually interpreted
by theoretical arguments suggesting that $D_0$ should be
proportional either to $T$ or to $\sqrt{T}$ \cite{Shelby77,Doremus},
namely the pre-exponential "constant" actually depends on
temperature.

Nevertheless, in regard to silica, Shelby and Keeton proposed
another interpretation, which appears to be closer to our
qualitative understanding of the properties of amorphous solids. In
fact, to explain the increase with temperature of the
pre-exponential factor for diffusion of He in \sil{}, they
hypothesized that in the amorphous matrix the activation energy must
be regarded as a randomized parameter, which statistically
fluctuates from site to site following a Gaussian distribution
centered on some value $\epsilon$ and with a width (standard
deviation) $\sigma$. Qualitatively, it can be argued that the local
diffusivity in the glass is necessarily nonuniform because the
moving species traverse interstitial voids of various shapes and
sizes.
\cite{ShelbyKeeton,GriscomJNCS84,KajiharaPRL02,KajiharaPRB06,BongiornoPRL01}
Hence, following Shelby and Keeton, the value of $D$ measured by the
classical experiment based on permeation through a membrane must be
actually interpreted as the \emph{mean} $D$ value on the
distribution of $E_a$, which is \cite{ShelbyKeeton,Shelby72}:
\begin{eqnarray}
  <D> & = & \int D_0\exp{\left(-\frac{E_a}{k_B
T}\right)}\times\frac1{\sigma\sqrt{2\pi}}exp\left[-{\frac{(E_a-\text{$<$$E_a$$>$})^2}{2\sigma^2}
}\right]dE_a \nonumber \\
   & = & D_0\exp{\left[\frac{\sigma^2}{2k_B^2T^2}\right]}\exp{\left(-\frac{\text{$<$$E_a$$>$}}{k_B
T}\right)}\label{distrd}
\end{eqnarray}
eq.~(\ref{distrd}) deviates from the Arrhenius behavior in that it
contains the additional $T$-dependent term
$\exp{[\sigma^2/(2k_B^2T^2)]}$, leading to an \emph{increase} of the
effective diffusion coefficient from the value \linebreak{}
$D_0\exp{(-\text{$<$$E_a$$>$}/k_B T)}$ corresponding to the mean
activation energy $<$$E_a$$>$. The authors found that
eq.~(\ref{distrd}) is able to reproduce a large set of experimental
data on He diffusion in \sil{} at 300\,K$<T<$1300\,K, with
$<$$E_a$$>$=0.33\,eV and $\sigma$=0.057\,eV. In particular, with
these parameters it is possible to reproduce quite well both
temperature intervals investigated by Swets \emph{et al.} (see
above). Similarly to the case of He, slightly temperature-dependent
values of the activation energy for molecular hydrogen or
deuterium\footnote{The dimension and the electronic structure of the
H$_2$ and D$_2$ molecules are essentially the same. As expected, the
two activation energies for diffusion are found to be almost
identical. In contrast, the pre-exponential factor $D_0$ scales as
m$^{-1/2}$.\cite{GriscomJNCS84,Shelby77,LeeJCP62,Doremus,Shewmon} }
diffusion in the interval 0.38\,eV\,--\,0.45\,eV have been measured
by several studies performed in different temperature ranges
\cite{LeeJCP62,LeeJCP63,PerkinsJPC71,Shelby77}. Actually, when one
considers not too wide temperature intervals (typically up to $\sim$
300\,K\,--\,500\,K wide), the expression (\ref{arrh}) remains a very
good approximation for many purposes and is commonly used in
reporting the diffusing parameters and in the analysis of diffusion
processes \cite{Doremus}. Even so, for the above discussed reasons
$D_0$ and $E_a$ are characterized by some "intrinsic" indeterminacy.
In regard to $E_a$, which influences the diffusion coefficient much
more strongly than $D_0$, these effects typically lead to
fluctuations within $\sim$10\% of the reported values. Hence, when
using literature diffusion parameters, one should always keep in
mind the specific temperature interval in which the values have been
measured. On the other hand, eq.~(\ref{distrd}) implies an important
step forward in the understanding of the diffusion process in an
amorphous matrix. Indeed, as it will be discussed in the following,
the introduction of distributed activation energies becomes
mandatory to describe the reaction kinetics of point defects with
diffusing molecules.

Once clarified the framework in which the description of diffusion
by eq.~(\ref{arrh}) is valid, we report in Table~(\ref{dc}) some
representative values of the diffusion parameters of common chemical
species in silica, as estimated by the work of several authors.
\begin{table}[h!]
\begin{center}
\begin{small}
\begin{tabular}{|l||c|c|l|c||p{4mm}r@{$\times$}l|c|}
    \hline
 & $D_0$[cm$^{2}$\,s$^{-1}$] & $E_a$[eV] & Reference & Range(K) & \multicolumn{3}{c|}{$D$(300K)[cm$^{2}$\,s$^{-1}$]} & $L$(h)[cm]\\
\hline \hline
H & 1.0$\times$10$^{-4}$ & 0.18 & \cite{GriscomJAP85} & 77-130 &  & 9.5 & 10$^{-8}$ & 4.5$\times$10$^{-2}$\\
H$_2$ & 5.7$\times$10$^{-4}$ & 0.45 & \cite{LeeJCP63} & 570-1270 & & 1.5 & 10$^{-11}$& 5.6$\times$10$^{-4}$\\
O$_2$ & 2.9$\times$10$^{-4}$ & 1.17 & \cite{Norton61} & 1220-1350 & & 5.9 & 10$^{-24}$& $<$10$^{-8}$\\
H$_2$O & 1.7$\times$10$^{-4}$ & 0.76 & \cite{DoremusJMR95} & 400-1500 & & 3.4 & 10$^{-17}$ & 8.6$\times$10$^{-7}$\\
He & 3.0$\times$10$^{-4}$ & 0.24 & \cite{SwetsJPC61} & 300-600 & & 2.8 & 10$^{-8}$ & 2.5$\times$10$^{-2}$\\
Ne & 5.1$\times$10$^{-4}$ & 0.41 & \cite{PerkinsJPC71} & 300-770 & & 6.7 & 10$^{-12}$ & 3.8$\times$10$^{-4}$\\
\hline
\end{tabular}
\end{small}
\end{center}
\caption{Diffusion parameters of some common chemical species in
amorphous silica. The column "Range" reports the temperature
intervals in which the diffusion parameters in the first two columns
were measured. $D$(300\,K) is the diffusion constant extrapolated at
room temperature. $L$(h) is the mean diffusion length in a time
interval of one hour, calculated from $D$(300\,K). }
 \label{dc}\end{table}From the reported parameters, we note that the species featuring the
lowest activation energies for diffusion are He, Ne, atomic and
molecular hydrogen. The mean diffusion length of these species at
$T$=300\,K on a time scale of a few hours is 10$^3$\,--\,10$^6$
times a typical interatomic distance: $\sim$5$\times$10$^{-8}$\,cm.
Since the noble gases are basically inert from the chemical point of
view, hydrogen can now be anticipated to be the diffusing species
most able to influence the properties of the material in experiments
performed at room temperature or below.

\section{Theoretical treatment of the reaction kinetics}\label{solutre}
\hspace{0.8cm}In this section, we introduce some theoretical
approaches that permit to describe chemical reaction kinetics in
\sil{}. Consider the following generic bimolecular reaction, in
which two species A and B combine to form C:
\begin{equation}
\text{A + B}\longrightarrow \text{C} \label{genericreaction}
\end{equation} Due to reaction (\ref{genericreaction}) the
concentrations [A], [B] and [C] depend on time. In many cases, the
time variations can be described by a simple second order rate
equation in which the time derivative of the concentrations (called
the \emph{reaction rate}) is proportional to the product of the
concentrations of the two reagents.\begin{equation}
\frac{d[\text{A}]}{dt}=\frac{d[\text{B}]}{dt}=-\frac{d[\text{C}]}{dt}=-k[\text{A}][\text{B}]
\label{rateconstant}
\end{equation} The constant $k$ is known as the \emph{rate
constant} of process~(\ref{genericreaction}), and can be estimated
experimentally by measuring the concentrations of the reagents as a
function of time.

Several theoretical treatments have been proposed to derive an
explicit expression for the rate constant, as a function of
microscopic parameters related to A and B, to be compared with
experimental data \cite{Atkinsbook,Waite57}. Before discussing some
of them, let us start with a qualitative discussion of the physical
factors determining the rate of reaction~(\ref{genericreaction}).
For the reaction to occur, the species A and B must first encounter
each other during their random diffusive motion, thereby forming an
(A--B) pair separated by a distance short enough (a few {\AA}) to
allow for the chemical interaction. Then, the fate of the encounter
can be one of the following two: A and B can either react forming C,
or diffuse away thus breaking the pair. In this sense, the overall
rate of the reaction depends on two factors: the diffusion constant,
which determines the rate at which the pairs form and break, and the
reaction rate, namely the probability per unit time that a pair
reacts resulting in the formation of C. Therefore, we can define two
typical time scales: the reaction time $\tau_R$, which is the
typical time required for a pair to react, provided that it is not
separated by diffusion, and the diffusion time $\tau_D$, which is
the lifetime of a pair before A and B diffuse away, provided that
they do not react. Let us suppose now for simplicity that A and B
interact only below a certain distance $r_0$. Then, the order of
magnitude of $\tau_D$ is the time required to diffuse out of a
sphere with radius $r_0$: $\tau_D\sim r_0^2/D$.

\begin{figure}[htb!]
\begin{center}
\includegraphics[width=.7\textwidth]{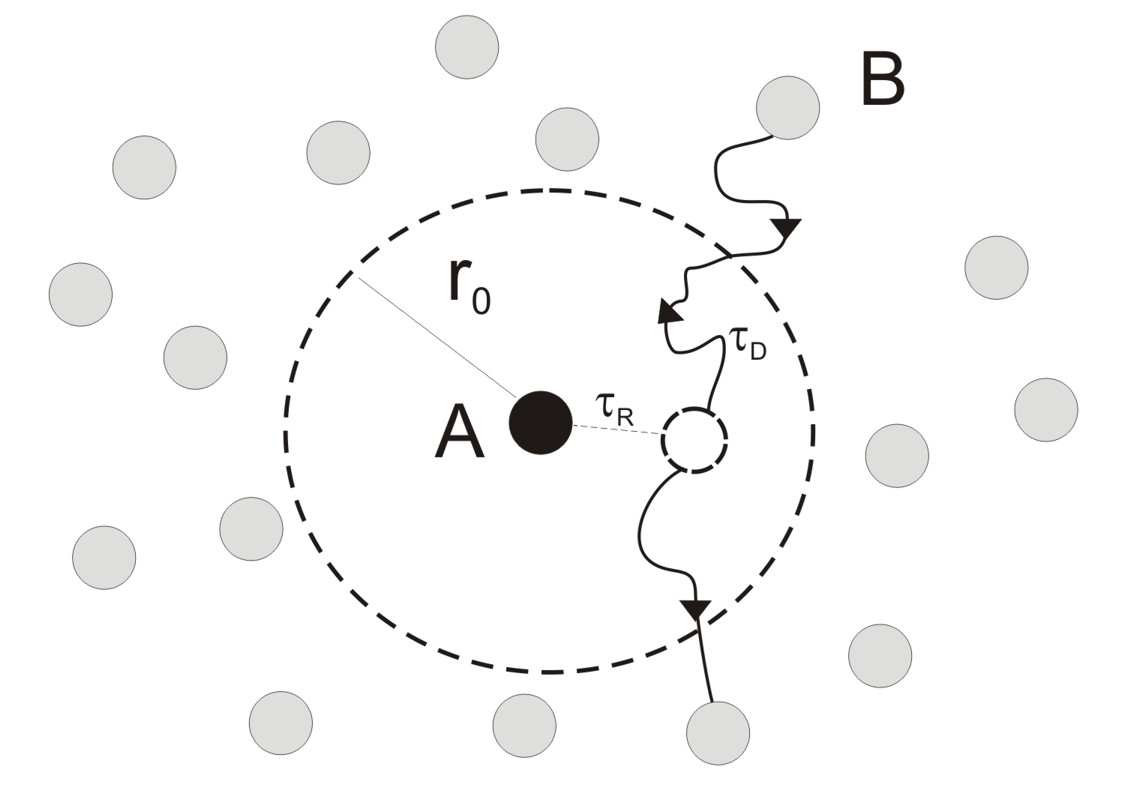}
\end{center}
\caption{Pictorial representation of a reaction between two species
A (supposed immobile for simplicity) and B (diffuser) within a
solid. The continuous line with arrows represents the random
diffusion trajectory of B. The typical time scale required for B to
move across the interaction sphere of radius $r_0$ is $\tau_D$,
while $\tau_R$ is the typical time required for reaction. }
\label{brownian}\end{figure}

Now, the reactions can be classified according to the relation
between the two time scales. Indeed, if $\tau_D\gg\tau_R$, almost
each encounter is successful in forming C: in fact, the diffusion is
so slow that the reagents, after an encounter, remain close for much
more time than typically required for the reaction. Hence, the
reaction is mainly controlled by the rate at which (A--B) pairs form
by diffusion, so as to say that the process is
\emph{diffusion-limited}. In the opposite case, if
$\tau_D\ll\tau_R$, the reagents need to meet many times before they
actually react. In this case, the process is said to be
\emph{reaction-limited} and is mainly controlled by the local
interaction between A and B rather than by the migration of the
species. Finally, it is worth noting that intermediate situations
are obviously possible. For reactions occurring in solution or in
solids, the diffusive motion is so slow that it often represent the
actual bottleneck for the process; therefore, most reactions in
solids involving mobile species are diffusion-limited processes
\cite{Atkinsbook,Waite57}.

To describe a diffusion-limited reaction, it is expected that the
details of the short-range interaction between the reagents are
unimportant. For this reason, one can try to describe the process on
the basis of a somewhat crude approximation first proposed by
Smoluchowski \cite{Smoluchowski}: the two species instantaneously
react when their distance falls below a given distance $r_0$, called
the \emph{capture radius}. From the mathematical point of view, this
represents a boundary condition for the diffusive motion, which
allows to model the occurrence of the reaction. On these grounds, we
are going now to derive an explicit expression for the reaction
constant of a diffusion-limited reaction. We start with introducing
a simple heuristic model called \emph{the trapping problem},
discussed in many standard textbooks\cite{Atkinsbook,Avraham}.
Suppose that in reaction (\ref{genericreaction}) A represents a
static point defect, whereas B is a mobile diffusing species. For
the sake of simplicity, we imagine a particle A as a perfect
spherical trap of radius $r_0$ centered at the origin of the axis
and surrounded by an infinite sea of diffusing particles B with
spatial density $\rho(r)$, which move following the Fick law, here
written in spherical coordinates:
\begin{equation}
J_r=-D\frac{\partial \rho(r)}{\partial r} \label{fickradial}
\end{equation} now, the total flux $\mathbb{J}$ of the species B through a sphere
of radius $r$ centered at the origin must be independent of $r$ in
stationary conditions, since the only sink for the diffusive motion
of B is represented by the spherical trap A:
\begin{equation}
\mathbb{J} =4\pi r^2J_r=-4 \pi D r^2\frac{\partial \rho(r)}{\partial
r}=constant \label{j}
\end{equation}$\mathbb{J}$ is a negative quantity representing
the variation per unit time of the
total number $N_B$ of B particles due to trapping.
\begin{equation}
\mathbb{J}=\frac{dN_B}{dt} \label{st1}
\end{equation}
The aforementioned hypothesis, i.e. particles B are instantaneously
trapped by A as soon as they arrive at $r_0$, can be incorporated in
the model as the condition $\rho(r_0)=0$ (Smoluchowski boundary
condition). Using this relation, we can now solve  eq.~(\ref{j}) for
$\rho(r)$:
\begin{equation}
\rho(r)=-\frac{\mathbb{J}}{4 \pi
D}\left(\frac{1}{r_0}-\frac{1}{r}\right)
\label{distributionparticles}
\end{equation} the last expression shows that in proximity of the
trap the concentration of B is depleted with respect to the value
$\rho_{\infty}$ at distances $r \gg r_0$. Besides, the unperturbed
value $\rho_{\infty}$ must be identified with the macroscopic
measured concentration of the chemical species B. If we take the
limit for $r\rightarrow\infty$ in the expression for $\rho(r)$, we
find $\mathbb{J}$:
\begin{equation}
\mathbb{J}=-4\pi r_0 D \rho_{\infty}=-4 \pi r_0 D [\text{B}]
\label{st2}
\end{equation}
This expression has been derived assuming a single trapping center
A. Therefore, $\mathbb{J}$ has to be multiplied for the number $N_A$
of A particles. Finally, since the macroscopic reaction rate is the
derivative of the concentration, rather than of the number of B
particles, $\mathbb{J}$ must also be scaled for the volume $V$ of
the sample. On this basis, starting from (\ref{st1}) and (\ref{st2})
we get:
\begin{equation}
\frac{d[\text{B}]}{dt}=V^{-1}\frac{dN_B}{dt}=\mathbb{J}\frac{N_A}{V}=-4\pi
r_0 D [\text{A}][\text{B}]
\end{equation}
and comparing with eq.~(\ref{rateconstant}), we finally obtain the
well known expression for the diffusion-limited reaction
rate\cite{Atkinsbook,Avraham}:
\begin{equation}
k_d=4\pi r_0 D \label{waiterateconstant}
\end{equation}
it is worth noting that the capture radius $r_0$ is expected to be
of the order of the interatomic distance (a few {\AA}) if the above
arguments make sense at all. As an example, for the case of a
process at room temperature in \sil{} limited by H$_2$ diffusion,
substituting in eq.~(\ref{waiterateconstant}) the value of
$D(300\,K)$ from Table (\ref{dc}), we estimate the reaction constant
to be of the order of: $4\pi r_0 D(300K)
\sim$10$^{-17}$\,cm$^{3}$\,s$^{-1}$.

It is clear that this very simple approach bears many limitations.
In particular, it does not take into account the removal of the
traps upon reaction, the occurrance of an initial arbitrary
non-stationary spatial distribution of the reagents and the
possibility that both A and B diffuse. In literature,
diffusion-limited reaction kinetics have been described by other
techniques that permit to take into account these effects. Among the
available approaches, we are going to focus on that proposed by
Waite, which is a theoretical treatment formulated in terms of the
pair probability densities of the reacting particles \cite{Waite57}.
Hypothesizing uniform and uncorrelated initial distributions of A
and B, and based upon the Smoluchowski boundary condition, Waite
found that the rate "constant" $k$ of
reaction~(\ref{genericreaction}) actually depends on time, and is
given by the following expression.
\begin{equation}
k_d=4\pi r_0 D \left(1+\frac{r_0}{\sqrt{\pi D t}} \right)
\label{waitet}
\end{equation}
Physically, the time-dependent term $\sqrt{\pi D t}$ accounts for a
transient regime during which the initial uniform distribution of A
and B, which also includes a fraction of particles that are
initially closer than the capture radius, spontaneously evolves
towards a situation in which the concentration of B particles is
depleted around A. After a time much longer than $r_0^{2}/D$, the
time-dependent term in (\ref{waitet}) becomes negligible, the
spatial distribution of reagents resembles
eq.~(\ref{distributionparticles}), and the rate constant coincides
\emph{de facto} with the expression~(\ref{waiterateconstant}).
Finally, eq.~(\ref{waitet}) is valid also if both species are
mobile, with the substitution $D=D_A+D_B$.

The treatment by Waite can be extended also to situations in which
the reaction is not purely diffusion-limited. To this aim, it may be
used another boundary condition, originally proposed by Collins
\cite{Collins}. The author suggested to define a interval of pair
separations, from $r_0-\Delta r$ to $r_0$, within which the reaction
is no longer controlled by diffusion, but follows a first order rate
equation. In other words
\begin{equation}
\frac{d[\text{C}]}{dt}=w[\text{(A--B)}]
\end{equation} where [(A--B)] is the concentration of (A--B) pairs
whose distance falls within the above defined interval, and $w$ is a
constant [s$^{-1}$]. On this basis, it can be derived a generalized
expression for the reaction constant:\cite{Waite57} similarly to
eq.~(\ref{waitet}), $k$ is found to be time-dependent but, once the
initial transient is completed, the following approximate
time-independent expression holds\cite{Waite57,Kuchinsky}:
\begin{equation}
k_r=4\pi r_0 D\frac{w}{w+D(r_{0}\Delta r)^{-1}}
\label{kwaitereaction}
\end{equation}
which generalizes eq.~(\ref{waiterateconstant}). This expression
allows to make quantitative the distinction among diffusion- and
reaction- limited processes, depending on the relation existing
between diffusion and reaction time scales \cite{Waite57,Kuchinsky}.
Indeed from (\ref{kwaitereaction}), when $w\gg D(r_{0}\Delta
r)^{-1}$, $k_r \sim k_d$ (compare with
eq.~(\ref{waiterateconstant})), and the reaction is a purely
diffusion-limited process; in particular, the diffusion-limited
reaction rate depends on $D$ but not on $w$ . In contrast, if $w\ll
D(r_{0}\Delta r)^{-1}$, one gets $k_r \sim 4 \pi r_0^{2} \Delta r
w$. In these conditions, the reaction constant depends on $w$ and
not on $D$ and the process is reaction-limited. If
$r_0$$\sim$$\Delta r$, the quantity $D(r_{0}\Delta r)^{-1}$ can be
interpreted as the inverse of the time necessary for the diffusing
species to move a distance $r_0$. Hence, the condition $w\gg
D(r_{0}\Delta r)^{-1}$ can be simply stated as follows: a reaction
is diffusion-limited if the time necessary for the mobile species to
diffuse away from the capture radius is much longer than the mean
time ($w^{-1}$) necessary for reaction. This confirms our previous
conclusions based on qualitative arguments. \subsubsection{Solving
the rate equations}\hspace{0.8cm}We are going now to briefly sketch
the structure of the solutions of the chemical rate equations. The
simplest situation is that of a single diffusion-limited bimolecular
reaction, described by the second order rate equation
(\ref{rateconstant}). The parameter $k$ will be supposed to be
independent of time, which is basically true apart from an initial
transient of duration $r_0^{2}/D$. In this case, it is easy to find
the following analytical solutions of eq.~\ref{rateconstant}. If the
initial concentration [A](0) is higher than [B](0):
\begin{equation}
[A](t) = \frac{\delta}{1-\lambda \exp{(-k \delta t)}}
\label{waitesolution1}
\end{equation}

\begin{equation}
[B](t) = \frac{\delta}{\lambda^{-1}\exp{(k \delta t)}-1}
 \label{waitesolution2}
\end{equation}
where $\delta=[A](0)-[B](0)$ and $\lambda=[B](0)/[A](0)$. For the
particular initial condition $[A](0)=[B](0)=c$, the solution assumes
a different form:
\begin{equation}
[A](t)=[B](t) = \frac{c}{1+kct}
 \label{waitesolution3}
\end{equation}
This initial condition, which may appear quite unlikely, can
actually be realized if $A$ and $B$ are introduced in the sample as
a consequence of a common mechanism. Both in
eq.~(\ref{waitesolution1}) and eq.~(\ref{waitesolution3}), the
typical time scale over which the reaction occurs is given by the
product of the reaction constant $k$ times a concentration, either
$\delta$ or $c$.

A more complex situation is that in which $n$ reactions proceed
together. In this case, a rate equation can be written for each
chemical species, containing a sum of terms each describing the
variation rate of the concentration of that species due to a given
reaction. In this way, one obtains a system of nonlinear
differential equations whose solutions can be compared with
experimental data. Some approximations can be often made which help
to simplify the system, typically based upon the orders of magnitude
of the reaction coefficients. To illustrate this approach, we
analyze a case that will be of much interest in the following,
namely the passivation of a defect X by molecular hydrogen in silica
at room temperature. To fix ideas, we can further specify X to be a
\emph{paramagnetic} defect: indeed, as molecular hydrogen is a very
stable molecule, usually it reacts efficiently only with
paramagnetic (rather than diamagnetic) centers, very reactive due to
the presence of an unpaired spin \cite{RadtzigJPC95,GriscomJNCS84}.
The treatment we are about to follow has been used in literature to
describe the reaction of NBOHC centers with H$_2$
\cite{GriscomJNCS84,KajiharaPRB06}, while in the following of this
work it will be applied to the passivation of $E'$. The generic
reaction with X and H$_2$ can be written
\begin{equation} \text{X + H$_2$} \stackrel{k_1}{\longrightarrow}
\text{X\,--\,H + H}\label{conh2}\end{equation} and results in the
conversion of the paramagnetic X into the diamagnetic X\,--\,H as
well as in the production of a free hydrogen atom H. Then, H
diffuses and can encounter another X, passivating
it:\begin{equation} \text{X + H} \stackrel{k_2}{\longrightarrow}
\text{X\,--\,H}\label{conh}\end{equation} but another possibility is
that H made available by (\ref{conh2}) meets another H and dimerizes
forming H$_2$.\begin{equation}\text{H + H}
\stackrel{k_3}{\longrightarrow}
\text{H$_2$}\label{dimerize}\end{equation} In order to find out the
time dependence of the concentrations of [X], [H$_2$] and [H], we
write down the system of rate equations derived from the three above
reactions:
\begin{equation}
\frac{d[\text{X}]}{dt}=-k_1[\text{X}][\text{H$_2$}]-k_2[\text{X}][\text{H}]
\label{ratex}
\end{equation}
\begin{equation}
\frac{d[\text{H$_2$}]}{dt}=-k_1[\text{X}][\text{H$_2$}]+k_3[\text{H}]^2
\label{rateh2}
\end{equation}
\begin{equation}
\frac{d[\text{H}]}{dt}=k_1[\text{X}][\text{H$_2$}]-k_2[\text{X}][\text{H}]-2k_3[\text{H}]^2
\end{equation}
where $k_1$, $k_2$, $k_3$ are the rate constants of reactions
(\ref{conh2}), (\ref{conh}), (\ref{dimerize}), respectively. The
three equations are self-explicatory generalizations of
eq.~(\ref{rateconstant}). Now, we start with proposing a qualitative
view of the solutions of this system. Let us suppose that the
initial concentrations of X and H$_2$ are comparable, being of order
$c$. Due to the reactions, the concentration of these two species
are progressively reduced until one of the two is exhausted. In
contrast, H plays quite a different role. Indeed, this species is an
\emph{intermediate} product, meaning that it is produced by reaction
(\ref{conh2}) and consumed by (\ref{conh}) and (\ref{dimerize}). In
addition, we observe that $k_1$ is a reaction constant limited by
the diffusion of \emph{molecular} hydrogen, whereas $k_2$ and $k_3$
are limited by the much faster \emph{atomic} hydrogen diffusion.
Based on the values of Table (\ref{dc}) and on the expression
(\ref{waiterateconstant}) for the diffusion-limited reaction
constant (where the order of magnitude of the capture radius is a
few {\AA}), it is evident that the following condition holds: $k_1
\ll k_2,k_3$. In other words, the typical time scale for H$_2$
diffusion and reaction is much longer than for H. From these
considerations, one can draw the following qualitative picture: H
that is "slowly" produced by reaction (\ref{conh2}) (controlled by
$k_1$) is very rapidly consumed by reactions (\ref{conh}) and
(\ref{dimerize}) ($k_2$, $k_3$); hence, H is expected to behave as a
\emph{transient} species, whose concentration always remains much
lower than H$_2$ and X: [H] $\ll$ [H$_2$],[X]. Even if an initial
anomalously high concentration of H is present, it is going to be
rapidly consumed by reactions (\ref{conh}) and (\ref{dimerize}), at
most in a time of $\sim k_2 c$, during which the much slower
reaction (\ref{conh2}) is actually frozen. Apart from this possible
fast transient, H just follows adiabatically the "slow" decrease of
the other two species, whose typical time scale is $\sim k_1 c$, but
keeping a much lower concentration. On this basis, one can make the
following approximation, called \emph{stationary state
approximation}:\cite{Atkinsbook,GriscomJAP85}
\begin{equation}
\frac{d[\text{H}]}{dt}\sim0\Longrightarrow
k_1[\text{X}][\text{H$_2$}]-k_2[\text{X}][\text{H}]-2k_3[\text{H}]^2\sim
0 \label{stationary}
\end{equation}
The stationary state approximation is widely applicable to reactions
involving a highly reactive intermediate product \cite{Atkinsbook}.
Now, since the concentration [X] is much higher than [H], and $k_2$
and $k_3$ are comparable, the following relation holds:
$k_3[\text{H}]^2 \ll k_2[\text{X}][\text{H}]$. Therefore, we can
neglect the quadratic term in [H] with respect to the linear one,
and from eq.~(\ref{stationary}) we finally find:
\begin{equation}
[\text{H}]\sim\frac{k_1}{k_2}[\text{H$_2$}] \label{hovsh2}
\end{equation}
Using eq.~(\ref{hovsh2}), the stationary state approximation allows
to eliminate the concentration [H] that can be simply expressed as a
function of [H$_2$]. It is worth noting that, since $k_1 \ll k_2$,
eq.~(\ref{hovsh2}) confirms \emph{a posteriori} that $[H]\ll [H_2]$,
so that the approximation gives self-consistent results. At
$T$=300\,K, from Table~(\ref{dc}) we estimate [H]/[H$_2$]$\sim
k_1/k_2\sim$10$^{-4}$: due to the very high diffusivity and
reactivity of H, the stable form of hydrogen at room temperature is
molecular rather than atomic.

Finally, we can substitute eq.~(\ref{hovsh2}) in eqs.~(\ref{ratex})
and~(\ref{rateh2}) and neglect again the quadratic term in the
latter, to get:
\begin{equation}
\frac{d[\text{X}]}{dt}=2\frac{d[\text{H$_2$}]}{dt}=-2k_1[\text{X}][\text{H$_2$}]
\label{ratefinale}
\end{equation}
the factor 2 has a very simple interpretation: each H$_2$ molecule
actually reacts with \emph{two} X centers, because the hydrogen atom
produced by reaction~(\ref{conh2}) readily recombines with another
X~(\ref{conh}). The expression (\ref{ratefinale}) represents now a
system of two equations in the form (\ref{rateconstant}), and can be
solved exactly to give [X](t). Using the expression
(\ref{waiterateconstant}) for $k_1$, indicating with $D_{H_2}$ the
diffusion constant of H$_2$ and with [X](0) and [H$_2$](0) the
initial concentrations, and substituting in (\ref{waitesolution1}),
we eventually obtain:
\begin{equation}
[\text{X}](t) =
\frac{([\text{X}](0)-2[\text{H$_2$}](0))[\text{X}](0)}{[\text{X}](0)-2[\text{H$_2$}](0)
\exp{ [-4 \pi r_0 D_{H_2} ([\text{X}](0)-2[\text{H$_2$}](0)) t]}}
\label{generich2decay}
\end{equation}from this expression, we see that [X](t) tends
asymptotically to [X]$_\infty$=[X](0)-2[H$_2$](0) (provided that
[X](0)$>$2[H$_2$](0)). The time constant of the exponential
appearing at the denominator, can be regarded as the typical
timescale $\tau$ in which the process occurs:
\begin{eqnarray}
\tau & \sim & [4 \pi r_0 D_{H_2} \{[\text{X}](0)-2[\text{H$_2$}](0)\}]^{-1} \nonumber \\
     &  =   & [4 \pi r_0 D_{H_2}[\text{X}]_\infty]^{-1} \nonumber \\
     & \sim & 10^{17}\text{cm}^{-3}\text{s}\cdot[\text{X}]_{\infty}^{-1}\qquad\text{(for H$_2$ at $T$=300\,K)} \label{timescale}
\end{eqnarray}using the values of Table \ref{dc} for the diffusion parameters of
H$_2$. As a final remark, we stress that even if in this particular
case we were able to find an analytical, albeit approximate,
expression for the time dependence of the concentrations of the
reacting species, in general the rate equation system can be solved
only numerically.

\section{Hydrogen in silica and its interaction with point defects}
\subsection{Forms of hydrogen in \texorpdfstring{\sil}{silica}} \hspace{0.8cm}The presence of hydrogen is so common in
silica that it can be barely considered as an impurity. It can be
found in \emph{free} form, as H atoms or H$_2$ molecules dissolved
in the matrix, or in \emph{bonded form}. Only in the first case it
is mobile and potentially able to diffuse in the matrix (provided
that the temperature is not too low). The most common bonded
configurations are Si\,--\,OH and Si\,--\,H groups. In the
following, we provide some more additional information about these
species and their detection techniques, and then we discuss in
detail the interaction of hydrogen with the two major point defects
in \sil{}, NBOHC and $E'$.

\paragraph{Si\,--\,OH:} Si\,--\,OH groups are the most known and common form
of bonded hydrogen in \sil{}. They can be incorporated in silica in
concentrations exceeding 10$^{20}$\,cm$^{-3}$, and they can be
detected both by IR absorption and Raman spectroscopy by the signal
around 3700\,cm$^{-1}$ associated to their stretching mode
\cite{Shelby94,SkujaSPIE01,Schmidt98}. Typical sensitivity of IR and
Raman (see footnote \ref{detlimit} of the previous chapter) allows
to measure a minimum concentration of Si\,--\,OH of $\sim$
10$^{17}$\,cm$^{-3}$. The IR absorption band associated to
Si\,--\,OH is detrimental to the telecommunications technologies,
since its first overtone at $\sim$1.4\,$\mu$m falls in the
wavelength range commonly used for transmitting signals in optical
fibers \cite{Nalwa,Erice,Doremus,Devinebook}. Silica samples are
commonly classified as "wet" or "dry" according to the high
(100\,--\,1000\,parts per million in weight) or low
(1\,--\,100\,ppm) concentration of Si\,--\,OH groups, which mainly
depends on the water content of the atmosphere in which the sample
is synthesized\cite{BruknerJNCS70,HetheringtonPCG65}.

An important property of this hydrogen-related center is that it
serves as a precursor for the NBOHC. Indeed, many experimental
evidences have demonstrated that irradiation can break the
oxygen-hydrogen bond in Si\,--\,OH generating NBOHC and atomic
hydrogen
\cite{GriscomJNCS84,KajiharaPRL02,KajiharaJNCS03,SkujaSPIE01,Erice}.
\begin{equation}
\equiv\text{Si\,--\,OH}\longrightarrow\equiv\text{Si\,--\,O}^{\bullet}
\text{ + H}
 \label{breakingoh}
\end{equation}
This process is observed under $\gamma$ or X-ray radiation, and can
be induced with high efficiency by F$_2$ laser radiation, which
photolyzes the bond by exciting the bonding-nonbonding electronic
transition of Si\--\,OH at E$>$7.4\,eV
\cite{KajiharaPRL02,KajiharaJNCS03,SkujaSPIE01}.

\paragraph{Si\,--\,H:} Si\,--\,H groups can form in reactions of hydrogen with
the silica network or with preexisting oxygen vacancies. They can be
detected by IR absorption or Raman measurements on their vibration
mode at about 2250\,cm$^{-1}$, with a typical sensitivity of $\sim$
10$^{17}$\,--\,10$^{18}$\,cm$^{-3}$ \cite{Shelby94,Schmidt98}. The
2250\,cm$^{-1}$ signal overlaps with a strong absorption due to an
intrinsic vibration mode of the Si\,--\,O\,--\,Si silica network. As
a consequence, the reliable observation of the Si\,--\,H signal by
IR absorption is usually possible only in difference spectra. Very
few experiments have investigated how the presence of hydrogen
incorporated in Si\,--\,H form depends on the manufacturing
procedure of \sil{}. As already discussed, it has been argued for a
long time that photo- (or radio-) induced breaking of the
silicon-hydrogen bond in Si\,--\,H generates $E'$ and H:
\begin{equation}
\equiv\text{Si\,--\,H}\longrightarrow\equiv\text{Si}^{\bullet}
\text{ + H}
 \label{breakingsih}
\end{equation}
for this reason, Si\,--\,H is generally regarded as an important
precursor for the paramagnetic
defect\cite{ImaiPRB91,SkujaSPIE01,SmithAO00,AfanasevJPCM00,NishikawaPRB90,NishikawaPRB93B,NishikawaPRB93,ImaiPRB93,ImaiJNCS94};
nevertheless, the understanding of process (\ref{breakingsih}) at
the moment is basically qualitative.

\paragraph{Other bonded forms:} Aside from Si\,--\,H and Si\,--\,OH, hydrogen
can be bonded in several other configurations, two examples of which
are the H(I) and the H(II) centers introduced in the previous
chapter. However, these structures usually appear only in irradiated
samples and store only a minor portion of the total hydrogen
population.
\paragraph{H:} Atomic hydrogen is very reactive, and
spontaneously combines with many point defects or dimerizes to form
H$_2$, thus being highly unstable except at low temperatures. For
this reason, at $T>$150\,K it exists only as a transient species,
generated by radiolysis or photolysis of Si\,--\,H or Si\,--\,OH
bonds or as a subproduct of the reaction of H$_2$ with defects. H
can be detected by ESR spectroscopy at $T<$150\,K by its typical
signal, which consists in a doublet split by 50mT due to the
hyperfine interaction of the electron and the proton spins
\cite{GriscomJNCS84,Abragam,Slichter}.
\paragraph{H$_2:$} The concentration of available solubility sites for
molecular hydrogen in silica glass at room temperature has been
estimated to be $\sim$10$^{21}$cm$^{-3}$, corresponding to a 4.6$\%$
molar concentration \cite{Shelby77}. H$_2$ in vacuum does not absorb
in the infrared due to its symmetry resulting in a zero dipole
moment \cite{Ottica}. In contrast, when incorporated in silica, the
structure of the molecule is slightly distorted, so as to activate a
weak IR absorption band at 4130--4140\,cm$^{-1}$ associated to its
fundamental stretching mode \cite{Doremus,Shelby94,Schmidt98}. The
vibration of the species can be also detected with higher
sensitivity by Raman spectroscopy, with a typical minimum detectable
concentration of 10$^{17}$\,--\,10$^{18}$\,cm$^{-3}$
\cite{Shelby94,Hartwig,Vandersteen,Schmidt98}. Several studies have
shown that usually H$_2$ is already present in variable
concentration in as-grown commercial silica materials
\cite{MorimotoJNCS,KajiharaPRL02}. As anticipated, due to its high
bond energy ($\sim$4.5\,eV) H$_2$ is a very stable molecule, which
usually reacts only with paramagnetic centers\footnote{It may occur,
however, that H$_2$ reacts efficiently with a diamagnetic center in
an excited electronic state, examples being the breaking of H$_2$
molecules by ODC(II) centers,\cite{RadtzigJPC95} or (possibly) by
GLPC centers in the S$_1$ state (see
\figurename~\ref{schemab}).\cite{MioJNCS06CONKADER}
}.\cite{RadtzigJPC95,GriscomJNCS84}

We have already mentioned several times that the presence of H$_2$
has important consequences on the response of silica to radiation,
partly as a consequence of its high diffusivity. Indeed, often its
concentration is increased by \emph{loading} techniques for several
technological purposes. One example is H$_2$-loading of Ge-doped
\sil{}, a procedure which, as we have seen, tends to enhance point
defect conversion processes leading to photosensitivity. On the
contrary, in \emph{pure} silica the presence of H$_2$ has in some
sense the opposite effect: it tends to passivate radiation-induced
defects. In particular, many experimental evidences have suggested
that the two main paramagnetic defects in \sil{}, the $E'$
($\equiv$Si$^{\bullet}$) and the NBOHC
($\equiv$Si\,--\,O$^{\bullet}$), can react with H$_2$ by the
following reactions:
\begin{equation} \label{reactnbohch2}
\equiv\text{Si\,--\,O}^{\bullet}\text{ + H$_2$ } \longrightarrow
\text{ }\equiv\text{Si\,--\,OH + H}\end{equation}
\begin{equation} \label{reacteh2}
\equiv\text{Si}^{\bullet}\text{ + H$_2$ } \longrightarrow \text{
}\equiv\text{Si\,--\,H + H}\end{equation} as discussed in the
previous section, atomic hydrogen produced at the right side of
these reactions may dimerize in H$_2$, or passivate another defect:
\begin{equation} \label{reactnbohch0}
\equiv\text{Si\,--\,O}^{\bullet}\text{ + H } \longrightarrow \text{
}\equiv\text{Si\,--\,OH}\end{equation}
\begin{equation} \label{reacteh0}
\equiv\text{Si}^{\bullet}\text{ + H } \longrightarrow \text{
}\equiv\text{Si\,--\,H}\end{equation}

In the presence of H$_2$, these reactions lead to a partial or
complete passivation of $E'$ and NBOHC produced by irradiation.
Hence, as the two defects absorb in the UV, hydrogen allows for a
recovery of the native transparency of the material compromised by
the presence of these centers. In this sense, loading with a high
concentration of H$_2$ is a standard technique used to make silica
materials more resistent to transparency loss upon exposure to
radiation, as common for example in multimode optical fibers to be
used to transmit deep UV (200\,nm\,--\,300\,nm) light. In the
following, we discuss what is known about the reaction properties of
these two defects with H$_2$.

\subsection{Reaction of NBOHC center with \texorpdfstring{H$_2$}{hydrogen}} 
\begin{figure}[htb!]
\begin{center}
%\makebox[\textwidth]{\framebox[5cm]{\rule{0pt}{5cm}}}
\includegraphics[width=.5\textwidth]{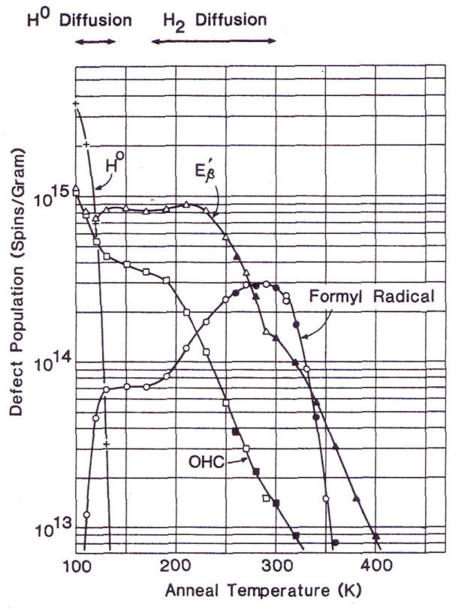}
\end{center}
\caption{Concentration of several paramagnetic defects during an
isochronous thermal treatment after X-ray irradiation at $T$=77\,K.
Figure adapted from Griscom\cite{GriscomJAP85}. }
\label{annealing2}\end{figure}

\hspace{0.8cm}One of the
first systematic investigations of the interaction between diffusing
hydrogen and point defects in \sil{} was carried out and reported by
Griscom in a 1984 paper\cite{GriscomJNCS84}. In that experiment, a
synthetic silica specimen was first exposed to X-ray radiation at
$T$=77\,K. Then, thermal treatments were performed at 77K$<T<$300K,
accompanied by ESR measurements, to investigate the annealing
properties of the generated defects. Data obtained during an
isochronous treatment are collected in \figurename~\ref{annealing2}.

The main experimental evidences are: \textbf{(i)} just after X-ray
irradiation NBOHC and H are detected in the sample in comparable
concentrations. \textbf{(ii)} Between 100\,K and 150\,K, H and NBOHC
decrease in concentration; in particular, the H signal completely
disappears. Also, it is observed the growth of two more signals ,
called the $E'_{\beta}$ and Formyl Radical (FR). The former is a
variety of $E'$ center already introduced in section \ref{strdef},
while the latter was identified as a HCO structure. \textbf{(iii)}
The concentrations of all the species remain relatively constant
until $\sim$200K is reached. \textbf{(iv)} Above 200\,K, a new stage
of the process begins: in particular it is observed a further anneal
of NBOHC and an increase of the FR. These results can be interpreted
as follows: X-ray irradiation at 77\,K generates NBOHC and H by
breaking the oxygen-hydrogen bond on pre-existing Si\,--\,OH
precursors. This result is particularly clear thanks to the direct
observation by ESR of both NBOHC and H, roughly in the same
concentration. Then, between $\sim$130\,K and 150\,K, \emph{atomic}
hydrogen diffusion is activated, so that H partly recombines with
NBOHC (Reaction~(\ref{reactnbohch0})) and partly dimerizes in H$_2$.
Above 200\,K, \emph{molecular} hydrogen diffusion is activated, and
the remaining portion of NBOHC are annealed by reaction with H$_2$
(Reaction~(\ref{reactnbohch2})). In this context, the observed
growth of the hydrogen-related Formyl Radical (HCO) plays an
important role: in fact, the FR is formed by reaction of (a portion
of) atomic hydrogen with pre-existing precursors in the CO form;
hence, the growth of FR represents an independent and clear
indication of the presence of diffusing hydrogen in the matrix,
which corroborates the attribution of the decay of NBOHC to
reactions with H and H$_2$. By demonstrating that NBOHC can be
passivated by reactions (\ref{reactnbohch2}) and
(\ref{reactnbohch0}), the experiment by Griscom was the first to
clearly evidence the role of diffusing hydrogeneous species in
annealing the defects generated at cryogenic temperatures.
Furthermore, it permitted to individuate the typical temperature
ranges for atomic ($T>$130\,K) and molecular ($T>$200\,K) hydrogen
diffusion in \sil{}.

\begin{figure}[htb!]
\begin{center}
%\makebox[\textwidth]{\framebox[5cm]{\rule{0pt}{5cm}}}
\includegraphics[width=.7\textwidth]{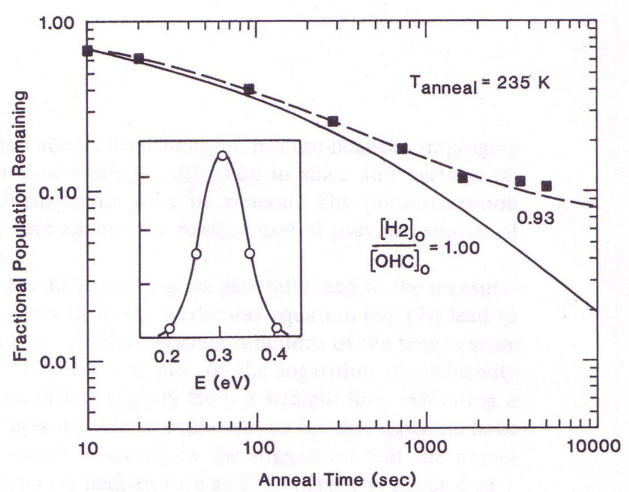}
\end{center}
\caption{Time dependence of the concentration of NBOHC induced by
irradiation at $T$=77\,K, as measured during an isothermal thermal
treatment at $T$=235\,K. The decay of NBOHC is due to reaction with
H$_2$. The fitting curves are obtained by a linear combination of
expressions (\ref{generich2decay}), weighted by a Gaussian
distribution (inset) of hydrogen diffusion activation energy. The
ratio between the initial concentrations of hydrogen and NBOHC is
used as a fitting parameter, with a best fit value of 0.93. Figure
adapted from Griscom\cite{GriscomJAP85}. }
\label{fitisocronogriscom}\end{figure}

Within this interpretation, the author made a step forward and tried
to fit the measured annealing kinetics on the basis of the Waite
model for diffusion-limited reactions, in the temperature range
($T$$>$200\,K) in which the annealing is due to reaction with
molecular hydrogen H$_2$. However, it was found that satisfactory
fits cannot be obtained by the expression~(\ref{generich2decay}) for
H$_2$ diffusion-limited kinetics. In contrast, a good fit could be
obtained only by a \emph{linear combination} of these expressions,
obtained by different values of $D_{H_2}$ weighted by a Gaussian
distribution of the activation energy for H$_2$ diffusion. This is
shown for an isothermal decay curve at $T$=235\,K in
\figurename~\ref{fitisocronogriscom}. This finding must be
interpreted as another manifestation of the existence of a
statistical distribution of diffusion activation energies,
consequent to the amorphous structure of silica; we have already
introduced this idea when discussing the non-Arrhenius behavior of
the diffusion constant $D$. In this sense, it is worth noting that
point defects like NBOHC may be considered an indirect tool to study
the properties of diffusion in glass.

Nonetheless, there is now a very important point to be stressed. We
have seen that the non-Arrhenius dependence of $D$ measured by
macroscopic diffusion experiments can be described by introducing a
Gaussian distribution in $E_a$ and identifying the measured value of
$D$ with the \emph{mean} diffusion constant calculated on the
distribution \cite{ShelbyKeeton,Shelby72}. On the contrary, when
fitting the data in \figurename~\ref{fitisocronogriscom}, the use of
a linear combination of expressions~(\ref{generich2decay}) is
\emph{not equivalent} to using a \emph{single} expression calculated
by simply substituting in (\ref{generich2decay}) the \emph{mean}
diffusion constant. Qualitatively, this suggests that the diffusing
species, in the typical experimental times, do not move enough to
experience the full range of activation energies, so as to be
characterizable by an "average diffusivity".\cite{GriscomJNCS84} In
contrast, one can imagine a set of H$_2$ molecules diffusing along
many independent and not crossing paths until encountering and
passivating a NBOHC, each path being characterized by a different
value for activation energy. As a consequence of this
interpretation, it was also suggested that the parameters of the
distribution of activation energy necessary to fit the reaction
kinetics may be slightly dependent on the explored concentration
range, which determines the mean free path of the diffusers
\cite{GriscomJNCS84}. This overall approach undoubtedly gives a
powerful insight on the diffusion in disordered solids and is quite
reasonable from a qualitative point of view; nevertheless, at the
moment it remains heuristic and lacks a solid theoretical basis
founded on a detailed microscopic modeling of the diffusion
phenomenon. Further experimental and theoretical studies are needed
to provide a thorough understanding of the problem.

Finally, we point out an important point that was left unsolved by
Griscom: the mean value $\sim$0.3\,eV of the activation energy
(inset of \figurename~\ref{fitisocronogriscom}), which was found by
fitting the reaction kinetics, is significantly lower than the
(mean) activation energy for diffusion of H$_2$ in \sil{} known from
classical diffusion studies, 0.38\,--\,0.45\,eV (Table \ref{dc}).
This poses a relevant problem in the interpretation of the results,
being apparently inconsistent with the attribution of the decay of
NBOHC to a diffusion-limited reaction with H$_2$.

Today the reaction between NBOHC and H$_2$, the role of hydrogen as
an important passivating agent in \sil{}, and the existence of a
site-to-site distribution of diffusion activation energies are well
established results. To overcome the limitations inherent in the
pioneering work by Griscom, the reaction properties of NBOHC with
hydrogen have been recently investigated again using a different
experimental approach, based on monitoring the NBOHC center by
observing its luminescence emission at 1.9\,eV. In detail, a pump
and probe PL technique has been applied to observe \emph{in situ}
the recombination of hydrogen and NBOHC centers, generated together
by breaking of Si\,--\,OH bonds (process \ref{breakingoh}) induced
by F$_2$ laser
photons.\cite{KajiharaPRL02,KajiharaAPL01,KajiharaPRB06,KajiharaJNCS03}
Similarly to the experiment by Griscom, by performing measurements
in different temperature ranges this experiment has permitted to
investigate both the reactions of NBOHC with H and H$_2$, the latter
species being formed by spontaneous dimerization of H atoms.

\begin{figure}[p!]
\begin{center}
%\makebox[\textwidth]{\framebox[5cm]{\rule{0pt}{5cm}}}
\includegraphics[width=.7\textwidth]{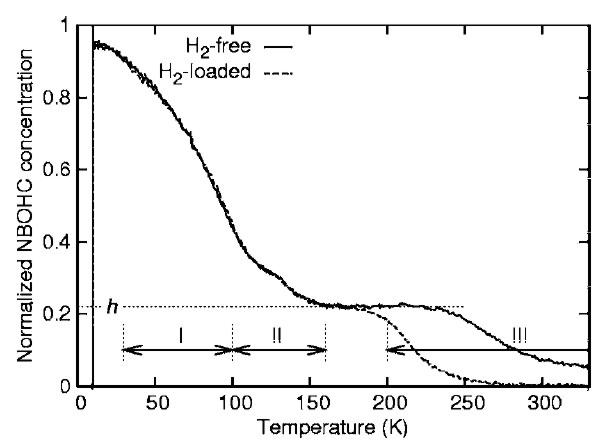}
\end{center}
\caption{Decay curves of NBOHC generated at 10\,K by a F$_2$ laser
pulse, as observed in a thermal annealing experiment in which the
sample was subjected to a progressive temperature increase from
10\,K with a constant rate of 3\,Kmin$^{-1}$. Results obtained in a
H$_2$-loaded material are compared with those in a sample initially
free of H$_2$. Figure adapted from Kajihara \emph{et
al.}\cite{KajiharaPRB06} } \label{isocronokaji}\end{figure}
\begin{figure}[p!]
\begin{center}
%\makebox[\textwidth]{\framebox[5cm]{\rule{0pt}{5cm}}}
\includegraphics[width=.7\textwidth]{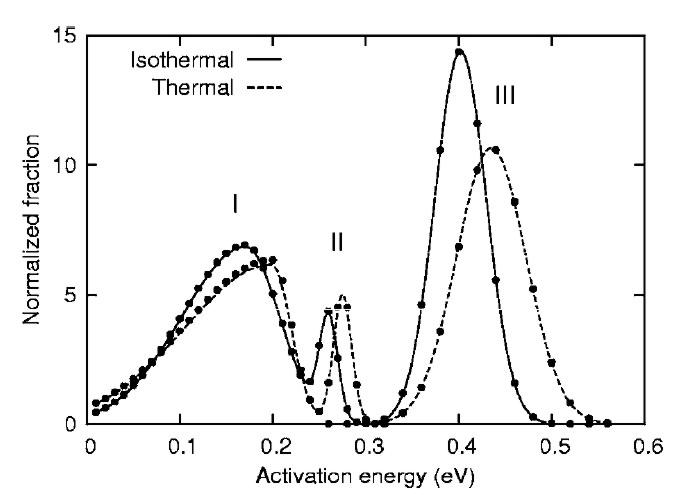}
\end{center}
\caption{Statistical distribution of activation energy for diffusion
of hydrogenous species, as obtained from the analysis of the
kinetics of isothermal and thermal (see
\figurename~\ref{isocronokaji}) decay curves. Figure adapted from
Kajihara \emph{et al.}\cite{KajiharaPRB06} }
\label{distridrogenokaji}\end{figure}

This investigation has permitted several steps forward in the
understanding of the problem. First, the study was performed
starting from very low (10\,K) temperatures, and it was found that
atomic H becomes mobile and starts to passivate NBOHC at
temperatures much lower than initially thought, i.e. $\sim$\,30\,K
(\figurename~\ref{isocronokaji}). Second, it was confirmed that the
kinetics of the reactions (\ref{reactnbohch2}) and
(\ref{reactnbohch0}) are not describable by ordinary
diffusion-limited reactions theory, but they can be accounted for by
including a distribution of diffusion activation energies. Within
this approach, from the accurate analysis of the measured thermal
and isochronous decay kinetics, it was possible to accurately
calculate the distribution of the activation energy $E_a$ for
diffusion of hydrogenous species in \sil{}, reported in
\figurename~\ref{distridrogenokaji}.
 Apart from some difference between the results found from isothermal
data with respect to isochronous data, we see that the distribution
of $E_a$ of \emph{atomic} H diffusion is strongly asymmetrical (peak
I), whereas that of H$_2$ (III) is well represented by a Gaussian.
An additional peak (peak II) is found that may be due to H trapped
within shallow traps in the silica matrix. The three peaks
correspond to temperature regions (\figurename~\ref{isocronokaji})
in which the diffusion of H and H$_2$ are thermally activated.
Moreover, the mean activation energy for H$_2$ diffusion is now
found to be $\sim$\,0.42\,eV, and the pre-exponential factor is
$\sim$\,8.5$\times$10$^{-5}$\,cm$^{2}$s$^{-1}$, both in a good
agreement with the values in Table~\ref{dc}. This demonstrates very
clearly that the reaction between NBOHC and H$_2$ is
diffusion-limited, solving the problem of the too small mean value
of $E_a\sim$0.3\,eV found by Griscom.\footnote{The discrepancy with
the results by Griscom was explained by hypothesizing the presence
in the first investigation of a certain amount of hydrogen already
dissolved in form of H$_2$ in the as-grown samples (as opposed to
hydrogen of radiolytic origin), and not taken into
account.\cite{KajiharaPRB06}. }

As a last remark we mention that, apart from the distribution of
activation energy, another model has been proposed in literature to
explain the "anomalous" decay kinetics in glasses, which as we have
seen cannot be fitted by standard chemical rate equations. The model
is based upon the idea of diffusion in fractal spaces. It is known
that fractals are geometrical structures characterized by a set of
properties in which they differ from "Euclidean" spaces, such as
self-similarity and fractional dimension. Fractals can derive either
from deterministic or from stochastic construction rules, and a
particular class of random fractals, the so-called percolation
cluster, is thought to well describe, at least from a qualitative
point of view, the structure of vitreous materials
\cite{GriscomPRB01,Avraham}. Now, it is known from simulations and
theoretical considerations that when diffusion takes place in a
fractal space, such as the percolation cluster, the simple Fick law
is not more valid. In particular, the relation
$L_d\propto$($D$t)$^{1/2}$ is substituted by
$L_d\propto$($D'$t)$^{\gamma}$, where $\gamma<$1/2; it can be shown
that this is formally equivalent to the introduction of a
time-dependent diffusion coefficient. This situation is usually
referred to as \emph{anomalous diffusion}. As a consequence, the
kinetics of diffusion-limited reactions is anomalous as well, and
can be described by rate equations similar to the ordinary but
having time-dependent rate "constants"
\cite{Avraham,GriscomPRB01,VerdiPRB93}. Therefore, in some works the
decay curves of point defects in silica have been fitted with
success by so-obtained fractal kinetic curves; the parameters were
found to be near to the values expected from the mathematical theory
of fractals \cite{GriscomPRB01,Avraham,VerdiPRB93}. This approach is
alternative to (but not incompatible with) the introduction of a
distribution of activation energies for diffusion. Also in this
case, further investigation is needed to found on a more rigorous
basis the fractal representation of the microscopic structure of
silica.

\subsection{Reaction of $E'$ center with \texorpdfstring{H$_2$}{hydrogen}} \label{reactewithh2section} \hspace{0.8cm}With respect
to the NBOHC, much less is known about the other basic
hydrogen-defect reaction in \sil{}, namely passivation of
 $E'$ by hydrogen (Reactions (\ref{reacteh2}) and (\ref{reacteh0})).
 We had previously mentioned this process when discussing the generation of $E'$ from Si\,--\,H precursors
(Reaction \ref{genEdaSi-HA}). In this case, hydrogen atoms are
generated as a co-product of $E'$. Therefore, H dimerizes in H$_2$
that recombines with $E'$ in the post-irradiation stage (Reaction
\ref{reacteh2}) causing the partial disappearance of E' centers in a
time scale of a few hours at room temperature
(\figurename~\ref{decaydaImai})
\cite{ImaiPRB91,CannasJNCS04,SmithAO00,KuzuuPRB91}. More in general,
hydrogen can also be not of radiolytic origin and react with $E'$
centers generated from other precursors. For example, passivation of
$E'$ has been observed to occur in silica films irradiated with
$\gamma$, UV or X-rays or in activated surfaces, after exposition of
the sample to gaseous H$_2$ in a vessel
\cite{RadtzigJPC95,BobyshevKK90,Vikhrev,ConleyAPL92,LiJNCS90,AfanasevJPCM00}.
Apart from being relevant for the general understanding of diffusion
and reaction dynamics in glass, reaction (\ref{reacteh2}) has a
notable practical importance as it allows to reduce the 5.8\,eV
absorption band of $E'$ detrimental for the UV transparency of the
glass. Even so, the main features of this process have been quite
debated in literature and at the moment a full understanding is
lacking.

First, although many experimental observations demonstrate that
reaction (\ref{reacteh2}) proceeds spontaneously at room
temperature, some theoretical calculations contrast with these
results, suggesting instead that the dissociation of a H$_2$
molecule on $E'$ is an endothermic process, so that $E'$ should not
be an efficient cracking center for H$_2$, differently from NBOHC
\cite{Erice,EdwardsJNCS94,VitielloJPCA00,PacchioniPRB98c}. Besides,
theoretical works and experimental results obtained in MOS systems
have suggested that the process can occur efficiently in the
\emph{opposite} direction, i.e. spontaneous reaction of H with
preexisting Si\,--\,H bonds de-passivates the dangling bond and
generates a H$_2$ molecule \cite{VitielloJPCA00,Lucovsky}.

Second, in experiments on $E'$ centers generated by laser
irradiation at room temperature, it was observed that the defects
partially decay in the post-irradiation stage, likely by reaction
with H$_2$. However, when trying to fit with
eq.~(\ref{generich2decay}) the time dependence of [$E'$], it was
found an unrealistically small capture radius of the order of
$\sim$10$^{-2}$\,{\AA} \cite{ImaiPRB91,CannasJNCS04}. This is
tantamount to saying that the kinetics is much slower than expected
for a H$_2$-diffusion-limited reaction. A possible interpretation of
this finding is that the passivation of $E'$ by H$_2$ is
reaction-limited rather than purely diffusion-limited, namely its
rate is significantly conditioned by the short-distance interaction
between the two species \cite{LiJNCS90,CannasJNCS04,Kuchinsky};
however, this is contrary to what occurs for NBOHC or for many other
reactions in solids, and has been questioned by several other
theoretical and experimental
works.\cite{KuzuuPRB91,EdwardsJNCS95,Mrstik,Waite57,Atkinsbook}

\begin{figure}[h!]
\begin{center}
%\makebox[\textwidth]{\framebox[5cm]{\rule{0pt}{5cm}}}
\includegraphics[width=.6\textwidth]{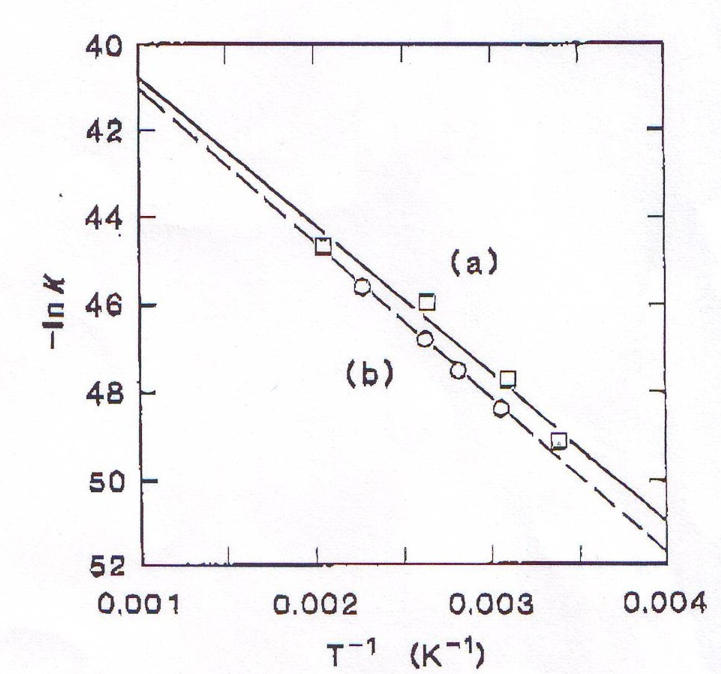}
\end{center}
\caption{Arrhenius plot of the reaction constant between $E'$
centers and H$_2$ in irradiated \sil{} films of different
thicknesses (450\,nm (a) and 100\,nm (b)) on a Si substrate. The
activation energy for reaction (\ref{reacteh2}) derived from these
data is 0.3\,eV. Figure taken from Li \emph{et al.}\cite{LiJNCS90} }
\label{lee}\end{figure}

From the experimental point of view, the character (diffusion- or
reaction- limited) of the reaction between $E'$ and H$_2$ could be
clarified by an accurate kinetic study able to estimate the reaction
rate as a function of temperature. Unfortunately, until now such
measurements have been carried out only for surface $E'$ centers
\cite{RadtzigJPC95}, or for $E'$ centers in thin silica films
\cite{LiJNCS90}, systems where the reaction can be directly
investigated by exposure of the surfaces to gaseous H$_2$ in a
vessel. These two works report disagreeing results: the measured
rates of reaction (\ref{reacteh2}) at $T$=300\,K differ by two
orders of magnitude, while the two activation energies are 0.4\,eV
and 0.3\,eV respectively\cite{RadtzigJPC95,LiJNCS90} (data for $E'$
in a thin silica film are reported in \figurename~\ref{lee}). In
regard to theoretical studies, the process was shown to require an
activation energy ranging from 0.2\,eV to 0.7\,eV depending on the
calculation methods
\cite{VitielloJPCA00,EdwardsJNCS94,EdwardsJNCS95,PacchioniPRB98c}.
For $E'$ centers in \emph{bulk} silica, the task of studying
experimentally the reaction becomes more difficult, since here the
concentration of available H$_2$ is not an external controllable
parameter depending on the pressure in the vessel, and diffusion of
H$_2$ in the bulk \sil{} matrix becomes a necessary step to bring
the reagents in contact. Furthermore, it is needed to measure
\emph{in situ} the kinetics of the transient defects generated by
irradiation in temperature-controlled experiments. For these
reasons, the thermal activation properties of the reaction between
$E'$ and H$_2$, as well as the limiting factor of its kinetics, have
still to be thoroughly investigated. Finally, no experimental data
exist about the influence on this process of the statistical
distribution of the activation energy that characterizes the
diffusion of H$_2$ in \sil{}. A contribution to the understanding of
these problems represents one of the most relevant results of this
PhD Thesis.

\chapter{Experimental techniques: a theoretical background}\label{mm1}
\hspace{0.8cm}The purpose of this chapter is to provide a basic
theoretical background for the interpretation of the parameters
deduced from the investigation of point defects in \sil{} by optical
and magnetic resonance spectroscopic techniques.
\section{Optical properties of a point defect}\label{oppd} 
\begin{figure}[htb!]
\begin{center}
%\makebox[\textwidth]{\framebox[5cm]{\rule{0pt}{5cm}}}
\includegraphics[width=.7\textwidth]{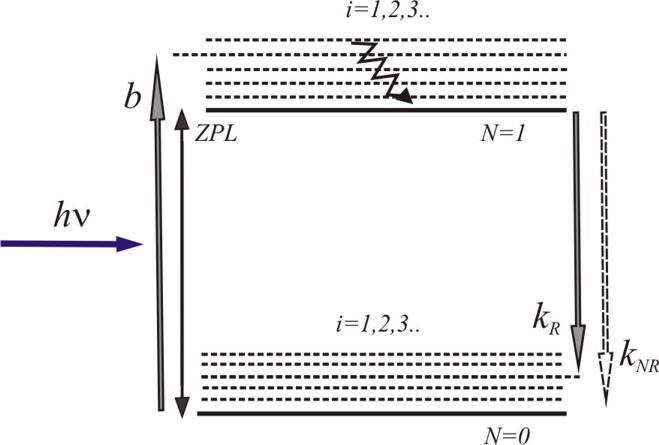}
\end{center}
\caption{Idealized electronic-vibrational level scheme of a two
levels point defect. The horizontal arrow represents an incoming
photon. The continuous arrow oriented upward represents one of the
possible absorption transitions (from the $(0,0)$ to the $(1,4)$
state). The continuous arrow oriented downward represents
spontaneous emission from $(1,0)$ to $(0,3)$. The dotted arrow
represents the non-radiative decay process. The double arrow
indicates the ZPL transition. The arrow within the $(1,j)$ levels
represents the internal relaxation
process.}\label{twolevelscheme}\end{figure}

\hspace{0.8cm}Optical
absorption spectroscopy investigates the properties of a point
defect embedded in a solid by the observation of the absorption
band(s) associated with the transitions of the center from the
electronic ground state to the excited state(s)\footnote{We refer
here only to electronic transitions, whose typical energies
(generally of the order of a few eV) fall in the range probed by UV
or visible optical absorption spectroscopy.}. Consider the simple
two-level scheme of a point defect represented in
\figurename~\ref{twolevelscheme}. To describe the optical properties
of this center, one must take into account the coupling between the
electronic and the vibrational degrees of freedom. Within the
Born-Oppenheimer and the Condon approximations, this can be done by
describing the state of the system as the (tensorial) product of an
electronic wave function and a nuclear wave function; on this basis,
the general state of the system is be characterized by a pair of
quantum numbers $(N,i)$, describing the state of the electronic and
vibrational subsystems respectively \cite{Erice,Nalwa}.

Suppose now that the system is initially in its electronic
\emph{and} vibrational ground state $(0,0)$, as expected in
particular at T=0. After optical excitation it can be promoted to
each of the vibrational sublevels $(1,i)$ within the excited
electronic state, the transition at lowest energy being that from
$(0,0)$ to $(1,0)$, called the \emph{zero phonon line} (ZPL). The
set of $(0,0)$$\rightarrow$$(1,i)$ combined electronic-vibrational
transitions (\emph{vibronic} transitions) is characterized by a
specific distribution of transition rates, which can be predicted
theoretically and depends on the extent to which the equilibrium
position of the nuclei in the excited electronic state differs from
the ground state. This problem is treated in many standard texts
\cite{Erice,Nalwa} and is not discussed in detail here. However, we
recall that: \textbf{(i)} very often the $(0,0)$$\rightarrow$$(1,i)$
vibronic transitions cannot be resolved in the optical absorption
spectrum, which therefore appears as a single bell-shaped broad
band\footnote{Aside from electron-phonon coupling, which leads to
so-called homogeneous broadening, in amorphous silica the absorption
bands of point defects are further broadened by inhomogeneity
effects.}, given by the envelope of many narrow subbands each due to
a given transition\cite{Erice,Nalwa}. We have already seen typical
examples of such broad absorption bands in
\figurename~\ref{inducedoa} for the case of $E'$ and NBOHC centers.
\textbf{(ii)} The energy E$_{OA}^{pk}$ at which falls the maximum of
the absorption band corresponds to the most probable vibronic
transition, which is generally higher than the "pure" ZPL (see
\figurename~\ref{twolevelscheme}).

If a light beam propagates along the $x$ direction within a sample
containing absorbing point defects, its intensity is progressively
reduced according to an exponential equation \cite{Erice,Nalwa}:
\begin{equation} I(\lambda,x)=I_0(\lambda)\exp{(-\alpha(\lambda)x)}
\end{equation}where $I(\lambda,x)$ is the intensity of the monochromatic
component of wavelength $\lambda$ in the position $x$. The last
equation defines implicitly the \emph{absorption coefficient}
$\alpha(\lambda)$, which characterizes the OA profile of the point
defect. The quantity that is directly measured by an absorption
spectrophometer is the \emph{optical density} $OD(\lambda)$ of the
sample. It is defined by:
\begin{equation}
OD(\lambda)=\log_{10}\frac{I_0(\lambda)}{I(\lambda,d)}
\end{equation}and is obtained by measuring for each $\lambda$ the light
intensity before ($I_0(\lambda)$) and after ($I(\lambda,d)$) the
sample. OD is related to $\alpha$ by the following relationship
\begin{equation}
\alpha=\ln{(10)}\frac{OD}{d}\sim2.303\frac{OD}{d}
\end{equation} where $d$ is the thickness of the sample. Finally, we
note that the absorption coefficient can be equivalently expressed
as $\alpha(E)$, namely as a function of the photon energy
$E=2\pi\hbar c/\lambda$, $\hbar$ and c being the Planck constant and
the speed of light respectively.

Consider a silica sample containing a single species of absorbing
point defects in concentration $\rho$. As anticipated, the
absorption profile $\alpha(E)$ due to the presence of the centers is
typically bell-shaped with a width between 0.1\,eV and 1\,eV. The
value of $\alpha$ at a given spectral position is proportional to
$\rho$ through the absorption cross section $\sigma(E)$
\cite{Erice,Nalwa}:
\begin{equation}
\alpha(E)=\sigma(E)\rho\label{relsa}
\end{equation}in turn, $\sigma(E)$ is connected to the Einstein
coefficient $b$ for absorption and stimulated emission\footnote{From
now on we suppose for simplicity that both the initial and the final
electronic state are non-degenerate. Generalizing the equations to
take into account electronic degeneracy is
straightforward.}:\cite{Erice,Nalwa,Ottica}
\begin{equation}
b=\frac{c}{2n\pi\hbar}\int\frac{\sigma(E)dE}{E}\label{bconsigma}
\end{equation}where $n$ is the refractive index of the medium in which the
defect is embedded. In the case of a very narrow absorption band,
the energy dependence of the denominator can be neglected, thus
giving the following approximate expression:
\begin{equation}
b\sim\frac{c}{2n\pi\hbar^2\omega}\int\sigma(E)dE\label{bconsigmaapprox}
\end{equation}which is equivalent to regarding the absorption process as a
simple transition between narrow, "atomic"-like levels separated by
an energy $\hbar\omega$.

The absorption properties of a point defect can be characterized
either by using the "local" (function of $E$) property $\sigma(E)$
or by a dimensionless integrated parameter known as the
\emph{oscillator strength}. The two are alternative but equivalent
ways to express the strength of the absorption signal arising from a
given concentration of the defect. In detail, the oscillator
strength $f$ of an electric dipole transition of frequency $\omega$
is given by \cite{Erice,Nalwa,Ottica}:
\begin{equation}f=\frac{2m\omega}{3\hbar e^2}
|<\psi_1|D|\psi_2>|^2\end{equation}where $m$ and $e$ are electron's
mass and charge, respectively, $D$ is the electric dipole moment
operator and $\psi_1$ and $\psi_2$ indicate the quantum initial and
final states of the transition. $f$ can be connected to the
\emph{integral} of the measured OA band. In fact, it can be shown
that \cite{Erice}:
\begin{equation}
\int\alpha(E)dE=\frac{\rho}{n}\left(\frac{E_{e}}{E_0}\right)^2\frac{2\pi^2e^2\hbar}{mc}f\label{sma}
\end{equation} where the $E_{e}/E_{0}$ term is
called the \emph{effective field correction}, and takes into account
the difference existing between the macroscopic electric field
$E_{0}$ and the \emph{local} microscopic field E$_{e}$ acting at the
position of the defect \cite{Erice,Ashcroft}. Several treatments of
the problem of the effective field are available in standard texts,
resulting in different mathematical expressions of the correction.
One of the most commonly used is called the Lorentz-Lorenz
correction: within this approach, and in the case of a Gaussian
absorption band, eq.~(\ref{sma}) can be approximated as follows
\cite{Erice}:
\begin{equation}
\rho
f\sim8.72\times10^{16}\frac{n}{(n^2+2)^2}\alpha_{max}\Delta\times[cm^{-2}\,eV^{-1}]
\end{equation} where $\alpha_{max}$ is the maximum amplitude of the band and $\Delta$
is the width (full width at half maximum) of the band.

The experimental estimate of the oscillator strength (or of the
cross section) of a point defect can be made using (\ref{sma}) (or
(\ref{relsa})) on the basis of the measured absorption band, only if
the concentration is already known from an independent approach. For
paramagnetic centers this is usually accomplished by ESR
measurements, which permit to estimate the absolute concentration of
the defects by comparison with a reference sample. For example, for
the 5.8\,eV absorption band of the $E'$ center it was determined
that the \emph{peak} absorption cross section is
$\sigma$(5.8\,eV)=6.4$\times$10$^{-17}$\,cm$^{2}$
\cite{CannasJNCS04,Nalwa}. In the case of photoluminescent centers,
the oscillator strength can be estimated from the knowledge of the
luminescence radiative decay lifetime (as explained below). If a
center is neither paramagnetic nor luminescent, the task becomes
much more complicated and must be solved indirectly, for example on
the basis of the observed conversion in another defect of known
concentration.

Another optical property of great importance in the study of point
defects is photoluminescence (PL), i.e. the process by which a
system excited by light with wavelength $\lambda_0$ emits light at
$\lambda_e>\lambda_0$ while decaying back to its ground state. Under
excitation with constant intensity, we indicate the spectral density
of the emitted light as
\begin{equation}
\frac{dI_e}{d\lambda_e}(\lambda_0,\lambda_e)
\end{equation}which is considered as a function of $\lambda_e$,
while $\lambda_0$ may be regarded as a parameter. A measurement of
the emitted spectral profile $dI_e/d\lambda_e$ (as a function of
$\lambda_e$) under excitation at a given $\lambda_0$ is called
\emph{emission} spectrum of the center. Not all the point defects
that feature a measurable absorption band decay by emitting
luminescence, but when this occurs, their study by PL spectroscopy
has some important advantages with respect to OA. In particular, PL
is more \emph{selective}, as it often allows to \emph{isolate} a
center whose absorption band overlaps to those arising from other
defects, based on the different emission properties.

To picture the physical processes determining the PL emission band,
let us consider again the two-level system of
\figurename~\ref{twolevelscheme}. Excitation to one of the $(1,j)$
states is followed by a rapid \emph{internal relaxation}, which
drives the system all the way down to the $(1,0)$ state (if at
T=0\,K), within a time scale comparable to that of nuclear
vibrations: 10$^{-12}$--10$^{-11}$s \cite{Harris}. Then, in general
the center relaxes back to the ground state either by
\emph{radiative} emission (photoluminescence) or by a
temperature-dependent \emph{non-radiative} process in which the
energy is dissipated by emission of phonons. The two processes can
be characterized by a radiative decay rate $k_R$ and a non-radiative
decay rate $k_{NR}$. The radiative decay is the origin of the
emission band, consisting in the combination of several
$(1,0)\rightarrow(0,j)$ vibronic transitions. The most probable of
such transitions determines the peak spectral position E$_{PL}^{pk}$
of emission. From \figurename~\ref{twolevelscheme} it is apparent
that E$_{PL}^{pk}$ must be \emph{lower} than the ZPL; Hence, it is
also lower than the spectral position of the absorption peak
E$_{OA}^{pk}$$>$ZPL. This results in a shift
(E$_{OA}^{pk}$-E$_{PL}^{pk}$) between the peak positions of
absorbtion and emission spectra, called \emph{Stokes shift}
\cite{Nalwa,Erice}. It can be demonstrated that the absorption and
emission spectra are approximately symmetric (\emph{mirror
symmetry}) with respect to the ZPL position \cite{Erice,Nalwa}.

When a defect emits by photoluminescence under excitation of
constant intensity within its absorption band (stationary
photoluminescence), the efficiency of the luminescence process can
be characterized by a parameter $\eta$ known as \emph{luminescence
quantum yield}, defined as the ratio between emitted photons and
absorbed photons. $\eta$ is determined by the competition between
radiative and non-radiative decay processes. For the case of
\figurename~\ref{twolevelscheme}, it can be easily demonstrated that
$\eta=k_R/(k_{NR}+k_R)$. Differently from stationary PL
measurements, in a \emph{time-resolved} PL measurement, it is
studied the time decay of the emitted light after an exciting light
\emph{pulse}. By this technique, it is possible to directly estimate
the decay time $\tau_e=(k_{NR}+k_R)^{-1}$ from the excited state. At
sufficiently low temperatures, the non-radiative decay channels are
usually quenched, i.e. $k_{NR}\ll k_{R}$, so that the measurement
directly yields the \emph{radiative} decay time $\tau=1/k_R$. The
importance of the knowledge of $\tau$ relies in the possibility of
calculating the oscillator strength $f$ of the center. To see this,
we observe that $1/\tau=k_R$ equals the $a$ Einstein coefficient for
spontaneous emission, which is connected to the $b$ coefficient by
the relation $a=2\hbar\omega^3n^3b\pi^{-1}c^{-3}$. Hence, combining
with (\ref{bconsigmaapprox}) and (\ref{relsa}), the following
relationship between $1/\tau$ and the absorption profile $\alpha(E)$
is derived:
\begin{equation}
\frac{\rho}{\tau}=\frac{n^2\omega^2}{\pi^2c^2\hbar}\int\alpha(E)dE
 \label{reltauf}
\end{equation} Based on the knowledge of $\tau$, if the absorption profile of the defects $\alpha(E)$ has been measured, this
equation can be used to find the concentration $\rho$, which in turn
can be inserted in eq.~(\ref{sma}) to estimate $f$. Nonetheless,
eq.~(\ref{reltauf}) has been deduced within the scheme of an
"atomic" two level system, in which there is no electron-phonon
coupling and absorption and emission occur at the same energy. For
point defects in solids, this treatment must be carried out
singularly for \emph{each} vibronic transition. Then, the radiative
transition rates must be integrated over the whole spectrum. On this
basis, one can generalize eq.~(\ref{reltauf}) to an equation
(F\"{o}rster equation) applicable to the case of point defects in
\sil{}\cite{Erice}:
\begin{equation}
\frac{\rho}{\tau}=\frac{n^2}{\pi^2c^2\hbar^3}\int(2E_0-E)^3\frac{\alpha(E)dE}{E}\label{Forster}
\end{equation}
where $E_0$ is the position of the ZPL, which can be estimated
experimentally on the basis of the mirror symmetry between emission
and absorption spectra.

\section{Magnetic resonance of a point defect} \label{mrpd}
 \hspace{0.8cm}Electron Spin Resonance (ESR) is the resonant
absorption of electromagnetic radiation by an electronic spin system
coupled to a static magnetic field, which takes place when the
frequency of the radiation matches one of the characteristic
transition frequencies of the system. For typical laboratory
magnetic fields of the order of 10$^3$\,G, the resonance frequency
$\omega_0$ falls in the microwave range (10$^{9}$--10$^{10}$\,Hz). A
typical ESR spectrometer explores the resonance condition by
exposing the sample to a fixed microwave frequency $\omega_0$ and
varying the static magnetic field, and not \emph{vice versa}, since
an accurate control of a varying frequency would be technically more
difficult. Moreover, to increase sensitivity, a modulation magnetic
field of frequency $\omega_{m}\ll\omega_0$ and small amplitude
$A_{m}$ is superimposed to the static field; hence, it is used the
so-called "lock-in" detection technique, which is selective with
respect to the frequency $\omega_{m}$ and sensitive to the phase of
the detected signal. As a consequence, it can be shown that the
spectrometer is sensitive to the \emph{first derivative of the
energy absorption curve}.

The ESR technique is applicable to a variety of systems, including
paramagnetic point defects, which possess a nonzero spin in their
lowest electronic state due to the presence of un unpaired electron.
We have already seen important examples of such defects in \sil{},
in particular the $E'$ and the NBOHC. The ESR technique possesses an
unsurpassed ability to provide detailed structural information about
the centers: in some sense, it can be argued that most of the
current knowledge about the structure of \emph{all} defects in
\sil{} is based, directly or indirectly, on the models of
paramagnetic defects elucidated by means of ESR investigation
\cite{Erice}.

From the theoretical standpoint, the interaction of an electronic
spin with its surroundings can be described by the introduction of
the \emph{spin hamiltonian} $\mathcal{H}_0$
\cite{Abragam,Erice,Slichter}. The basic term of the spin
hamiltonian is the Zeeman contribution, which accounts for the
interaction of the magnetic moment associated to the spin
$\overrightarrow{S}$ with the magnetic field. For a field of
intensity $B_z$ directed along the $z$ axis of the laboratory frame:
\begin{equation}
\mathcal{H}_0=-\overrightarrow{\mu}\cdot\overrightarrow{B}=g \beta
\overrightarrow{S}\cdot\overrightarrow{B}=g \beta S_z B_z
\label{hamiltonian}
\end{equation}
where $\beta=|e|\hbar/(2mc)$ is the Bohr magneton and $g$ is the
spectroscopic splitting factor (the gyromagnetic ratio). This
Hamiltonian gives rise to two electronic levels, from which it is
found the following resonance condition:
\begin{equation}
B_z=\frac{\hbar\omega_0}{g\beta}\label{resonanceB}
\end{equation}
hence, the quantity $g$ can be measured from eq.~(\ref{resonanceB}),
based upon the observation of the field position at which the
resonance occurs.

A thorough description of the magnetic resonance phenomenon in a
solid requires not only to find the positions of the resonance
line(s), but also to deal with the interaction \emph{dynamics} of
the spin with the magnetic field. To this end, it is mandatory to
take into account the effects of the spin-lattice coupling, allowing
for the dissipation of energy, and of the spin-spin interaction,
which lead to a homogeneous broadening of the resonance line
\cite{Abragam,Erice,Slichter}. Without entering into the details, we
recall that this problem was treated by Bloch, who proposed a set of
equations able to describe the interaction of the macroscopic
magnetization $\overrightarrow{M}$ of a solid containing
paramagnetic centers with a time-dependent magnetic field. In
particular, the spin-lattice and spin-spin interactions were dealt
with by the introduction of two phenomenological relaxation times
$T_1$ and $T_2$, respectively. In the presence of the static field
$B_z$ and of a perpendicular microwave field $B_x$ oscillating at
the frequency $\omega_0$, if we define a complex magnetic
susceptivity $\chi$:
\begin{equation}M_x=\chi B_x e^{j\omega_0 t}\end{equation}then the
energy absorption of the system is connected to the imaginary part
$\chi''$ of the susceptivity. $\chi''$ can be equivalently
considered as a function of $\omega_0$ or $B_z$; as expected from
(\ref{resonanceB}), it can be demonstrated that $\chi''(B_z)$ shows
a resonance for $B_z=\hbar\omega_0/(g\beta)$ described by the
following equation\cite{Abragam,Erice,Slichter}:
\begin{equation}
\chi''(B_z)=\frac{1}{2}\gamma M_z^0 T_2 \frac{1}{1+T_2^2(\gamma
B_z-\omega_0)^2+\frac{1}{4}\gamma^2B_x^2T_1T_2}
\label{blochsolution}
\end{equation}where $M_z^0$ is the $z$
magnetization at thermal equilibrium, proportional to the number of
spins $N$, and $\gamma=g\beta/\hbar$. As apparent from
eq.~(\ref{blochsolution}), the shape and the intensity of the
resonance generally depend on the incident microwave power through
the term in $B_x^2$ appearing on the denominator. However, a
particular condition (called \emph{non-saturation condition}) is
realized when $\gamma^2B_x^2T_1T_2\ll 1$, which physically means
that the incident microwave power is low enough to allow for the
relaxation channels to efficiently dissipate the absorbed energy.
When the non-saturation condition is verified, the term in $B_x^2$
can be neglected from eq.~(\ref{blochsolution}), and the resonance
line acquires a Lorentzian shape independent of incident power and
with a $(T_2\gamma)^{-1}$ width.

As described in more detail in the next chapter, the measured
\emph{ESR spectrum} $S(B_z)$ is proportional to the derivative of
the absorption curve multiplied by the amplitude $B_x$ of the
microwave field and the modulation amplitude $A_m$ defined at the
beginning of this section:
\begin{equation}
S(B_z)\propto\frac{d\chi''}{dB_z}B_xA_{m}\label{esrmeasured}
\end{equation}
Hence, substituting $\chi''$ from (\ref{blochsolution}), the
\emph{doubly-integrated} signal gives the following quantity
$\Gamma$.
\begin{equation}
\Gamma=\int_{-\infty}^{+\infty}dy\int_{-\infty}^{y}dB_z\frac{d\chi''}{dB_z}B_xA_{m}=\frac{\pi
M_z^0A_{m}}{2}\frac{B_x}{\sqrt{(1+\frac{1}{4}\gamma^2B_x^2T_1T_2)}}\label{integralega}
\end{equation}If we define $\Gamma'=\Gamma/(B_xA_{m})$, far from saturation ($\gamma^2B_x^2T_1T_2\ll 1$), the last equation
can be approximated as $\Gamma'\sim \pi M_z^0/2$. Therefore, the
doubly-integrated signal normalized for the microwave ($B_x$) and
the modulation ($A_m$) amplitudes, $\Gamma'$, is proportional to the
number of spins $N$ through $M_z^0$, so as to represent a
\emph{relative} measurement of $N$. $\Gamma'$ can be converted to an
\emph{absolute} concentration measurement by comparison with a
reference sample in which the concentration of a given defect is
known by an independent technique. This interpretation of $\Gamma'$
is lost when the transition is saturated, because in this case (see
(\ref{integralega})) the proportionality factor depends on T$_1$ and
T$_2$, making impossible a comparison between different centers or
samples.

For a free electron, $g=g_e=2.0023$; in the case of a point defect
in a solid, the situation becomes more complicated because of the
admixture of \emph{angular} momentum into the spin ground state of
the center. In particular, this leads to the \emph{anisotropy} of
the Zeeman interaction, which is accounted for by promoting $g$ to a
\emph{tensorial} quantity $\underline{g}$, whose principal values
$g_1$,$g_2$,$g_3$ generally differ from $g_e$. Hence, the
hamiltonian assumes the following form
\cite{Abragam,Erice,Slichter}:
\begin{equation}
\mathcal{H}_0=\beta
\overrightarrow{B}\cdot\underline{g}\cdot\overrightarrow{S}
\end{equation}and the resonance condition becomes:
\begin{equation}
B_z=\frac{\hbar\omega_0}{\beta\sqrt{g_1^2cos^2{\theta_1}+g_2^2cos^2{\theta_2}+g_3^2cos^2{\theta_3}}}
\label{resonanceani}
\end{equation}where $\theta_i$ are the angles between the direction of the magnetic
field (i.e. the $z$ axis of the laboratory frame), fixed by the
geometry of the spectrometer, and the principal axes of the
$\underline{g}$ tensor, dictated by the local symmetry properties of
the defect.

\begin{figure}[htb!]
\begin{center}
%\makebox[\textwidth]{\framebox[5cm]{\rule{0pt}{5cm}}}
\includegraphics[width=.85\textwidth]{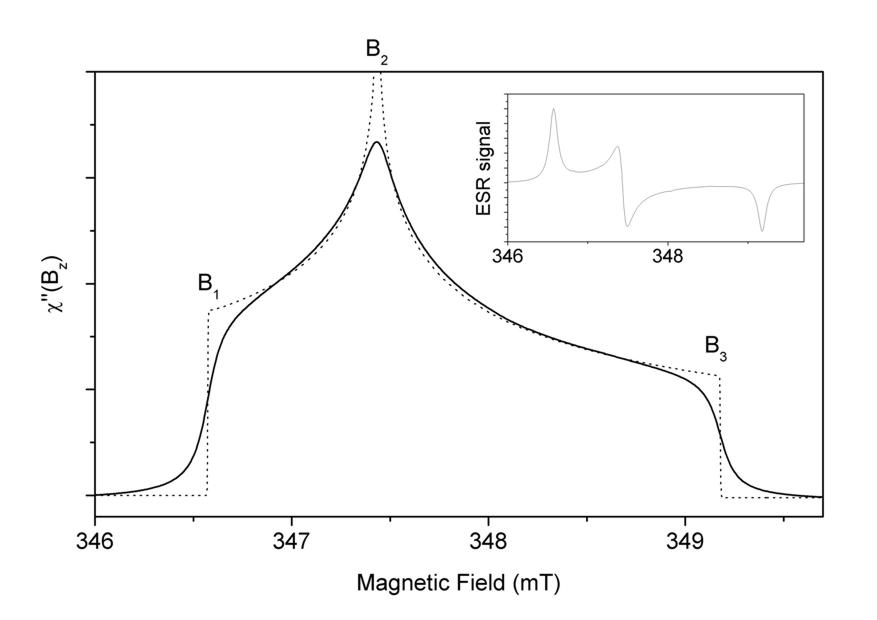}
\end{center}
\caption{Simulated ESR absorption curve ($\chi''$(B$_z$)) for a
system of randomly oriented defects in a crystal powder. The values
$B_i$ are defined as $\hbar\omega/\beta g_i$, where $g_i$ are the
principal axes of the tensor (g$_1$$\neq$ g$_2$$\neq$ g$_3$). The
dotted line represents the distribution of the resonance fields in
the powdered sample due to the random orientations of the spins. The
continuous line represents the ESR absorption curve, calculated by
convoluting the dotted line with a (much narrower) Lorentzian
lineshape. The ESR lineshape that would be observed in this case by
a measurement is reported in the inset and corresponds to the
derivative of the absorption curve. The case represented here is the
most general; in several cases of interest the actual signal appears
simpler thanks to the axial symmetry ($g_2$=$g_3$) possessed by many
defects of interest (such as the $E'$ center).}
\label{lineshapeESR}\end{figure}

If we execute a measurement on a \emph{powdered} crystal sample, in
which the defects are randomly oriented in all directions, the
observed resonance line results to be the envelope of many narrow
Lorentzian lines at the positions given by (\ref{resonanceani}) for
all the possible values of the random $\theta_i$. This gives rise to
the so-called \emph{powder lineshape}. From the mathematical point
of view, the isotropic distribution of the defect orientations can
be represented by statistical distributions of the three angles
$\theta_i$ proportional to $\sin(\theta_i)$; then, one has to use
eq.~(\ref{resonanceani}) to find the corresponding distribution of
$g=\sqrt{g_1^2cos^2{\theta_1}+g_2^2cos^2{\theta_2}+g_3^2cos^2{\theta_3}}$,
which must finally be convoluted with the single-packet lineshape
given by eq.~(\ref{blochsolution}) to obtain the overall resonance
curve\cite{Erice}. A typical result is sketched in
\figurename~\ref{lineshapeESR}: it is worth noting that from the
observation of the powder lineshape it is still possible to extract
the three principal values $g_i$, which correspond to specific field
positions detectable by inspection of the curve.
 Finally, when one considers defects
in an \emph{amorphous} matrix, another effect must be taken into
account: also within a set of centers all sharing the same
orientation with respect to the magnetic field, an intrinsic
site-to-site variation of the principal values of $\underline{g}$ is
expected as a consequence of the disordered nature of the amorphous
solid. In this sense, the $g_i$ are to be regarded as statistically
distributed, this leading to a further broadening of the resonance
line \cite{Erice}.

When a nucleus with nonzero spin $I$ is positioned near the
electronic spin, the \emph{hyperfine interaction} between the two
magnetic moments gives rise to another energy term that must be
included to the spin hamiltonian. In general, the hyperfine
interaction is accounted for by a \emph{hyperfine tensor}
$\underline{A}$, which contains an anisotropic portion due to the
dipolar spin-spin interaction, and an isotropic portion due to the
so-called \emph{Fermi contact term} \cite{Abragam,Erice,Slichter}.
Often, the contact term prevails, so that in first approximation the
interaction is completely described by a scalar $A_0\propto
g_eg_N\beta\beta_N|\psi(0)|^2$, where $g_N$ and $\beta_N$ are the
nuclear gyromagnetic ratio and the nuclear Bohr magneton
respectively, and $|\psi(0)|^2$ is the probability of finding the
electron at the position of the nucleus. The spin hamiltonian
assumes the following form:
\begin{equation}
\mathcal{H}_0= g \beta S_z B_z + A_0
\overrightarrow{I}\cdot\overrightarrow{S}
\end{equation} where for simplicity it was neglected the anisotropy
of $\underline{g}$, as well as the (usually minor) nuclear Zeeman
term, which describes the interaction between $\overrightarrow{I}$
and $\overrightarrow{B}$. Although the treatment proposed here is
actually too simplified to thoroughly describe several practical
cases of interest, it permits to evidence the main consequence of an
hyperfine interaction on the observed ESR spectrum: a splitting of
the resonance line in a multiplet of $2I+1$ lines, corresponding to
the possible values of the quantum number $M_I$ of the $z$ component
of $I$. At the first order in $A_0$, the positions of the resonances
are:
\begin{equation}
B_{z}=\frac{\hbar\omega_0}{g\beta}+\frac{A_0}{g\beta}M_I\qquad with
\qquad M_I=-I, -I+1,\ldots,I-1,I\label{hfsplitting}
\end{equation}

When an ESR signal comprises an observable hyperfine interaction,
this results to be a very powerful instrument to elucidate the
structure of the defect. Indeed, for different isotopes of the same
atom, the number of lines in the multiplet depends on the nuclear
spin $I$, and their separation parameter $A_0$ is proportional to
the nuclear magnetic moment: in this way, the nucleus responsible
for the hyperfine interaction can be unambiguously identified by
experimentally observing how the splitting (\ref{hfsplitting})
varies in an isotopically-enriched sample \cite{Erice}. For
instance, by substitution of hydrogen (I=1/2) with deuterium (I=1),
it was possible to clarify that the characteristic hyperfine
splitting of the ESR signals of the H(I) (=Si$^{\bullet}$\,--\,H)
and H(II) (=Ge$^{\bullet}$\,--\,H) centers in \sil{} is a
consequence of the interaction of the unpaired electron with a
hydrogen nucleus, thereby demonstrating in a straightforward way the
presence of this impurity in the chemical structure of the defect
\cite{RadtzigKK79,RadtzigPSS86,TsaiJNCS87,Vitko}. Moreover, it can
be demonstrated that in the case of \emph{non-isotropic} hyperfine
interaction, the parameters of the $\underline{A}$ tensor convey
detailed information on the wave function of the electron and its
symmetry properties, which can be extracted on the basis of simple
models. For instance, by this method it was demonstrated that the
unpaired electron of the $E'$ center resides in a sp$^3$ orbital of
a 3-fold coordinated Si atom, thus giving valuable information on
the structure of the defect \cite{Erice}.

It is worth noting that also taking into account anisotropy,
site-to-site inhomogeneity, and hyperfine interactions, it remains
valid the property that the above defined parameter $\Gamma'$
(doubly-integrated intensity of the ESR spectrum far from
saturation, divided by $B_x$ and $A_m$) is \emph{proportional to the
number $N$ of paramagnetic centers}. In this sense, if a reference
sample is available, ESR may be used to provide a measurement of the
absolute concentration $\rho=N/V$ of every paramagnetic defect,
where $V$ is the volume of the sample.

\part{Experiments and Results}
\chapter{Materials and experimental setups}\label{mm2}
This chapter is concerned with the description of the materials,
instruments, and setups, used to perform the experiments discussed
in the rest of the work.
\section{Silica samples}\label{samples}\hspace{0.8cm}
As discussed in Part I of this Thesis, the effects induced by
irradiation on \sil{} are often found to be strongly dependent on
the manufacturing procedure of the material. In fact, the history of
the sample determines the nature and the concentration of
pre-existing defects, which may get involved in the processes
triggered by radiation. For this reason, in this section we provide
information on the samples used in this work.

The specimens involved in our experiments were all of commercial
origin, produced by \emph{Heraeus QuarzGlas} and \emph{Quartz} \&
\emph{Silice}. An important advantage of commercial samples is that
their properties are highly reproducible, since they are
manufactured by standardized industrial techniques. Even though
commercial silica specimens have been investigated for a long time,
in the following chapters we are going to see that significant new
information can still be extracted by them by careful experimental
work. Commercial samples are usually classified in four categories
following the scheme described
below.\cite{HetheringtonPCG65,BruknerJNCS70}

\noindent\textbf{Natural dry (Type I):} Samples produced from
natural quartz powder, which is melted by an electric arc in an
inert atmosphere and then cooled down to get the amorphous material.
These materials feature relatively low (or the order of $\sim$10
Parts Per Million [ppm] in weight) concentrations of Si\,--\,OH
impurities, which is the reason for the name "dry". Usually they
contain also significant concentrations ($>$1\,ppm) of extrinsic
impurities, mainly metallic, already present in the starting
material (quartz). Ge, Al, and alkali are the most common.

\noindent\textbf{Natural wet (Type II):} Similar to natural dry, but
the melting is performed by a H$_2$/O$_2$ flame. Due to the
composition of the flame, a higher concentration of Si\,--\,OH
groups ($\sim$100ppm) is incorporated in these materials than in dry
silicas. Also here extrinsic impurities are usually present in
significant concentrations.

\noindent\textbf{Synthetic wet (Type III):} Produced by oxidation of
suitable compounds of Si (typically SiCl$_4$) in a H$_2$/O$_2$
flame. So-produced silica contains a high concentration of
Si\,--\,OH due to the hydrogen present in the flame. Synthetic
\sil{} contains lower concentrations of extrinsic elements with
respect to natural samples due to the higher purity of the starting
compounds.

\noindent\textbf{Synthetic dry (Type IV):} Produced by oxidation of
SiCl$_4$ in a water-free-plasma so as to obtain a low concentration
of Si\,--\,OH. For what concerns extrinsic elements, these materials
usually retain relatively high concentrations of chlorine
([Cl]$\sim$100\,ppm).

In our experiments we used representatives of each of the four types
of \sil. In Table \ref{materials} are reported the commercial
nicknames of the used samples and the content of some impurities
\cite{Heraeus,QeS,AgnelloPHD}.
\begin{table}[h!]
\begin{center}
\begin{small}
\begin{tabular}{|c|c||c|c|c|c|}
    \hline
Name & Nickname & Type & TMI[ppm] & [Si\,--\,OH][cm$^{-3}$] & [Ge][cm$^{-3}$] \\
\hline \hline
Infrasil 301 & I301 & I & $\sim$20& $6.2\times$10$^{17}$ & $<$0.8$\times$10$^{16}$  \\
Silica EQ 906 & Q906 & I & $\sim$25 &  $1.6\times$10$^{18}$ & $(1.4\pm0.3)\times$10$^{16}$ \\
Herasil 1 & HER1 & II & $\sim$20& $1.2\times$10$^{19}$ & $(1.6\pm0.3)\times$10$^{16}$  \\
Suprasil 1 & S1 & III & $<$1& 7.8$\times$10$^{19}$& $<$10$^{15}$  \\
Suprasil 300 & S300 & IV &$<$1 &$<$8$\times$10$^{16}$& $<$10$^{15}$  \\
\hline
\end{tabular}
\end{small}
\end{center}
\caption{Materials used in our experiments, representative of the
four categories (defined above) of commercial \sil{}. Ppm stands for
Parts Per Million (in weight) while TMI stands for Total Metallic
Impurities. Silica Q906 is produced by \emph{Quartz and Silice},
while the other 4 materials are produced by \emph{Heraeus} }
 \label{materials}\end{table}

 The reported total concentrations of both metallic impurities
and Si\,--\,OH are nominal values provided by the manufacturer
\cite{Heraeus,QeS}, but the latter were subsequently verified by IR
spectroscopy \cite{AgnelloPHD}. The concentrations of Ge were
estimated in previous works\cite{AgnelloPHD} by using the
\emph{neutron activation} technique: the samples are bombarded with
fast neutrons in a nuclear reactor, and Ge impurities become
unstable isotopes by trapping neutrons\footnote{In particular,
$^{75}$Ge and $^{77}$Ge are formed by neutron trapping on the
naturally occurring isotopes $^{74}$Ge and $^{76}$Ge, respectively.
\cite{AgnelloPHD} }; then, the presence and the concentration of Ge
is inferred by observing the emission of $\gamma$ radiation at a
characteristic energy that is peculiar of the nuclear decay of a Ge
isotope.

The samples were received in slabs of sizes
5$\times$50$\times$1\,mm$^3$ (or 5$\times$50$\times$2\,mm$^3$),
optically polished only on the largest surfaces, of sizes
5$\times$50\,mm$^2$, apart from HER1 samples, which were optically
polished on all surfaces. Prior to any experiment, they were cut in
pieces of size 5$\times$5$\times$1\,mm$^3$ or
5$\times$5$\times$2\,mm$^3$. All the specimens presented no ESR
signals before irradiation, as checked by preliminary measurements.
Actually, for reasons that will be clear in the following, most of
the results presented in this Thesis were obtained from experiments
on \emph{natural} silica samples. These materials will be also
equivalently indicated with the common expression \emph{fused
silica}.
\section{Irradiations and \emph{in situ} optical measurements}
\begin{figure}[p!]
\begin{center}
%\makebox[\textwidth]{\framebox[5cm]{\rule{0pt}{5cm}}}
\includegraphics[width=1.3\textwidth,angle=90]{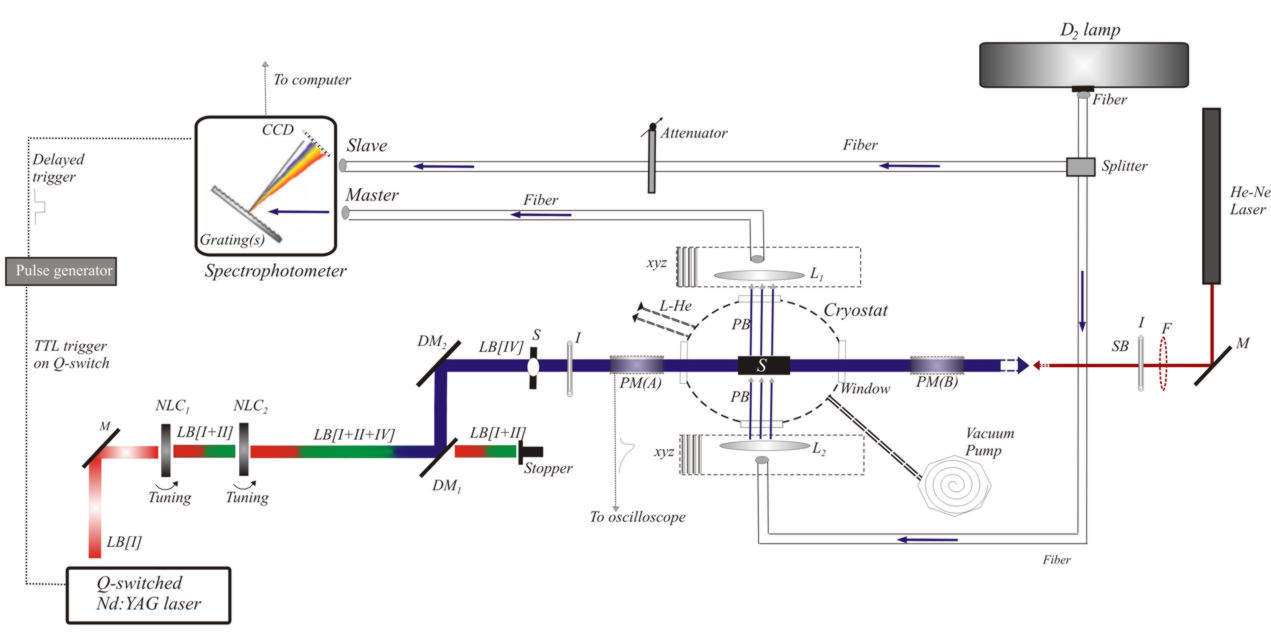}
\end{center}
\caption{Scheme of the experimental station used for Nd:YAG laser
irradiations and \emph{in situ} optical absorption measurements.
M=Mirror, I=Iris, F=Filter, S=Shutter, NLC = nonlinear crystal, DM =
dichroic mirror, LB=laser beam, PM = power meter, S=Sample, PB =
probe beam, L= Lens, xyz = micrometric translation stage}
\label{SchemaAvantes}\end{figure} \hspace{0.8cm}To investigate the
effects of laser irradiations on \sil{}, contextually to this work
it was set up an \emph{ad hoc} experimental system suitable to
perform \emph{in situ} optical absorption (OA) measurements during
laser irradiations. A scheme of the experimental station is reported
in \figurename~\ref{SchemaAvantes}. The main components are a Nd:YAG
laser system, an optical fiber spectrophotometer and a cryostat,
plus a variety of accessory elements, all mounted on a standard
optical table.

\subsection{The Nd:YAG laser
system} \hspace{0.8cm}The Q-switched Nd:YAG laser (Quanta System
SYL-201) emits pulses of 5\,ns duration and wavelength
$\lambda_0$=1064\,nm with a 1-20\,Hz repetition rate. The active
medium is a Nd:YAG rod 10\,mm long and 7\,mm in radius, pumped by a
pulsed Xe discharge lamp. To achieve Q-switching, the laser cavity
is engineered in such a way that the quality factor Q remains very
low for most of the time, and is increased only when a strong
electric pulse is applied to an electro-optic crystal (Pockel cell,
PC) within the cavity, which acts like a switch. When the PC is not
polarized, Q is low and a population inversion is obtained by
pumping, without any laser oscillations. Then, when the PC is
activated, the sudden increase of the Q factor causes the onset of
laser action with a strong initial inversion, resulting in the build
up of an intense pulse that in a few ns dissipates all the energy
stored in the active medium. More details on the principle of
Q-switching used to obtain short laser pulses can be found in many
bibliographic references.\cite{Haken,Saleh,Costa}.

A nonlinear KD$^*$P (KH$_2$PO$_4$) birefringent crystal (NLC$_1$ in
\figurename~\ref{SchemaAvantes}) is used to perform frequency
up-conversion, generating a second harmonic signal at
$\lambda_0/2$=532\,nm. A further BBO ($\beta$-BaB$_2$O$_4$) crystal
(NLC$_2$) is used to generate the fourth harmonic at
$\lambda_0/4$=266\,nm, corresponding to a photon energy of 4.7\,eV.
The nonlinear conversion process critically depends on the relative
orientation of the polarization axis of the incident beam and the
NLC$_1$ and NLC$_2$ axes. The maximum efficiency is obtained when a
condition known as \emph{phase matching} is verified, which assures
that the phase velocities of the frequency-doubled and the
fundamental waves are the same within the crystal \cite{Harper}. For
this reason, the laser includes a system that allows to finely
rotate (\emph{tune}) separately each of the nonlinear crystals in
order to maximize the intensity either of the II or of the IV
harmonic beams. The maximum laser energies per pulse are 600\,mJ,
280\,mJ and 65\,mJ, for the I, II, IV harmonic beams respectively.
The pulse energy can be controlled by varying the voltage applied on
the pumping lamp.

In all the experiments reported in this work, only the IV harmonic
UV laser beam was used to irradiate the silica samples. Indeed,
preliminary data had shown that the IR I-harmonic and the visible II
harmonic have negligible effects on the as-grown samples. Since the
output of the laser comprises the I, II and the IV harmonic, a pair
of dichroic mirrors (DM$_1$ and DM$_2$) is used to reflect
selectively the IV harmonic (LB[IV]) towards the sample position.
The diameter $d_b$ of LB[IV] can be regulated by an iris (I) from
(1.0$\pm$0.1)\,mm up to (6.0$\pm$1.0)\,mm, and an electronic shutter
(S) is present on the beam path, permitting to start and end the
irradiation session from remote. The beam is unfocused and has an
uniform intensity profile, as inferred by verifying that the pulse
energy measured after the iris is proportional to $d_b^2$
\cite{Costa}.

The irradiation beam enters the cryostat and hits the sample (S) on
one of its minor surfaces (5\,mm$\times$1\,mm). A secondary
continuous He:Ne laser is positioned in such a way that it shares a
portion of the optical path of the primary beam; its beam (SB) is
used to check the alignment of the sample with the 4.7\,eV laser
beam.

The energy of the laser pulses is measured either with a bolometer
or with a pyroelectric detector connected to an oscilloscope, with
consistent results. The bolometer measures the mean incident power
of the Nd:YAG IV harmonic beam, based on the heating produced upon
absorption of the laser light. The pyroelectric detector responds to
each laser shot giving in output a short ($\mu$s) electric pulse,
whose amplitude is proportional to the energy of the shot; in
particular, this detector allows to estimate pulse-to-pulse
intensity fluctuations. Each of the two power meters can be
positioned either before the sample (PM(A)), to measure the laser
intensity before an irradiation begins, or after the sample (PM(B)),
to check the stability of laser intensity during long irradiations.
The accuracy of pulse energy measurements with the pyroelectric
detector is $\sim$5\%.

\subsection{The optical fiber spectrophotometer}
\hspace{0.8cm}Optical absorption spectra are performed with an
optical fiber AVANTES S2000 spectrophotometer. The optical source is
a deuterium lamp that injects light into an optical fiber that
splits up in two channels, referred to as \emph{Master} and
\emph{Slave}, each 2\,m long from source to detector. The optical
fibers are multimode pure silica core/F$_2$-doped silica cladding
with diameter of 200\,$\mu$m. They are loaded with H$_2$ to better
resist to the prolonged exposure to UV light without being
deteriorated.

The light carried by the \emph{master} channel gets out of the fiber
and is used as the probe beam (PB). The PB is collimated by a lens
(L$_1$) so as to have a $\sim$3\,mm diameter; then, it passes trough
the cryostat windows and arrives on the larger surface of the sample
(5$\times$5\,mm$^{2}$). The typical intensity of the PB is
$\sim$2$\mu$W. The transmitted portion of the PB is collected from a
second lens (L$_2$) coupled to another fiber that brings it to the
detector. The two lenses are mounted on independent micrometric
positioning controls (xyz), which permit both to control the
alignment of the PB to the sample and to optimize the collection
efficiency after the sample. The \emph{slave} channel passes through
a variable attenuator, after which it goes to the detector. Since
the slave channel does not traverse the sample, it can be used to
correct experimental data for the temporal drift of the lamp when
monitoring on the master channel the kinetics of laser-induced OA.

The detector consists in a 1200\,lines/mm grating with blaze at
300\,nm, dispersing on a 2048 channels Charge Coupled Device (CCD)
array. For the two channels are used two different gratings,
virtually identical, but a single CCD detector. The latter is coated
with a fluorescent compound ("lumogen" coating) to enhance its
sensitivity in the UV. The instrument works in the 200\,nm--500\,nm
range with a spectral resolution of 5\,nm. It is worth noting that
such a relatively low resolution is usually sufficient to study the
absorption bands of point defects in \sil{}, usually very broadened
by both homogeneous and inhomogeneous effects.\cite{Erice} The
detector features a minimum integration time of 3\,ms. Including
also the time required to transmit to the acquisition system, the
instrument takes about 20\,ms to perform a complete measurement of
the intensity profile $I(\lambda)$ of the light carried by each
channel.

We describe now the \textbf{experimental procedure} used to perform
time dependent absorbtion measurements \emph{in situ} during and
after the end of a laser irradiation. First, the D$_2$ lamp is
disconnected from the fibers and it is acquired a \emph{dark}
reference signal on both channels: $D^M(\lambda)$ (master) and
$D^S(\lambda)$ (slave). Then, the lamp is connected again; at a time
$t_0$ just before the irradiation begins, a reference signal is
acquired for both channels, $I^M(\lambda,t_0)$ (master) and
$I^S(\lambda,t_0)$ (slave). Finally, the Nd:YAG laser is turned on
(or the shutter S is open) and the irradiation begins. Using an
electronic signal provided by the laser concurrently to each
Q-switching, a delayed pulse is produced that triggers the
spectrophotometer so as to perform measurements only during the time
span separating one laser pulse and the successive one
(\emph{interpulse}). In this way, it is avoided that scattered laser
light may be detected by the CCD altering the measurements. Defining
$I^M(\lambda,t)$ (master) and $I^S(\lambda,t)$ the signals acquired
at time $t$ by the detector, the \emph{difference absorbtion induced
in the sample by the irradiation} after $t$ is calculated as:
\begin{equation}
\Delta OD(t)=\log_{10}{\left(
\frac{I^M(\lambda,t_0)-D^M(\lambda)}{I^M(\lambda,t)-D^M(\lambda)}
\right)}-\log_{10}{\left(\frac{I^S(\lambda,t_0)-D^S(\lambda)}{I^S(\lambda,t)-D^S(\lambda)}\right)}\label{avantesabs}
\end{equation}this equation is valid in the approximation in which the dark signal
does not depend on time, which results to be a very good one. In
eq.~(\ref{avantesabs}), the first term already represents the
difference absorption induced in the sample between $t_0$ and $t$.
The subtraction of the second term, formally identical apart from
being estimated from the slave channel, allows to correct for the
temporal drift of the lamp and/or the detector; the correction is
necessary since this is a \emph{single beam} system in which the
acquisition of the reference is performed only once, at the time
$t_0$. It is worth noting that the validity of this procedure is
necessarily limited by the fact that the drift of the two channels
are not, in general, perfectly identical. For this reason, the
stability of the \emph{corrected} $\Delta OD$ calculated by
eq.~(\ref{avantesabs}) was checked by a test experiment: a
not-irradiated sample was monitored for many hours while keeping the
laser off; in this way, it was obtained that $\Delta
OD(t)<5\times$10$^{-4}$OD/hour.

The system can be used also to perform standard \emph{ex situ}
absolute absorption measurements, by the following procedure. Two
signals are acquired without ($I_0(\lambda)$) and with
($I(\lambda)$) the sample in the probe beam; hence, the absorption
profile of the specimen is given by
\begin{equation}
OD=\log_{10}{\left(\frac{I_0(\lambda)-D(\lambda)}{I(\lambda)-D(\lambda)}\right)}
\end{equation}
Finally, an additional spectrophotometer (JASCO-V560) was used to
perform \emph{ex situ} optical absorption measurements on the
irradiated samples after removal from the irradiation site. The
instrument is a traditional double beam spectrophotometer based upon
a D$_2$ discharge lamp source, a photomultiplier tube detector, and
a double monochromator (two gratings with 1200\,lines/mm) on the
excitation side, which allows to reduce stray light to 0.0003\%.
Measures with this instruments were performed with a spectral
resolution of 2\,nm. The optical absorption profiles obtained with
the JASCO spectrophotometer are consistent with those detected with
the optical fiber instrument.

\subsection{The cryostat} \hspace{0.8cm}To perform
temperature-controlled experiments, the samples are placed in a
continuous liquid Helium flow cryostat (Optistat CF-V) produced by
Oxford Instruments working between 4\,K and 500\,K. The cryostat
mounts four synthetic silica windows, reasonably transparent to the
UV laser and the probe beams. A two stage rotary/turbomolecular pump
(Leybold Vacuum PT 50) is used to achieve high vacuum in the
cryostat by overnight pumping (base pressure $\sim$10$^{-6}$\,mbar).
Helium is picked up from a storage dewar by a standard transfer tube
and delivered within the cryostat just above the sample holder.
Thermal equilibrium within 1\,K at the working temperature is
achieved by a Oxford-ITC503 instrument, which controls the He flow
and the electric current input on a resistor, positioned near the
sample holder as well, and acting as a heating element. The
experiments started after a delay of $\sim$1hour after reaching each
nominal operative temperature, to allow for thermal equilibrium. The
cryostat is equipped with an aluminum radiation shield that grants a
better thermal isolation of the sample. In addition, in this context
the use of the radiation shield is important also to reduce
undesired condensation of the residual gases present in the vacuum
chamber on the sample surfaces. Indeed, insufficiently clean
surfaces may compromise the results of the experiments because
laser-induced removal of the condensed film can result in fake
negative absorption signals during \emph{in situ} OA measurements
under Nd:YAG irradiation.
\section{Photoluminescence measurements} \subsection{The instrument}
\begin{figure}[htb!]
\begin{center}
%\makebox[\textwidth]{\framebox[5cm]{\rule{0pt}{5cm}}}
\includegraphics[width=.9\textwidth]{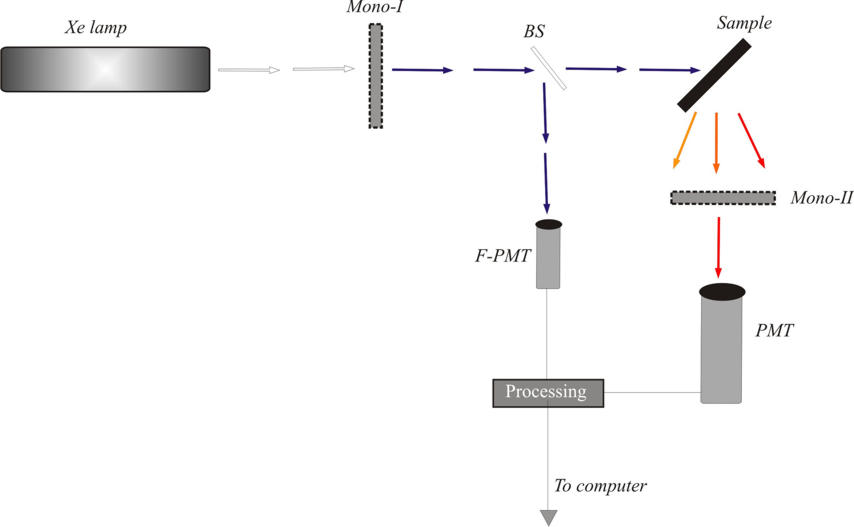}
\end{center}
\caption{Idealized scheme of the spectrofluorometer JASCO-FP 6500
}\label{schemaPL}\end{figure}

\hspace{0.8cm}Stationary PL measurements were carried out by a JASCO
FP-6500 spectrofluorometer, whose scheme is reported in
\figurename~\ref{schemaPL}. The light emitted by the excitation
source (a 150W Xenon discharge lamp) is dispersed by a monochromator
(MONO-I), based on a grating with 1800\,lines/mm, which permits to
select the excitation wavelength $\lambda_{0}$ with variable
bandwidth $\Delta\lambda_{0}$. The monochromatized excitation light
is then directed on the sample, which is positioned in a standard
45$^{\circ}$ backscattering geometrical configuration. The emitted
luminescence signal is collected by a second monochromator (MONO-II,
1800\,lines/mm) which selects the emission bandwidth $\lambda_{e}$
with variable bandwidth $\Delta\lambda_{e}$. Finally, the emitted
photons are detected by a phomultiplier (PMT) giving an output
signal $S_R(\lambda_{0},\lambda_{e})$. The instrument includes a
feedback system that corrects the measurements for the temporal
fluctuations of the source intensity: to this end, a beam splitter
(BM) separates a portion of the excitation light from the main beam
and directs it on a secondary PMT (F-PMT), which measures its
intensity; hence, the signal detected by the primary PMT is
automatically rescaled by the so-obtained reference signal.

Two basic types of measurements are possible: \textbf{(i)} the
\emph{emission} spectrum, in which $S_R(\lambda_{0},\lambda_{e})$ is
measured as a function of $\lambda_{e}$ for fixed $\lambda_{0}$.
This type of acquisition aims to measure the spectral shape and
intensity of the band emitted by the center while decaying from the
upper electronic state (see \figurename~\ref{twolevelscheme}).
\textbf{(ii)} The \emph{excitation} spectrum, in which
$S_R(\lambda_{0},\lambda_{e})$ is measured as a function of
$\lambda_{0}$ for fixed $\lambda_{e}$, representing a measurement of
the efficiency of the emission process in dependence of the
excitation wavelength.

 \subsection{Correction procedures}
\hspace{0.8cm}Luminescence measurements require specific correction
procedures before they can be related to physically meaningful
quantities. Let us start with considering a thin sample of width
$dx$ containing a concentration $\rho$ of identical\footnote{in the
following treatment we are going to suppose a population of
identical homogeneous defects, thereby neglecting inhomogeneity
effects, common in amorphous silica
\cite{CannasPHD,CannizzoPHD,LeonePRB99,AgnelloPRB03b}. However, this
approximation does not alter the main conclusions of this
subsection.} luminescent centers. We express as
$\Omega=\Delta\lambda_{0}\times [dI_{0}(\lambda_{0})/d\lambda_{0}]$
the excitation intensity, meaning the number of photons from the
source illuminating the sample per unit time. If
$\sigma(\lambda_{0})$ is the absorption cross section of the
defects, the number of absorbed photons per unit time is given by
$\rho\Omega\sigma(\lambda_{0})dx$. Therefore, recalling the
definition of the luminescence quantum yield (section \ref{oppd})
$\eta$, the number of photons emitted per unit time by the sample
with wavelength between $\lambda_{e}$ and
$\lambda_{e}+\Delta\lambda_{e}$ (i.e. the spectral density of the
emitted light) is given by:
\begin{equation}
\frac{dI_{e}}{d\lambda_{e}}(\lambda_0,\lambda_e)\Delta\lambda_{e}=\rho\Omega\Delta\lambda_{e}\eta\sigma(\lambda_{0})\Phi(\lambda_{e})dx
\label{thinemission}
\end{equation}
where $\Phi(\lambda_{e})$ represents the normalized emission
lineshape function, which for simplicity we have supposed to be
independent of $\lambda_{0}$.

For a sample of finite thickness $d$, we substitute $\Omega$ in the
last equation by $\Omega=\Omega_0\exp{[-\alpha(\lambda_0)x]}$,
namely the dependence of the excitation intensity on the position
$x$ through the sample. Here, $\alpha(\lambda_0)$ is the
\emph{overall} absorption coefficient, to which in general may
contribute also other centers. Then, we integrate on $x$. In this
way we obtain the generalization of eq.~(\ref{thinemission}):
\begin{equation}
\frac{dI_{e}}{d\lambda_{e}}(\lambda_0,\lambda_e)\Delta\lambda_{e}=\Omega_0\Delta\lambda_{e}\eta\left[1-\exp{(-\alpha(\lambda_{0})
d)}\right]\frac{\sigma(\lambda_0)\rho}{\alpha(\lambda_0)}\Phi(\lambda_{e})\label{emitted}
\end{equation}

Now, the measured \emph{raw} signal $S_R(\lambda_{0},\lambda_{e})$
is proportional to eq.~(\ref{emitted}) multiplied by the spectral
response $R_{d}(\lambda_{e})$ of the detecting system (MONO-II +
PMT). Furthermore, the action of the feedback system is to
substitute $\Omega_0$ (which could be time-dependent due to
fluctuations in the lamp) with a term $R_{f}(\lambda_{0})$
expressing the spectral response of the BS plus the F-PMT.
Eventually we get:
\begin{equation}
S_R(\lambda_{0},\lambda_{e})\propto\Delta\lambda_{e}\eta\left[1-\exp{(-\alpha(\lambda_{0})
d)}\right]\frac{\sigma(\lambda_0)\rho}{\alpha(\lambda_0)}\Phi(\lambda_{e})\frac{R_{d}(\lambda_{e})}{R_{f}(\lambda_{0})}\label{raw}
\end{equation}

Therefore, to obtain from the raw signal a meaningful physical
quantity, it is necessary to divide for $R_{d}(\lambda_{e})$ and
multiply for $R_{f}(\lambda_{0})$.\footnote{If one is interested
only in the \emph{lineshape}, it is sufficient to divide for
$R_{d}(\lambda_{e})$ the raw emission spectra and to multiply for
$R_{f}(\lambda_{0})$ the raw excitation spectra} These two
corrections \emph{were performed for all the luminescence data
reported in this Thesis}, based upon the measurement of $R_d$ and
$R_f$ performed as follows: $R_{f}(\lambda_{0})$ can be estimated
(apart from an arbitrary factor) by performing an excitation
spectrum on a sample whose luminescent centers feature high
absorption ($\sigma(\lambda_{0})\rho d\gg1$) and a quantum yield
independent of $\lambda_0$. In fact, in this case the term within
squared parentheses in (\ref{raw}) may be approximated to unity;
moreover, if no other absorbing centers are present in the sample,
$\sigma(\lambda_0)\rho/\alpha(\lambda_0)=1$. In these conditions,
eq.~(\ref{raw}) becomes (retaining only the terms that depend on
$\lambda_{0}$):
\begin{equation}
S(\lambda_{0})\propto\frac{1}{R_{f}(\lambda_{0})}
\end{equation}thereby allowing to estimate $R_f$ from an excitation
spectrum performed in these conditions. For this purpose it was used
a sample of Rhodamine B (in glycerol), which emits at 640\,nm and
can be excited from 220\,nm to 600\,nm with a constant (near to 1)
quantum yield \cite{Luminescence}. To measure $R_{d}(\lambda_{e})$,
a mirror was put in the instrument, redirecting the excitation light
to the detecting system. Then, an acquisition (\emph{synchronous}
spectrum) was performed while keeping $\lambda_{0}=\lambda_{e}$. If
we assume that the reflection efficiency of the mirror is
independent of $\lambda$, it is easy to see that the signal measured
in this conditions can be expressed by:
\begin{equation}
S(\lambda)\propto\frac{R_{d}(\lambda)}{R_{f}(\lambda)}
\end{equation}from which it can be derived $R_{d}(\lambda)$ if
$R_{f}(\lambda)$ has already been measured.

Using these two procedures to eliminate $R_d$ and $R_f$ from
eq.~(\ref{raw}), the corrected signal $S_C(\lambda_0,\lambda_e)$ is
finally given by:
\begin{equation}
S_C(\lambda_0,\lambda_{e})=y\Delta\lambda_e\eta\left[1-\exp{(-\alpha(\lambda_{0})
d)}\right]\frac{\sigma(\lambda_0)\rho}{\alpha(\lambda_0)}\Phi(\lambda_{e})\label{correctedlambda}
\end{equation}where the proportionality constant $y$ accounts for
geometrical factors that result in a partial collection of the light
emitted by the sample.

When the absorbance of the sample at the excitation wavelength is
low ($\alpha(\lambda_{0})d\ll1$), the last equation can be
approximated as follows:
\begin{equation}
S_C(\lambda_{0},\lambda_{e})=y\Delta\lambda_e\eta\sigma(\lambda_{0})
\rho d\Phi(\lambda_{e})\label{lowabsorptionpl}
\end{equation}As apparent from eq.~(\ref{lowabsorptionpl}), in conditions of low absorbance the
intensity of the corrected signal \emph{is proportional to the
concentration} $\rho$ of the luminescent centers. It is worth noting
that this is untrue in the general case (\ref{correctedlambda})
because $\rho$ indirectly contributes also to $\alpha(\lambda_0)$
appearing in the exponential term.

Finally, let us focus now on an emission measurement, where
$\lambda_0$ is kept fixed and $\lambda_e$ is scanned. As the only
term in eq.~(\ref{lowabsorptionpl}) depending from $\lambda_e$ is
$\Phi(\lambda_e)$, we see that after the correction procedure, the
\emph{shape} of the corrected emission spectrum resembles the
spectral density of the emitted light. Furthermore, comparing with
eq.~(\ref{correctedlambda}) we observe that this property in itself
does not depend on the low absorbance hypothesis.\footnote{Also, the
corrected excitation spectrum reproduces the shape of the absorption
spectrum; this property, however, is true \emph{only} in the case of
low absorbance.} The precision of the measurement of $S_C$ is
$\sim$10\%, being mainly limited by the repeatability of the
mounting conditions. From the physical point of view, it is usually
preferable to report the spectral density of the emission bands as a
function of photon energy $E_e=hc/\lambda_e$, instead of using the
wavelength $\lambda_e$. To this purpose, the following
transformation must be applied:
\begin{equation}
\frac{dI_{e}}{dE_{e}}=\frac{dI_{e}}{d\lambda_{e}}\left|\frac{d\lambda_{e}}{dE_{e}}
\right|\propto\lambda^2\frac{dI_{e}}{d\lambda_{e}}
\label{passarelambdaE}
\end{equation}
Due to (\ref{passarelambdaE}), all the emission spectra reported in
this Thesis (after scaling for $R_d(\lambda_e)$) were further
corrected by multiplication for $\lambda_e^2$, before changing the
independent variable to $E_e$. In literature this is usually called
correction for the \emph{dispersion} of the detecting system.
\subsection{Time-resolved photoluminescence measurements}
\hspace{0.8cm}The luminescence decay measurements reported in this
work were performed at the I-beamline of SUPERLUMI station at DESY,
Hamburg, under excitation by 130\,ps synchrotron radiation pulses.
The excitation pulses were monochromatized with a bandwidth of
0.3\,nm. The emitted light was acquired by a photomultiplier
(Hamamatsu R2059) with a time resolution of 0.02\,ns and an emission
bandwidth of 5\,nm. Temperature was varied from 8\,K to 300\,K by a
Helium-based continuous flow cryostat.
\section{Electron spin resonance measurements} \subsection{The instrument} 
\begin{figure}[htb!]
\begin{center}
\includegraphics[width=.9\textwidth]{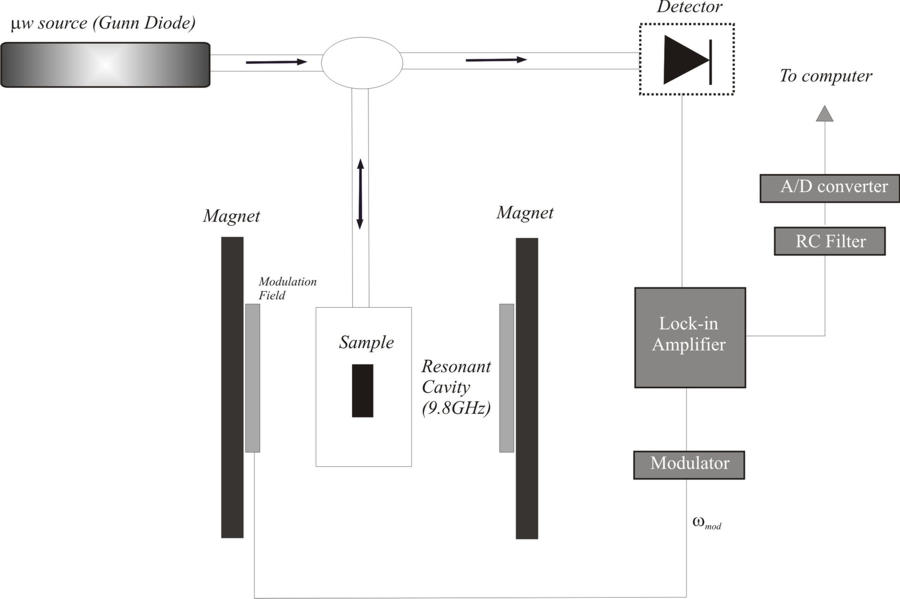}
\end{center}
\caption{Idealized scheme of the Bruker EMX electron spin resonance
spectrometer including only the most important elements.}
\label{ESRscheme}\end{figure}

\hspace{0.8cm} The ESR measurements reported in this
Thesis were performed by a Bruker EMX spectrometer working at
$\omega_0$=9.8\,GHz. In \figurename~\ref{ESRscheme} it is reported a
simplified scheme of the instrument. The sample is positioned in a
resonant cavity, fed by a waveguide transporting microwaves produced
by a Gunn Diode source. A variable attenuator permits to regulate
the actual power $P_i$ incident to the cavity from a maximum value
200\,mW down to 200\,nW; In this way, $P_i$ is usually chosen so as
to avoid the saturation of the observed magnetic resonance
transition.\footnote{In the following, we are going to report the
ESR spectra of $E'$ and H(II) centers (see chapter \ref{backI}).
Previous studies have demonstrated that the maximum not-saturating
powers for these two defects are 3\,mW and 8$\times$10$^{-4}$\,mW
respectively \cite{AgnelloPHD}. }
 The microwaves arriving on the entrance of the cavity are
partially reflected and partially transmitted. The reflected power
$P_R$ is measured by a detector that gives a current signal $I$
proportional to the square root of $P_R$. Indicating with $B_R$ the
magnetic field amplitude of the reflected microwaves, we have:
$P_R\propto B_R^2$, so that $I\propto B_R$. The cavity is positioned
within the polar expansions of an electromagnet, which permits to
generate a static magnetic field $B_z$ up to 10$^4$\,G. The magnetic
field $B_i$ of the incident microwave field is perpendicular to
$B_z$.

 In the typical measurement scheme, a
static magnetic field interval of width $\Delta B_z^{sw}=B_z^{max} -
B_z^{min}$ is swept in a time $T_{sw}$. At some value of the
magnetic field the onset of the magnetic resonance condition causes
the absorption of microwaves by the sample. It can be demonstrated
that this leads to a variation $\Delta P_R\propto\Delta B_R^2$ of
the reflected power, with $\Delta B_R\propto B_i\chi''$. Hence, we
have $\Delta I\propto\Delta B_R\propto B_i \chi''$, so that the
variation of the output detector current allows to measure the
quantity $\chi''B_i$. We take into account now the effects of
modulation: to the static magnetic field is superimposed a
modulation field with frequency $\omega_m$ and amplitude $A_m$. As a
consequence of the modulation of $B_z$, $\chi''(B_z)$ is modulated
as well. A necessary condition to avoid distortion in the measured
signal is that $A_{m}$ is chosen to be significantly smaller than
the resonance linewidth. In these conditions, the following relation
holds:
\begin{eqnarray}
 \Delta I\propto B_i\chi''(B_z) & = & B_i\chi''[B_z^0+A_m\sin(\omega_m
t)] \nonumber \\
  & \sim & B_i\chi''(B_z^0)+B_i\frac{d\chi''}{dB_z}(B_z^0)A_msin(\omega_m
t)\label{modulation}
\end{eqnarray}Therefore, the detector signal $I$ has a component oscillating
with frequency $\omega_m$ that is selectively detected by the
\emph{lock-in} acquisition system (see chapter \ref{mm1}), thus
allowing for an increasing sensitivity if compared with detection of
the non-modulated portion. Therefore, selecting in
(\ref{modulation}) only the component at $\omega_m$, we conclude
that the revealed signal is proportional to
\begin{equation}
 \frac{d\chi''}{dB_z}B_iA_m
\end{equation} which is equivalent to eq.~(\ref{esrmeasured}) because the
oscillating field in the cavity, $B_x$, is proportional to the
incident amplitude $B_i$. Finally, to reduce the signal to noise
ratio, the last stage of the instrument comprises a RC filter, whose
integration time $\tau_{RC}$ can be regulated by the user, and must
verify the following relation
\begin{equation}
\tau_{RC}\ll T_{sw} \frac{\Delta B_{pp}}{\Delta B_z^{sw}}
\end{equation} where $\Delta B_{pp}$ is the width of the resonance line.
This condition prevents from filtering out part of the signal
together with the noise, thereby avoiding distortion of the measured
lineshape.

\subsection{Absolute concentration measurements} \hspace{0.8cm}\label{acm}The
precision of the \emph{relative} concentration measurement $\Gamma'$
(defined in section \ref{mrpd}) in our experimental system, and in
the typical acquisition conditions used throughout this work, is
$\sim$10\%, including both the effects of noise and of repeatability
of the mounting conditions\footnote{The 10\% estimate refers to an
ESR signal acquired with a reasonable signal/noise ratio, as it is
the case for all the spectra analyzed in this work. When the signal
is very noisy due to a very low concentration of defects, the
precision can significantly drop below this value.}. Then, as
anticipated in chapter \ref{mm1}, to convert this result to an
\emph{absolute} concentration measurement it is needed a reference
sample. To this purpose, in this work it was used a specimen where
the absolute number of $E'$ centers (purposely generated by $\gamma$
irradiation) was known by spin-echo measurements
\cite{AgnelloPHD,Echo}.

The expression \emph{ spin-echo} indicates the characteristic
response of the transverse (perpendicular to $\overrightarrow{B}$)
magnetization ($\overrightarrow{M_\bot}$) of a spin system submerged
in a static magnetic field $\overrightarrow{B}$, to a sequence of
two appropriately engineered microwave \emph{pulses} separated by an
interpulse time $\Delta$t. The phenomenon consists in a partial
recovery after the second pulse of the coherence between the spins,
which had been lost due to the inhomogeneous spreading of their
resonance frequencies causing the decay of
$\overrightarrow{M_\bot}$; as a consequence of the recovery, after a
time $\Delta$t from the second pulse it is observed a temporary
increase (echo) of $\overrightarrow{M_\bot}$
\cite{Slichter,AgnelloPHD,Echo}. By an appropriate experimental
setup suitable to perform \emph{transient} ESR
spectroscopy,\cite{AgnelloPHD,Echo} it is possible to measure the
intensity of the echo signal as a function of $\Delta$t, so as to
study in particular its typical decay dynamics at low temperatures.
A contribution to the decay arises from the spin-spin magnetic
dipole interaction, whose strength is proportional to $d^{-3}$,
where $d$ is the mean spin-spin distance. Hence, the measurement of
the decay time of the spin-echo signal permits to find $d$ and the
mean concentration $<\rho>=d^{-3}$.\cite{AgnelloPHD,Echo}

The accuracy of the absolute concentration measurements obtained by
ESR, based on comparison with the spin-echo reference sample, is
estimated as 20\%. However, we stress that this error never
explicitly appears when reporting in the following the uncertainties
on the concentration measurements; indeed, the stated errors
represent only the $\sim$10\% relative precision of the estimates.
The reason for this choice is that the 20\% uncertainty coming from
spin-echo plays the role of a \emph{systematic} error that affects
in the same way all the concentration measurements reported
throughout the Thesis.

\section{Raman measurements}
The Raman effect is the anelastic scattering of light (by a molecule
or a point defect) due to emission or absorption of a vibrational
quantum \cite{Raman,Ottica}. If $E_i$ is the energy of the incident
photons, scattering at $E_s<E_i$ implies the excitation of a
vibrational mode of energy $\hbar\omega=E_i-E_s$. A Raman spectrum
consists of a plot of the scattered intensity as a function of
$\omega$. This spectroscopy allows to probe the vibrational modes of
a molecule or a point defect, sometimes bearing some advantages with
respect to common IR spectroscopy. For example, it can happen that a
vibrational mode is Raman-active but non IR-active, or \emph{vice
versa} \cite{Raman,Ottica}. In this Thesis we are going to report
Raman measurements aimed to detect the Si\,--\,H vibration mode at
2250\,cm$^{-1}$, which is difficult to see in IR due to the overlap
with an intrinsic vibration of the \sil{} matrix, but is more easily
detectable by Raman spectroscopy\cite{Shelby94,Schmidt98}.
Measurements were performed by the group led by Prof. Y. Ouerdane at
the TSI laboratory - Universit\'{e} Jean Monnet - Saint-Etienne
(France), using an Argon laser source ($\lambda$=488\,nm) with
P=1--2\,W intensity and a photomultiplier detector.

\chapter{Effects induced on silica by 4.7eV laser radiation}\label{overview}
This is the first of the five chapters in which the report and the
discussion of our experimental results is organized. Here we
describe and discuss qualitatively the basic defect-related
phenomenology induced by 4.7\,eV pulsed laser irradiation on
amorphous silica, as observed by the combined use of several
spectroscopic techniques.

\section{Introduction}\hspace{0.8cm}As discussed in detail in the introductory
chapters, laser irradiation of amorphous silica triggers a complex
landscape of processes manifesting themselves in alterations of the
macroscopic properties of the material, and often related to
laser-induced generation and transformation processes of point
defects. In particular, many studies have investigated the
generation of the dangling silicon bond ($E'$) and of the dangling
oxygen bond (NBOHC), namely two of the basic intrinsic point defects
in \sil{}, while other works have focused on the effects of laser
radiation on the defects related to Germanium impurities. A last
class of processes of relevant scientific and technological interest
is that of point defect conversions driven by diffusion of mobile
chemical species; in fact, these effects may strongly condition the
response of the material to irradiation, and some of their features
are a fingerprint of the disordered structure of the amorphous
solid. Despite the great amount of work devoted to understand this
wide class of processes, the current understanding of many important
aspects is mainly qualitative, and several relevant questions remain
unanswered.

Starting from this chapter, we present the results of a series of
experiments investigating the effects of pulsed UV laser irradiation
on amorphous silica. In particular, we are going to show that in a
subclass of silica materials, fused silica, 4.7\,eV laser photons
triggers an articulated landscape of point defect conversion
processes, which involve both intrinsic and extrinsic point defects,
and whose features permit to use this material as a model system to
investigate some important aspects of laser-induced and
diffusion-driven effects in \sil{}. In this
chapter\cite{MioJNCS03,MioJNCS04,MioPSS05,MioJNCS05EPRIMO,MioJNCS06EPRIMO},
we basically present the phenomenology observed in our experiments,
which is then discussed on a qualitative basis in order to introduce
the main problems that we are going to thoroughly investigate in the
rest of the work.

\section{Optical properties of the as-grown samples}
\hspace{0.8cm}The experiments described throughout this Thesis were
carried out on 4 types of silica samples (Table \ref{materials}).
The purpose of this section is to introduce the optical properties
of the as-grown materials.
\begin{figure}[htb!]
\begin{center}
\includegraphics[width=.8\textwidth]{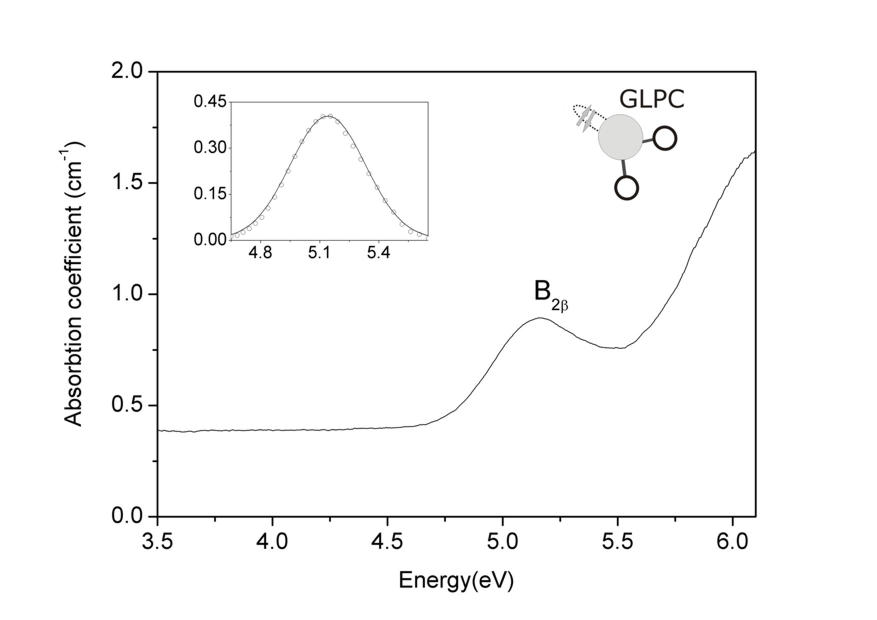}
\end{center}
\caption{Absorption spectrum of an as-grown natural dry I301 silica
sample. The main detected signal is the B$_{2\beta}$ band associated
to the twofold coordinated Ge center, =Ge$^{\bullet\bullet}$ (GLPC),
represented in the upper right corner. After proper baseline
subtraction (the background is representable as the tail of a band
peaked at E$>$6\,eV), the B$_{2\beta}$ band is well reproduced by a
Gaussian shape (inset)} \label{i301preabs}\end{figure}

\begin{figure}[htb!]
\begin{center}
\includegraphics[width=.8\textwidth]{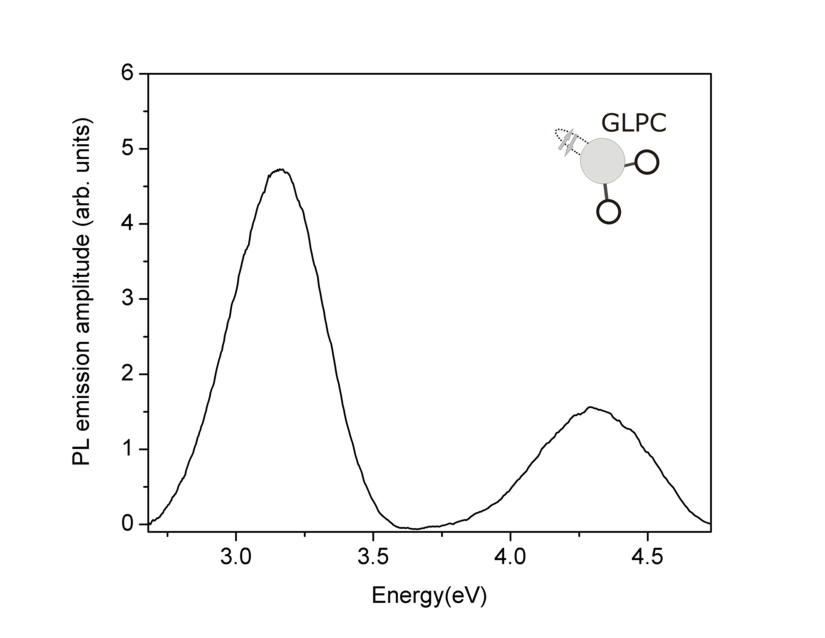}
\end{center}
\caption{PL emission spectrum detected in an as-grown natural silica
sample under excitation at 5.0\,eV with 3\,nm excitation and 3\,nm
emission bandwidths. The signal is due to the GLPC center
represented in the upper right corner } \label{PLH1}\end{figure}

A typical absorption spectrum of a \emph{natural dry} \sil{} sample
in the UV range prior to any treatment, is reported in
\figurename~\ref{i301preabs}. The main detected signal is a band
(B$_{2\beta}$ band\cite{Tohmon}) peaked at (5.13$\pm$0.02)\,eV with
(0.45$\pm$0.03)\,eV Full Width at Half Maximum (FWHM). The area of
the band is (0.19$\pm$0.02)\,cm$^{-1}$eV both for the I301 and Q906
natural dry samples. An analogous absorption profile is detected in
as grown \emph{natural wet} HER1 samples, where the intensity of the
B$_{2\beta}$ band is (0.13$\pm$0.02)\,cm$^{-1}$eV.

By exciting the B$_{2\beta}$ absorption, we detect a
photoluminescence (PL) emission signal consisting in a band peaked
at (3.14$\pm$0.02)\,eV with FWHM=(0.42$\pm$0.02)\,eV and a band
peaked at (4.28$\pm$0.02)\,eV with FWHM=(0.46$\pm$0.02)\,eV
(\figurename~\ref{PLH1}). In addition, the excitation spectrum (not
reported) of both PL signals closely resembles the 5.1\,eV
absorption band in \figurename~\ref{i301preabs}
\cite{CannizzoPHD,CannasPHD,Nalwa,AgnelloPHD}.

In previous studies it was found a linear correlation between the
intensities of the three bands (absorption and the two emissions),
valid in a large number of commercial natural silica
materials\cite{LeonePRB99}. This evidence strongly suggests the
overall optical activity to be due to a single defect. Then, from
comparison of the spectroscopic features of the bands with
literature, this optical activity can be ascribed to the Germanium
Lone Pair Center (GLPC), consisting in a two-fold coordinated Ge
impurity (=Ge$^{\bullet\bullet}$)
\cite{Skuja84,SkujaJNCS92,Nalwa,CannasPHD,CannizzoPHD}. The presence
of this defect in the as-grown material is consistent with the
independent observation of Ge impurities in natural silica by
neutron activation measurements (see section \ref{samples} and
references therein).\footnote{To avoid confusion, it is worth
clarifying a subtle difference in the notation we are using here
with respect to that of subsection \ref{Gedefsection}, where
Ge-related defects were introduced. Indeed, we had originally used
the symbol "B$_{2\beta}$" to indicate generically the $\sim$5.1\,eV
OA commonly detected in as-grown Ge-doped silica materials; in
general, other defects besides GLPC (e.g. the Ge-related neutral
oxygen vacancy, NOV) may contribute to this absorption. Now, since
the as-grown optical activity of natural silica materials can be
completely ascribed to GLPC, here and in the following we are going
to use the symbol B$_{2\beta}$ to indicate more specifically only
the absorption band at 5.13\,eV of the GLPC center. }
\begin{figure}[htb!]
\begin{center}
\includegraphics[width=.6\textwidth]{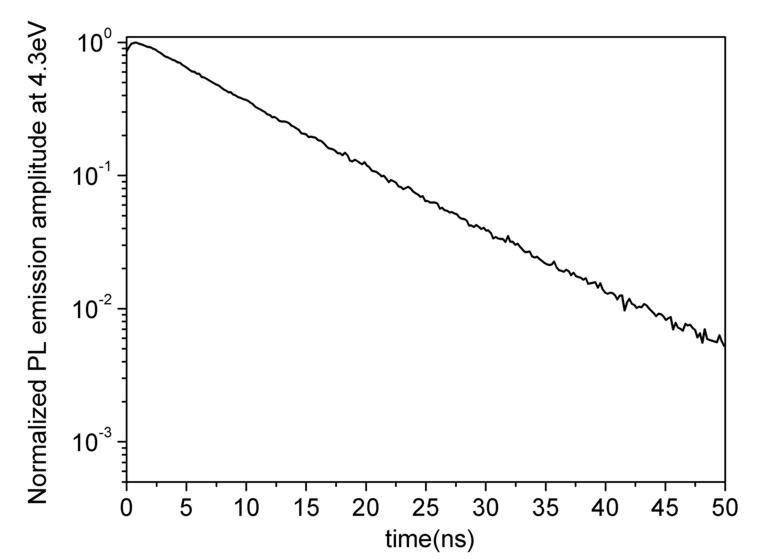}
\end{center}
\caption{Decay curve of PL emission at 4.3\,eV (maximum of the
emission band) measured at 10\,K after excitation with a synchrotron
radiation pulse monochromatized at 5.00\,eV (near the maximum of the
excitation band).} \label{decayDESY}\end{figure}

As already discussed in the introduction (see
\figurename~\ref{schemab}), the 4.3\,eV and 3.1\,eV PL emissions are
associated to the decay from the excited singlet (S$_1$) and triplet
(T$_1$) electronic states to the ground singlet (S$_0$) state
respectively. The luminescence activity of GLPC is further
characterized by its radiative emission lifetime $\tau$=7.8\,ns from
S$_1$; $\tau$ is estimated by measuring at T=10\,K the emission
decay curve (\figurename~\ref{decayDESY}) after excitation with
pulsed synchrotron radiation at 5.00\,eV. It is worth noting that
performing this measurement at low temperature is mandatory so as to
quench the non-radiative decay channel from S$_1$, which at higher
temperatures alters the observed lifetime \cite{AgnelloPRB03b}.

Differently from natural silica, the native absorption profile of
synthetic dry and wet \sil{} samples in the same spectral region
does not show any measurable absorption band. This is not
surprising, given the lower concentration of impurities typical of
synthetic \sil{} with respect to natural silica (chapter \ref{mm2}).

\section{\emph{In situ} observation of the generation and decay of $E'$ center}
\hspace{0.8cm}One of the main techniques employed in this work to
investigate the effects of laser irradiation on \sil{} is \emph{in
situ} optical absorption spectroscopy, carried out by the
experimental apparatus described in detail in chapter \ref{mm2}.
\begin{figure}[h!]
\begin{center}
\includegraphics[width=.6\textwidth]{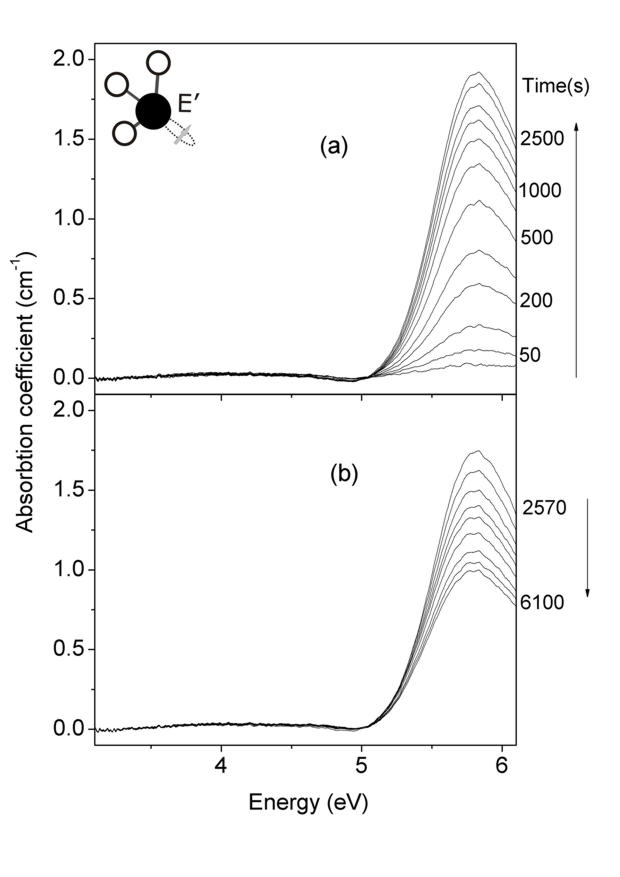}
\end{center}
\caption{Induced OA measured \emph{in situ} at different times
during (a) and after (b) an irradiation session with 2520 laser
pulses (4.7\,eV photon energy, 5\,ns pulsewidth, 1\,Hz repetition
rate, 40\,mJcm$^{-2}$ energy density per pulse) of a natural dry
I301 silica sample. The observed 5.8\,eV absorption band is due to
the $E'$ center (represented in the upper left corner of panel (a))
induced in the sample by irradiation.}
\label{spettrinsitui301}\end{figure} In a representative experiment,
an as-grown sample is irradiated at room temperature by 4.7\,eV
pulsed (5\,ns pulsewidth) radiation from a frequency-quadruplied
Nd:YAG laser, using a 1\,Hz repetition rate and a 40\,mJcm$^{-2}$
energy density per pulse. The total duration of the irradiation
session is 2520\,s. During and for a few $\sim$10$^3$\,s after
irradiation, the absorption profile induced in the UV spectral range
is monitored by the optical fiber spectrophotometer. In particular,
during the irradiation session 10-20 spectra are collected and
averaged during each interpulse time interval. The spectra are
corrected for the temporal drift of the lamp using the second
reference channel. Hence, we calculate the difference spectra with
respect to the native absorption profile of
\figurename~\ref{i301preabs}, at different times during and after
the end of the exposure session: the results are reported in
\figurename~\ref{spettrinsitui301}.

The main detected signal is the band centered at 5.81$\pm$0.02\,eV
with 0.71$\pm$0.03\,eV FWHM which, as widely known, is associated to
one of the fundamental defects in \sil{}: the silicon dangling bond,
known as $E'$ center\cite{SkujaSPIE01,Nalwa,Erice,Devinebook}
($\equiv$Si$^{\bullet}$, see chapter \ref{backI}). The peak
amplitude of the induced band grows up to (1.90$\pm$0.02)cm$^{-1}$
during the irradiation session (panel (a)); then, as soon as the
laser is switched off, we observe that the signal begins
spontaneously to decrease with time (panel (b)), its reduction being
$\sim$40\% in the first 3600\,s. Many works in literature have
discussed the generation of $E'$ under laser irradiation, suggesting
a variety of possible mechanisms, discussed in detail in subsection
\ref{geneprimosection}. Elucidating the specific process that is
active in our case is one of the main topics of the next chapter.
For the moment, we proceed to discuss the main features of the
process, as apparent from experimental observations.

\begin{figure}[p!]
\begin{center}
\includegraphics[width=.7\textwidth]{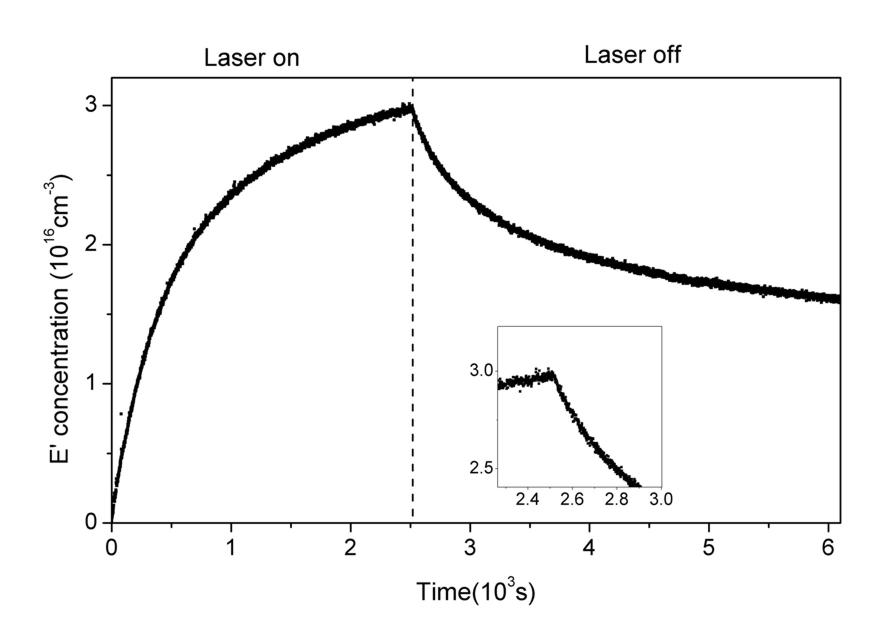}
\end{center}
\caption{Typical kinetics of [$E'$] in a laser-irradiated I301
natural dry \sil{} sample, as calculated from the induced absorption
profile measured \emph{in situ}
(\figurename~\ref{spettrinsitui301}). Inset: zoom at the end of the
irradiation session.} \label{kini301}\end{figure}
\begin{figure}[p!]
\begin{center}
\includegraphics[width=.7\textwidth]{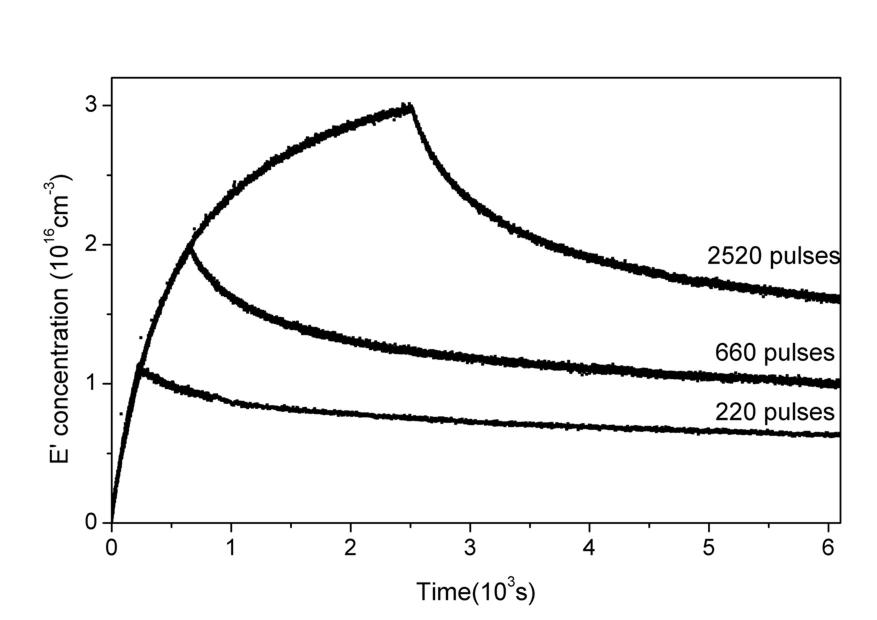}
\end{center}
\caption{Three kinetics of [$E'$] measured during and after
irradiation of I301 natural dry silica with different numbers of
laser pulses. } \label{severali301}\end{figure}

Within experimental error, both the growth and the decay of the
signal take place without changes in shape. Hence, from the peak
amplitude $\alpha$(5.8\,eV) of the band and the known peak
absorption cross section
$\sigma$(5.8\,eV)=6.4$\times$10$^{-17}$cm$^{2}$
\cite{Nalwa,CannasJNCS04}, we can estimate the concentration of the
defects: [$E'$]=$\alpha\sigma^{-1}$, which is plotted in
\figurename~\ref{kini301} as a function of time. In this particular
case, during irradiation [$E'$] grows up to
2.9$\times$10$^{16}$\,cm$^{-3}$.
\begin{figure}[h!]
\begin{center}
\includegraphics[width=.7\textwidth]{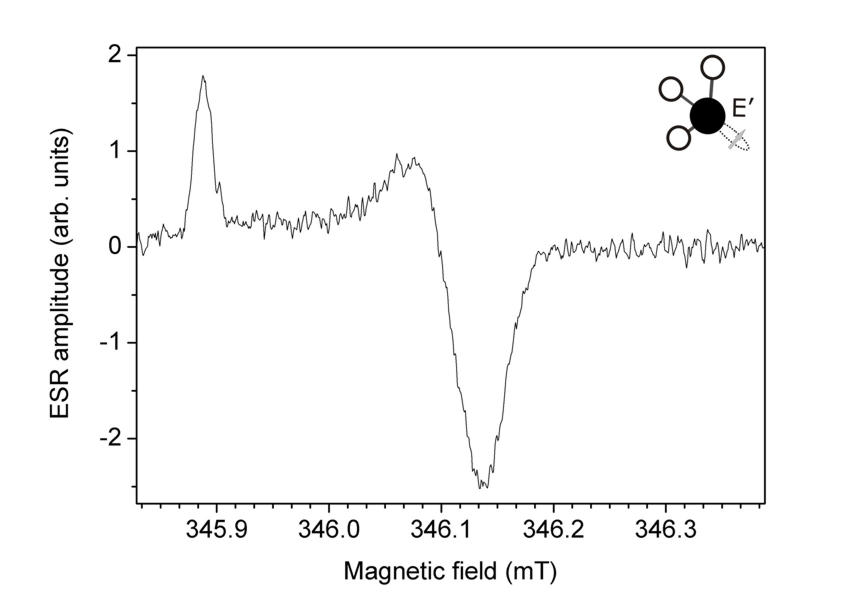}
\end{center}
\caption{ESR signal of $E'$ center (represented in the upper right
corner), as detected on the sample of \figurename~\ref{kini301} a
few days after the end of exposure. The signal was detected with a
0.01mT modulation amplitude and with a
not-saturating\cite{AgnelloPHD} 8$\times$10$^{-4}$\,mW microwave
power.} \label{centroEepr}\end{figure} As soon as the laser is
switched off (t=2520\,s), we observe a dramatic change of slope
characterizing the beginning of $E'$ decay, as evidenced in the
inset. While the growth stage of the curve can be obviously measured
\emph{in situ} once for all for a given type of \sil{} and laser
intensity, the post-irradiation decay stage of the kinetics depends
on the number of pulses after which the irradiation session is
interrupted. This is shown in \figurename~\ref{severali301}, where
we compare three kinetics of [$E'$] induced by irradiation with
different total numbers of pulses. From the same Figure, the degree
of overlap of the growth stages of the three curves evidences the
high repeatability of the \emph{in situ} OA measurement.

\begin{figure}[h!]
\begin{center}
\includegraphics[width=.7\textwidth]{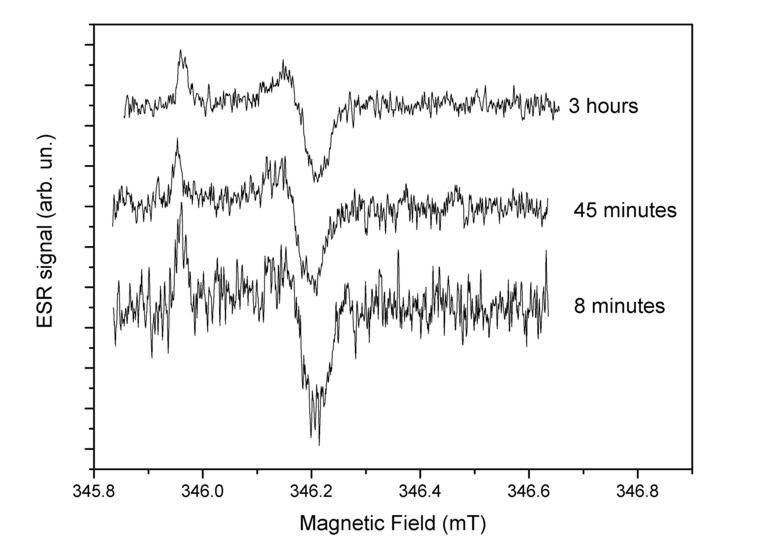}
\end{center}
\caption{ESR signal of $E'$ center detected in a Q906 sample at
different delays after the end of laser irradiation. The three
spectra have been normalized with respect to the acquisition
parameters so that their intensities are proportional to the
concentrations of $E'$; after normalization, the curves have been
vertically shifted to avoid overlap. The signal is much noisier than
\figurename~\ref{centroEepr} because of the lower (but progressively
increasing) integration time used in these measurements, as
mandatory to follow the time dependence of the signal.}
\label{pikesr}\end{figure}

The presence of $E'$ in the irradiated sample is also independently
confirmed by \emph{ex situ} ESR measurements carried out starting
from $\sim$5$\times$10$^2$\,s after the end of exposure, which show
the typical signal of the paramagnetic center. An example of the
characteristic ESR lineshape of $E'$ acquired in optimal conditions
in a (natural dry) laser-irradiated specimen is reported in
\figurename~\ref{centroEepr}. Also, by measurements at different
delays from the end of exposure, the intensity of the ESR signal is
found to decay with time (\figurename~\ref{pikesr}), consistently
with the results coming from optical measurements.

\begin{figure}[h!]
\begin{center}
\includegraphics[width=.9\textwidth]{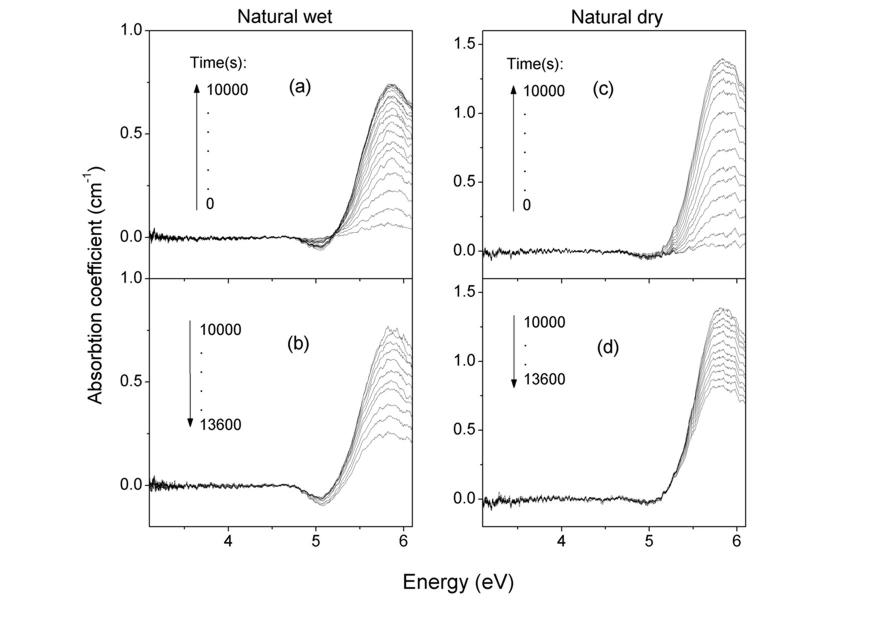}
\end{center}
\caption{Induced OA measured \emph{in situ} during and after a 10000
pulses irradiation with 40\,mJcm$^{-2}$ energy density per pulse and
1\,Hz repetition rate in wet (HER1) and dry (Q906) natural silica}
\label{cindryewet}\end{figure}
\begin{figure}[h!]
\begin{center}
\includegraphics[width=.75\textwidth]{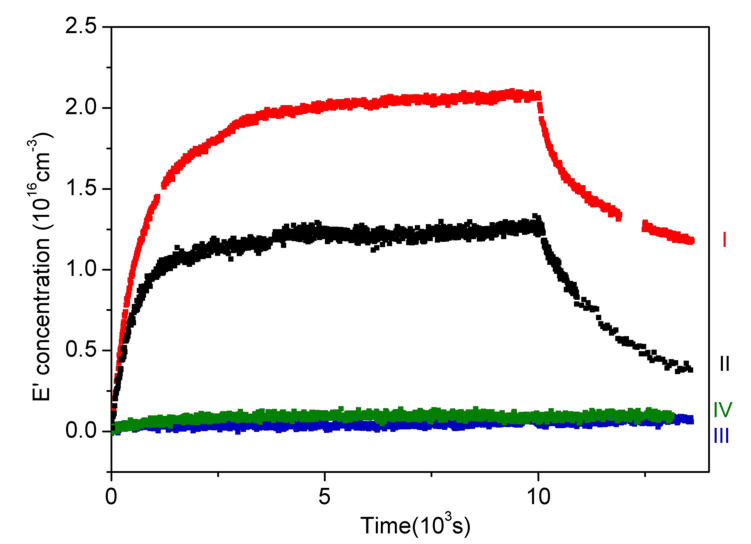}
\end{center}
\caption{Kinetics of $E'$ concentration measured \emph{in situ}
during and for 1 hour after the end of a 10000 pulses laser
irradiation with 40\,mJcm$^{-2}$ energy density per pulse and 1\,Hz
repetition rate in the four varieties of \sil{}: natural dry (I,
red), natural wet (II, black), synthetic wet (III, blue) and
synthetic dry (IV, green).} \label{cinmater}\end{figure}

From \figurename~\ref{kini301} it is apparent that the decay of $E'$
is still in progress after 1 hour from the end of irradiation. While
\emph{in situ} measurement are the only way to investigate the
\emph{growth} stage of the kinetics as well as the first
$\sim$10$^3$\,s of the decay, they are not suitable to follow the
post-irradiation kinetics for more than $\sim$10$^{4}$\,s, because
they are performed with a single-beam system (see chapter
\ref{mm2}). For this reason, the most reliable estimate of the
asymptotic stationary value [$E'$]$_\infty$ of [$E'$] at the end of
the decay process, comes from \emph{ex situ} ESR or OA measurements
performed for a few days after irradiation. The two techniques yield
consistent results, but ESR allows for a higher precision. In this
way, from the intensity of the ESR signal of
\figurename~\ref{centroEepr}, it was determined that
[$E'$]$_\infty$=(1.0$\pm$0.1)$\times$10$^{16}$\,cm$^{-3}$ for the
kinetics in \figurename~\ref{kini301}. This asymptotic value is
reached within a few $\sim$10$^5$\,s from irradiation, after which
the concentration of $E'$ was observed to remain stationary within
experimental error on a timescale of (at least) several
months.\footnote{For the meaning of the uncertainty affecting
[$E'$]$_{\infty}$, and any other concentration estimate obtained by
ESR, please refer to subsection \ref{acm}.
}$^,$\footnote{\label{fnsyst}The absorption cross section of $E'$,
$\sigma$, was estimated from the slope of the linear correlation
between $\alpha$(5.8\,eV) and the concentration measured by
ESR.\cite{CannasJNCS04,Nalwa} Since the latter depends on the value
of [$E'$] in the spin-echo reference sample (see subsection
\ref{acm}), $\sigma$ is affected by the same 20\% error featured by
any concentration measurement derived by ESR throughout this work.
On the other hand, due to the systematic nature of this uncertainty,
it does not affect the agreement between the concentrations
calculated from ESR and OA data, which are consistent within the
repeatability of the respective measurements.}

Similar growth and decay kinetics of the $E'$ absorption band are
observed in all the irradiated natural \sil{} materials. In
\figurename~\ref{cindryewet} we compare the time dependencies of the
induced absorption profile, as observed during irradiation and for
the first hour of the post-irradiation stage in a wet (panels (a)
and (b)) and dry (panels (c) and (d)) natural silica sample
irradiated with 10000 laser pulses with 40\,mJcm$^{-2}$ energy
density per pulse and 1\,Hz repetition rate.
 The kinetics of [$E'$] calculated
from these data are reported in \figurename~\ref{cinmater} (kinetics
I, dry and II, wet), and compared with the result of the same
experiment performed in synthetic dry (III) and wet (IV) silica
samples.

\begin{figure}[h!]
\begin{center}
\includegraphics[width=.7\textwidth]{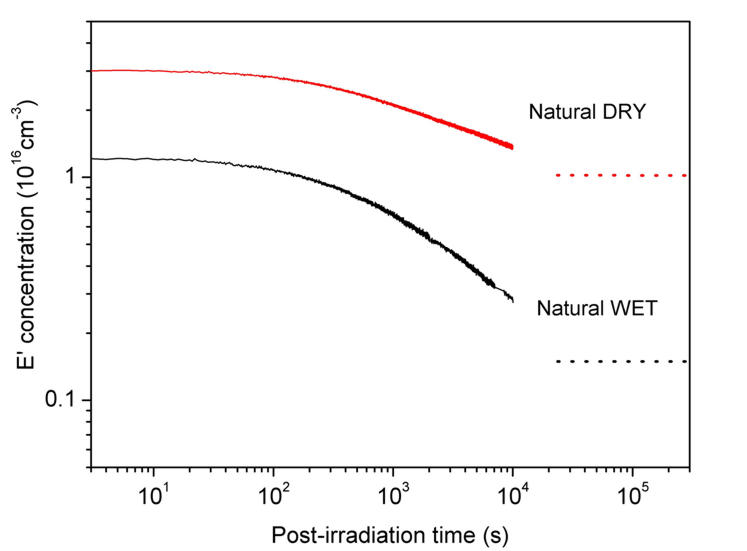}
\end{center}
\caption{Post-irradiation kinetics of [$E'$] measured \emph{in situ}
after the end of laser irradiation on dry and wet natural silica
samples. Dotted lines: asymptotic stationary values of [$E'$]
measured a few days after the end of the irradiation session by ESR.
The origin of the time scale corresponds here to the end of
exposure.} \label{compdrywetcontrollarevasintotico}\end{figure}
These data permit to discuss the different response of the four
\sil{} varieties to 4.7\,eV laser photons. First, generation of $E'$
is observed only in natural silica, whereas laser irradiation
results to be \emph{ineffective on synthetic materials}, (within
$\sim$5$\times$10$^{14}$\,cm$^{-3}$), at least at the explored laser
intensities, which are unable to induce any detectable absorption
band in these specimens. Second, both in wet and dry natural silica
the $E'$ centers grow during irradiation until they reach a
saturation value after a certain number of pulses, and decay after
the laser is switched off, but the kinetics differ in two important
aspects: \textbf{(i)} more defects are induced in the dry materials
for a given irradiation dose (values of curve I are higher than
those of II). \textbf{(ii)} The decay is more effective in wet
silica, where the reduction of [$E'$] in the first hour is already
70\%, to be compared with a 40\% reduction in the dry material.
Further info on this latter point comes from
\figurename~\ref{compdrywetcontrollarevasintotico}, where two
representative post-irradiation kinetics, as observed in dry or wet
silica, are reported in a logarithmic scale, and compared with the
asymptotic stationary concentration values obtained by ESR
measurements a few days after the end of exposure. In the plot, the
origin of the time scale has been redefined to correspond to the
\emph{end} of the irradiation session.

 We see that in the wet sample the post-irradiation decay anneals about 90\% of the
$E'$ centers that had initially been induced by laser exposure,
resulting in a stationary concentration of
$\sim$1.5$\times$10$^{15}$\,cm$^{-3}$. In contrast, in dry natural
silica the portion of annealed centers is lower (67\%) and the
concentration of stationary centers $\sim$10$^{16}$\,cm$^{-3}$ one
order of magnitude higher. These are general features of all the
laser-induced kinetics we have observed. Wet natural silica appears
to be a remarkable material for what concerns post-irradiation
effects, since the $E'$ induced by laser exposure are \emph{almost
completely transient}. In contrast, the dry material differs in that
a significant concentration of residual $E'$ are still present after
the decay is completed. Finally, we can characterize the decay of
$E'$ by a typical time scale, defined as the time necessary to
achieve half of the total decrease. From data, it can be estimated
that this time is close to 10$^3$\,s for both kinetics, although in
both cases appreciable concentration variations can be
experimentally observed for a few days after irradiation.

\section{Response of Ge-related defects to irradiation}
\hspace{0.8cm}By a closer look to the induced absorption spectra
measured \emph{in situ} (\figurename s~\ref{spettrinsitui301} and
\ref{cindryewet}) we observe that, in addition to the 5.8\,eV band,
it is detected a weak negative component near $\sim$5\,eV.

\begin{figure}[h!]
\begin{center}
\includegraphics[width=.75\textwidth]{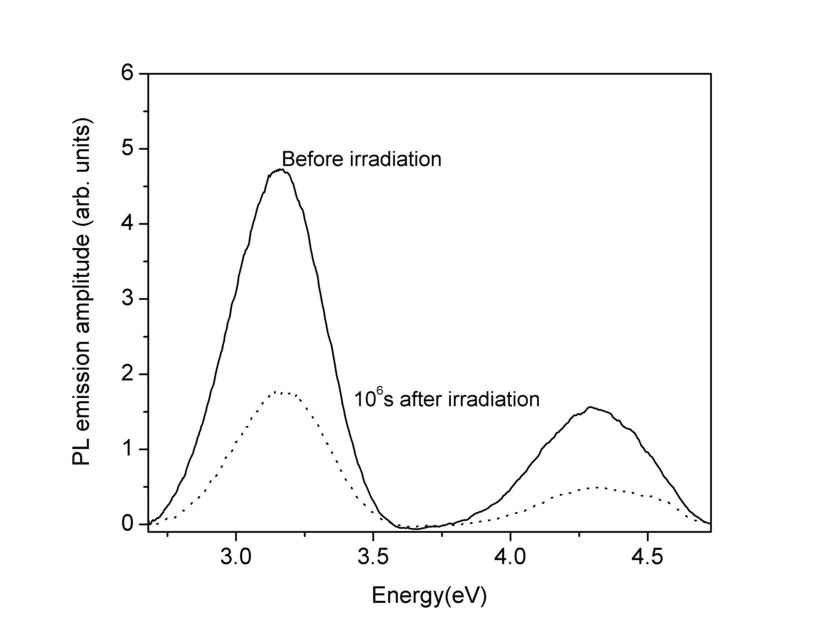}
\end{center}
\caption{Typical PL emission spectrum of GLPC center excited at
5.0\,eV in an as-grown natural silica material (continous line,
already reported in \figurename~\ref{PLH1}) and in the same sample a
few days after being exposed to 2000 laser pulses with
40\,mJcm$^{-2}$ energy density per pulse and 1\,Hz repetition rate
(dashed line).} \label{bleaching}\end{figure}
\begin{figure}[p!]
\begin{center}
\includegraphics[width=.8\textwidth]{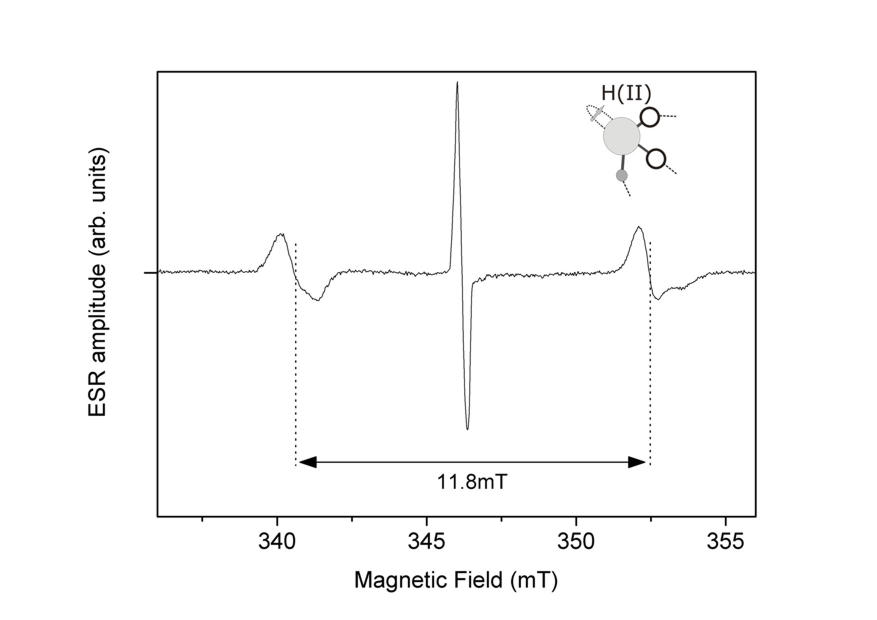}
\end{center}
\caption{Typical ESR spectrum of a laser-irradiated natural dry
silica sample, measured in the post-irradiation stage. The signal
was acquired with a 3\,mW microwave power and a 4\,G modulation
amplitude. The 11.8\,mT doublet is due to the Ge-related H(II)
center (=Ge$^{\bullet}$\,--\,H), represented in the upper right
corner. } \label{esrspettrolargo}\end{figure}
\begin{figure}[p!]
\begin{center}
\includegraphics[width=.6\textwidth]{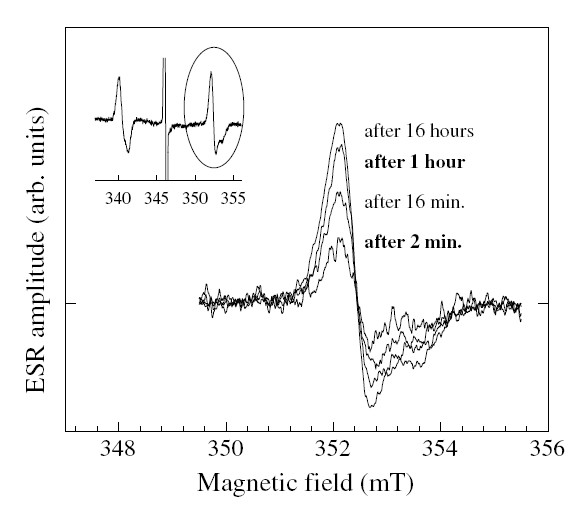}
\end{center}
\caption{High field component of the 11.8\,mT ESR doublet of the
H(II) center (as evidenced in the inset) as detected at different
delays after the end of laser exposure. The signal was acquired
using a 3\,mW non-saturating\cite{AgnelloPHD} microwave power and a
4\,G modulation field. Figure taken from Cannas \emph{et
al.}\cite{MioJNCS03} } \label{crescitadoppiettohii}\end{figure}

Due to the presence in the as-grown absorption profile
 (\figurename~\ref{i301preabs}) of the B$_{2\beta}$ band, peaked at 5.1\,eV, the
negative contribution in the difference spectra may be explained as
an intensity reduction of this signal. In turn, this finding
evidences a partial conversion of the pre-existing Ge-related GLPC
centers responsible for the band. Since this signal is partially
concealed by the much more intense 5.8\,eV component, the conversion
of GLPC is more conveniently investigated by luminescence
measurements. In fact, in \figurename~\ref{bleaching} we show the PL
signal of GLPC in an as-grown sample and in an irradiated specimen,
as detected a few days after exposure. It is apparent that laser
irradiation induces a reduction (\emph{bleaching}) of the native PL
activity, which confirms the occurrence of laser-induced conversion
processes transforming the GLPC in other defects. However, we stress
that at this stage no information is available on the
\emph{kinetics} of the bleaching process.

\begin{figure}[h!]
\begin{center}
\includegraphics[width=.8\textwidth]{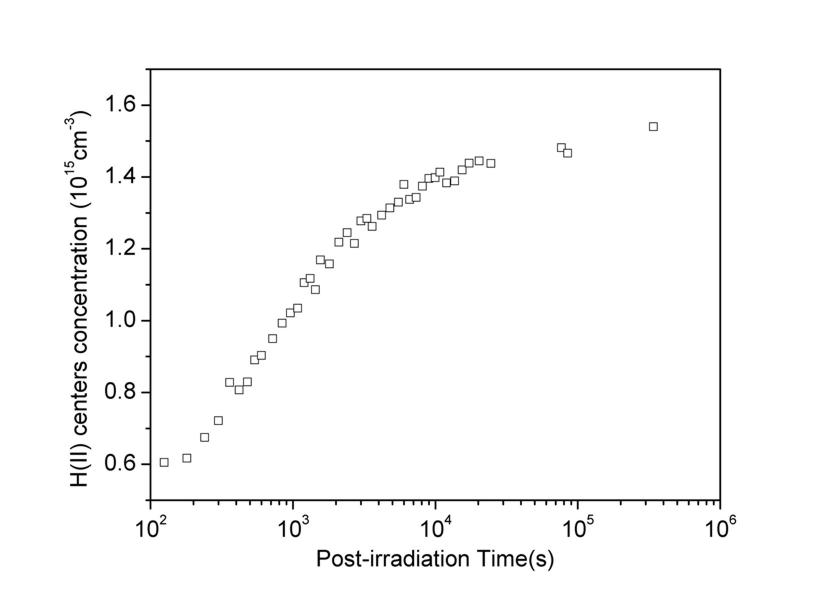}
\end{center}
\caption{Post-irradiation kinetics of H(II) centers in an irradiated
Q906 sample, after irradiation with 10000 laser pulses with
40\,mJcm$^{-2}$ energy density per pulse and 1\,Hz repetition rate,
as calculated from the time dependence of their 11.8\,mT doublet.
t=0 represents the end of the irradiation session.}
\label{pikhiiq90610000}\end{figure}
\begin{figure}[p!]
\begin{center}
\includegraphics[width=.75\textwidth]{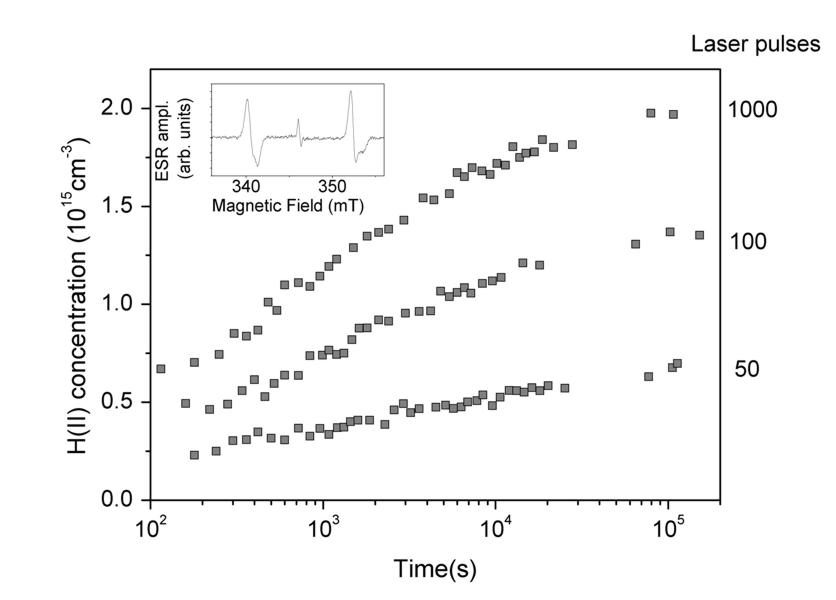}
\end{center}
\caption{Post-irradiation kinetics of H(II) centers observed in HER1
wet natural \sil{} after exposure of several as-grown samples to
different numbers of laser pulses with 40\,mJcm$^{-2}$ energy
density per pulse and 1\,Hz repetition rate. Inset: typical ESR
spectrum showing the 11.8\,mT doublet of H(II) centers.}
\label{variecinetichehii}\end{figure}
\begin{figure}[p!]
\begin{center}
\includegraphics[width=.75\textwidth]{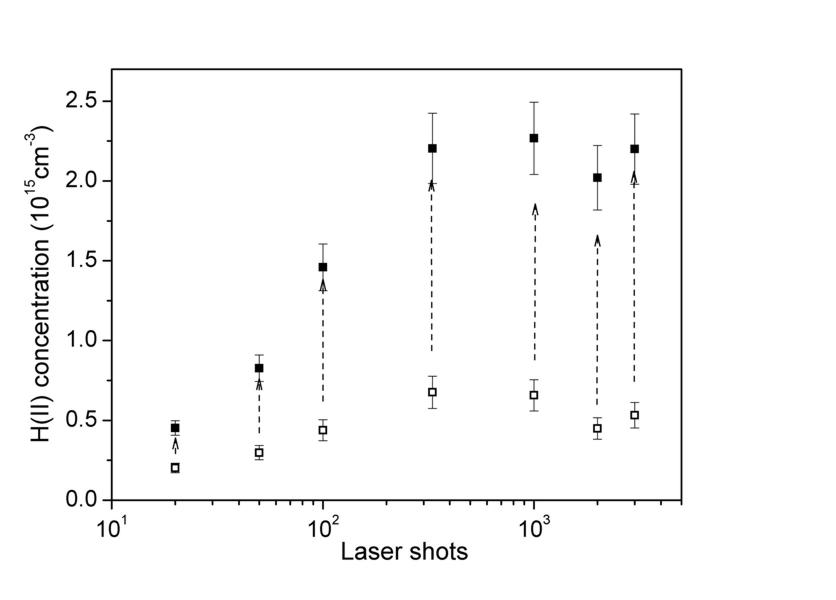}
\end{center}
\caption{Dose-dependence of the concentrations measured a few
minutes after the end of irradiation (empty symbols) and at the end
of the post-irradiation kinetics (full symbols) for HER1 natural wet
\sil{} irradiated with with 40\,mJcm$^{-2}$ energy density per pulse
and 1\,Hz repetition rate.} \label{frommintomaxhii}\end{figure}

Another process related to Germanium impurities is evidenced by ESR.
In detail, if an as-grown natural silica sample is laser-irradiated,
and we perform in the post-irradiation stage an \emph{ex situ} ESR
measurement on a wide magnetic field region, we detect a typical
signal that is reported in \figurename~\ref{esrspettrolargo}. The
strong component near 346\,mT is the $E'$ signal in
\figurename~\ref{centroEepr}, which here appears very distorted due
to the high modulation field and microwave power being used in the
acquisition\footnote{In this acquisition, the instrumental
parameters were optimized to detect the 11.8\,mT doublet, which
features a saturation power\cite{AgnelloPHD} and a linewidth much
larger than $E'$.}. In addition, the spectrum evidences a doublet
split by 11.8\,mT, which by comparison with literature can be
attributed to the Ge-related H(II)
center\cite{Vitko,RadtzigPSS86,PacchioniPRB98}.

The H(II) center consists in a Ge bonded to two oxygen atoms and one
hydrogen, and hosting an unpaired electron (=Ge$^{\bullet}$\,--\,H);
the doublet structure of its ESR signal arises from the hyperfine
interaction between the electron and the proton spins. If the ESR
measurements are repeated at different delays from the end of
exposure, two effects are observed: \textbf{(i)} the progressive
decrease of the $E'$ signal, above mentioned, consistent with the
post-irradiation decay of the 5.8\,eV band observed by \emph{in
situ} OA, and \textbf{(ii)} a progressive \emph{increase} of the
intensity of the 11.8\,mT doublet, as shown in
\figurename~\ref{crescitadoppiettohii} for one of the two components
of the signal.

This observation demonstrates a growth of the concentration of H(II)
centers taking place in the post-irradiation stage,
\emph{simultaneously to the decay of $E'$}. From the intensity of
the doublet, measured\footnote{We stress that the signal of H(II) is
monitored as a function of post-irradiation time without removing
the sample from the ESR spectrometer. This allows to measure the
kinetics of [H(II)] with a higher precision, it being not limited by
the repeatability of the mounting conditions. } at different delays
from the end of exposure, we calculate the concentration of the
defects, which is plotted in \figurename~\ref{pikhiiq90610000} as a
function of time for a natural dry sample irradiated with 10000
laser pulses. The concentration a few minutes after the end of
exposure (\emph{initial} concentration) is
$\sim$6$\times$10$^{14}$\,cm$^{-3}$, from which it grows up to
$\sim$1.5$\times$10$^{15}$\,cm$^{-3}$ (\emph{stationary}
concentration) measured after a few days. A similar kinetics is
observed in all the irradiated dry and wet natural silica specimens.
In all samples, the stationary concentration of H(II) is always
[H(II)]$<$2$\times$10$^{15}$\,cm$^{-3}$. Similarly to $E'$, also the
growth of H(II) can be characterized by a typical time scale,
defined as the time necessary to achieve half of the total
\emph{growth}. From inspection of \figurename~\ref{pikhiiq90610000}
one can easily estimate that this time is $\sim$10$^3$\,s,
comparable to the time scale of $E'$ decay.

To show the dose-dependence of the H(II) generation process, we
report in \figurename~\ref{variecinetichehii} their growth kinetics
measured after exposure to different numbers of pulses in natural
wet \sil{} samples.
 From each kinetics we extract
the initial and stationary concentrations, which are reported in
\figurename~\ref{frommintomaxhii} as a function of irradiation dose
(number of pulses). We see that the stationary concentration
increases with dose, and after $\sim$300 pulses it becomes invariant
with increasing number of pulses at
(2.1$\pm$0.2)$\times$10$^{15}$cm$^{-3}$. The initial concentrations
show a similar dose-dependence but are always 3-4 times smaller than
stationary concentrations, independently of dose. The results of
\figurename~\ref{frommintomaxhii} are representative of all natural
silica samples investigated.

Finally, no ESR signals are detected in irradiated synthetic \sil{}
specimens, consistently with the absence of any induced absorption
signal in \emph{in situ} measurements. In particular, in regard to
the H(II) doublet, it is worth noting that its absence is a
necessary consequence of the lack of Ge impurities in synthetic
\sil{}, differently from the case of natural specimens.

\section{Discussion}\label{modelpiktutti}
\hspace{0.8cm}We have shown that one of the main features of natural
\sil{} exposed to pulsed 4.7\,eV laser irradiation is that the
induced $E'$ centers are unstable and decay in the post-irradiation
stage. As reviewed in chapter \ref{backII}, in literature
post-irradiation effects have often been attributed to the diffusion
in \sil{} of mobile species able to react with point defects causing
their conversion in other centers. Among the many species whose
diffusion in \sil{} was evidenced experimentally, only hydrogen
(aside from chemically inert noble gases, see Table \ref{dc}) is
able to readily diffuse at room temperature, where the experiments
presented up to this point were carried out. Moreover, at room
temperature hydrogen in \sil{} is found only in \emph{molecular}
H$_2$ form, because H diffuses so fast and is so reactive that it
exists only as a transient species, which rapidly recombines with
reactive defects or dimerizes forming H$_2$. On this basis, it is
natural to hypothesize that the post-irradiation decay of $E'$
($\equiv$Si$^{\bullet}$) is due to reaction with H$_2$:
\begin{equation} \label{reacteh2reprep}
\equiv\text{Si}^{\bullet}\text{ + H$_2$ } \longrightarrow \text{
}\equiv\text{Si\,--\,H + H}\end{equation}H produced at the right
side of the reaction rapidly dimerizes again in H$_2$ or passivates
another $E'$:
\begin{equation}\label{reacteh0rep} \equiv\text{Si}^{\bullet}\text{
+ H } \longrightarrow \text{ }\equiv\text{Si\,--\,H}\end{equation}
In literature, the reaction of $E'$ with H$_2$ has been
experimentally observed in several works both for $E'$ in bulk
\sil{} and on silica surfaces (see subsection
\ref{reactewithh2section}). The overall kinetics of
(\ref{reacteh2reprep}) and (\ref{reacteh0rep}) is basically driven
by the former reaction, because diffusion of H is so fast that the
latter follows adiabatically the slow concentration variations of
$E'$ driven by H$_2$ diffusion (see chapter \ref{backII}).
\begin{figure}[htb!]
\begin{center}
\includegraphics[width=.8\textwidth]{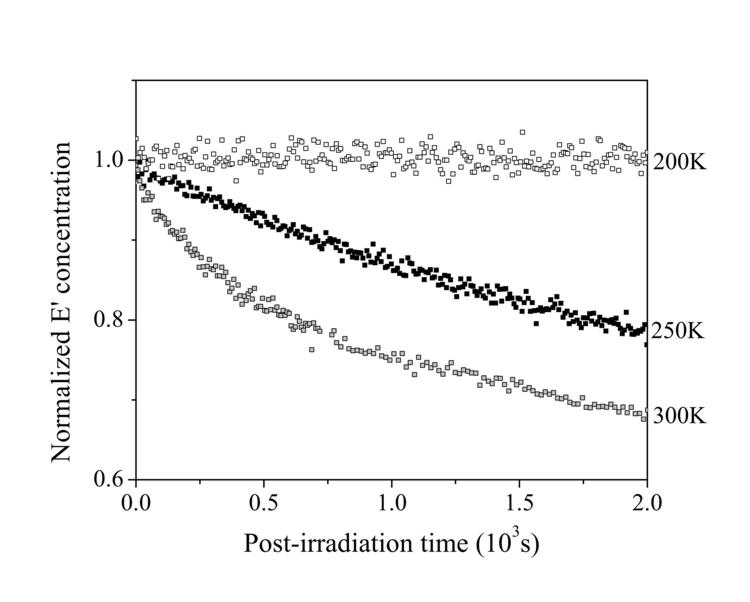}
\end{center}
\caption{Post-irradiation kinetics of $E'$ after irradiation of a
natural dry silica sample with 2000 laser pulses with
40\,mJcm$^{-2}$ energy density per pulse and 1\,Hz repetition rate,
performed at three different temperatures. The curves are normalized
to the concentration measured at the end of exposure.}
\label{temperature}\end{figure} The analysis of the temperature
dependence is useful to verify if the post-irradiation decay of $E'$
is consistent with this interpretation scheme. To this purpose we
measured \emph{in situ} the kinetics of $E'$ during and after
irradiation with 2000 laser pulses performed at several temperatures
in the 200\,K$<T<$300\,K interval. Each irradiation experiment
consisted in the exposure of the sample to laser radiation at a
given temperature and in the observation of the post-irradiation
kinetics of $E'$ at the same temperature. Here we focus only on the
post-irradiation (decay) stage of the kinetics, some of which are
reported in \figurename~\ref{temperature} after normalization to the
concentrations of $E'$ at the end of the exposure session. The
decrease becomes progressively slower on decreasing $T$, and it is
absent within experimental uncertainty at $T$=200\,K. These data
demonstrate the post-irradiation decay to be a thermally activated
process, frozen at T$_0$$\sim$200\,K. This finding is consistent
with literature
studies\cite{GriscomJNCS84,KajiharaPRL02,KajiharaPRB06} where
diffusion of H$_2$ in silica has been characterized by the same
threshold temperature, thus supporting our interpretation
(\ref{reacteh2reprep}).

It is interesting to briefly discuss this process on the basis of
the Waite theory of diffusion-limited reactions (section
\ref{solutre}), which allows to estimate from literature parameters
of H$_2$ the \emph{expected} time scale of $E'$ anneal at room
temperature: applying eq.~(\ref{timescale}) to the kinetics of $E'$
in dry silica of \figurename~\ref{compdrywetcontrollarevasintotico},
we obtain:
\begin{equation}
\tau\sim
10^{17}\text{cm}^{-3}\text{s}\cdot[E']_{\infty}^{-1}\sim10\text{\,s}
\end{equation}which is roughly two orders of magnitude lower that
the 10$^3$\,s characterizing the experimental decay. On the one
hand, this means that our kinetics are actually \emph{compatible}
with the idea of a process driven by H$_2$ diffusion, as the
experimental reaction rate does not \emph{overcome} the maximum
possible value consistent with the mobility of H$_2$ in silica at
$T$=300\,K. It is easy to see that a similar check excludes that the
process may be driven by O$_2$ or H$_2$O diffusion, much slower than
H$_2$ (Table \ref{dc}). On the other hand, the difference between
the observed and the predicted decay time scales indicates that a
physical reason, unaccounted for by purely diffusion-limited
reactions theory, slows down the overall reaction rate. This is
consistent with the unrealistically small capture radius found in
previous works when trying to fit the post-irradiation kinetics
measured \emph{ex situ} on the basis of the theoretical predictions
obtained by the Waite theory \cite{ImaiPRB91,CannasJNCS04}. The
reason of this discrepancy will be clarified in chapter \ref{PRL}.
Finally, it is worth noting that at this stage we have not yet
addressed the problem of the \emph{origin} of mobile H$_2$: it may
be already present in the as-grown samples due to the manufacturing
procedure, or be induced radiolytically by laser-induced rupture of
pre-existing precursors, such as Si\,--\,OH or Si\,--\,H bonds.

\vspace{1.5cm}Aside from the above considerations, there is another
totally independent line of reasoning that strongly suggests
hydrogen to be at the basis of the post-irradiation effects. In
fact, ESR measurements evidence the growth of H(II) centers
occurring concurrently to the decay of $E'$
(\figurename~\ref{pikhiiq90610000}). The features of the
paramagnetic signal of H(II) (see \figurename~\ref{esrspettrolargo})
are an unambiguous \emph{fingerprint of the presence of a hydrogen
atom} in its structure, as was proved by studies on isotopically
enriched\footnote{In detail, by substitution of hydrogen with
deuterium, the 11.8\,mT doublet was shown to become a
\emph{triplet}, and its separation to scale (as expected) as the
ratio of deuteron and proton magnetic moments.\cite{Vitko} }
samples, which allowed to determine the microscopic structure
(=Ge$^{\bullet}$\,--\,H) of the defect \cite{Vitko}. Consequently,
the post-irradiation growth of the H(II) center is most likely
attributable to \emph{trapping of a hydrogen atom} on a suitable
pre-existing Ge precursor. Now, given the structure of the H(II),
the precursor is clearly expected to be (=Ge$^{\bullet\bullet}$),
namely the GLPC center, as also put forward in previous works on
$\gamma$-irradiated silica or on surface defects
\cite{RadtzigPSS86,AgnelloPRB00,SkujaReview98,AgnelloPHD}:
\begin{equation} \label{formHII}
=\text{Ge}^{\bullet\bullet}\text{ + H} \longrightarrow \text{
}=\text{Ge$^{\bullet}$\,--\,H}\end{equation}the attribution of the
post-irradiation kinetics of H(II) to process (\ref{formHII}) is
also consistent with the observed reduction of the GLPC typical
optical activity (\figurename~\ref{bleaching}), even if it would be
necessary to prove that the \emph{kinetics} of GLPC bleaching is
correlated with H(II) growth.

Reaction (\ref{formHII}) necessarily requires \emph{atomic} hydrogen
because, as widely known, H$_2$ does not react spontaneously with
\emph{diamagnetic} defects, being a very stable molecule with a high
bond energy ($\sim$4.5\,eV). On the contrary, the reaction with
(some) \emph{paramagnetic} centers is possible because the presence
of an unpaired electron confers to these defects a much higher
reactivity \cite{RadtzigJPC95,GriscomJNCS84}. Hence, since the
stable form of hydrogen in \sil{} at room temperature is H$_2$, for
reaction (\ref{formHII}) to be possible, it is needed a paramagnetic
defect that reacts with H$_2$ thus acting as a \emph{cracking
center}, which makes available H to be trapped at the GLPC site.
Given the post-irradiation decay of $E'$, it is clear now that $E'$
is the main candidate for the role of the cracking
center.\footnote{A further indication of the role of $E'$ is the
absence here of the other basic defect in silica able to serve as a
H$_2$ cracking center, namely the NBOHC
\cite{GriscomJNCS84,KajiharaPRL02,KajiharaPRB06}. Indeed, the
typical 4.8\,eV absorption band of NBOHC is not observed by \emph{in
situ} OA. This point is further discussed in the next chapter.}
\begin{figure}[h!]
\begin{center}
\includegraphics[width=.55\textwidth]{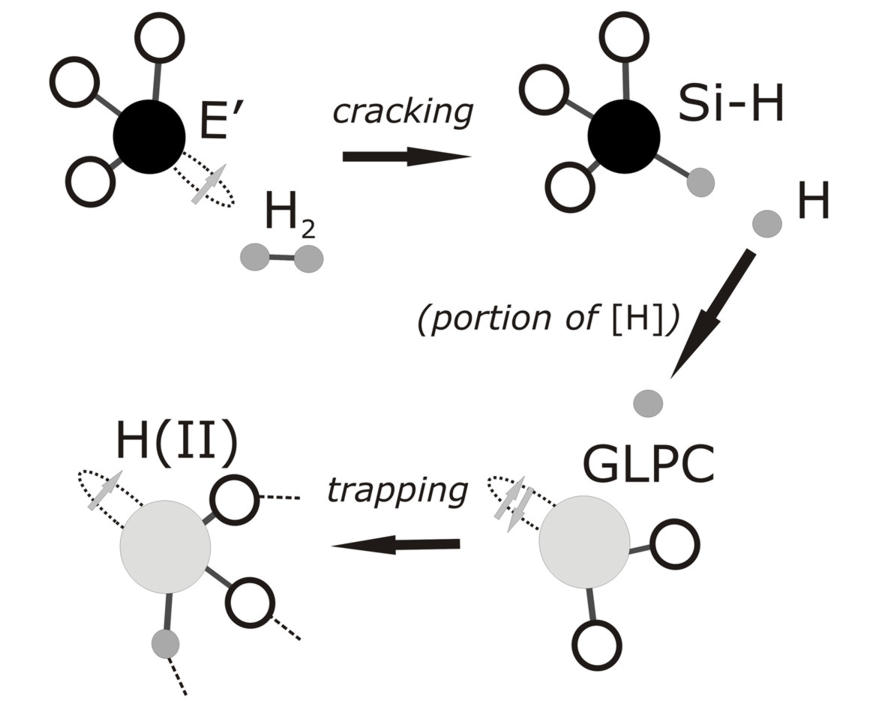}
\end{center}
\caption{Pictorial representation of post-irradiation processes in
natural silica. Diffusing H$_2$ reacts with $E'$ center: this leads
to passivation of the defect and produces free atomic hydrogen H. A
portion of the population of H migrates until encountering a GLPC
center, where it is trapped producing the H(II) center. The
remaining portion of H (not represented) passivates another $E'$ or
dimerizes in H$_2$. } \label{schemamodello}\end{figure}

So, in this scheme, displayed in \figurename~\ref{schemamodello},
diffusing H$_2$ reacts with $E'$ by (\ref{reacteh2reprep}) causing
the post-irradiation decay of this paramagnetic center. \emph{A
portion} of the population of H produced at the right side of
(\ref{reacteh2reprep}) is then captured by GLPC (Reaction
\ref{formHII}) thereby causing the simultaneous growth of the H(II)
center. This interpretation is also corroborated by the comparable
time scales ($\sim$10$^3$\,s) of the $E'$ decay and the H(II) growth
kinetics. Comparing the typical H(II) concentration
(\figurename~\ref{pikhiiq90610000})
$\sim$2$\times$10$^{15}$\,cm$^{-3}$ with the entity of the total
post-irradiation decrease of $E'$
(\figurename~\ref{compdrywetcontrollarevasintotico})
$\sim$10$^{16}$\,cm$^{-3}$, it appears that hydrogen trapping by
GLPC (\ref{formHII}) can be regarded as a \emph{secondary} reaction,
which consumes no more than 10-20\% of the available hydrogen atoms,
most of which react with $E'$ by (\ref{reacteh2reprep}) and
(\ref{reacteh0rep}). On the other hand, it is clear that the
importance of H(II) in the present context goes much further. In
fact, the observation by ESR of this hydrogen-related center whose
concentration increases in time, naturally leads to attribute the
post-irradiation decay of $E'$ to H$_2$, ruling out \emph{a priori}
other possibilities, like electron/hole detrapping or diffusion of
other mobile species. Then, H(II) may be considered \emph{a probe of
the presence of mobile hydrogen}, allowing in some sense to overcome
the difficulty of a direct observation of H$_2$.

Although the present qualitative considerations offer quite
convincing reasons to attribute the post-irradiation effects to the
presence of mobile H$_2$, the problem of verifying in more detail
the quantitative consistency of this model with the observed
kinetics will be discussed again in greater detail in chapters
\ref{PRB} and \ref{PRL}.

\section{Conclusions}
\hspace{0.8cm}Exposure of natural silica to 4.7\,eV pulsed laser
radiation induces the generation of $E'$ centers. The defects are
unstable and decay in the post-irradiation stage in a typical time
scale of the order of 10$^3$\,s. The $E'$ is not induced in
synthetic silica materials. ESR measurements show that a Ge-related
paramagnetic center, the H(II), grows in the post-irradiation stage
concurrently to the decay of $E'$. Since the structure of the H(II)
can be unambigously identified by ESR, this center may be considered
as a probe of the presence of mobile hydrogen, so as to attribute
both the decay of $E'$ and the growth of H(II) to diffusion of H$_2$
in the glass after the end of exposure. In this scheme, $E'$ decays
by reaction with H$_2$, and a portion of the H made available by
this process is then captured on pre-existing GLPC centers so as to
form H(II). Our interpretation is also confirmed qualitatively by
the temperature dependence of the $E'$ decay and by the observed
reduction of the PL optical activity of GLPC upon irradiation.

\chapter{Generation of $E'$ center}\label{JPCMs} In this
chapter we discuss in more detail the \emph{in situ} kinetics of the
$E'$ center, with the purpose to understand the generation mechanism
of the defect, as well as the origin of the hydrogen responsible for
its post-irradiation decay.
\section{Introduction}
\hspace{0.8cm} The experimental results presented in the previous
chapter demonstrate the ability of 4.7\,eV laser light to induce at
room temperature the generation of $E'$ in natural \sil{} materials.
Besides, the induced defects are unstable and decay in the
post-irradiation stage, supposedly due to reaction with diffusing
H$_2$. The decay is particularly effective in wet natural silica,
where it anneals almost completely the $E'$ induced by irradiation.
However, many questions remained open such as the generation
mechanism of the $E'$ center and the origin of hydrogen responsible
for the post-irradiation decay.

The passivation of $E'$ by H$_2$ is one of the most important
processes among the several defect conversions in silica induced by
hydrogen, quite common even well below room temperature due to the
high mobility and reactivity of H and H$_2$. A careful study of the
reaction dynamics of defects with diffusing species requires
spectroscopic techniques suitable to probe \emph{in situ} their
concentration changes. So far, \emph{in situ} photoluminescence
measurements have been used to clarify the generation and the decay
of another defect of fundamental interest, the non bridging oxygen
hole center (NBOHC) induced by photolysis of Si\,--\,OH bond
\cite{KajiharaAPL01,KajiharaPRL02,KajiharaPRB06}. In contrast, the
current understanding of the generation of $E'$ centers by UV laser
(reviewed in subsection \ref{geneprimosection}) is mainly founded on
\emph{ex situ} ESR and OA measurements. Hence, even though the
passivation of $E'$ by H$_2$ (reviewed in subsection
\ref{reactewithh2section}) has been repeatedly observed in
literature, the interplay between the photo-induced creation of $E'$
and its decay due to reaction with mobile hydrogen is not well
understood. Our \emph{in situ} technique is based upon the the
observation of the complete time-dependent absorption profile in the
UV\cite{MioPSS06INSITU,MioJPCM05,MioJNCS05EPRIMO,MioJPCM06,MioJNCS06EPRIMO},
differently from previous works, which reported only the absorption
at a fixed wavelength \cite{LeclercJNCS92,MarshallJNCS97,SmithAO00}.
In addition to the possibility of monitoring the kinetics of induced
absorption, another advantage of our approach is the possibility of
clearly determine which transient absorbing centers are induced (and
which are not), provided that their absorption line shapes can be
unambiguously identified. In this
chapter\cite{MioJPCM05,MioJPCM06,MioJNCS06EPRIMO,MioJNCS05EPRIMO}
the \emph{in situ} OA technique is applied to perform a
comprehensive investigation of the processes controlling the
generation and decay dynamics of $E'$ centers. Our discussion starts
from wet fused quartz, which is a remarkable material to study
transient transmittance losses; in fact, as already shown, in these
glasses UV absorption due to $E'$ centers is effective mainly during
laser exposure, while \emph{transparency is almost completely
recovered in the post-irradiation stage}. In the last part of the
chapter we are going to extend our considerations to the other
varieties of \sil{}.
\section{$E'$ center in natural wet \texorpdfstring{\sil}{silica}}\label{sectionwet}
\subsection{Experiment} The experiments reported in this section were performed
on HER1 \emph{natural wet} \sil{} samples (see Table
\ref{materials}), of sizes 5$\times$5$\times$1\,mm$^3$ and optically
polished on all surfaces. The specimens were irradiated at room
temperature (T$_0$=300\,K) with 4.7\,eV pulsed laser radiation
perpendicularly to one of the minor surfaces. We verified that
during laser irradiation the temperature of the samples did not vary
significantly from T$_0$. It was used a repetition rate of the laser
pulses of 1\,Hz, corresponding to an interpulse time of $\Delta
t$=1\,s. The diameter $2r$ of the laser beam was (6.0$\pm$0.1)\,mm.
Since the intensity profile of the laser beam is uniform (see
chapter \ref{mm2}), the ratio of pulse energy to the beam section
($\pi r^2$) and duration ($\tau$=5\,ns) gives the (\emph{mean})
laser peak intensity $\Lambda$. We performed several irradiation
sessions on different \emph{virgin} samples at different laser pulse
energies, from 3.7\,mJ to 27\,mJ. These values correspond to peak
intensities $\Lambda$ from (2.6$\pm$0.2)$\times$10$^6$\,Wcm$^{-2}$
to (19$\pm$1)$\times$10$^6$\,Wcm$^{-2}$, respectively.

During each irradiation session (consisting in a few thousand
pulses), we measured \emph{in situ} the absorption profile induced
in the sample. As described in the previous chapter, these
measurements yield the kinetics of the OA on a time scale longer
than $\Delta t$. For some of the irradiations, the measurements were
carried on at the same rate (1 OA spectrum per second) also for a
few hours in the post-irradiation stage, so as to follow the decay
of the induced absorption profile. Our investigation was completed
by ESR spectra performed for a few days after the end of
irradiation.

\subsection{Results} 
\begin{figure}[htb!]
\begin{center}
\includegraphics[width=\textwidth]{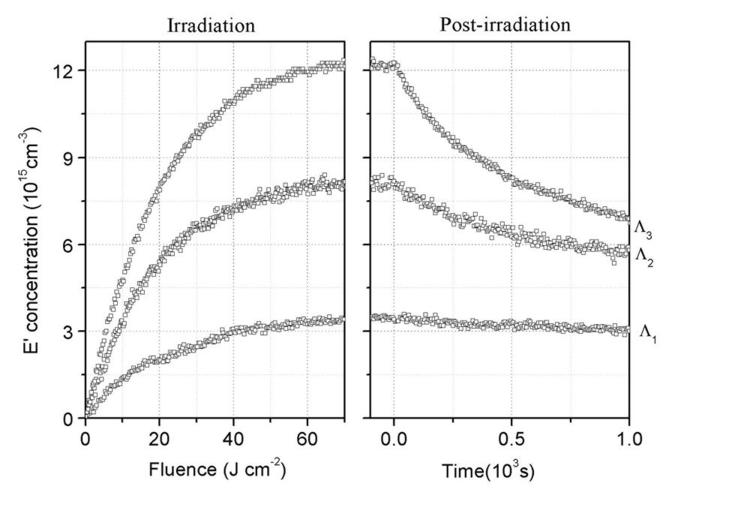}
\end{center}
\caption{Three representative kinetics of induced absorption
measured \emph{in situ} during and after pulsed UV laser irradiation
of wet natural silica at different laser intensity levels. The
origin of the time scale corresponds to the end of the irradiation
session. The parameters $\Lambda_i$ are defined in the text. For
graphical reasons, not all datapoints are plotted.}
\label{compakin}\end{figure}

\hspace{0.8cm}As already discussed, the main signal observed in the
induced OA spectrum measured \emph{in situ} during and after laser
irradiation of wet natural silica is the 5.8\,eV band due to the
$E'$ centers; from the knowledge of the peak absorption cross
section of the defects, we estimated their concentration [$E'$] as a
function of time. In \figurename~\ref{compakin} are reported three
representative kinetics observed upon exposure of the specimens to
radiation with the peak laser intensities
$\Lambda_1$=2.6$\times$10$^6$\,$Wcm^{-2}$,
$\Lambda_2$=4.8$\times$10$^6$\,$Wcm^{-2}$, and
$\Lambda_3$=12$\times$10$^6$\,$Wcm^{-2}$.
 For practical reasons, the
\emph{irradiation stage} of the kinetics is plotted as a function of
laser fluence $\Phi=\Lambda \tau t \Delta t^{-1}$, while data for
the first 10$^{3}$\,s of the \emph{post-irradiation stage} are
plotted versus time, $t=0$ corresponding to the end of the
irradiation session.

During irradiation, [$E'$] saturates after $\sim$ 70\,J$cm^{-2}$ to
a constant value [$E'$]$_S$ that depends on intensity. For instance,
[$E'$]$_S$=1.2$\times$10$^{16}$\,cm$^{-3}$ during irradiation with
$\Lambda_3$=12$\times$10$^6$\,$Wcm^{-2}$. As soon as the laser is
switched off, the defects begin to decay. The decay appears to be
progressively faster with increasing $\Lambda$; indeed, the
concentration of the $E'$ generated with intensity $\Lambda_3$ is
reduced of $\sim$40\% after 10$^3$\,s, whereas for
$\Lambda=\Lambda_1$ the decrease is $\sim$15\% in the same time
interval. However, ESR measurements performed for a few days after
irradiation confirm the result of the previous chapter, namely that
the $E'$ centers are \emph{almost completely transient} since $[E']$
tends to an asymptotic value lower than 20\% of the maximum
concentration.

\begin{figure}[p!]
\begin{center}
\includegraphics[width=.7\textwidth]{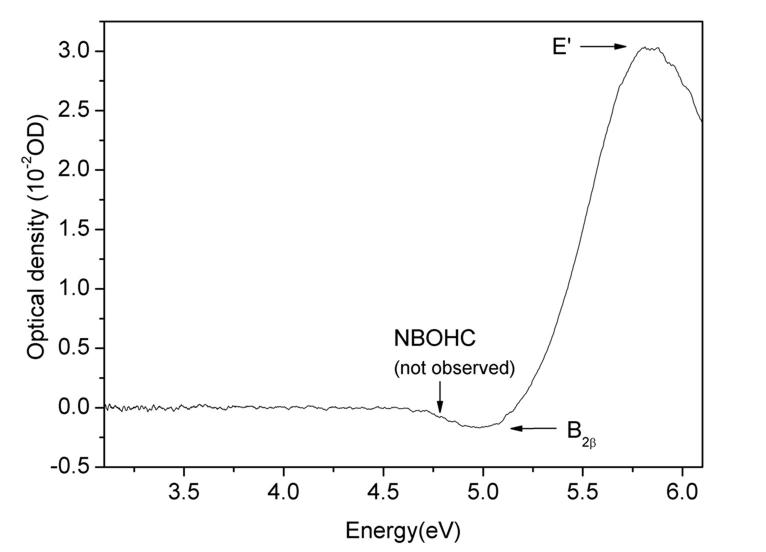}
\end{center}
\caption{Typical UV profile of the transient OA induced by 4.7\,eV
pulsed laser irradiation in wet natural \sil{}.}
\label{profile}\end{figure}

We discuss now in more detail the typical OA profile induced in wet
natural silica, as reported in \figurename~\ref{profile}.
\begin{figure}[p!]
\begin{center}
\includegraphics[width=.7\textwidth]{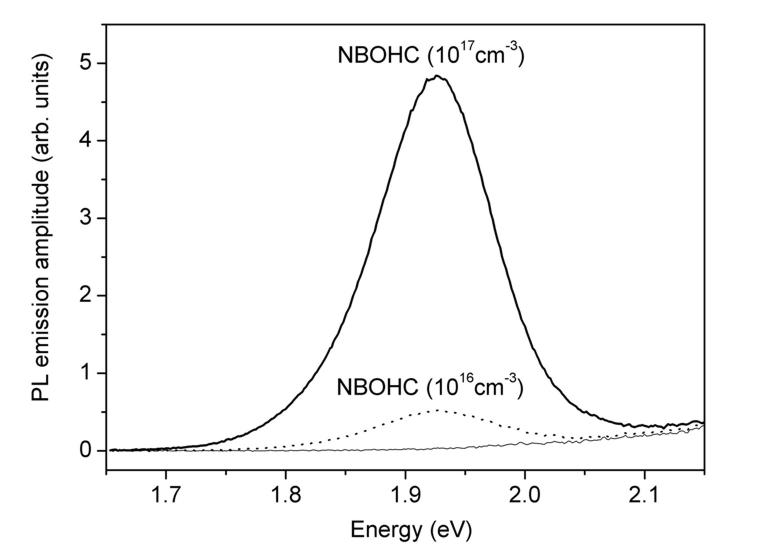}
\end{center}
\caption{Photoluminescence emission spectra excited at 4.8\,eV and
measured with an emission bandwidth of 10\,nm in a
$\gamma$-irradiated synthetic silica sample that contains
10$^{17}$\,cm$^{-3}$ NBOHC centers (spectrum G, thick continous
line), and in a laser-irradiated wet natural silica sample (L,
continous line). The dotted line is a simulation of the signal that
would be expected for 10$^{16}$\,cm$^{-3}$ NBOHC; it was obtained by
adding G/10 to L.} \label{plshow}\end{figure}Apart from the
Gaussian-shaped band centered at (5.84$\pm$0.03)\,eV with
\linebreak{} FWHM=(0.70$\pm$0.04)\,eV associated with the $E'$
centers, we observe the small negative component at 5.1\,eV, already
attributed to conversion of GLPC centers pre-existing in natural
\sil{}. Furthermore, no other measurable absorption bands are
present in the spectrum: in particular, we stress that there is no
evidence of the typical 4.8\,eV band associated with NBOHC centers
($\equiv$ Si\,--\,O$^\bullet$, see chapter \ref{backI}) within our
sensitivity, which is $\sim$0.02cm$^{-1}$ in optimal conditions
(\figurename~\ref{profile}). The oscillator strength of the NBOHC
centers was estimated to be in the interval 0.03\,--\,0.05 on the
basis of the decay time of their luminescence band at 1.9\,eV
\cite{SkujaSPIE01,CannasPRB04,VaccaroJNCS}. Hence, the sensitivity
limit can be converted by eq.~(\ref{sma}) to a \emph{concentration}
upper limit for NBOHC of 3\,--\,5$\times$10$^{15}$cm$^{-3}$.

In this way, the \emph{in situ} technique allows to exclude that
NBOHC centers are generated in concentrations comparable to $E'$ by
IV harmonic Nd:YAG radiation, at least at the investigated intensity
levels. To further corroborate this conclusion, we performed
photo-luminescence measurements in laser-irradiated wet and dry
natural silica samples a few days after the end of irradiation. In
fact, it is known that the NBOHC feature a PL emission bands
centered at 1.9\,eV under excitation at 4.8\,eV, which was not
observed in the measured emission spectra
(\figurename~\ref{plshow}). For comparison, it is shown the 1.9\,eV
emission band as measured in a $\gamma$-irradiated synthetic \sil{}
sample, which corresponds to a concentration of NBOHC of
$\sim$10$^{17}$\,cm$^{-3}$ as deduced from the intensity of the OA
band at 4.8\,eV. Since the intensity of the PL emission signal is
proportional to the concentration (in conditions of low absorbance,
see chapter \ref{mm2}), these data confirm that the concentration of
NBOHC in laser-irradiated samples is lower than
$\sim$2$\times$10$^{15}$\,cm$^{-3}$. However, it is worth noting
that the \emph{in situ} technique yields a stronger information than
PL measurements, since it excludes also possible \emph{transient}
NBOHC centers. We stress that the absence of NBOHC is going to be a
very important point in what follows.
\subsection{Discussion I: Precursor of $E'$ center}
We are going to show now that the analysis of \emph{in situ} OA
measurements permits to infer important information about the
generation mechanism of the $E'$ center under Nd:YAG laser
irradiation. In particular, we begin our discussion with the
analysis of the \emph{decay} stage of the process.

A thorough study of the decay of $E'$ due to reaction with mobile
hydrogen, must be performed on the basis of a set of chemical rate
equations, comprehending also the secondary trapping of atomic
hydrogen on preexisting twofold coordinated Ge (Reaction
\ref{formHII}). This problem will be dealt with in a following
chapter, while here we propose an analysis founded on the properties
of the \emph{first stage} of the decay. Besides, we neglect here for
simplicity the secondary reaction, which involves only a minor
portion of hydrogen, and we assume in the following that H$_2$ is
involved only the reaction with $E'$ center, also because it is
absent the NBOHC center that would react with a portion of the
available hydrogen \cite{GriscomJNCS84,KajiharaPRB06}.

\begin{figure}[htb!]
\begin{center}
\includegraphics[width=.7\textwidth]{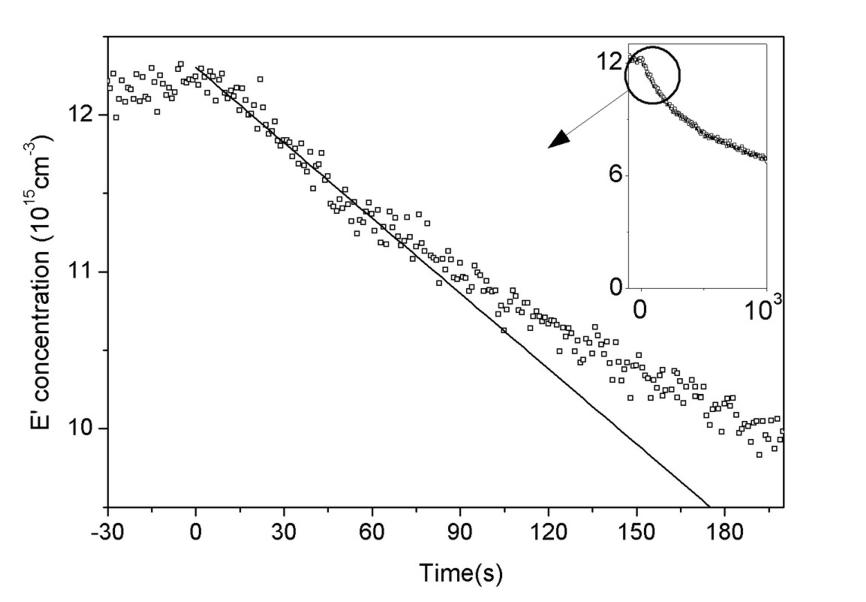}
\end{center}
\caption{Zoom of the first portion of the decay curve $\Lambda_3$ of
the right panel of \figurename~\ref{compakin}, and reported again in
the inset. The continuous line is obtained by a linear least-square
fit of the data in the first 50\,s.} \label{zoomdecay}\end{figure}

To characterize the first stage of the decay, we can determine by a
linear fit in the first $\sim$50\,s of the post-irradiation stage
the \emph{initial decay slope} $d[E']/dt(t=0)$, as shown in
\figurename~\ref{zoomdecay} for the kinetics at \linebreak{}
$\Lambda$=12$\times$10$^6$\,$Wcm^{-2}$.
Then, the slope is normalized dividing by the concentration
$[E']_{S}$, measured at the beginning of the decay as well, so as to
give the following parameter $\Gamma$:
\begin{equation}
\Gamma=-\frac{1}{[E']_{S}}\frac{d[E']}{dt}(t=0)
\end{equation}which represents the probability per unit time that an
$E'$ center disappears, estimated immediately after the end of
irradiation. The importance of the parameter $\Gamma$ derives from
the following considerations. We know that the post-irradiation
annealing of $E'$ ($\equiv$Si$^{\bullet}$) in the present
experimental conditions is due to the reaction with mobile H$_2$,
which we rewrite again here:
\begin{equation}
\label{reacteh2rep} \equiv\text{Si}^{\bullet}\text{ + H$_2$ }
\longrightarrow \text{ }\equiv\text{Si\,--\,H + H}
\end{equation}where H produced at the right side may passivate
another $E'$ or dimerize in H$_2$. As described in section
\ref{solutre}, the concentration variations due to reaction
(\ref{reacteh2rep}) are accounted for by the following rate
equation, valid in the stationary-state approximation
\cite{Atkinsbook,GriscomJAP85}:
\begin{equation}
\frac{d[\text{$E'$}]}{dt}=2\frac{d[\text{H$_2$}]}{dt}=-2k_0[\text{$E'$}][\text{H$_2$}]\label{rateeq}
\end{equation}where $k_0$ is the reaction constant between $E'$ and molecular hydrogen, and
the factor 2 derives from the fact that one H$_2$ passivates
\emph{two} $E'$. Hence, evaluating eq.~(\ref{rateeq}) at the end of
irradiation and dividing both members by [$E'$], we get:
\begin{equation}
\Gamma=-\frac{1}{[\text{$E'$}]}\frac{d[\text{$E'$}]}{dt}(t=0)=2k_0[\text{H$_2$}](t=0)
\end{equation}which means that the above defined parameter $\Gamma$
is proportional to the concentration [H$_2$]$_{S}$=[H$_2$](t=0) of
molecular hydrogen present in the sample at the end of irradiation.
In other words, $\Gamma$ can be considered as an \emph{indirect
measure} of [H$_2$]$_{S}$.

\begin{figure}[htb!]
\begin{center}
\includegraphics[width=.8\textwidth]{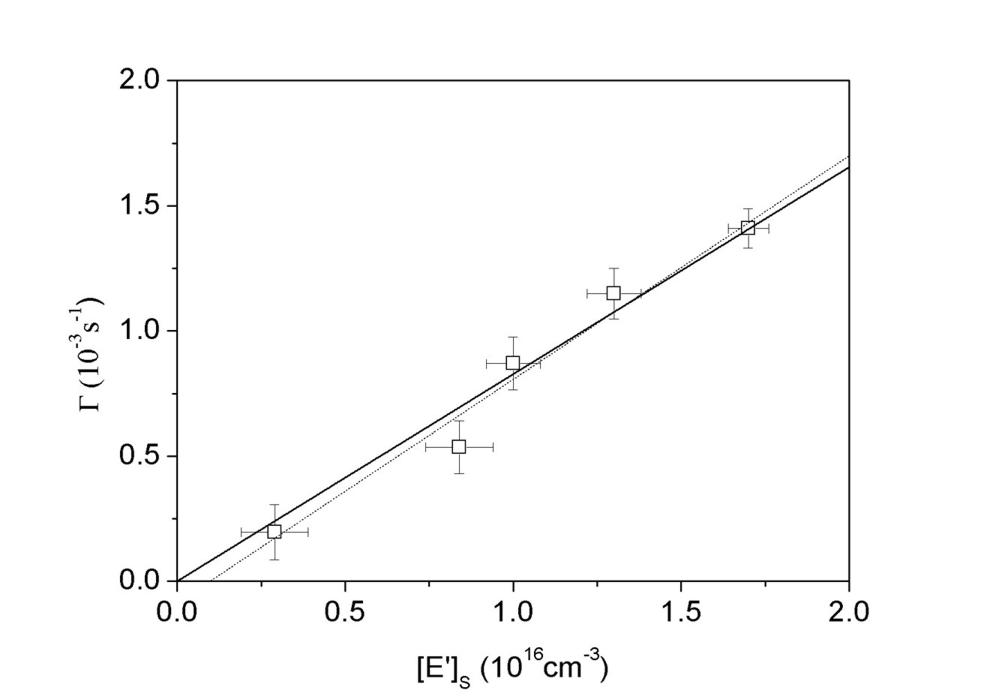}
\end{center}
\caption{Parameter $\Gamma$, proportional to the concentration of
molecular hydrogen at t=0, as a function of the concentration of
$E'$ measured at the same time instant. The continuous line is a
least-square fit of the data with the function y=$\alpha$x, from
which we find:
$\alpha$=(8.3$\pm$0.8)$\times$10$^{-20}$\,cm$^{3}$s$^{-1}$.
Alternatively, the dotted line is a least-square fit with the
function y=$\alpha$x+$\beta$; in this case, the $\beta$ coefficient
obtained from the fit is consistent with zero within the error of
the fitting procedure. } \label{gammavseprimo}\end{figure}
Therefore, in \figurename~\ref{gammavseprimo} we analyze how
$\Gamma$ is related to the concentration of $E'$ at t=0,
[$E$]$_{S}$. To this purpose, $\Gamma$ is plotted as a function of
[$E$]$_{S}$ for 5 points obtained at different power levels
$\Lambda$. We stress that at this stage $\Lambda$ serves only as a
parameter useful to span the concentration of E'. We see that the
plot evidences a linear dependence y=$\alpha$x with a slope
$\alpha$=8.3$\times$10$^{-20}$\,cm$^{3}$s$^{-1}$. This implies that
[$E'$]$_{S}$ is \emph{proportional} to [H$_2$]$_{S}$.

This observation has very important implications for two of the
questions we posed at the beginning of this chapter, i.e. the origin
of $E'$ and of mobile H$_2$. Let us start with discussing H$_2$. In
general, molecular hydrogen in a \sil{} sample can be either already
present in the material prior to any treatment, or be generated
radiolytically by the irradiation itself; in both cases it can then
get involved in post-irradiation processes. However, data in
\figurename~\ref{gammavseprimo} allow to exclude that H$_2$
available for reaction with $E'$ is already present in the samples
before exposure, since in that case [H$_2$]$_{S}$ would be
independent of irradiation, resulting in a constant value of
$\Gamma$ regardless of [$E'$]$_{S}$. Then hydrogen has a
\emph{radiolytic origin}.

To make a step forward, we first observe that the rate equation
(\ref{rateeq}) conserves the difference [$E'$]-2[H$_2$].\footnote{In
fact, from (\ref{rateeq}): d/dt([$E'$]-2[H$_2$])=0 } As a
consequence, the total variation of the concentration of $E'$ in the
post-irradiation stage is at most the double of the amount of
available hydrogen [H$_2$]$_{S}$:
\begin{equation}
\Delta[E']=2\Delta[\text{H$_2$}]<2[\text{H$_2$}]_{S}.\label{conditionh2}
\end{equation}

Now, we observe that radiolytic hydrogen must be generated by
breaking of a suitable precursor in which H is stored in bonded form
in the as-grown material. In particular, we have seen in chapter
\ref{backII} that the two main forms of bonded hydrogen in \sil{}
are Si\,--\,OH and Si\,--\,H, \cite{Shelby94,SkujaSPIE01,Schmidt98}
whose UV-induced breaking can release atomic H, which dimerizes to
form H$_2$. However, photo-induced breaking of Si\,--\,OH generates
NBOHC ($\equiv$Si\,--\,O$^{\bullet}$) as a co-product of H,
according to the following
reaction\cite{GriscomJNCS84,KajiharaPRL02,KajiharaJNCS03,SkujaSPIE01,Erice}:
\begin{equation}
\equiv\text{Si\,--\,OH}\stackrel{h\nu}{\longrightarrow}\equiv\text{Si\,--\,O}^{\bullet}
\text{ + H}
 \label{breakingohrep}
\end{equation}The absence of the 4.8\,eV band in the \emph{in situ}
absorption profile, which we have seen to lead to the the absence of
NBOHC within a few 10$^{15}$cm$^{-3}$, indicates that the photolysis
of Si\,--\,OH \emph{cannot account} for the available [H$_2$]$_S$,
expected by eq.~(\ref{conditionh2}) to be at least one-half of
[$E'$]$_{S}$ since $E'$ centers decay almost completely in the
post-irradiation stage. Therefore, we can exclude process
(\ref{breakingohrep}) as a possible source of radiolytic
hydrogen.\footnote{Our results suggest that 4.7\,eV laser is unable
to break efficiently Si\,--\,OH bonds, differently from 7.9\,eV
radiation which has been demonstrated to generate NBOHC from
Si\,--\,OH with high efficiency \cite{KajiharaPRL02}. Our finding is
consistent with data by Nishikawa \emph{et al.}, according to which
Si\,--\,H breaking is much more efficient than Si\,--\,OH breaking
under 6.4\,eV laser irradiation
\cite{NishikawaPRB93B,NishikawaPRB90}. It is possible, however, that
the lack of observation of NBOHC under 4.7\,eV laser derives from a
much higher reactivity of NBOHC (compared to $E'$) with H$_2$, which
prevents the defect from significantly growing during the
irradiation session, due to fast recombination with hydrogen during
each interpulse time span. The different reaction properties with
H$_2$ of NBOHC and $E'$ are further discussed in chapter \ref{PRL}.
}

Additionally, the linear correlation between [H$_2$]$_{S}$ and
[$E'$]$_{S}$ of \figurename~\ref{gammavseprimo} strongly suggests
that the generation processes of the two species are not
independent. Based on this observations, the simplest model is that
\emph{hydrogen and $E'$ are formed from a common precursor}, i.e.
the Si\,--\,H group, whose dissociation produces $E'$ centers and H
in the same amount:
\begin{equation}
\text{Si\,--\,H
}\stackrel{h\nu}{\longrightarrow}\equiv\text{Si$^{\bullet}$ +
H}\label{generation}
\end{equation}where the produced H is supposed to rapidly diffuse (at T=300\,K) and
dimerize forming H$_2$. Indeed, as a consequence of this process,
[H$_2$]$_{S}$=[$E'$]$_{S}$/2, leading to
\begin{equation}
 \Gamma = 2k_0[\text{H$_2$}]_S=k_0[E']_{S}\label{gammareltoeprimo}
\end{equation}
in agreement with the results in \figurename~\ref{gammavseprimo}. In
this model, the slope
$\alpha$=8.3$\times$10$^{-20}$\,cm$^{3}$s$^{-1}$ of
\figurename~\ref{gammavseprimo} equals the reaction constant $k_0$
between $E'$ and H$_2$; this is corroborated by its close agreement
with the value k=(8.4$\pm$0.5)$\times$10$^{-20}$cm$^3$s$^{-1}$,
which was estimated by fitting the post-irradiation kinetics of $E'$
by a second-order kinetic curve (eq.~(\ref{generich2decay}) with
$k=4\pi r_0 D_{H_2}$) derived from the Waite
theory\cite{CannasJNCS04,Waite57}. Aside from the linear correlation
of \figurename~\ref{gammavseprimo}, some more considerations further
support the attribution to Si\,--\,H of the role of precursor of
$E'$ centers in wet natural silica:

\noindent\textbf{A.}\hspace{0.8cm}This model explains why in the wet
natural silica samples the $E'$ are almost \emph{completely}
transient, independently of laser intensity at least in the
investigated range. In fact, process (\ref{generation}) produces
$E'$ and hydrogen \emph{exactly in the same amount}, which recombine
in the post-irradiation stage so as to completely anneal the effects
of exposure. In this sense, the transient nature of $E'$ may be
considered an \emph{intrinsic} property of its very generation
process. The fact that ESR measurements a few days after irradiation
show a residual small concentration ($\sim$10--20\% of [$E'$]$_S$)
of $E'$ can be interpreted as a consequence of the minor reaction
(\ref{formHII}), which causes a small portion of the total hydrogen
population to escape from recombination with the $E'$.

\begin{figure}
\begin{center}
\includegraphics[width=.9\textwidth]{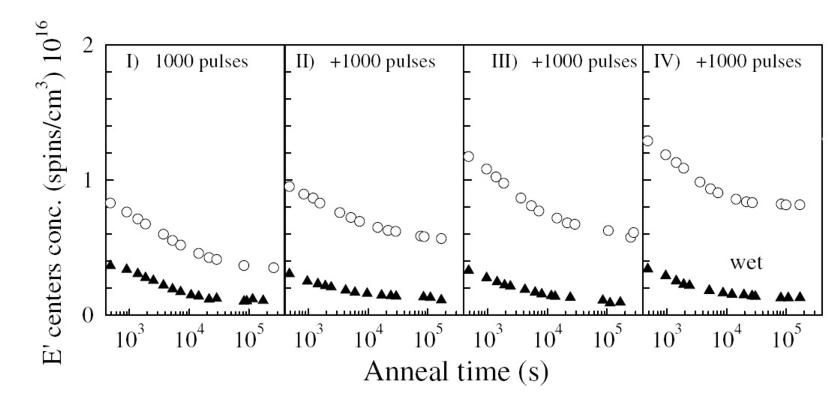}
\end{center}
\caption{Post-irradiation kinetics of $E'$ measured upon a sequence
of identical irradiation sessions consisting in 1000 laser pulses
with 1\,Hz repetition rate, separated by a $\sim$10$^6$\,s
post-irradiation stage. Full and empty symbols represent wet and dry
natural silica, respectively. In wet natural silica, the decay
curves repeat themselves identically after each exposure. Figure
adapted from Cannas \emph{et al.}\cite{CannasJNCS04} }
\label{cinrip}\end{figure}

\noindent\textbf{B.}\hspace{0.8cm}A consequence of the precursor
Si\,-–\,H hypothesis is that the irradiated sample virtually returns
to the same condition as the virgin material after the recombination
of $E'$ and H$_2$ in the post-irradiation stage is completed. Hence,
the material is expected to show an "elastic" response to
irradiation, meaning that if the sample is irradiated once more, the
growth and annealing of $E'$ should repeat themselves with the same
kinetics as observed after the first exposure. This prediction was
confirmed, at least in regard to the post-irradiation stage, by a
multiple-irradiation experiment recently performed on wet fused
quartz (\figurename~\ref{cinrip})\cite{CannasJNCS04}, thereby
supporting the model proposed here.

\noindent\textbf{C.}\hspace{0.8cm}Finally, further support to the
Si\,--\,H precursor model comes from the critical analysis of the
other possible generation models, as proposed in literature and
reviewed in detail in subsection \ref{geneprimosection}. In general,
generation of $E'$ centers under sub-bandgap radiation has been
hypothesized to occur either by conversion of pre-existing
precursors, such as oxygen deficient centers (ODC(I) and ODC(II)),
strained Si\,-–\,O\,-–\,Si bonds and impurity bonds (Si\,-–\,H,
Si\,-–\,Cl), or by intrinsic non-radiative decay of self trapped
excitons (STEs), which has been observed only under femtosecond
laser radiation or highly focused excimer laser radiation. In the
present experiment, where the radiation source was a non-focused ns
laser, we expect non-radiative decay of STEs to be inefficient, and
the defects to be generated from precursors. The optical properties
of the as-grown samples permit to go further. In fact, the typical
7.6\,eV band of the ODC(I) (corresponding to an oxygen vacancy
$\equiv$\,Si\,--\,Si\,$\equiv$\cite{SkujaReview98}) is not observed
by spectrophotometric measurements in the vacuum UV (VUV) range on
the as-grown natural wet samples \cite{CannasJNCS01}. On the basis
of the known oscillator strength \cite{SkujaSPIE01}, this fixes a
concentration limit of $\sim$10$^{16}$\,cm$^{-3}$ for this defect.
For what concerns the ODC(II) (whose models are a twofold
coordinated silicon, =Si$^{\bullet\bullet}$, or an \emph{unrelaxed}
oxygen vacancy\cite{SkujaReview98}), its typical optical activity,
consisting in an absorption band peaked at 5.0\,eV exciting two
emission bands at 2.7\,eV and
4.4\,eV\cite{SkujaSPIE01,SkujaReview98,SkujaJNCS92}, was not
observed in the as-grown materials. From the lack of these signals
we infer that ODC(II) is absent within $\sim$10$^{15}$\,cm$^{-3}$.
So, the concentration of both varieties of ODCs is too small to
account for the observed values of [$E'$]$_{S}$
(\figurename~\ref{gammavseprimo}) and also this precursor can be
excluded. Finally, the \emph{strained}
$\equiv$Si\,-–\,O\,-–\,Si$\equiv$ \emph{bonds} are excluded as well,
since their photolysis has been proposed in literature to generate
$E'$--NBOHC pairs\cite{KajiharaJNCS03,DonadioPRB05,HosonoPRL01}, but
NBOHC is not observed here. In conclusion, by comparison with
literature, we deduce by exclusion that the most plausible
precursors \emph{a priori} are extrinsic impurity bonds like
Si\,-–\,H , Si\,-–\,Cl and Si\,--\,F. It is clear that among the
three only the Si\,-–\,H is consistent with data in
\figurename~\ref{gammavseprimo} and with the post-irradiation decay
of $E'$, as it allows to explain the origin of hydrogen as a
byproduct of $E'$ generation. A further comparison with literature
is proposed at the end of this section.

\begin{figure*}[t!]
\begin{center}
\fcolorbox{postitA}{postitA}{\begin{minipage}[c]{.95\textwidth}\linespread{1.2}\hrule\vspace{0.2cm}
\textbf{\textsc{Summary I. }}\textsl{\textcolor{postitC}{Section
\ref{sectionwet} is concerned with the generation of $E'$ centers
under 4.7\,eV laser irradiation in wet natural silica. To study this
process we have measured the growth and decay kinetics of $E'$
induced by irradiation at different laser intensities
(\figurename~\ref{compakin}). Our initial aim was to answer to two
apparently unrelated questions, namely the origin of $E'$ and of
hydrogen responsible for the post-irradiation decay of the defect.
Eventually, our experimental observations have led to the unexpected
conclusion that the two questions have in natural wet silica a
common answer, namely the two species both derive from a common
generation mechanism, the breaking of pre-existing Si\,--\,H
precursors, eq.~(\ref{generation}). In particular, this model is
suggested by the linear correlation between the concentrations of
$E'$ and H$_2$ at the end of irradiation
(\figurename~\ref{gammavseprimo}), where the latter is indirectly
estimated from the decay rate of $E'$.
\vspace{0.2cm}}\hrule}\end{minipage}}
\end{center}
\end{figure*}

Before going on with the discussion, it might be worth clarifying
why our treatment was based upon an indirect approach, i.e. the use
of the kinetic parameter $\Gamma$ proportional to [H$_2$] at the end
of the irradiation session: indeed, the \emph{direct} measurement of
[H$_2$] would require a system able to perform \emph{in situ}, and
on a time scale of $\sim$1\,s, a Raman measurement sufficiently
sensitive to appreciate a concentration of
$\sim$10$^{16}$\,cm$^{-3}$ of the mobile species. Raman techniques,
also when performed \emph{ex situ}, are quite insensitive to such a
low concentration of H$_2$ so that an indirect approach results to
be much more feasible.

\subsection{Discussion II: Generation mechanism}
Up to now, we have provided experimental evidence that the dominant
generation mechanism of $E'$ in wet fused quartz under Nd:YAG laser
radiation is the breaking of preexisting Si\,--\,H precursors.
Moreover, in the irradiated samples no other OA signal significantly
overlaps with the 5.8\,eV band (\figurename~\ref{profile}). In this
sense, this appears to be a material of choice to isolate and study
selectively the generation process (\ref{generation}), in order to
obtain further information on the underlying mechanism and on its
dynamics. This is the purpose of this and the next subsection, which
are mainly concerned with the growth stage of the kinetics in
\figurename~\ref{compakin}. One of the basic features of the
kinetics in \figurename~\ref{compakin}, is the presence of a
saturation tendency after a sufficiently high number of pulses. In
addition, the saturated concentration [$E'$]$_{S}$ varies with
incident power. This simple observation has an important
consequence: the saturation during UV irradiation is not ascribable
to exhaustion of the Si\,--\,H precursors, whose initial
concentration in the as-grown material would determine [$E'$]$_{S}$
regardless of the irradiation conditions. Then, the saturation must
arise from the reaching of an equilibrium between two competitive
process, the photo-induced generation and a concurrent depletion
mechanism of the induced $E'$ population. Now, we know that after
the laser is switched off, reaction (\ref{reacteh2rep}) causes the
decay of almost all the defects initially produced; besides, the
typical time scales of growth and decay ($\sim$10$^3$\,s) are
comparable. Due to the behavior of $E'$ in the post-irradiation
stage, we argue that reaction (\ref{reacteh2rep}) is also effective
in the interpulse time span, and is a possible candidate for the
depletion mechanism. So, even the growth kinetics of the defects is
conditioned by the interplay between pulse-induced generation
(\ref{generation}) and interpulse annealing (\ref{reacteh2rep}).
\begin{figure}[htb!]
\begin{center}
\includegraphics[width=.7\textwidth]{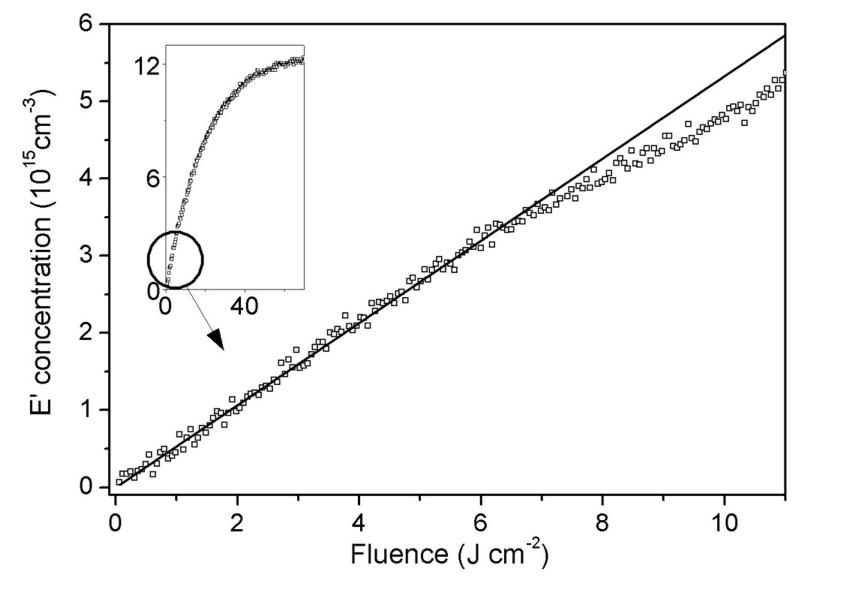}
\end{center}
\caption{Zoom of the first portion of the growth curve $\Lambda_3$
of the left panel of \figurename~\ref{compakin}, and reported again
in the inset. The continuous line is obtained by a linear
least-square fit of the data in the first $\sim$3\,Jcm$^{-2}$.}
\label{zoomsalita}\end{figure}

\begin{figure}[htb!]
\begin{center}
\includegraphics[width=.9\textwidth]{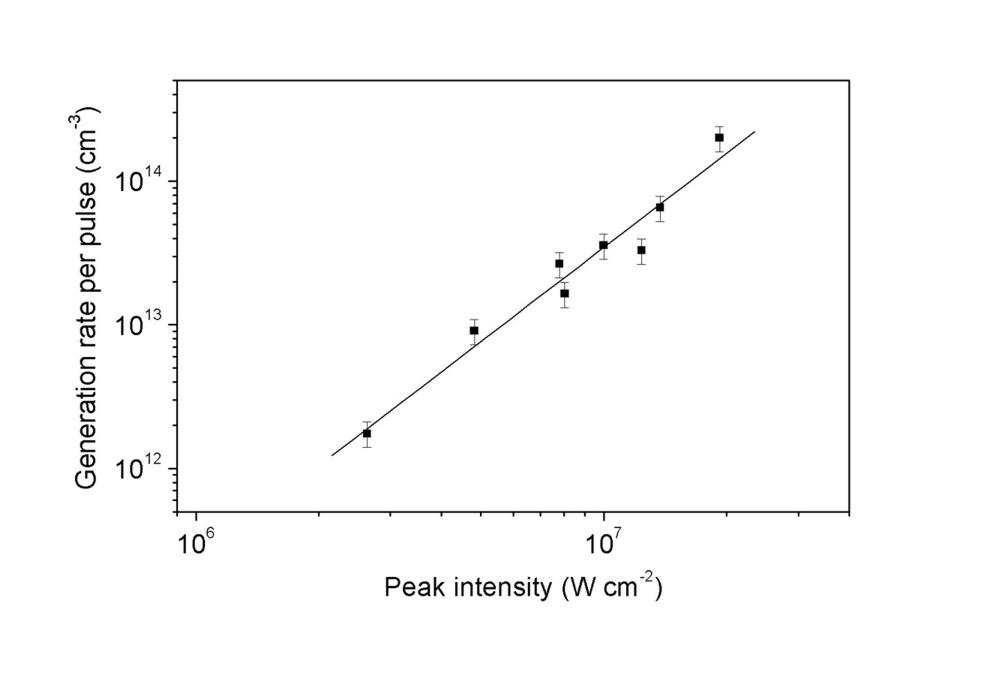}
\end{center}
\caption{Generation rate per pulse $R$ as a function of laser peak
intensity $\Lambda$. The continuous line in the plot is a
least-square fit with the function $R=a\Lambda^b$. In addition to
the five irradiations in \figurename~\ref{gammavseprimo}, three more
exposures were performed to check the repeatability of the results
and to extend the explored range of $\Lambda$.}
\label{rvslambda}\end{figure}

Pulse-induced generation can be isolated by looking at the first
stages of the growth of $E'$, when the growth curve is still
approximately linear because the annealing process is still too slow
to compete with the generation rate. So, in
\figurename~\ref{zoomsalita}, we report a zoom of one the kinetics
in \figurename~\ref{compakin} comprising the first
$\sim$11\,Jcm$^{-2}$. By a linear fit, we can determine the initial
growing slope d[$E'$]/d$\Phi$($\Phi$=0). Then, we convert the
so-obtained values to generation rates \emph{per pulse} $R$:
\begin{equation}
R=\frac{d[E']}{dN}(N=0)=\Lambda\tau\frac{d[E']}{d\Phi}(\Phi=0)
\end{equation}where the laser pulse duration $\tau$ is 5\,ns. The quantity
$R$ is finally plotted in \figurename~\ref{rvslambda} as a function
of the peak laser intensity $\Lambda$.  By a least-square fit of
these data with the function $R=a\Lambda^b$ we obtain
$b$=2.2$\pm$0.2; this means that the behavior of $R$ is consistent
with a quadratic dependence from $\Lambda$.
 The data in
\figurename~\ref{rvslambda} allow to address an important feature of
the laser-induced breaking process of Si\,--\,H (\ref{generation}).
Indeed, the quadratic dependence of $R$ on peak laser intensity
demonstrates that \emph{two-photon processes} are involved in $E'$
generation.

The UV absorption properties of Si\,--\,H are basically unknown at
the moment \cite{SkujaSPIE01}; however, we can say that a two-photon
mechanism is qualitatively consistent with current, though
incomplete, knowledge on Si\,--\,H group: indeed, this center does
not show any measurable OA at energies below the silica bandgap, and
its lowest transition was predicted by theoretical calculations to
be at $\sim$9\,eV, leading to an anti-bonding state
\cite{Robertson88}. Hence, the simplest $E'$ generation mechanism
consistent with present results is two photon absorption by
Si\,--\,H leading to the excited state with consequent breaking of
the bond. Yet, other nonlinear processes are conceivable, such as
production of excitons by two photon absorption followed by
non-radiative decay on Si\,--\,H. In this sense, our work provides
valuable experimental support to \emph{ab initio} theoretical
calculations on the absorption properties of the hydrogen-related
precursor.

As reviewed in detail in subsection \ref{geneprimosection}, several
works in literature have discussed the generation mechanisms of $E'$
in \sil{} under laser radiation, distinguishing between single- and
multi-photon processes. Nevertheless, an important distinction must
be made at this point: indeed, most of these investigations dealt
with \emph{permanent} defects, while only a few have directly
observed the transient $E'$ centers originating from the Si\,--\,H
precursor \cite{LeclercJNCS92,SmithAO00,ShimboJJAP01}. The reason
for this fact is clear: due to the transient nature of the generated
$E'$, the comprehensive study of process (\ref{generation}) requires
\emph{in situ} measurements, which have only been widely available
for a few years and are the most appropriate way to find out whether
UV-induced Si\,–-\,H breaking is a single- or two-photon process.

Comparing now the present results with previous \emph{in situ}
studies, we see that only in two papers
\cite{LeclercJNCS92,ShimboJJAP01} was the dependence on laser energy
density of the initial generation rate of transient $E'$ centers
studied in synthetic \sil{} exposed to KrF (5.0\,eV) or ArF
(6.4\,eV) laser radiation; it was reported to be linear
\cite{LeclercJNCS92} and sublinear \cite{ShimboJJAP01}. These
results are in disagreement with the quadratic dependence found
here, and also difficult to reconcile with the theoretical
predictions on UV absorption properties of Si\,--\,H
\cite{Robertson88}. At the moment the reason for this discrepancy is
unclear, but it may be related to differences in the materials
employed, leading to the activation of different mechanisms leading
to Si\,-–\,H breaking, or to the coexistence of (transient) $E'$
centers not arising from Si\,-–\,H, whereas the present samples
permit a selective observation of process (\ref{generation}).

Finally, we point out that the information derived here may also be
relevant for the understanding of other systems, such as Si/SiO$_2$
interfaces, where Si\,--\,H breakage is a very common process
resulting in the formation of P$_b$-type interface centers (whose
structure is Si$_3$$\equiv$Si$^{\bullet}$ at the (111)Si/SiO$_2$
interface), this being an important mechanism of degradation of
microelectronics devices \cite{Poindexterreview,Browerreview}. In
particular, under laser radiation, breakage of Si\,--\,H at
Si/SiO$_2$ interfaces has been observed to occur either by a
photothermal mechanism or by direct photolysis\cite{PuselPRL}. In
these systems, theoretical work has fixed the bonding-nonbonding
electronic transition of Si\,--\,H to be at 8.5\,eV
\cite{BeckerPRL}, not far from the value of 9\,eV found in
\sil{}\cite{Robertson88}; consistently, Si\,--\,H photolysis was
observed to occur by single-photon absorption of F$_2$ laser
(7.9\,eV) radiation\cite{PuselPRL}. Taking into account the lower
laser energy used in the present work, it appears that the
photochemical Si\,--\,H breaking process may be quite similar in the
two systems. On the other hand, breaking of Si\,--\,H under less
energetic photons occurs efficiently by a \emph{photothermal}
mechanism peculiar to the Si/SiO$_2$ system, in which the Si\,--\,H
bond is broken by providing the $\sim$2.6\,eV dissociation energy
via heating due to strong absorption of the laser light by the
silicon substrate\cite{PuselPRL,BrowerPRB90,TuttlePRB99}.
\subsection{Discussion III: Generation kinetics}
We now proceed to address the issue of quantitatively modeling the
growth kinetics of the defects. As above discussed, the growth of
$E'$ concentration is conditioned by the interplay between
two-photon laser-induced generation, and concurrent annealing due to
reaction (\ref{reacteh2rep}). In this scheme, the kinetics of [$E'$]
on the scale of many laser pulses should be described by the
following rate equation, provided that N is approximately treated as
a continuous variable:
\begin{equation}
\frac{d[E']}{dN}\sim\frac{\Delta[E']}{\Delta N}\sim R-2k_0\Delta
t[E'][\text{H$_2$}]\label{re1}
\end{equation}where $R$ is the generation rate plotted in \figurename~\ref{rvslambda},
which equals the initial growth slope of the kinetics, while the
negative term accounts for the decrease of [$E'$] during the
$\Delta$t = 1\,s interpulse due to annealing with H$_2$. Moreover,
the reaction constant k$_0$ can be fixed to the value obtained by
the fit in \figurename~\ref{gammavseprimo}. Then, \textbf{(i)} due
to the correlated generation of $E'$ and hydrogen of
eq.~(\ref{generation}), \textbf{(ii)} supposing fast dimerization of
H after creation, and \textbf{(iii)} based also on the absence of
dissolved H prior to laser exposure, we have [H$_2$]=[$E'$]/2.
Substituting in eq.~(\ref{re1}) we obtain:
\begin{equation}
\frac{d[E']}{dN}\sim R-k_0\Delta t[E']^2\label{re2}
\end{equation}

We found that this rate equation is in disagreement with the
experimental kinetics, being in particular incapable of predicting
the observed saturation values of [$E'$]. In fact, the saturation
concentration of [$E'$], found from eq.~(\ref{re2}) when
d[$E'$]/dN=0, is given by [$E'$]$_{S}$ = (R/k$_0$$\Delta$t)$^{1/2}$;
for instance we obtain [$E'$]$_{S}$ =
2.0$\times$10$^{16}$\,cm$^{-3}$ for the kinetics at
$\Lambda$=12$\times$10$^6$\,Wcm$^{-2}$, which is higher than the
actual [$E'$]$_{S}$ = 1.2$\times$10$^{16}$\,cm$^{-3}$ in
\figurename~\ref{compakin}. A similar situation is found for all the
kinetics. More to the point, we verified that the agreement with
experimental data is not improved by taking into account the
statistical distribution of the diffusion parameters of H$_2$, which
has been proposed in literature to take into account the amorphous
structure of the silica matrix
\cite{GriscomJNCS84,KajiharaPRB06,KajiharaPRL02,ShelbyKeeton}. This
implies that annealing driven by H$_2$-diffusion (reaction
\ref{reacteh2rep}) significantly slows down the growth of the
defects, but alone is insufficient to explain the observed
saturation concentrations, which are still lower than predicted; for
this reason, an additional negative term has to be added to the rate
equation.

\begin{figure}[htb!]
\begin{center}
\includegraphics[width=.9\textwidth]{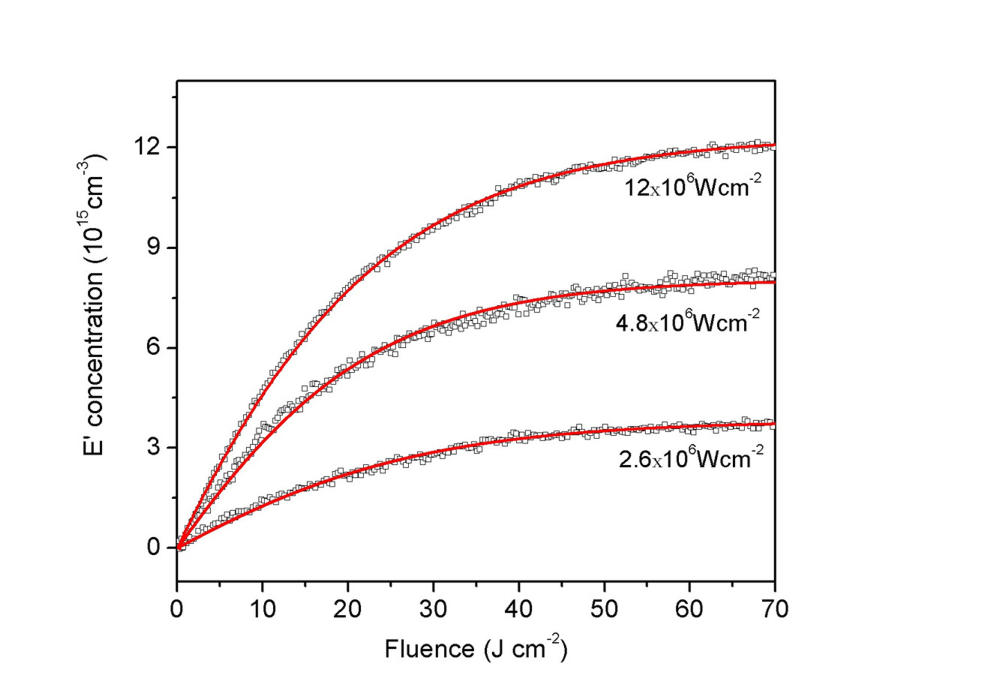}
\end{center}
\caption{Growth kinetics of $E'$ from \figurename~\ref{compakin}
fitted by eq.~(\ref{solutionre}) with respect to the free parameter
$\alpha$.} \label{fits}\end{figure} On this basis, we found
empirically that all the kinetics can be satisfactorily fitted by
adding a linear term to eq.~(\ref{re2}) as follows:
\begin{equation}
\frac{d[E']}{dN}= R(1-\alpha[E'])-k_0\Delta t[E']^2\label{re3}
\end{equation}this being equivalent to supposing a (linear) concentration
dependence of the generation rate. The analytical solution of
eq.~(\ref{re3}) with the initial condition [$E'$]($N$=0)=0, and
changing variable back to fluence ($\Phi$=$\Lambda\tau N$) is the
following:
\begin{equation}
[E'(\Phi)]=\frac{\gamma}{k_0\Delta
t}\left[\tanh\left(\frac{\gamma}{\Lambda
\tau}\Phi+\beta\right)-\tanh\beta \right] \label{solutionre}
\end{equation}
where \begin{equation*} \gamma=\frac{\alpha R}{2}
\sqrt{1+4\frac{k_0\Delta t}{\alpha^2 R}}\qquad \text{and} \qquad
\tanh\beta=\frac{\alpha R}{2\gamma}
\end{equation*}
Hence, the growth stage of the three representative kinetic curves
from \figurename~\ref{compakin} was fitted with the function
(\ref{solutionre}), depending on the free parameter $\alpha$. The
results are reported in \figurename~\ref{fits}: it is seen that the
predicted curves are consistent with experimental data. Similar
results are obtained in all Nd:YAG irradiation experiments on wet
natural silica. The values of $\alpha$ from the fits all fall in the
interval (5.7$\pm$1.4)$\times$10$^{-17}$\,cm$^{3}$.

We discuss now the possible interpretations of the linear term
$-R\alpha[E']$. The rate equation (\ref{re1}) has been written in
the approximation of an infinite population of Si\,--\,H. Actually,
even if the saturation of [$E'$] is due to the equilibrium between
generation and annealing (as above discussed), it is anyway possible
that the progressive reduction of the precursors has to be accounted
for in the generation term. In this respect, we note that
eq.~(\ref{re3}) can be rewritten as follows
\begin{equation}
\frac{d[E']}{dN}= R'(\alpha^{-1}-[E'])-k_0\Delta t[E']^2\label{re4}
\end{equation}where $R'=R\alpha$; this equation can be seen as the generalization of (\ref{re1}) in the case of a finite
concentration $\alpha^{-1}$ of pre-existing Si\,--\,H. This
interpretation also seems to be corroborated by the fact that
$\alpha^{-1}\sim$1.8$\times$10$^{16}$\,cm$^{-3}$ is found by the
fits to be reasonably independent of $\Lambda$. Hence, to better
understand this point, we compared this predicted value of the
Si\,--\,H concentration in the as-grown specimens with the results
of Raman measurements.

\begin{figure}[htb!]
\begin{center}
\includegraphics[width=.9\textwidth]{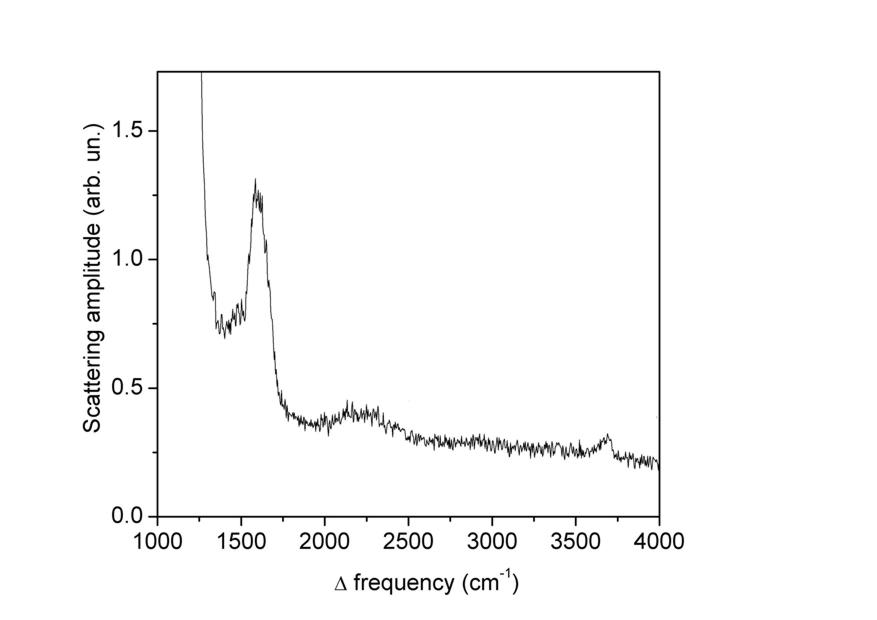}
\end{center}
\caption{Raman spectrum of an as-grown wet natural silica sample.
The signal between 2000\,cm$^{-1}$ and 2500\,cm$^{-1}$ supposedly
comprises a contribution (expected at 2250\,cm$^{-1}$) due to the
Si\,--\,H centers, while the signal near 3700\,cm$^{-1}$ is due to
Si\,--\,OH impurities.} \label{ramanprofile}\end{figure}

A typical Raman spectrum of a HER1 sample is shown in
\figurename~\ref{ramanprofile}. The spectrum shows a weak and broad
signal between 2000\,--\,2500\,cm$^{-1}$, in the same spectral
region in which it is known to peak the typical 2250\,cm$^{-1}$
signal of Si\,--\,H groups in \sil{}.\cite{Shelby94,Schmidt98}
However, the width of the signal in \figurename~\ref{ramanprofile}
appears to be significantly larger than that reported in literature
for Si\,--\,H ($<$100\,cm$^{-1}$). Hence, it is likely that other
components aside from Si\,--\,H contribute to the measured signal, a
possibility being the intrinsic vibrational peak of the silica
matrix whose frequency is reported in literature to fall near the
Si\,--\,H signal.\cite{Schmidt98,Shelby94}. Hence, our present data
permit only to fix a superior limit for [Si\,--\,H]. To this
purpose, if one tentatively ascribes all the amplitude of the signal
in \figurename~\ref{ramanprofile} to Si\,--\,H, the corresponding
concentration of the defect can be calculated by comparison with
literature data and exploiting the signal (out of scale in
\figurename~\ref{ramanprofile}) near 800\,cm$^{-1}$ as a benchmark
to take into account different instrumental sensitivity
\cite{Schmidt98}. In this way we obtain
[Si\,--\,H]$\sim$5$\times$10$^{17}$\,cm$^{-3}$. This value is more
than one order of magnitude higher than $\alpha^{-1}$. If taken
literally, this estimate suggests that the linear term cannot arise
from the finite number of precursors, which is much higher than the
maximum concentration of attainable [$E'$]. We cannot exclude,
however, a much smaller Si\,--\,H signal concealed under the broad
component observed in \figurename~\ref{ramanprofile}. While the
present data do not allow to provide a definite answer to this
point, we discuss in the following other two possible explanations
of the linear term.

One of them may be the reaction of [$E'$] with molecular hydrogen
already dissolved in the sample before laser exposure, in
concentration [H$_2$]$_0$. In fact, in this case the relation
[H$_2$]=[$E'$]/2 deriving from (\ref{generation}) becomes
[H$_2$]=[H$_2$]$_0$+[E']/2, so that eq.~(\ref{re2}) is generalized
as follows:
\begin{equation}
\frac{d[E']}{dN}= R-2k_0\Delta t[\text{H$_2$}]_0[E']-k_0\Delta
t[E']^2\label{re4}
\end{equation}giving rise to a linear term (whose coefficient should be independent of laser intensity).

Now, as already observed, the linear correlation between $E'$ and
H$_2$ at the end of exposure (\figurename~\ref{gammavseprimo})
excludes the presence of H$_2$ in significant concentrations prior
to irradiation. To make quantitative this argument, we note that in
presence of [H$_2$]$_0$, eq.~(\ref{gammareltoeprimo}) would be
modified as follows:
\begin{equation}
\Gamma=2k_0[\text{H$_2$}]_0+k_0[E']_S
\end{equation}introducing a nonzero intercept in \figurename~\ref{gammavseprimo}.
On this basis, from the data one can fix a limit for [H$_2$]$_0$:
[H$_2$]$_0$$<$10$^{15}$cm$^{-3}$, which in turn gives the following
estimate of the linear coefficient,
2k$_0$$\Delta$t[H$_2$]$_0$$\sim$10$^{-4}$. However, this is between
one and two orders of magnitude smaller than the values of $R\alpha$
coming from the fits (except at the two lowest laser intensities
used). So, this mechanism at the basis of the linear term can be
excluded.

Finally, another qualitative interpretation can be tentatively
proposed, based on a more accurate discussion of the generation
process (\ref{generation}): in general, a H produced by
(\ref{generation}) diffuses in the matrix and may experience two
different fates: it can either meet another H and dimerize in H$_2$
or it can come across an $E'$ and passivate it; for this reason,
aside from the slow annealing due to reaction (\ref{reacteh2rep}),
the $E'$ centers undergo a much faster decay (FD) due to
recombination with a portion of the H population made available by
each pulse. In addition, the portion of H involved in the FD is
expected to increase with $E'$ concentration, which enhances the
probability of encountering an $E'$ before meeting another H. A FD
stage with similar features was directly observed \emph{in situ} for
NBOHC produced by F$_2$ laser irradiation at T$_0$=300\,K, and
occurs on a typical timescale which is shorter than the 1\,s
interpulse\cite{KajiharaAPL01}. Besides, the FD is not accounted for
in the stationary-state approximation for H, which considers only
the H produced at the right side of eq.~(\ref{reacteh2rep}): in
other words, during the irradiation the stationary-state
approximation is expected to fail just after each pulse due to the
excess H produced by (\ref{reacteh2rep}) which is still to dimerize.
From the experimental standpoint, since the FD cannot be directly
observed with the time resolution available here, it is incorporated
\emph{de facto} in the generation term, which must actually be
interpreted as the net concentration of [$E'$] generated by
(\ref{generation}) and surviving fast recombination with H. Now,
given that the portion of H quickly recombining with [$E'$] must
increase with $E'$ concentration as the irradiation session
progresses, we expect a consequent reduction of the generation rate
from its initial value $R$, which to a first approximation can be
represented by a linear term in [$E'$], as in (\ref{re3}).

We acknowledge that understanding the origin of the linear term is
not an easy problem, and that more measurements are surely needed to
conclusively solve it. However, it is worth pointing out the main
merits of the simple model proposed here: it allows to describe the
kinetics and the saturation of $E'$ as a consequence of
hydrogen-related reactions, and it is able to reproduce independent
datasets coming from several irradiation sessions with only one free
parameter.

\begin{figure*}[t!]
\begin{center}
\fcolorbox{postitA}{postitA}{\begin{minipage}[c]{.95\textwidth}\linespread{1.2}\hrule\vspace{0.2cm}
\textbf{\textsc{Summary II. }}\textsl{\textcolor{postitC}{In wet
natural silica, the dependence of the generation rate of $E'$
centers from laser intensity is found to be quadratic
(\figurename~\ref{rvslambda}). This suggests that two-photon
processes are involved in the photochemical generation mechanism of
$E'$, eq.~(\ref{generation}). Moreover, a peculiar feature of the
generation process of $E'$ from Si\,--\,H is to produce mobile
hydrogen together with $E'$ centers. As a consequence, the growth of
$E'$ during irradiation is limited by the simultaneous occurrence of
reaction (\ref{reacteh2rep}). This leads to a characteristic growth
curve (\figurename~\ref{fits}) which can be satisfactorily modeled
by a suitable rate equation, eq.~(\ref{re3}). The ability of
eq.~(\ref{re3}) to fit the growth stage of the kinetics of $E'$
centers is a further, albeit indirect, confirmation of the validity
of our interpretation eq.~(\ref{generation}) of the generation
process of the defect. \vspace{0.2cm}}\hrule}\end{minipage}}
\end{center}
\end{figure*}

\subsection{Previous literature on the \texorpdfstring{Si\,--\,H}{Si-H}
generation model} In a sense, the Si\,--\,H model provides a
particularly straightforward interpretation of the effects of pulsed
Nd:YAG laser irradiation in natural wet \sil{}: $E'$ and H are
generated together by laser light and recombine in the
post-irradiation stage, so that the matrix virtually returns to the
native state after the decay is completed. Also, this process
appears the analogue on Si\,--\,H to the generation of NBOHC from
Si\,--\,OH groups (Reaction \ref{breakingohrep}), quite clearly
demonstrated in several works
\cite{GriscomJNCS84,KajiharaPRL02,KajiharaJNCS03}.

Generation of $E'$ from Si\,--\,H under laser irradiation has been
repeatedly proposed in literature (see subsection
\ref{geneprimosection}), most of the time on the basis of
qualitative considerations. A convincing evidence of this process
has been provided by Imai \emph{et al.}, who irradiated an
oxygen-deficient sample preliminary treated at high temperature in
H$_2$ atmosphere, so as to strongly increase the concentration of
Si\,--\,H by the reaction:
\begin{equation} \label{genSi-HdaSiSIrep}
\equiv\text{Si\,--\,Si}\equiv\text{ + H$_2$ } \longrightarrow \text{
}\equiv\text{Si\,--H + H\,--Si}\equiv\end{equation}
 As a consequence of reaction (\ref{genSi-HdaSiSIrep}), Si\,--\,H becomes one of the dominant
impurities in the material. Then, Imai \emph{et al.} reported that
after treatment in H$_2$, the generation efficiency of $E'$ is
strongly increased, supposedly because Si\,--\,H serves as a
precursor for the paramagnetic center. One of the possible
generation mechanism of $E'$ proposed by Imai \emph{et al.} to apply
in this case is:
\begin{equation} \label{genEdaSi-HArep}
\equiv\text{Si\,--H + H\,--Si}\equiv\text{ }\longrightarrow \text{
}\equiv\text{Si}^{\bullet}\text{ H\,--Si}\equiv\text{ +
$\frac{1}{2}$ H$_2$}\end{equation} where the first term at the right
side is to be considered an $E'$ as it contains the
($\equiv$Si$^{\bullet}$) structure. To further confirm the role of
Si\,--\,H as a precursor, Imai \emph{et al} also observed that
\emph{a portion} of the paramagnetic centers decay in the
post-irradiation stage, supposedly due to reaction with H$_2$. The
circumstance that the decay was not complete indicates either that
other precursors still contributed to the generation of $E'$ or that
a portion of H$_2$ was involved in a reaction with another
paramagnetic defect.

Some interesting observations arise from comparison of our results
with those by Imai \emph{et al}, obtained on samples in which the
Si\,--\,H precursor should be present mainly in the form predicted
by eq.~(\ref{genSi-HdaSiSIrep}), i.e. a \emph{hydrogen-decorated
vacancy}. Indeed, in general it is conceivable for Si\,--\,H to
exist also as an isolated defect, depending on the reactions that
control the incorporation of such impurity during the manufacturing
procedure of the material.\footnote{A further possibility is that
Si\,--\,H defects are coupled to nearby Si\,--\,OH, if both are
formed by reaction of H$_2$ with the silica matrix within the
atmosphere in which the material is produced.\cite{Doremus,Shelby94}
} For the moment, our results do not allow to understand the
prevalent form in which Si\,--\,H is incorporated in our samples;
more in general, these considerations suggest the need of further
studies to understand if different varieties of Si\,--\,H actually
exist in \sil{} and whether their UV-induced breaking process are
different. However, it is worth noting the following: if the
H-decorated vacancy happened to be the most common form of Si\,--\,H
in \sil{} also in the case of \emph{non}-H$_2$-loaded samples (such
as our materials), then it is possible to argue that the generation
process eq.~(\ref{generation}) proposed in this chapter should be
re-interpreted as eq.~(\ref{genEdaSi-HArep}). This would not
invalidate the essence of our conclusions, but it would give further
information on the structure of the $E'$ center produced in our
experiments. In fact, while the $E'$ produced from an isolated
Si\,--\,H is simply an (isolated) silicon dangling bond
($\equiv$Si$^{\bullet}$), the defect at the right side of
eq.~(\ref{genEdaSi-HArep}) comprises a nearby Si\,--\,H group,
thereby coinciding with what in literature has been generally called
$E'$$_{\beta}$\footnote{The $E'$$_{\beta}$ was originally introduced
as the reaction product between an oxygen vacancy and a hydrogen
atom, giving the same result as reaction (\ref{genEdaSi-HArep})
\cite{GriscomJNCS84,Erice}. Actually its microscopic model
presupposes a structural relaxation (not represented in
(\ref{genEdaSi-HArep})) at the end of which the unpaired electron
points away from the former vacancy. }\cite{GriscomJNCS84,Erice}.
The $E'$$_{\beta}$ is distinguishable in principle from the other
varieties of $E'$ by ESR, as it features a slightly different
$\underline{g}$ tensor \cite{Erice}. Unfortunately, present ESR data
are not accurate enough to operate such a distinction. We note,
however, that this is an interesting hypothesis that deserves
further investigation, because it implies that the $E'$$_{\beta}$
absorbs at 5.8\,eV, contrary to the fact that a 5.4\,eV band has
been provisionally ascribed to the defect\footnote{However, we
stress that the identification of our $E'$ with the $E'$$_{\beta}$
is not a necessary consequence of the decorated vacancy hypothesis;
indeed, Imai \emph{et al.} proposed an alternative photochemical
reaction, eq.~(\ref{genEdaSi-HB}), which generates the usual
$E'$$_\gamma$. }\cite{SkujaReview98}.

After Imai \emph{et al.}, only a few researchers improved the
experimental approach by using \emph{in situ} optical absorption to
investigate laser-induced generation of $E'$. In particular, Smith
\emph{et al.}\cite{SmithAO00} proposed that laser-induced $E'$ are
unstable only in samples containing free dissolved H$_2$ prior to
irradiation. Their conclusion is apparently at variance with the
Si\,--\,H model, in which H$_2$ should be always available at the
end of irradiation because it is produced radiolytically. In
contrast, in the pioneering work by Leclerc \emph{et al.}
\cite{LeclercJNCS92}, it was already proposed that generation of
transient $E'$ occurs by breakage of Si\,--\,H, but the detailed
description of the generation process involved a two-step mechanism,
in which $E'$ are first produced by unknown precursors, then
converted to an unknown $\overline{E'}$ (later identified with
Si\,--\,H), which successively serves as an efficient precursor for
$E'$ for successive generation upon repeated irradiations of the
same sample. Finally, as already discussed, disagreeing results were
reported in literature when trying to apply \emph{in situ}
measurements to find out if transient $E'$ generation occurs by a
single- or by a two- photon process.

Present results were obtained in a non-H$_2$-loaded silica sample
and improve the understanding of this problems in many aspects.
Despite previous works had proposed Si\,--\,H breaking to be one of
the possible channels contributing to $E'$ generation in
H$_2$-loaded materials, where Si\,--\,H is one of the prevalent
impurities, our evidences suggest this to be a mechanism that can be
dominant in standard non-treated samples. Also, the post-irradiation
instability that leads to the almost complete disappearance of the
induced defects, is now proposed as an \emph{intrinsic} feature of
the $E'$ centers inherent to their very generation process, not
being related to hydrogen dissolved prior to laser exposure.
Furthermore, we have provided rigorous proofs basically leaving
generation of Si\,--\,H as the only possible generation mechanism of
$E'$ in our experimental conditions. In contrast with previous
results, we have obtained a direct proof of a two photon generation
mechanism for Si\,--\,H breaking, consistently with theoretical and
experimental expectations on the UV absorption properties of the
Si\,--\,H groups. Finally, for the first time we have addressed the
issue of quantitatively modeling the growth kinetics of the defects
with a suitable rate equations model.

In this sense, \emph{in situ} OA spectroscopy demonstrates its
usefulness, as it has permitted to study quantitatively the first
stage of the decay, thus yielding the correlation of
\figurename~\ref{gammavseprimo}, and the first stage of the growth,
demonstrating a two-photon mechanism for Si\,--\,H breaking.
Moreover, only the observation of the complete absorption profile
has allowed to exclude the generation of (transient) NBOHC, and in
turn to exclude laser-induced breaking of Si\,--\,OH.

From a wider point of view, it is expected that the generation
processes induced by laser exposure can be triggered as well by
ionizing radiation. Consistently, the formation of $E'$ from
Si\,--\,H precursors has been proposed in literature also as a
mechanism activated by $\gamma$ or $\beta$
irradiation\cite{ImaiPRB93,ImaiJNCS94,Erice,Devinebook}. In
particular, a recent investigation has supported this generation
model on the basis of an accurate study of the ESR lineshape of
$\gamma$-induced $E'$ centers, carried out in several commercial
\sil{} samples including those used in the present
work\cite{BuscarinoTESI}. In this context, we remark that also in
the case of $\gamma$ irradiation, the study of the \emph{in situ}
kinetics and of the post-irradiation stability of $E'$ may allow to
infer additional information complementary to what is known from the
\emph{ex situ} investigations.

As a final remark we add that the present results, together with all
the others in literature promoting the Si\,--\,H generation model,
suggest an important hint on the still open problem of understanding
the nature of the 5.8\,eV transition leading to the $E'$ absorption
band. In fact, as anticipated in the introduction, some theoretical
calculations \cite{PacchioniPRB98b,PacchioniPRL98} have suggested
the band to be due to a \emph{charge transfer} transition from the
silicon dangling bond ($\equiv$Si$^{\bullet}$) to the charged
Si$^{+}$, which is expected in front of it when the $E'$ is
generated by ionization of a vacancy:\begin{equation}
\label{genEdasisirep} \equiv\text{Si\,--\,Si}\equiv\text{ }
\longrightarrow \text{ }\equiv\text{Si}^{\bullet}\text{
 $^{+}$}\text{Si}\equiv\end{equation} the $E'$ at the right side
of eq.~(\ref{genEdasisirep}) is what we have defined
\emph{vacancy}-$E'$. On the other hand, this interpretation of the
OA band has been questioned by other authors\cite{OReillyPRB83}, and
calculations have also predicted another weaker absorption, falling
in the same energy region but due to a transition entirely comprised
within the basic $\equiv$Si$^{\bullet}$ moiety of the defect
\cite{SkujaReview98,PacchioniPRB98b}. In this context, our results
strongly suggest that an $E'$ absorbing in the "standard" 5.8\,eV
band may come from Si\,--\,H breakage. So-produced $E'$ is expected
to be structurally different from \emph{vacancy}-$E'$, not
containing the Si$^{+}$ portion and being more similar to an
isolated dangling bond. Hence, the lack of the Si$^{+}$ fragment
excludes the charge transfer model, and we conclude that our results
point towards the \emph{intra}-dangling-bond interpretation of the
the 5.8\,eV electronic transition.

\section{$E'$ center in other types of \texorpdfstring{\sil}{silica}}
\subsection{Dry natural silica}\label{dns}
\hspace{0.8cm}We have seen in the previous chapter that the
generation of $E'$ centers under 4.7\,eV pulsed laser irradiation
occurs as well in \emph{dry} natural silica materials, and its
efficiency is comparable with \emph{wet} natural silica. Moveover,
also in the dry materials the $E'$ centers are unstable in the
post-irradiation stage due to reaction with mobile H$_2$.
Nonetheless, there is an important difference between the two cases.
In fact, while the $E'$ generated in the wet samples are almost
completely annealed after the end of exposure by reaction
(\ref{reacteh2rep}), in the dry samples the passivation is
incomplete, meaning that only a fraction of the $E'$ are annealed
and a residual concentration remains in the sample after that the
reaction is completed. This in turn implies that the concentration
[H$_2$]$_S$ of hydrogen available at the end of irradiation must be
insufficient to completely cancel the defects, i.e.
[H$_2$]$_S<$[$E'$]$_S$/2, differently from the wet samples where, as
a consequence of the generation process (\ref{generation}), we have
[H$_2$]=[$E'$]/2.
\begin{figure}[h!]
\begin{center}
\includegraphics[width=.9\textwidth]{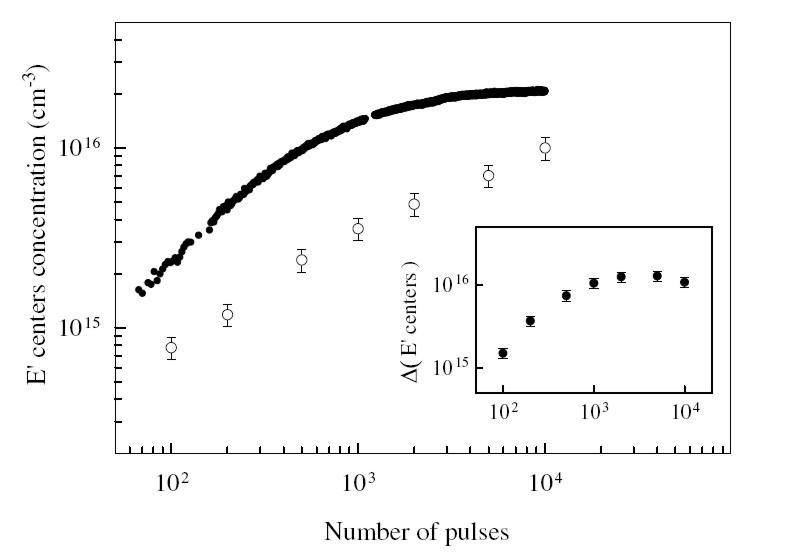}
\end{center}
\caption{Concentration of $E'$ measured \emph{in situ} during laser
irradiation (full symbols), and stationary concentrations measured
with ESR at the end of the post-irradiation annealing process after
exposure to different laser of pulses (empty symbols). Inset:
difference between the two curves, representing the post-irradiation
reduction of the $E'$. Figure taken from Messina \emph{et
al.}\cite{MioJNCS05EPRIMO} } \label{q906differenza}\end{figure}

To better understand this issue, we performed the following
experiment on natural dry Q906 samples. A set of virgin samples were
irradiated with different numbers of pulses ranging from 100 to
10000, using a pulse repetition rate of 1\,Hz and an energy density
of 40\,mJcm$^{-2}$ per pulse. For each sample, we waited a few days
after the end of the irradiation session, so as to be sure that the
post-irradiation kinetics was completed; then, the stationary
concentration [$E'$]$_\infty$ was estimated by ESR. On another
virgin sample it was measured \emph{in situ} the growth curve of
$E'$ during irradiation in the same conditions. In
\figurename~\ref{q906differenza} the results of the two measurements
are compared. These data show that the stationary concentration
[$E'$]$_\infty$ follows a sub-linear dependence on the number of
received pulses, very different from the behavior of the
\emph{transient} concentration measured \emph{in situ}, and
different also from wet natural silica, where there is no
accumulation of $E'$ with dose, as they are almost completely
annealed after the end of irradiation independently of the received
number of pulses. In addition, the difference $\Delta [E']$ between
transient and stationary concentration, reported in the inset,
depends on the irradiation dose as well. Analogous results are
obtained on I301 dry fused silica samples.

Now, the post-irradiation decrease $\Delta [E']$, equals two times
[H$_2$]$_S$, so as to represent another indirect measurement of the
amount of available hydrogen at the end of exposure. In this sense,
the results in \figurename~\ref{q906differenza} demonstrate that
\emph{also} in natural dry silica hydrogen has a \emph{radiolytic
origin}, since its concentration grows with dose. Furthermore, the
same considerations that demonstrate that NBOHC are not generated in
significant (comparable to $E'$) concentrations in wet samples apply
also here, where the typical induced OA profiles (\figurename
s~\ref{spettrinsitui301} and \ref{cinmater}) do not contain any
appreciable 4.8\,eV band and PL measurements do not detect the
characteristic 1.9\,eV emission.

Based upon these considerations and on comparison with wet samples,
the simplest interpretation is the following. Also in dry samples,
process (\ref{generation}) is the most likely source of the hydrogen
responsible for the decay of $E'$, and generates a \emph{portion} of
the total $E'$ population. However, in dry glass a second formation
channel of $E'$ apart from (\ref{generation}) is active: as a
consequence, concentration of generated $E'$ is higher than
[H$_2$]$_S$/2, so that available molecular hydrogen is insufficient
to completely cancel the induced defects.\footnote{We have performed
Raman measurements also on dry fused silica samples: they show a
signal similar to that in \figurename~\ref{ramanprofile}, which
supposedly comprises the contribution at 2250\,cm$^{-1}$ due to the
Si\,--\,H groups. In this case, however, the presence of Si\,--\,H
in the I301 glass, in concentrations exceeding 10$^{18}$ cm$^{-3}$,
was reported also by the manufacturer of the material \cite{Kuen}.
This supports the role of Si\,--\,H as a precursor of $E'$ also in
dry silica, and suggests that the Si\,--\,H signal significantly
contributes to the broad component between 2000\,--\,2500\,cm$^{-1}$
observed in our Raman spectra. } In this respect it is worth
stressing that the entity of the post-irradiation decay of $E'$ is
comparable in the two materials (confront
\figurename~\ref{q906differenza} with the typical concentrations of
$E'$ in wet, as in \figurename~\ref{compakin}, or see
\figurename~\ref{cinmater}). This suggests that dry and wet silica,
although differing of more than one order of magnitude for what
concerns the concentration of Si\,--\,OH impurities (Table
\ref{materials}), seem to be unexpectedly similar in the content of
Si\,--\,H groups.

Within the proposed scheme, the population of $E'$ centers consists
in two portions: the defects $E'$$_H$ coming from Si\,--\,H, plus a
contribution $E'$$_X$ arising from the second formation process.
Since the concentration of hydrogen made available by each laser
pulse is one-to-one correlated to the $E'$$_H$ component, the
stationary concentration [$E'$]$_\infty$ in
\figurename~\ref{generation} is related only to the presence of
$E'$$_X$: more precisely, [$E'$]$_\infty$ equals the total
concentration of defects generated from the precursor $X$ during the
irradiation session\footnote{This does not imply, however, that
[$E'$]$_\infty$ equals the concentration of $E'$$_X$ \emph{at the
end of exposure}. In fact, since both $E'$$_X$ and $E'$$_H$ react
with H$_2$ (see later), [$E'$$_X$] at the end of exposure is already
depleted of the portion of defects which have been passivated by
H$_2$ during the irradiation session. What is true is only that
[$E'$]$_\infty$ (times the volume of the sample) equals the number
of generation \emph{events} from the precursor $X$ which have
occurred during all the irradiation session. }. Dry and wet
materials also differ in the behavior under repeated irradiations:
in fact, while wet samples appear to be "elastic" upon
re-irradiation, in the dry glass (\figurename~\ref{cinrip}) the
presence of the additional $E'$$_X$ component leads to the
progressive accumulation of the defects with increasing number of
exposures. At the moment it is not possible to identify the
generation process of $E'$$_X$, although the presence of oxygen
deficient centers (Si\,--\,Si vacancies) in dry fused
glass\cite{CannasJNCS01} suggests this defect to be a possible
precursor. More studies may help to clarify this point.

\begin{figure}[h!]
\begin{center}
\includegraphics[width=0.9\textwidth]{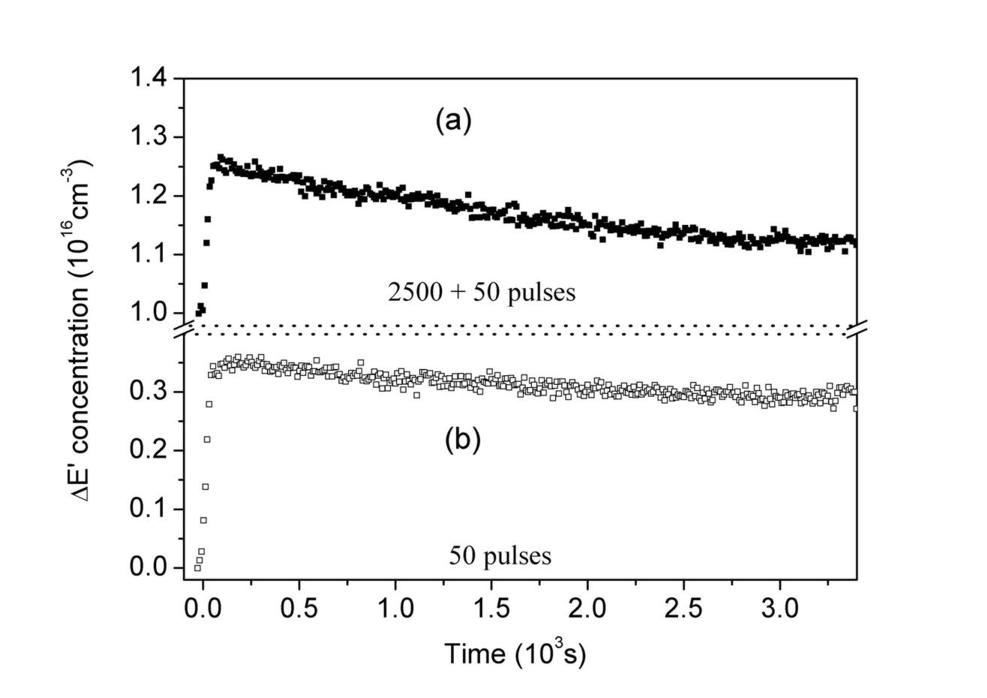}
\end{center}
\caption{Panel (b): Kinetics of [$E'$] induced on an as-grown I301
sample by irradiation with 50 pulses. Panel (a): the same
experiment, but performed on a sample which already contained
[$E'$]$\sim$10$^{16}$cm$^{-3}$ due to a previous irradiation.}
\label{risolutivoi301} \end{figure}

A last subtle question must be addressed to have a comprehensive
scheme of the $E'$ generation-annealing dynamics in these materials.
Indeed, in the case of dry natural silica, it is not obvious \emph{a
priori} if reaction with H$_2$ involves \emph{all} E' centers
generated by irradiation or only the portion coming from the Si-H
precursor. To clarify this issue, we performed the following
repeated irradiation experiment. An I301 natural dry sample was
preliminarily irradiated with N$_1$=2500 laser pulses of
70\,mJcm$^{-2}$ energy density. After the post-irradiation kinetics
is completed, we measured with ESR that
[$E'$]$_\infty$(N$_1$)=(1.0$\pm$0.1)$\times$10$^{16}$\,cm$^{-3}$.
Then, the sample was irradiated a second time with 50 laser pulses.
The concentration kinetics of $E'$ during and after this second
exposure is reported in panel (a) of
\figurename~\ref{risolutivoi301}. For comparison, we report in panel
(b) the result of the same 50 pulses irradiation on an as-grown
sample.
 The two kinetics are different;
indeed, on the as-grown sample, 50 pulses induce the generation of
[$E'$]$_S$(50)=(0.36$\pm$0.04)$\times$10$^{16}$cm$^{-3}$ defects,
$\sim$17\% of which decay after 1 hour of the post-irradiation
stage; in contrast, the re-irradiation of the sample which had
already received 2500 pulses induces a smaller variation of [$E'$]
from [$E'$]$_\infty$(N$_1$):
(0.25$\pm$0.02)$\times$10$^{16}$cm$^{-3}$; most important, the decay
of the induced defects is faster ($\sim$55\% after 1 hour).

To discuss this experimental result, let us consider again the
expression for the reaction rate (\ref{rateeq}). If only the $E'_H$
centers were reactive with H$_2$, [$E'$] on the right side of
(\ref{rateeq}) should be identified with [$E'$$_H$] and the decay
kinetics would be independent of the previous history of the sample,
not influenced by other contributions to the \emph{total} $E'$
concentration. Results in \figurename~\ref{risolutivoi301} are at
variance with this picture, as the decay of the defects induced by
50 pulses is accelerated when the sample, prior to irradiation,
already hosts a concentration [$E'$]$_\infty$(N$_1$) due to a
previous exposure. This is a consequence of the increase of the
right side of (\ref{rateeq}) due to the addition of
[$E'$]$_\infty$(N$_1$) to the total population of reacting centers.
Then we conclude that \emph{all} $E'$ centers participate in the
reaction of H$_2$ independently of their origin.

\subsection{Synthetic silica}\hspace{0.8cm}As described in the previous chapter, no $E'$ centers
generation is observed in synthetic silica upon laser irradiation,
at least within the same intensity range used in the experiments on
natural silica. Within the interpretation proposed so far, the
simplest explanation of this result would be the absence of
Si\,--\,H precursors in synthetic \sil{}. Nonetheless, this seems
somewhat unlikely, since we have seen that both dry and wet natural
\sil{} contain appreciable concentrations of Si\,--\,H, which must
be incorporated during the manufacturing process, and is not
expected to depend on properties of the starting material (quartz
\emph{vs} SiCl$_4$). Raman measurements aimed to investigate the
presence of Si\,--\,H in synthetic S1 and S300 samples (see Table
\ref{materials}) were not able to observe any signal in the
2000\,--\,2500\,cm$^{-1}$ region, because of the presence of an
unknown luminescence signal which made impossible the observation of
the Raman scattering in this spectral range. Finally, differently
from the case of Si\,--\,OH no comprehensive literature data are
available to clarify how the concentration of Si\,--\,H depends on
the type of silica.

Another possible explanation is that $E'$ are not efficiently
generated due to the presence of pre-existing H$_2$ already
dissolved in the as-grown samples in high concentration. In fact, in
this case the generated $E'$ would be immediately annealed by
reaction with H$_2$ so as to prevent their growth. This could be the
case for synthetic \emph{wet} materials, which several authors have
claimed to be impregnated with high (up to
$\sim$10$^{18}$\,cm$^{-3}$) concentrations of hydrogen molecules
even when as-grown \cite{KajiharaPRL02,MorimotoJNCS}. This
hypothesis need to be verified by applying to the samples a
degassing procedure prior to irradiation, or by studying the effects
of irradiation at low ($<$200\,K) temperatures, so as to block
molecular hydrogen mobility.

\begin{figure*}[t!]
\begin{center}
\fcolorbox{postitA}{postitA}{\begin{minipage}[c]{.95\textwidth}\linespread{1.2}\hrule\vspace{0.2cm}
\textbf{\textsc{Summary III. }}\textsl{\textcolor{postitC}{After
discussing the generation of $E'$ in wet natural silica, the last
part of this chapter discusses the experimental results found on
other varieties of amorphous \sil{}. The kinetics of $E'$ centers in
dry natural silica are found to be different from wet natural
silica, in that the post-irradiation decay of the defects induced in
the dry specimens is only partial
(\figurename~\ref{q906differenza}). This can be explained by
supposing a second generation channel of $E'$ active in dry silica
concurrently to Si\,--\,H breaking, so that the concentration of
hydrogen available for reaction (\ref{reacteh2rep}) is always
smaller than that of $E'$ centers. Finally, 4.7\,eV laser
irradiation appears to be basically unable to generate $E'$ centers
in synthetic silica; several interpretations of this finding are
conceivable and discussed in the text.
\vspace{0.2cm}}\hrule}\end{minipage}}
\end{center}
\end{figure*}

Finally, we should consider the possibility of extrinsic impurities
in the material playing some indirect role in $E'$ generation. Even
if the (main) precursor of the paramagnetic center is Si\,--\,H,
impurities may for example serve as intermediate states in \emph{two
step absorption} assisting the generation of hole-electron pairs,
which then could form excitons whose non-radiative decay on the
Si\,--\,H precursor would give the $E'$. In conclusion, the lack of
observation of $E'$ by 4.7\,eV pulsed laser radiation in synthetic
silica remains here as an open issue that requires further
investigation.
\section{Conclusions} \hspace{0.8cm}We
studied the generation and annealing dynamics of $E'$ centers
induced by 4.7\,eV pulsed laser irradiation in \sil{}, mainly on the
basis of \emph{in situ} optical absorption spectroscopy. We first
discussed the process as observed in wet natural silica. The
experimental results indicate that hydrogen responsible for the
decay of $E'$ in the post-irradiation stage has a radiolytic origin.
Furthermore, the generation processes of $E'$ and hydrogen are
correlated, consistently with a model in which the two species are
formed by photo-induced breaking of a common precursor Si\,-–\,H.
The dependence of the initial generation rate on laser intensity is
quadratic, demonstrating a two-photon mechanism for $E'$ generation.
The kinetics and the saturation of the process are the result of
competition between the action of radiation and the annealing of
$E'$ due to reaction with H$_2$. On this basis, a rate equation
model was proposed and tested against experimental data. In dry
natural silica, also a second generation process contributes to the
total concentration of the defects; both types of E' centers
participate in the reaction with H$_2$. The possible reasons of the
lack of $E'$ in irradiated synthetic silicas are discussed from a
qualitative point of view. These results prove the usefulness of
\emph{in situ} time-resolved detection of absorption spectra to
perform comprehensive studies on transient point defect conversion
processes and their effect on the transparency of optical materials
during UV exposure.

\chapter{Post-irradiation conversion processes at room
temperature}\label{PRB} This chapter focuses on the kinetics of
point defects in natural \sil{} after the end of laser irradiation
at room temperature. The results, obtained exclusively by \emph{ex
situ} experimental techniques, provide a deeper understanding of the
processes involving Ge-related impurities and of the kinetics driven
by diffusion and reaction of H$_2$.
\section{Introduction}
\hspace{0.8cm}In the previous chapter we have carried out a detailed
analysis of the \emph{in situ} kinetics of $E'$, permitting in
particular to clarify several aspects of its generation mechanism
and growth kinetics. We make now a step backwards, returning to the
interpretation of post-irradiation processes that in chapter
\ref{overview} was proposed mostly on the basis on qualitative
considerations. Indeed, in our initial discussion, one of the most
convincing reasons to interpret the decay of $E'$ as a consequence
of reaction with mobile H$_2$ has been the observation of the
simultaneous growth of the Ge-related H(II) centers
(=Ge$^{\bullet}$-H), attributed to H trapping on pre-existing GLPC
(=Ge$^{\bullet\bullet}$) impurities:
\begin{equation}\label{trappingsuglpcrep}
=\text{Ge}^{\bullet\bullet}\text{ + H} \longrightarrow\text{
}=\text{Ge$^{\bullet}$\,--\,H}
\end{equation}where atomic hydrogen at the left side of reaction
(\ref{trappingsuglpcrep}) was supposed to be made available by
cracking of H$_2$ on $E'$ centers. Hence, the conversion of GLPC in
H(II) played the central role of \emph{probing} the presence of
diffusing hydrogen, strongly suggesting its involvement also in the
decay of $E'$.

Nevertheless, several questions were left open by our discussion in
chapter \ref{overview}: \textbf{(i)} to convincingly prove our
interpretation of the post-irradiation processes we still have to
demonstrate that the time dependencies of GLPC, H(II) and $E'$ can
be quantitatively accounted for by a chemical kinetics model
describing the diffusion and reactions of mobile H$_2$ and H.
\textbf{(ii)} Although process (\ref{trappingsuglpcrep}) very
reasonably explains the growth of H(II), to prove it we still need
to evidence an anticorrelation between the reduction of
GLPC\footnote{which can be monitored by the reduction
(\emph{bleaching}) of the typical optical activity of GLPC centers,
consisting in the B$_{2\beta}$ absorption band at 5.1\,eV exciting
the two emissions at 3.1\,eV and 4.3\,eV \cite{SkujaJNCS92}. See
\figurename~\ref{bleaching}. } and the growth of H(II).
\textbf{(iii)} The reason why the concentration of H(II) remains
very low \emph{during} irradiation is still to be elucidated.
\textbf{(iv)} Finally, it is interesting to find out if GLPC
subjected to laser irradiation undergoes other conversion processes
parallel to (\ref{trappingsuglpcrep}). For these reasons, in this
chapter\cite{MioPRB05,MioJNCS04,MioJNCS05HII} we extend the
discussion of chapter \ref{overview} on the post-irradiation
processes observed in natural silica, with the aim to clarify these
unsolved problems.

Aside from the present relation with the kinetics of $E'$ center,
there is also another reason motivating us to further investigate
the conversion process of GLPC into H(II). In fact, we recall from
subsection \ref{Gedefsection} that the laser-induced conversion
processes of Ge-related defects in \sil{} currently attract a strong
research interest because of a variety of properties featured by
Ge-doped silica under laser exposure, e.g. photosensitivity and
second harmonic generation, quite relevant for applications and
partly ascribable to Ge-related defects. The existing experimental
investigations on these topics have been invariably carried out on
samples of Ge-doped silica (with typical Ge concentrations of a few
\%), material of choice in the fabrication of optical fibers.
As-grown Ge-doped \sil{} usually features an OA band at
$\sim$5.1\,eV: the defects contributing to this signal are widely
believed to behave as efficient precursors readily convertible in
paramagnetic centers by laser radiation. Nonetheless, while this
absorption band seems to play a central role in laser-induced
processes in Ge-doped silica, the current understanding of the
microscopic conversion mechanisms is still incomplete. Also, the
presence of H$_2$ has been demonstrated to enhance Ge-related
conversion processes and photosensitivity, but the underlying
microscopic interactions between Ge defects and H$_2$ are unclear to
a large extent. Finally, scanty literature data exist on the
kinetics of these phenomena, since most of the studies have been
carried out by investigating only the stationary defect
concentrations before and after laser irradiation.

In this context, natural \sil{} is a potentially interesting
material since it contains Ge in concentrations (a few ppm, see
Table \ref{materials}) much lower than Ge-doped silica. In this
sense, Ge atoms are incorporated as \emph{dispersed} impurities, as
opposed to the high concentrations deliberately obtained by doping
procedures. Hence, one can wonder whether laser-induced conversion
processes of Ge-related defects in these conditions are different
from those observed in Ge-doped silica, and it is possible that such
a simple model system may allow to isolate some specific Ge-related
process within the complex landscape emerging from literature.
Another important difference exists between Ge impurities in natural
silica and Ge-doped \sil{}. Indeed, it was demonstrated that in
as-grown natural \sil{} the majority (50\%--100\%) of Ge impurities
are incorporated in twofold coordinated form
\cite{Grandi,LeonePRB99,CannasPHD,Skuja84,SkujaJNCS92}, i.e. as GLPC
centers (=Ge$^{\bullet\bullet}$). This situation is quite different
from Ge-doped \sil{}, where the portion of Ge atoms in the
two-coordinated configuration is much lower (10$^{-4}$--10$^{-2}$),
while significant portions of Ge are present in the three- and
four-fold coordinated arrangements or as Ge
clusters\cite{HosonoPRB92,Grandi}. The reasons why Ge impurities in
\sil{} prevalently choose the GLPC arrangement only when their total
concentration is low are not completely clear at the moment. Even
so, this circumstance can be exploited for our purposes: indeed,
natural silica can be regarded as a material of choice to
investigate \emph{selectively} the conversion processes of the GLPC
center elicited by UV radiation. The experiments presented in this
chapter aim also to investigate this point.

\section{Conversion processes of GLPC center}
\subsection{Results}
\subsubsection{ESR measurements}
 \hspace{0.8cm}
As-grown dry Q906 and wet HER1 samples were irradiated at room
temperature with 2000 pulses of 4.7\,eV laser radiation with
40\,mJcm$^{-2}$ energy density per pulse and 1\,Hz repetition rate.
\begin{figure}[h!]
\begin{center}
\includegraphics[width=.8\textwidth]{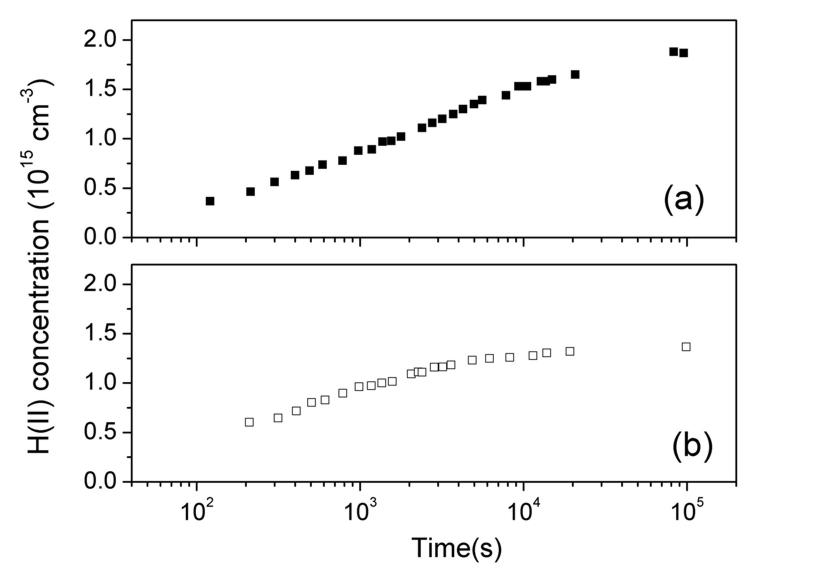}
\end{center}
\caption{Post-irradiation kinetics of H(II) centers induced in wet
(a) and dry (b) natural \sil{} by 2000 4.7\,eV laser pulses with
40\,mJcm$^{-2}$ energy density per pulse and 1\,Hz repetition rate
at room temperature. The origin of the time scale corresponds to the
end of irradiation.} \label{PIKhiimaterials}\end{figure}

As already described, by ESR measurements performed in the
post-irradiation stage we are able to detect the 11.8\,mT doublet
typical of the H(II) center. From the time dependence of the
intensity of the doublet, we calculated the two kinetics [H(II)](t),
which are reported in \figurename~\ref{PIKhiimaterials}.
\begin{figure}[h!]
\begin{center}
\includegraphics[width=.8\textwidth]{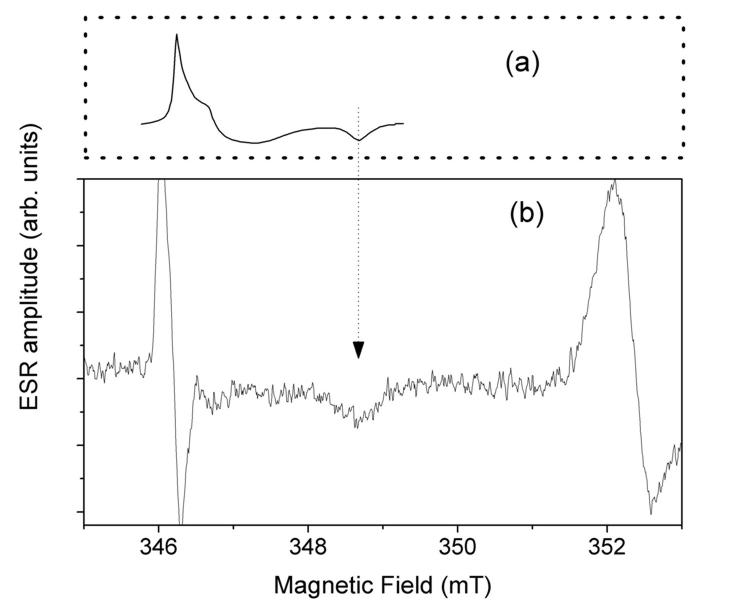}
\end{center}
\caption{ESR spectrum (b) of a natural dry \sil{} a few days after
exposure to 500 laser pulses with 40\,mJcm$^{-2}$ energy density per
pulse and 1\,Hz repetition rate. The signal was acquired with a 2G
modulation amplitude and a P=6mW microwave power. Panel (a):
simulated Ge(2) lineshape taken from Friebele \emph{et
al.}\cite{FriebeleJAP74} with purpose of comparison.}
\label{spettroge2}\end{figure} In both materials we observed the
usual post-irradiation growth of [H(II)], which increases from the
initial values of (3.7$\pm$0.4)$\times$10$^{14}$\,cm$^{-3}$ (wet)
and (6.0$\pm$0.6)$\times$10$^{14}$\,cm$^{-3}$ (dry), measured at
t$\sim$10$^2$\,s from the end of irradiation, to the stationary
values of (1.9$\pm$0.2)$\times$10$^{15}$\,cm$^{-3}$ (wet) and
(1.4$\pm$0.1)$\times$10$^{15}$\,cm$^{-3}$ (dry), measured at
t$\sim$10$^5$\,s.

We have seen in chapter \ref{overview} that the two main signals
detected by ESR in irradiated natural silica are H(II) and $E'$.
Aside from these two, a further very weak component
(\figurename~\ref{spettroge2}) is detected, which is known in
literature as Ge(2) center. In particular, the characteristic
negative peak of the Ge(2) ESR signal is found in our spectra in the
expected spectral position, being a clear fingerprint of the
presence of this defect \cite{FriebeleJAP74}. Ge(2) is a
paramagnetic Ge-related center, whose microscopic structure is
currently debated to be either a ionized twofold coordinated Ge
(=Ge$^{\bullet}$)\cite{AnoikinSLC91,NeustrevJPCM94,FujimakiPRB98,YamaguchiPRB02},
or an electron trapped at a GeO$_4$ site (GeO$_4$$^{\bullet}$)$^{-}$
\cite{KawazoeJNCS85,NishiiPRB95,NishiiPRB99,FriebeleinDIG86,TsaiDDD}.
 Even if the signal is very
weak and partially superimposed to the much stronger $E'$ resonance,
it is possible to approximately assess the concentration of Ge(2) by
comparing the negative peak with a reference lineshape available in
literature\cite{FriebeleJAP74}; by this procedure we obtain:
[Ge(2)]$\sim$2$\times$10$^{14}$\,cm$^{-3}$. The concentration of
Ge(2) never exceeds this order of magnitude in all the investigated
samples. H(II) and Ge(2) are the only Ge-related defects observed by
ESR in irradiated natural \sil{}. In particular, we do not detect
either the Ge\,--$E'$
($\equiv$Ge$^\bullet$)\cite{TsaiJAP87,PurcellPCG69,FriebeleJAP74,TamuraPRB04,PacchioniPRB00},
or the Ge(1) ((GeO$_4$$^{\bullet}$)$^{-}$)
centers\cite{KawazoeJNCS85,NeustrevJPCM94,FriebeleinDIG86,TsaiDDD,PacchioniPRB00},
both commonly found in Ge-doped silica exposed to laser radiation .
\subsubsection{PL measurements}\label{pikglpcsection}
\hspace{0.8cm}We already reported in chapter \ref{overview} that
laser irradiation induces the bleaching of the native optical
activity of the GLPC, although no information was provided on the
\emph{kinetics} of this bleaching process. However, if H(II) are
generated by (\ref{trappingsuglpcrep}), it is expected that the
reduction of the GLPC occurs in the post-irradiation stage, i.e.
simultaneously to the observed growth of H(II). To investigate this
issue, we carried out other two irradiation experiments on as-grown
wet and dry natural \sil{} specimens in the same conditions as those
of \figurename~\ref{PIKhiimaterials}. PL measurements under lamp
excitation at 5.0\,eV were performed on both samples, before and at
different delays (10$^2$–-10$^5$)\,s from the end of irradiation. In
particular, to monitor the intensity variations of the PL signal of
GLPC in the post-irradiation stage, $\Delta$PL, the irradiated
specimens were positioned into the sample chamber of the
spectrofluorometer about 10$^2$\,s after exposure, after which they
were kept in place and measured for 10$^4$–-10$^5$\,s; with this
choice, the precision of $\Delta$PL is not limited by repeatability
of the mounting conditions and increases to $\sim$1\% of PL
intensity.

\begin{figure}[h!]
\begin{center}
\includegraphics[width=.7\textwidth]{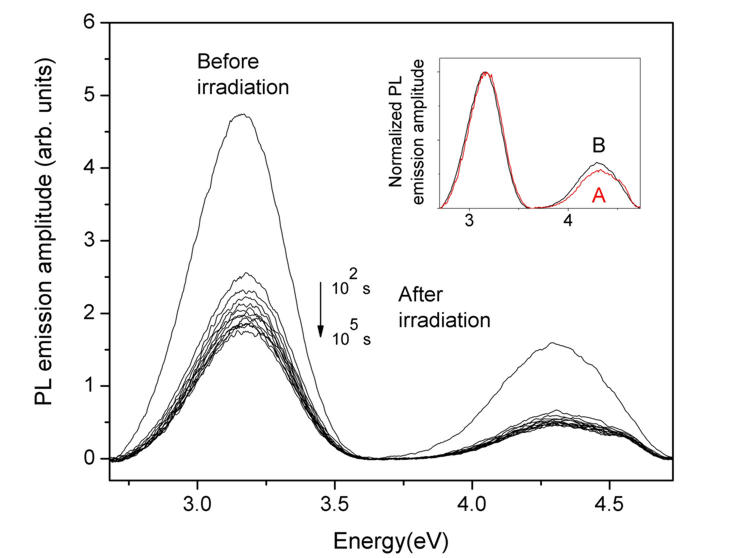}
\end{center}
\caption{PL signal of GLPC before irradiation and at several times
from 10$^2$\,s up to 10$^5$\,s after irradiation, all detected in a
HER1 natural wet sample irradiated with 2000 laser pulses with
40\,mJcm$^{-2}$ energy density per pulse and 1\,Hz repetition rate.
Inset: normalized spectra before (B) and after (A) irradiation.}
\label{pikglpc1}\end{figure}

The detected emission spectra consist in the 3.1\,eV and 4.3\,eV
bands. We found that the signal intensity reduction induced by
irradiation occurs in two clearly distinguishable stages (see
\figurename~\ref{pikglpc1}): \textbf{(i)} during illumination, an
intensity reduction of 50\% in wet and 15\% in dry silica takes
place, as we observe by comparing the as-grown PL spectrum with the
first detected after exposure at t$\sim$10$^2$\,s \textbf{(ii)}
after the end of irradiation, the PL intensity further decreases in
time, as apparent from the spectra measured in the wet specimen at
different delays from the end of exposure. An analogous result was
obtained on the dry material. In both cases, measures were continued
until a constant PL intensity was reached within experimental error.
Finally, we observed that the bleaching is accompanied by a small
alteration of the ratio between the 3.1\,eV and 4.3\,eV PL bands of
GLPC, as shown in the inset of \figurename~\ref{pikglpc1}.

\begin{figure}[t!]
\begin{center}
\includegraphics[width=.8\textwidth]{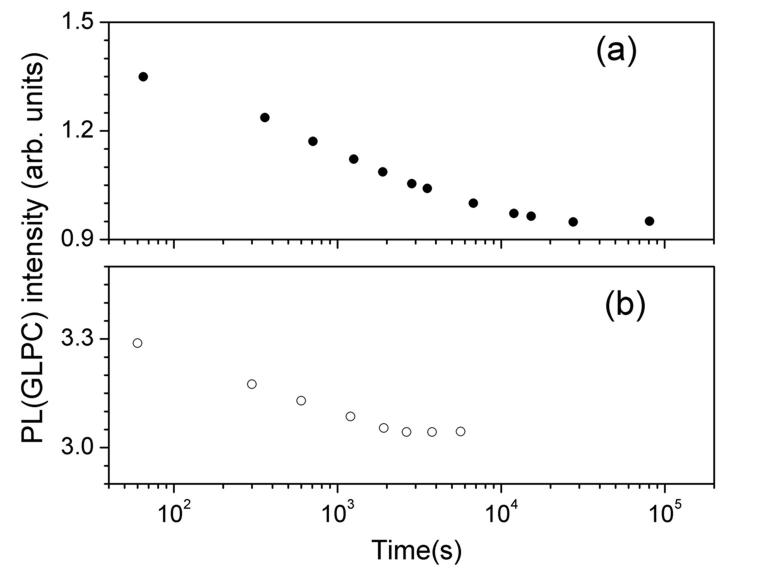}
\end{center}
\caption{Post-irradiation kinetics of the intensity of the PL signal
of GLPC after exposure of a wet (a) or a dry (b) natural \sil{}
sample to 2000 laser pulses with 40\,mJcm$^{-2}$ energy density per
pulse and 1\,Hz repetition rate at room temperature.}
\label{PIKglpcmaterials}\end{figure} The post-irradiation kinetics
of the GLPC is summarized in \figurename~\ref{PIKglpcmaterials},
where the integrated intensity PL(GLPC) of the signal is plotted for
the two materials against time. Luminescence intensity decreases of
0.40$\pm$0.01 arb.units from 65\,s to 8$\times$10$^4$\,s in the wet
sample. In the dry specimen, the decrease is 0.24$\pm$0.03 arb.units
from 60\,s to 6$\times$10$^3$\,s.

\subsubsection{Repeated irradiations}
\hspace{0.8cm}To deeper analyze the relationship between H(II) and
GLPC, we investigated the concentration variations of both defects
under repeated irradiations. In general, this approach is useful in
that it allows to observe the interaction of already formed defects
with laser light; besides, the application of multiple exposures
separated by dark intervals reproduces a situation often encountered
in optical applications of amorphous silica.
\begin{figure}[h!]
\begin{center}
\includegraphics[width=.9\textwidth]{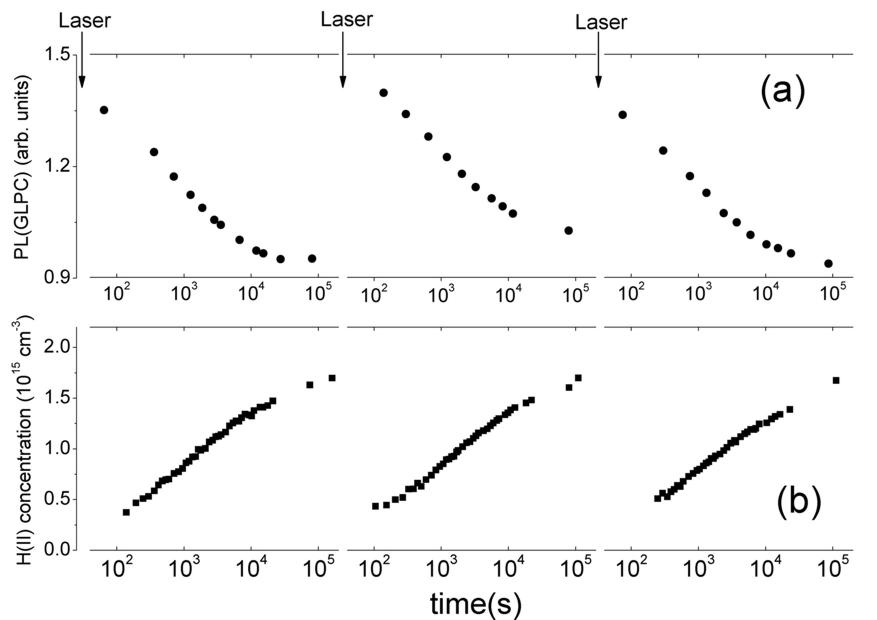}
\end{center}
\caption{PL signal of GLPC (a) and concentration of H(II) (b)
induced in a wet natural \sil{} sample by a cycle of 3 irradiations
of 2000 laser pulses with 40\,mJcm$^{-2}$ energy density per pulse
and 1\,Hz repetition rate. Each arrow represents an irradiation
session. The irradiation sessions are separated by a waiting time of
$\sim$10$^6$\,s to allow the post-irradiation kinetics to be
completed. } \label{repeated}\end{figure} In this case, an
experiment was performed in which a natural wet specimen was
irradiated 3 times with 2000 laser pulses; after each exposure, the
post-irradiation kinetics of PL(GLPC) was measured until completion.
Results are shown in \figurename~\ref{repeated}-(a). On a second
sample subjected to the same irradiation sequence, the
post-irradiation kinetics of H(II) centers was measured after each
exposure (\figurename~\ref{repeated}-(b)). As apparent from
experimental data, each exposure destroys most of H(II) that had
formed upon the previous illumination, their concentration
decreasing approximately to the same value as immediately after the
previous irradiation; simultaneously, we observe a rebuild of
luminescence intensity to approximately the same value found at the
same time after the previous exposure. After every re-irradiation,
the sample \emph{loses memory} of its previous history, meaning that
both PL(GLPC) and [H(II)] repeat again the same decrease/growth
kinetics.

We stress that the repeatable decay and recovery cycles of GLPC
observed upon multiple irradiations involve only the portion
bleached in the post-irradiation stage, whereas the reduction
observed \emph{during} exposure occurs irreversibly only during the
earliest irradiation. However, we clarify that the term
"irreversibly" is used here only in relation to the effects of
further irradiations. Indeed, data obtained by a thermal treatment
experiment show that the "irreversible" portion of the bleaching can
actually be reversed by heating the irradiated material at
300C$^{\circ}$ for 3 hours. This aspect is of no concern here and
will be not further discussed.

\subsection{Discussion I: Correlation between GLPC and H(II) centers}
\hspace{0.8cm}Being the optical density of our samples at 5.0 eV
smaller than 0.02 (see for example \figurename~\ref{i301preabs}),
PL(GLPC) is proportional to the concentration of the twofold
coordinated center (chapter \ref{mm2}). Hence, the bleaching induced
by irradiation is a manifestation of conversion processes triggered
by UV exposure, which transform the diamagnetic center in other
defects. Results in \figurename s~\ref{pikglpc1} and
\ref{PIKglpcmaterials}, allow to isolate two different stages of
GLPC conversion:  the "irreversible" stage, occurring only once
during the earliest irradiation, and that taking place after each
irradiation. In particular, the second stage was expected from our
preliminary discussion, since it occurs simultaneously to the growth
of H(II) consistently with reaction (\ref{trappingsuglpcrep}). Now,
we have to find out if a quantitative anti-correlation between the
two effects exists \emph{and} if the amount of the absolute
concentration growth of H(II) is compatible with the absolute
concentration decrease of GLPC.

To this aim, we begin with converting the PL intensity PL(GLPC) to
an absolute concentration measurement [GLPC]=$\xi$PL(GLPC). The
coefficient $\xi$ can be estimated from the radiative emission
lifetime $\tau$=7.8\,ns of the defect from the excited singlet state
(see \figurename~\ref{decayDESY}) in two steps. First, by using the
F\"{o}rster equation (\ref{Forster}), we can calculate from $\tau$
the oscillator strength\footnote{As a result of this calculation,
using the Lorentz-Lorenz effective field correction we get an
oscillator strength of: f$\sim$0.07. We stress, however, that the
value of $\xi'$ does \emph{not} depend on the expression chosen for
the effective field correction.} of GLPC, or equivalently the
proportionality coefficient $\xi'$ between [GLPC] and the area
A(B$_{2\beta}$) of the \emph{absorption} band at 5.1\,eV of the
defect, [GLPC]=$\xi'$A(B$_{2\beta}$). Hence, we find $\xi$ by
multiplying $\xi'$ for the known constant ratio between the
intensities A(B$_{2\beta}$) and PL(GLPC) of absorption and emission
signals\cite{LeonePRB99}. In this way we estimate\footnote{It is
worth noting that the value of $\xi$ (as well as PL(GLPC), but
differently from $\xi'$) does not possess an absolute meaning. In
fact, $\xi$ is a function of the instrumental parameters used to
acquire the luminescence spectrum. The present calculation refers to
the acquisition conditions used for all the PL data reported in this
chapter.}: $\xi$=(4.4$\pm$0.7)$\times$10$^{15}$cm$^{-3}$.

Being now able to transform PL intensities in absolute
concentrations, we first discuss the irreversible conversion of GLPC
occurring during irradiation. Using $\xi$, we estimate the
concentrations of the centers converted in this stage,
$\Delta_0$=(6.0$\pm$0.9)$\times$10$^{15}$\,cm$^{-3}$ (wet) and
$\Delta_0$=(2.7$\pm$0.4)$\times$10$^{15}$\,cm$^{-3}$ (dry). Present
data do not allow to clarify in what is transformed this portion of
GLPC, thereby leaving open this specific issue. Nonetheless, we can
rule out H(II) and Ge(2), whose concentration (at t=0) are both too
small to account for $\Delta_0$, and Ge\,--$E'$ and Ge(1), which are
absent in the exposed specimen within the EPR sensitivity of
$\sim$2$\times$10$^{14}$\,cm$^{-3}$. Then, we infer that during
irradiation a portion of GLPC is most likely converted to some
unknown \emph{diamagnetic} center that happens to be virtually
invisible at this concentration.

\begin{figure}[h!]
\begin{center}
\includegraphics[width=.9\textwidth]{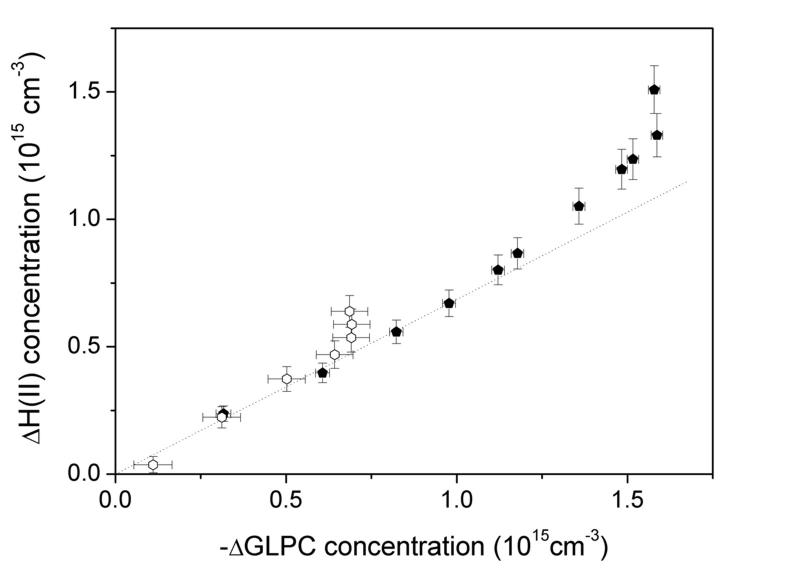}
\end{center}
\caption{Correlation plot between the increase of [H(II)] and the
decrease of [GLPC], measured from t$\sim$10$^2$ from the end of
irradiation. Full (empty) symbols correspond to wet (dry) natural
\sil{}.} \label{correlationgraph}\end{figure}

Hence, we proceed to examine the relation between the
post-irradiation decay of GLPC and the simultaneous growth of H(II).
To this aim, we plot in \figurename~\ref{correlationgraph} the
increase $\Delta$[H(II)] of H(II) concentration as a function of the
decrease of [GLPC], -$\Delta$[GLPC]=-$\xi\Delta$PL(GLPC). For each
of the two materials, $\Delta$[H(II)] and $\Delta$[GLPC] were
calculated from the same time instant t$_0$; ESR data used to
calculate $\Delta$[H(II)] were obtained by extrapolation at the same
time instants at which luminescence spectra had been acquired. For
practical reasons, t$_0$ was chosen to be the time at which the
first ESR spectrum was acquired, i.e. the x-coordinate of the first
point in \figurename~\ref{PIKhiimaterials} (t$_0$$\sim$10$^{2}$ for
the wet sample and t$_0$$\sim$2$\times$10$^{2}$ for the dry sample).

 We see that data from
both materials fall on a single line for short times, whereas for
long times they tend to depart from the line towards the upper
semiplane. In the short time region, H(II) and GLPC are indeed
anti-correlated, with a correlation coefficient independent of the
material and represented by the slope of the line, $S\sim$~0.7, as
estimated by a best fit procedure on the first points (corresponding
to t$<$2$\times$10$^3$\,s). This value of $S$, which is founded on
two completely independent concentration
measurements\footnote{\label{acmII}In contrast to the case discussed
in footnote \ref{fnsyst} of Chapter \ref{overview}, the errors on
the concentration measurements derived from PL and ESR are
completely independent. Then, the accuracy (rather than the
repeatability only) of these two concentration estimates becomes
important when assessing the closeness of the parameter $S$ to
unity. In this sense, we recall that the uncertainty on [H(II)]
measured by ESR is $\sim$20\%. As for the concentration of GLPC
derived from PL data, it is more difficult to obtain a precise
estimate of the experimental error due to the relative complexity of
the procedure; however, it can be argued that also in this case the
uncertainty is at least (another) 20\,--\,30\%}, can be considered
to be in a reasonable agreement with unity for our present purposes.
Also, the circumstance that data obtained on two different materials
fall on a single line strengthens the result.

Therefore, the following interpretation of
 \figurename~\ref{correlationgraph} is proposed: the linear
relationship approximately valid at short times represents a
one-to-one conversion between GLPC and H(II) centers by process
(\ref{trappingsuglpcrep}), whereas the deviations from linear
correlation indicate that H(II) centers are generated also by a
second channel prevailing over the main reaction
(\ref{trappingsuglpcrep}) at long times. In both samples, this
second generation channel accounts for $\sim$30\% of the total
$\Delta$[H(II)]. Since we detect a low concentration of Ge(2)
centers after irradiation (\figurename~\ref{spettroge2}), if we
assume the model of Ge(2) as a ionized twofold coordinated Ge
impurity (=Ge$^{\bullet}$), a possible mechanism producing the
portion of H(II) not anti-correlated to GLPC may be the successive H
and electron trapping on Ge(2), as proposed by Fujimaki \emph{et al}
\cite{FujimakiPRB99}.
\begin{equation}
=\text{Ge}^{\bullet}\text{ + H + $e^-$ }\longrightarrow\text{
=Ge$^{\bullet}$\,--\,H}
\end{equation}

However, it may be possible that the deviations from a straight line
of \figurename~\ref{correlationgraph} arise from a not-perfect
proportionality between PL(GLPC) and [GLPC], due to inhomogeneities
in the optical properties of the GLPC defect embedded in the
disordered solid
\cite{CannasPHD,CannizzoPHD,LeonePRB99,AgnelloPRB03b,MioPSS06Simone}.
Indeed, it is conceivable that both the oscillator strength and the
reactivity of GLPC with H fluctuate within the population of defects
due to site-to-site inhomogeneity. Suppose now that the two
statistical distributions are inter-correlated\footnote{this may
occur for example if the two distributions actually reflect the
randomization of a single "structural" parameter (bond length,
angle, etc..) that varies from site to site, and on which both
quantities depend. It is relatively common in \sil{} to find
situations in which such cross-correlations effects alter the
perfect proportionality between two parameters associated to the
same defect, and expected to be proportional in absence of
inhomogeneity effects \cite{SkujaReview98}. }: in this scheme, the
progress of reaction (\ref{trappingsuglpcrep}) is expected to lead
to a variation of the \emph{mean} value of $\xi$, above supposed for
simplicity to be unique, which could explain the deviations from a
perfect linear correlation. The possible influence of such effects
in the present context is also suggested by the laser-induced
variation of the ratio between the 3.1\,eV and the 4.3\,eV PL bands
in the PL signal (inset of \figurename~\ref{pikglpc1}). This
evidence indicates a variation of the (mean) value of the
intersystem crossing (ISC) rate (see the level scheme of GLPC in
\figurename~\ref{schemab}); this parameter in GLPC has been clearly
proved to be affected by inhomogeneity effects\cite{LeonePRB99}.
Hence, similarly to ISC, it is likely that inhomogeneity may affect
also other parameters controlling the luminescence intensity, such
as the PL quantum yield or the oscillator strength (see chapter
\ref{mm1}).

Even though a definite interpretation cannot be proposed for the
deviations of \figurename~\ref{correlationgraph}, the correlation at
short times, found for both materials and with a common slope $S$
near unity, is sufficient to conclude that our results are in a good
agreement with the hypothesized model that ascribes the formation of
H(II) to process (\ref{trappingsuglpcrep}). It is worth noting that
our present approach significantly differs from the customary way in
which this kind of problems are addressed in literature. In fact, a
correlation between \emph{relative} concentration measurements (i.e.
the intensities of spectroscopic signals) is often considered
sufficient to confirm a supposed conversion mechanism between a
diamagnetic precursor and a paramagnetic center. Then, since it is
usually easier to obtain an \emph{absolute} concentration estimate
by ESR measurements than by optical data, the slope of the
correlation is exploited to "calibrate" the signal of the
diamagnetic center to its absolute concentration. In contrast, we
have proceeded here in a more rigorous way. Starting from two
independent concentration measurements, each potentially affected by
several sources of experimental uncertainty (see footnote
\ref{acmII}), we have verified not only the existence of a linear
correlation, but also that the linear coefficient, when expressed as
an absolute dimensionless quantity, is sufficiently close to one for
the conversion process (\ref{trappingsuglpcrep}) to be reliable.

\begin{figure*}[t!]
\begin{center}
\fcolorbox{postitA}{postitA}{\begin{minipage}[c]{.95\textwidth}\linespread{1.2}\hrule\vspace{0.2cm}
\textbf{\textsc{Summary I. }}\textsl{\textcolor{postitC}{Aside from
the generation of $E'$ centers discussed in the previous chapter,
4.7\,eV laser irradiation induces in natural silica conversion
processes of defects related to the Ge impurities present in low
concentration in this material. In particular, in the
post-irradiation stage GLPC defects (monitored by PL spectroscopy,
\figurename~\ref{PIKglpcmaterials}) are converted in H(II) centers
(monitored by ESR, \figurename~\ref{PIKhiimaterials}) by hydrogen
trapping, eq.~(\ref{trappingsuglpcrep}), as proved by the
anti-correlated concentration variations of the two defects
(\figurename~\ref{correlationgraph}). The conversion of GLPC into
H(II) can be temporarily reversed by re-irradiating the material
(\figurename~\ref{repeated}). Indeed, re-irradiation causes the
photo-induced destruction of the H(II) centers,
(\figurename~\ref{destructionhii}) and
eq.~(\ref{hiidestructionreaction}), which are back-converted to
GLPC. Natural silica permits a selective study of the interplay
between GLPC and H(II) centers triggered by UV radiation,
differently from Ge-doped materials where several conversion
processes of Ge-related defects have been reported to be active at
the same time. \vspace{0.2cm}}\hrule}\end{minipage}}
\end{center}
\end{figure*}

\subsubsection{Comparison with literature on conversion processes of Ge-related
defects} \hspace{0.8cm}As reviewed in subsection \ref{Gedefsection},
literature studies on conversion process of Ge-related centers are
usually carried out on Ge-doped silica, with a typical Ge impurity
concentration of the order of a few \%. Most of these works report a
reduction of the pre-existing $\sim$5.1\,eV band accompanied by the
generation of Ge(1), Ge(2) and Ge\,--$E'$ centers
\cite{HosonoPRB92,FujimakiPRB98,NishiiPRB95,
NishiiAPL94,TakahashiAO03,NeustrevJPCM94,SakohOE03}. Several models
have been proposed to explain the formation of these paramagnetic
centers, often on the basis of observed correlations between the
stationary concentrations of the induced defects or between the
reduction of the native OA and the concentration of the induced
defects (see subsection \ref{convgeiilabel}-B). Present results show
that Ge-related processes induced by laser irradiation in natural
silica, where the concentration of Ge is only a few ppm, are
remarkably different. In fact, we did not observe either Ge(1) or
Ge–-\,$E'$ centers. Ge(1) has been proposed to arise from trapping
of an electron on 4-fold coordinated Ge precursors
\cite{FujimakiPRB96,FujimakiPRB98,FujimakiPRB99,NishiiPRB95,HosonoPRB96,NishiiPRB99},
while Ge--\,$E'$ is supposed to be induced by ionization of
pre-existing Neutral Oxygen Vacancies (NOV) on Ge, in which the Ge
atom is 3-fold coordinated \cite{HosonoPRB92}. While the NOV is
believed to contribute to the $\sim$5.1\,eV absorption without
emitting luminescence \cite{HosonoPRB92}, no measurable optical
activity below bandgap has been ascribed to the 4-fold coordinated
Ge precursor of Ge(1).

On this basis, the lack of Ge(1) and Ge\,--$E'$ in irradiated
natural silica is likely due to the lower concentration of 4-fold
and 3-fold coordinated Ge precursors with respect to Ge-doped glass,
as already anticipated in the introduction of this chapter. Two
considerations support this idea: \textbf{(i)} The close resemblance
in natural silica between the PLE spectrum of the 3.1\,eV and
4.3\,eV bands and the 5.1\,eV absorption profile, as well as the
strong linear correlation between the three bands, exclude the
presence of the NOV in this material in concentration comparable to
GLPC. \textbf{(ii)} The concentration of native two-fold coordinated
GLPC centers, calculated using the conversion ratio $\xi$ between PL
intensity and GLPC concentration estimated above, is
1.2$\times$10$^{16}$ in wet specimens. This corresponds to about
75\% of the total Ge content estimated from neutron activation
measurements (see section \ref{samples} and references therein). The
ratio [GLPC]/[Ge]$\sim$0.75, is to be compared with Ge-doped samples
where it was measured to be [GLPC]/[Ge]$\sim$10$^{-4}$--10$^{-2}$
\cite{Grandi,HosonoPRB92}. Such a difference confirms the much lower
concentration of 3- and 4-fold Ge precursors in natural \sil{},
where Germanium is almost completely arranged as GLPC. The reasons
of the difference between the prevalent arrangements of the Ge
impurities in the two materials are not completely understood at the
moment.

In this sense, our results can be regarded as a selective
investigation of the laser-induced conversion processes of the GLPC
centers, very difficult to isolate in Ge-doped silica due to the
presence of NOVs and other configurations of Ge. Our results
strongly suggest that a significant portion of GLPC can be converted
to a diamagnetic center, at variance with the common practice in
literature to ascribe the conversion of GLPC to the formation of
paramagnetic signals only, mainly the Ge(2)
\cite{FujimakiPRB96,FujimakiPRB98,FujimakiPRB99}. Consistently with
our view, in recent times the conversion of GLPC in another
diamagnetic defect has been proposed by a few other experimental and
computational works \cite{TakahashiJAP02,UchinoPRL00,UchinoPRB02}.
In our system, the prevalent paramagnetic Ge-related center induced
by irradiation is H(II), formed by process
(\ref{trappingsuglpcrep}). The role of H(II) as one of the main
products of GLPC conversion was evidenced in $\gamma$-irradiated
natural silica \cite{AgnelloPHD,AgnelloPRB00}, even if no evidence
was found of other parallel conversion processes of GLPC. In
contrast, in laser-irradiated Ge-doped silica the generation of
H(II) has been reported only as a minor process occurring in
H$_2$-loaded samples, and being accompanied by the disappearance of
Ge(2) \cite{FujimakiPRB99}. This further difference with our result
could be due to the absence of available mobile H$_2$ in common
irradiated Ge-doped \sil{} samples, whereas natural silica contains
an intrinsic source of radiolytic hydrogen (supposedly Si\,--\,H) so
as to allow for the growth of H(II) (and possibly cause the
passivation of Ge(2)).

As a final remark, we stress that the present results are founded on
the choice of studying the \emph{time-dependence} of the Ge-related
defects. \emph{A posteriori}, we can say this choice to be mandatory
in order to isolate process (\ref{trappingsuglpcrep}) from the
conversion channels active during exposure and possibly at long
times. Our approach is innovative in that it differs from the common
practice in literature of studying (and trying to correlate) only
the \emph{stationary} concentrations of the defects. In this sense,
we propose that also in Ge-doped materials or in other systems,
kinetic investigations may help to clarify the picture of
laser-induced conversion processes.

\subsection{Discussion II: Correlation under repeated irradiations and photo-induced decay of H(II) center}
The data in \figurename~\ref{repeated} show evidence that the two
defects have an anticorrelated behavior also under repeated
irradiations. In fact, each exposure causes a photo-induced decay of
the H(II) grown during the last post-irradiation kinetics, and
simultaneously restores the GLPC. This finding indicates that H(II)
centers, generated by reaction (\ref{trappingsuglpcrep}) activated
by irradiation, are also \emph{reduced} by laser exposure. This
observation allows us to understand the feature of the growth
kinetics of H(II) center, that are formed mostly in the
post-irradiation stage rather than during it, so that the final
concentrations are always 3-4 times higher than initial
concentrations (chapter \ref{overview}). In fact, this property can
be now explained as follows: the growth of H(II) \emph{during}
irradiation by reaction (\ref{trappingsuglpcrep}) is inhibited
because of the competition with the photo-induced decay of the
centers.
\begin{figure}[h!]
\begin{center}
\includegraphics[width=.9\textwidth]{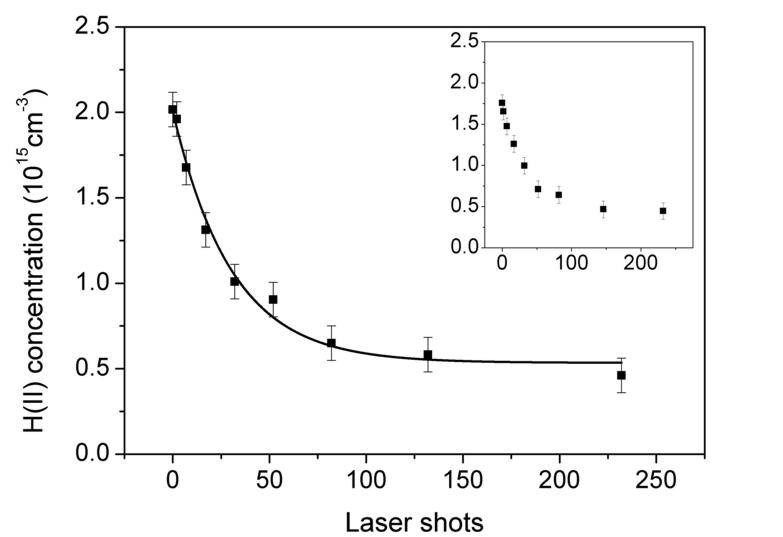}
\end{center}
\caption{Variations of [H(II)] induced in a natural wet sample by
reirradiation after preliminary exposure to 2000 laser shots.
Continuous line: least-square fit by an exponential function. Inset:
result of the same experiment on a dry natural sample.}
\label{destructionhii}\end{figure} Also, it is interesting to note
the similarity between the present finding and the photo-induced
decay of H(I) (=Si$^{\bullet}$\,--\,H) center observed by Radtsig et
al \cite{RadtzigJPC95}; H(I) and H(II) both belong to an
isoelectronic series of defects, localized on a Si, Ge, Sn atom,
which are known to have similar formation and spectroscopic
properties \cite{SkujaJNCS92,PacchioniPRB98}. From this point of
view, we suggest that the isoelectronic defects share also similar
photo-induced decay properties.

The re-growth of GLPC concurrent to the disappearance of H(II),
combined with the structural relationship existing between H(II)
(=Ge$^{\bullet}$\,--\,H) and GLPC (=Ge$^{\bullet\bullet}$), suggest
the following microscopic mechanism responsible for the
photo-induced decay of H(II):
\begin{equation}
\text{=Ge$^{\bullet}$\,--\,H }\stackrel{h\nu}{\longrightarrow}\text{
=Ge$^{\bullet\bullet}$ + H}\label{hiidestructionreaction}
\end{equation}
namely, exposure of H(II) to 4.7\,eV photons causes detaching of the
hydrogen atom from the Ge\,--\,H bond reconstructing the precursor
GLPC. We found that H(II) can be destroyed also by exposure of the
sample to light from a Xe lamp. Due to the much lower peak intensity
with respect to laser, this observation suggests process
(\ref{hiidestructionreaction}) to occur by single-photon absorption
at the defect site.\footnote{On the other hand, this leads to the
necessity of some precautions when measuring the post-irradiation
kinetics of the GLPC (subsection \ref{pikglpcsection}); indeed, to
prevent H(II) from being destroyed during PL measurements due to the
excitation light, in all PL measurements reported above we used a
high scan speed (500\,nm min$^{-1}$) to reduce as much as possible
the exposure time of the sample to the Xe lamp. We verified that
this choice avoids the photo-decay of H(II) as well as any
appreciable distortion of the experimentally observed PL kinetics.}
A further experiment was carried out to measure the cross section of
process (\ref{hiidestructionreaction}): we started with a natural
wet sample exposed to an irradiation dose large enough (2000 pulses,
see \figurename~\ref{frommintomaxhii}) to produce the maximum H(II)
concentration. Such a dose is referred to as a \emph{high-dose}.
After waiting the H(II) post-irradiation kinetics to be completed,
we measured [H(II)]=(2.0$\pm$0.2)$\times$10$^{15}$cm$^{-3}$. Then,
we irradiated again the specimen with an increasing number of pulses
of $W$=40\,mJcm$^{-2}$ energy density per pulse. We found that after
each exposure, the defect concentration remains almost invariant
(within 10\%) during the time interval required for the ESR
measurement ($\sim$500\,s). Hence, between successive exposures we
measured the defect concentration, obtaining the results in
\figurename~\ref{destructionhii}. Consistently with
\figurename~\ref{repeated}, we observed that the new irradiation
results in the destruction of H(II) generated by the first dose,
their concentration decreasing to $\sim$25\% of the initial value
after $\sim$250 laser pulses. An identical effect was observed also
on natural dry samples (inset). The same response of H(II) embedded
in the two materials agrees with our interpretation in which the
decay is due to the direct absorption of UV light at the defect site
(process \ref{hiidestructionreaction}). After the whole 250 pulses
sequence, [H(II)] increases again but does not recover its initial
magnitude within the investigated time scale: their concentration
increases to (1.4$\pm$0.1)$\times$10$^{15}$\,cm$^{-3}$ in
2$\times$10$^6$\,s. This behavior is different from the observed
repeatability of the kinetics after 2000 pulses
(\figurename~\ref{repeated}); in other words, only a high-dose
irradiation is able to cause a memory loss by the sample. Finally we
note that the concentration of $\sim$5$\times$10$^{14}$cm\,$^{-3}$
to which [H(II)] tends at the end of the 250 pulses sequence, is in
good agreement with the value observed in \figurename~\ref{repeated}
just after each re-irradiation. It is reasonable to assume that this
concentration represents an equilibrium between the photo-induced
decay of H(II) and the growth of the defect due to reaction
(\ref{trappingsuglpcrep}) taking place during each interpulse time
span.

The H(II) reduction with the number of pulses N is fitted by an
exponential function:
\begin{equation}
 [\text{H(II)}]=A_1\exp{\left(-\frac{N}{N_0}\right)}+A_2
\end{equation}
 where $N_0$=30$\pm$3, $A_1$=(1.5$\pm$0.1)$\times$10$^{15}$\,cm$^{-3}$,
 $A_2$=(0.5$\pm$0.1)$\times$10$^{15}$\,cm$^{-3}$. From these data, we calculate
the cross section of process (\ref{hiidestructionreaction}),
$\sigma_D=N_0^{-1}(h\nu/W)$=(6.2$\pm$0.6)$\times$10$^{-19}$\,cm$^{2}$.

Given that the H(II) is able to absorb 4.7\,eV light (though
destructively), at this point one could wonder why this does not
correspond to a OA signal near 4.7\,eV, present in the observed
absorption profiles due to the presence of the defect. In this
sense, we note that the H(II) concentration,
H(II)=2.1$\times$10$^{15}$cm$^{-3}$, leads to the following estimate
of the 4.7\,eV absorption coefficient of the defects,
$\alpha$=$\sigma_D$[H(II)]=1.3$\times$10$^{-3}$\,cm$^{-1}$, if we
suppose that each absorbed photon results in the destruction of a
center. Therefore, the presence of the much stronger B$_{2\beta}$
band overlapping in same spectral region prevents us from detecting
the anticipated signal that could be ascribed to H(II). On the other
hand, as carried out in a recent experiment\cite{Origlio}, a tunable
UV radiation source can be used to measure $\sigma_D$ as a function
of wavelength; this method allows to reconstruct indirectly the
absorption profile of the H(II) center, thus overcoming the
difficulty of directly observing the weak OA signal of the defect by
a standard spectrophotometric approach.

Finally, we briefly comment the observed repeatability of the
post-irradiation kinetics. The kinetics of H(II) and GLPC centers
repeat themselves after each \emph{high-dose} exposure cycle
(\figurename~\ref{repeated}). In principle, the post-irradiation
kinetics is determined by the concentrations at t=0 of all the
centers involved in the reactions that induce the growth of H(II).
In the model we have proposed starting from section
\ref{modelpiktutti}, these defects are GLPC, hydrogen and $E'$, the
latter playing the role of a cracking center for H$_2$. So, the
memory loss implies that each 2000 pulses re-irradiation is able to
reset the concentrations of H(II), GLPC, $E'$ and hydrogen to fixed
concentration values independent of the previous history of the
sample. The first step to achieve this effect is process
(\ref{hiidestructionreaction}), which destroys the H(II) and
rebuilds their precursors GLPC, this being completed just after the
first 250 shots (\figurename~\ref{destructionhii}). As for the $E'$
centers, in wet natural silica their concentration is almost null at
the beginning of each exposure, as the post-irradiation annealing
leads to the almost complete disappearance of the defect; moreover,
also $E'$ is consistently found to repeat "elastically" the same
decay kinetics after each high dose re-irradiation (see
\figurename~\ref{cinrip}). The combination of these findings suggest
the following picture for wet natural silica: in a 2000 pulses
exposure $E'$, H(II), GLPC and hydrogen reach some equilibrium
concentrations regardless of the previous history of the sample,
thereby leading to repeatability of the post-irradiation kinetics.
This scheme is also consistent with the observed \emph{in situ}
kinetics of $E'$ in natural wet silica (\figurename~\ref{cinmater}),
where 2000 pulses (with the same laser intensity 40\,mJcm$^{-2}$ and
repetition rate 1\,Hz used here) were found to (almost) saturate
[$E'$].

Apparently, this scheme does not apply to the dry material, where it
was observed a progressive accumulation of $E'$ upon successive
irradiations (\figurename~\ref{cinrip}) and 2000 pulses are not
enough to saturate [$E'$] (\figurename~\ref{cinmater}). In
principle, this could alter the post-irradiation kinetics leading to
a not perfect repeatability. Likely, the influence of the slowly
increasing concentration of $E'$ from an irradiation to the
successive one is negligible within the experimental conditions
explored so far.

\section{Modeling the reaction kinetics}\label{modelingkin}
According to our results, the post-irradiation processes in natural
\sil{} after 4.7\,eV irradiation are the result of the following
system of reactions, involving $E'$ ($\equiv$Si$^{\bullet}$), GLPC
(=Ge$^{\bullet\bullet}$), and giving Si\,--\,H and H(II)
(=Ge$^{\bullet}$\,--\,H) as the final results:
\begin{align} \label{systemsofreactions}
 \equiv\text{Si}^{\bullet}\text{ + H$_2$ } &
\stackrel{k_0}{\longrightarrow} \text{ }\equiv\text{Si\,--\,H +
H} \nonumber \\
 \equiv\text{Si}^{\bullet}\text{ + H }
& \stackrel{k_1}{\longrightarrow} \text{ }\equiv\text{Si\,--\,H}
 \\ =\text{Ge}^{\bullet\bullet}\text{ + H}
&\stackrel{k_2}{\longrightarrow} \text{
}=\text{Ge$^{\bullet}$\,--\,H} \nonumber \end{align} However, it is
still necessary to find out if the model inherent in
(\ref{systemsofreactions}) is able to account for the measured time
dependence of the concentrations of all the involved species. To
this aim, we start with writing down the chemical rate equations
governing the kinetics of (\ref{systemsofreactions}):
\begin{equation}\label{systemnossa}
\left\{
\begin{array}{l}
\displaystyle{\frac{d[\text{$E'$}]}{dt}}=-k_0[\text{$E'$}][\text{H$_2$}]-k_1[\text{$E'$}][\text{H}]
\\[2ex] \displaystyle{\frac{d[\text{H$_2$}]}{dt}}=-k_0[\text{$E'$}][\text{H$_2$}]
\\[2ex] \displaystyle{\frac{d[\text{GLPC}]}{dt}}=-k_2[\text{GLPC}][\text{H}]
\\[2ex] \displaystyle{\frac{d[\text{H}]}{dt}}=k_0[\text{$E'$}][\text{H$_2$}]-k_1[\text{$E'$}][\text{H}]-k_2[\text{GLPC}][\text{H}]
\end{array}
\right.
\end{equation}

Based on the fact that H is an intermediate reaction product, and
the reaction constants driven by the fast diffusion of H are much
larger than that depending on H$_2$ diffusion ($k_1$,$k_2\gg k_0$),
we can apply to this system the \emph{stationary state
approximation} \cite{Atkinsbook,GriscomJAP85}. As described in
detail in chapter \ref{backII}, this consists in assuming that both
the concentration of H and its time derivative remain negligible
during the progress of the kinetics, so as to set:
\begin{equation}
\displaystyle{\frac{d[\text{H}]}{dt}}=k_0[\text{$E'$}][\text{H$_2$}]-k_1[\text{$E'$}][\text{H}]-k_2[\text{GLPC}][\text{H}]\sim0
\end{equation}this equation can be used to express [H] as a
function of the other concentrations, and eliminate it from system
(\ref{systemsofreactions}). In this way, we obtain:
\begin{equation}\label{systemssa}
\left\{
\begin{array}{l}
\displaystyle{\frac{d[\text{$E'$}]}{dt}}=
-k_0[\text{$E'$}][\text{H$_2$}]\left(1+\frac{1}{1+\zeta\displaystyle{\frac{[\text{GLPC}]}{[\text{$E'$}]}}}\right)
\\[2ex] \displaystyle{\frac{d[\text{GLPC}]}{dt}}=-\displaystyle{\frac{d[\text{H(II)}]}{dt}}=
-k_0[\text{$E'$}][\text{H$_2$}]\left(1-\frac{1}{1+\zeta\displaystyle{\frac{[\text{GLPC}]}{[\text{$E'$}]}}}\right)
\\[2ex] \displaystyle{\frac{d[\text{H$_2$}]}{dt}}=-k_0[\text{$E'$}][\text{H$_2$}]
\end{array}
\right.
\end{equation}where $\zeta=k_2/k_1$ is a parameter that controls the
ratio between the amount of H captured by GLPC and that captured by
$E'$. This ratio is usually referred to as the \emph{branching
ratio} of the last two reactions in (\ref{systemsofreactions}).
System (\ref{systemssa}) explicitly shows that the main parameter
controlling the overall reaction rate is $k_0$, namely the reaction
constant between H$_2$ and $E'$: the whole post-irradiation stage of
the processes is driven by H$_2$ diffusion.

Before discussing the fit of experimental data with the rate
equations (\ref{systemssa}), we recall that chemical reaction
constants such as $k_0$ typically depend on temperature according to
the Arrhenius equation \cite{Atkinsbook,Doremus}:
\begin{equation}
k_0(\epsilon)=A\exp{\left(-\frac{\epsilon}{k_BT}\right)}
\label{rcgenericarrh}
\end{equation}where $A$ and $\epsilon$ are referred to as pre-exponential factor and activation energy for the
reaction of $E'$ with H$_2$, respectively. In the particular case of
a \emph{diffusion-limited reaction}\footnote{In a diffusion-limited
reaction (see chapter \ref{backII}), the diffusion of the mobile
species (rather than the local interaction with the other reagent)
is the bottleneck in determining the overall kinetics of the
process. This is regarded as the most common situation for reactions
in solids, where diffusion is quite a slow process
\cite{Doremus,Atkinsbook,GriscomJNCS84,KajiharaPRL02,KajiharaPRB06}.
}, the theoretical model by Waite (chapter \ref{backII}) can be
applied. In this case the value of the reaction constant $k_0$ is
predicted to be (from eq.~\ref{waiterateconstant}):
\begin{equation}\label{rcwaiterep}
k_0=4\pi r_0D_{0L}\exp(-E(\text{H$_2$})/k_BT)
\end{equation}
In (\ref{rcwaiterep}), $D=D_{0L}\exp(-$E$(\text{H$_2$})/k_BT)$ is
the diffusion constant of H$_2$ in silica glass, where \linebreak
$D_{0L}$=5.65$\times$10$^{-4}$\,cm$^{2}$s$^{-1}$ and
$E$(H$_2$)=0.45\,eV, as reported in literature from macroscopic
diffusion experiments of H$_2$ in \sil{}
\cite{LeeJCP63,GriscomJNCS84,Doremus}. The parameter $r_0$ is the
capture radius of the defect, i.e. a distance of the order of a few
10$^{-8}$\,cm under which $E'$ is supposed to react instantaneously
with H$_2$. Comparing (\ref{rcgenericarrh}) with (\ref{rcwaiterep}),
we see that within the Waite model for diffusion-limited reactions,
the activation energy $\epsilon$ coincides with that for H$_2$
diffusion in \sil{}, i.e. $E$(H$_2$), while the value for the
pre-exponential factor $A$ is:
\begin{equation}
A_W=4\pi r_0 D_{0L} \label{expressionA}
\end{equation}

\begin{figure}[t!]
\begin{center}
\includegraphics[width=.9\textwidth]{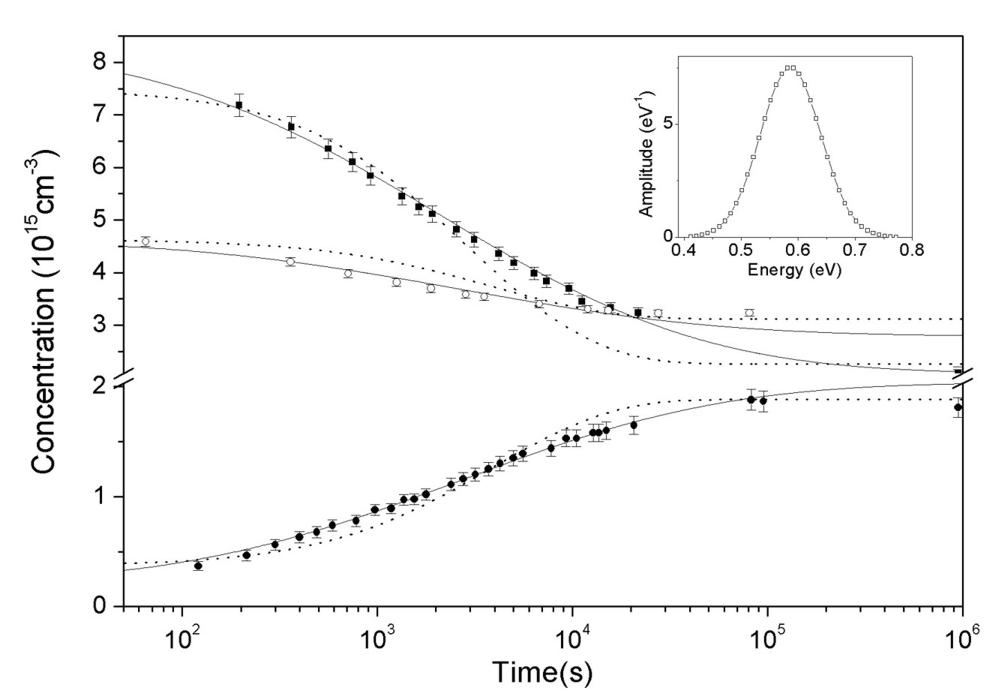}
\end{center}
\caption{Concentrations of $E'$ (squares), GLPC (empty circles) and
H(II) (circles) in a natural wet sample after exposure to 2000 laser
shots with 40\,mJcm$^{-2}$ energy density per pulse and 1\,Hz
repetition rate. Dotted lines are numerical solutions of system
\ref{systemssa}. Solid lines take also into account the statistical
distribution of the activation energy $\epsilon$ (represented in the
inset) which controls by (\ref{rcgenericarrh}) the reaction constant
between $E'$ and H$_2$.} \label{fitmatlab}\end{figure}

We measured the post-irradiation decay kinetics of $E'$ in a natural
wet \sil{} HER1 sample subjected to 2000 laser pulses, from the
amplitude of the 5.8\,eV OA band and using the known value of the
peak absorption cross section of the paramagnetic center. In this
context it becomes important to follow the kinetics until completion
(for a few days); hence, we monitored $E'$ by \emph{ex situ} OA
measurements performed with the JASCO spectrophotometer. In fact, as
already mentioned, the \emph{in situ} approach is useful to follow
the first $\sim$10$^4$\,s of the decay, but not suitable to perform
measurements on a very extensive time range. An obvious drawback of
the present choice is the loss of information on the first
$\sim$10$^2$\,s of the kinetics, which is a necessary compromise for
the present purposes. The results are reported in
\figurename~\ref{fitmatlab} as full square points. In the same graph
are reported again the kinetics of [H(II)] from
\figurename~\ref{PIKhiimaterials}, and of [GLPC], calculated from
the data in \figurename~\ref{PIKglpcmaterials} using the conversion
coefficient $\xi$. Then, the solutions of system (\ref{systemssa}),
calculated numerically, were fitted to the experimental datasets. In
the fitting procedure, the initial concentrations of $E'$, GLPC and
H(II) were constrained to the values obtained by extrapolating the
experimental curves at t=0,
[$E'$](t=0)=(8.6$\pm$0.5)$\times$10$^{15}$\,cm$^{-3}$,
[GLPC](t=0)=(4.7$\pm$0.2)$\times$10$^{15}$\,cm$^{-3}$ and
[H(II)](t=0)=(2.0$\pm$0.2)$\times$10$^{14}$\,cm$^{-3}$. Hence, the
fitting parameters that remain to be determined are [H$_2$](t=0),
$\zeta$ and $k_0$; the best fit values of the first two were found
to be [H$_2$](t=0)=(4.1$\pm$±0.3)$\times$10$^{15}$\,cm$^{-3}$ and
$\zeta$=(1.1$\pm$0.2).

\begin{figure*}[t!]
\begin{center}
\fcolorbox{postitA}{postitA}{\begin{minipage}[c]{.95\textwidth}\linespread{1.2}\hrule\vspace{0.2cm}
\textbf{\textsc{Summary II. }}\textsl{\textcolor{postitC}{In the
last part of this chapter we deal with the problem of fitting with a
suitable rate equation model the kinetics of $E'$, GLPC and H(II)
centers, observed after the end of laser irradiation and due to the
reactions eqs.~(\ref{systemsofreactions}) driven by diffusion of
H$_2$. Consistently with previous findings on other point defects
embedded in the disordered silica matrix, we show that our task can
be accomplished only by incorporating in the mathematical treatment
(eqs.~(\ref{systemssa})) a statistical distribution of the
activation energy $\epsilon$ which controls the reaction constant of
$E'$ with H$_2$. By a fit procedure (\figurename~\ref{fitmatlab}) we
are able to estimate the width of this (gaussian) distribution to be
FWHM=(0.12$\pm$0.02)\,eV. \vspace{0.2cm}}\hrule}\end{minipage}}
\end{center}
\end{figure*}

For what concerns $k_0$, a more complex picture emerges. In fact, as
already known from
literature,\cite{GriscomJNCS84,KajiharaPRL02,KajiharaPRB06} we found
that a good fit to the data on all the 10$^2$–-10$^6$\,s time scale
can be achieved only by introducing a \emph{statistical
distribution} of the activation energy $\epsilon$. This is done by
fitting experimental data with a \emph{linear combination of
solutions} of (\ref{systemssa}): individual solutions are first
calculated for different values of $\epsilon$, resulting in
correspondent values of $k_0$ from (\ref{rcgenericarrh}); then, the
weight of each solution in the linear combination equals the weight
of the associated value of $\epsilon$ within the distribution curve
of the activation energy. Such a situation is referred to by saying
that the kinetics of point defects in glasses are "anomalous". This
result is commonly interpreted as a consequence of the amorphous
nature of silica, which manifests itself in a statistical
distribution of activation energies, this being a fingerprint of
site-to-site inhomogeneity of the \sil{} matrix. A comment is
mandatory before applying this procedure to the current case. In
fact, present data at the single temperature T=300\,K do not allow
to estimate \emph{separately} $\epsilon$ and $A$, since only $k_0$
appears in the rate equations. This limitation will be overcome in
the next chapter on the basis of the study of the temperature
dependence of $k_0$. In the present context, before using the
statistical distribution of $\epsilon$ to calculate the
correspondent values of $k_0$ from (\ref{rcgenericarrh}), it is
first necessary to \emph{choose} a value of the pre-exponential
factor $A$. As the simplest hypothesis, to this purpose we used the
expression (\ref{expressionA}) valid within the Waite theory of
diffusion-limited reactions, with a capture radius arbitrarily
chosen to be $r_0$=5$\times$10$^{-8}$\,cm. Of course, it must be
kept in mind that the distribution of $\epsilon$ could be
\emph{shifted}, giving in particular a different mean value, by a
different choice of $A$ leading to the same distribution of $k_0$.

 Based on
these premises, we show as dotted lines in
\figurename~\ref{fitmatlab} the typical solutions of
eqs.~(\ref{systemssa}) obtained with a single value of $\epsilon$,
which manifestly fail to reproduce the shape of the experimental
kinetics. In contrast, we found that an excellent agreement (solid
curves) is attained introducing a Gaussian distribution of
$\epsilon$ with mean $<$$\epsilon$$>$=(0.59$\pm$0.01)\,eV and FWHM
$\Delta \epsilon$=(0.12$\pm$0.02)\,eV. The finding that all three
independent experimental datasets can be fitted at once for a
suitable choice of parameters, is a clear proof of the validity of
the chemical model (\ref{systemsofreactions}) hypothesized to
explain the post-irradiation processes, thus accomplishing one of
the goals that we wished to fulfill at the beginning of this
chapter. This is particularly true if we consider the approximations
implicit in eqs.~(\ref{systemssa}), such as neglecting every other
generation channel of H(II) and the stationary state approximation.
We propose in the following some comments on the values of the
fitting parameters.

\paragraph{I.}As explained above, present data do not allow to separate the
contributions of $A$ and $\epsilon$ to the reaction constant $k_0$,
and in particular the distribution of $\epsilon$ could be
\emph{shifted}, giving a different $<$$\epsilon$$>$, by a different
choice of $A$ leading to the same distribution of $k_0$. Hence,
apart from $k_0$ only the parameter $\Delta
\epsilon$=(0.12$\pm$0.02)\,eV has an \emph{absolute} meaning, and it
can be said to represent a measure of the effect of inhomogeneity on
the reaction properties of $E'$ with H$_2$.

\paragraph{II.}Nevertheless, the circumstance that $<$$\epsilon$$>$=0.59\,eV is
higher than the $E$(H$_2$)=0.45\,eV value reported for H$_2$
diffusion in \sil{}, contrary to what should be expected from
(\ref{rcwaiterep}), is not to be underestimated. In fact, although
$<$$\epsilon$$>$ does not possess an absolute meaning, this
disagreement can be restated saying that the value of the reaction
constant $k_0$($<$$\epsilon$$>$), given by (\ref{rcgenericarrh})
with the value (\ref{expressionA}) for the pre-exponential factor
$A$, is \emph{lower} than expected (eq.~(\ref{rcwaiterep})) from the
Waite model for a diffusion-limited reaction. Hence, it
\emph{unambiguously} indicates the inadequacy of the Waite theory to
describe the reaction between $E'$ and H$_2$, in the sense that the
present reaction is much \emph{slower} than expected for a purely
diffusion-limited process.\footnote{To see this point: we could have
chosen $A$ so as to \emph{fix} the mean activation energy
$<$$\epsilon$$>$ to $E$(H$_2$), but this would have not reconciled
the results of the fit with the Waite model; indeed, notwithstanding
the agreement of $\epsilon$, in this case the chosen $A$ would have
been much lower than given by (\ref{expressionA}). } Such a result
is analogous to the comparison proposed in chapter \ref{overview},
in which the typical decay time of $E'$ was found to be
\emph{longer} than expected for a purely diffusion-limited reaction
with H$_2$. These findings (as well as a few other similar results
in literature \cite{ImaiPRB91,CannasJNCS04}) suggest the idea that
the passivation of $E'$ by H$_2$ is also \emph{reaction-limited}, so
as to yield an effective value of $\epsilon$ \emph{lower} than
$E$(H$_2$) because $k_0$ incorporates somehow also the activation
energy \emph{for reaction}.

\paragraph{III.}In regard to the statistical distribution of the activation
energy, the situation we have found for $E'$ is subtly different
from previous works, in which a similar "anomalous" decay kinetics
was evidenced for the decay of NBOHC centers. In fact, differently
from the present case, the mean value of $\epsilon$ for NBOHC was
found close to $E$(H$_2$) \cite{KajiharaPRL02,KajiharaPRB06}. This
permitted to conclude that the reaction of NBOHC with H$_2$ is
diffusion-limited and, most important, to identify $<$$\epsilon$$>$
coming from the fit with the activation energy for H$_2$ diffusion,
so as to interpret the very existence of a randomization of
$<$$\epsilon$$>$ as a property \emph{inherent to diffusion} in
silica. Consistently with this view, macroscopic diffusion
experiments directly showed an anomalous temperature-dependence of
the diffusion constant of some species in \sil{}, which can be
explained on the basis of a distribution of the activation energy
for diffusion \cite{ShelbyKeeton}. In comparison, even if the
present results extend from NBOHC to $E'$ the observation of the
influence of inhomogeneity on the reaction properties with H$_2$, at
this stage it is not completely clear how to interpret from a
physical point of view the distribution of $\epsilon$ introduced to
treat the kinetics. All these problems will be dealt with in the
next chapter.

\paragraph{IV.}In some works in literature it was used a
\emph{flat} statistical distribution of $\epsilon$ to describe
anomalous reaction kinetics in glass, the advantage being the
possibility of finding an analytical expression for the kinetics in
some cases \cite{BobyshevKK90}. Anyway, more recent studies on the
reaction between NBOHC and H$_2$ have confirmed to a large extent
the applicability of a Gaussian distribution of $\epsilon$ to the
modeling of both isothermal and isochronous reaction kinetics
\cite{KajiharaPRB06}. We observe here that, due to the heuristic
nature of the approach itself of fitting the kinetics with a linear
combination of solutions of the rate equation system, it is
difficult to predict \emph{a priori} if the correct form of the
distribution should be flat or Gaussian. Moreover, this detail is
not particularly important to our purposes; in fact, the main
results of our line of reasoning (the demonstration that model
(\ref{systemsofreactions}) is consistent with experimental data, and
the evidence of an anomalous kinetics) are not influenced by this
choice, which could only bear consequences on the precise estimate
of $\Delta \epsilon$.

\paragraph{V.}The concentration of H$_2$ at t=0 found from the fit is approximately
one half of $E'$ at the same time instant. This is consistent with
the model for the generation of $E'$ proposed in chapter \ref{JPCMs}
for wet natural silica: $E'$ and H are generated during irradiation
from the common precursor Si\,--\,H. Also, the present fit procedure
quantitatively confirms that the small residual concentration of
$E'$ ($\sim$2$\times$10$^{15}$cm$^{-3}$) still remaining at the end
of the post-irradiation stage in wet samples is due to the portion
of hydrogen which escapes recombination with $E'$ and is trapped on
GLPC.

\paragraph{VI.}Finally, we note that a value of $\zeta$ of the order of unity is to be
expected if the reactions of $E'$ and GLPC with \emph{atomic}
hydrogen  are diffusion-limited. In fact, in the framework of the
Waite model\cite{Waite57}, $\zeta$ should equal the ratio of the
capture radii of GLPC and $E'$ for H, which both should be of the
order of an atomic dimension if the model is applicable at all.

\section{Conclusions}
We investigated the post-irradiation kinetics of point defects
induced at room temperature in natural \sil{} by 4.7\,eV laser
irradiation, with particular attention to the conversion processes
of Ge-related defects. The analysis of the time dependence of the PL
signal of the GLPC center permits to isolate the post-irradiation
stage of its conversion process, which is ascribed to trapping of H
at the defect site leading to the generation of H(II) center, on the
basis of the anti-correlated concentration variations of the two
species. This process can be reversed by a second laser exposure,
which destroys H(II) and restores the precursor GLPC. We provide a
measure of the cross section of photo-induced rupture of H(II):
$\sigma_D$=6.2$\times$10$^{-19}$\,cm$^{2}$. Multiple high-dose
irradiations result in the repetition of the same post-irradiation
kinetics of H(II) centers after each exposure. This effect is
achieved by the photo-decomposition of previously formed defects and
by the ability of each high dose exposure to reset to fixed values
the concentrations of the centers involved in the post-irradiation
phenomena. Atomic hydrogen to be trapped at the GLPC site is
produced by breaking of diffusing H$_2$ on $E'$ centers.
Consistently, the time dependence of $E'$, H(II), and GLPC
concentrations can be fitted by a suitable set of coupled rate
equations describing the chemical reactions occurring in the
post-irradiation stage. The kinetics of the three species are mainly
controlled by the reaction between $E'$ and H$_2$ and their features
suggest a statistical distribution of the activation energy
controlling the reaction constant. The results of the fit suggest
that the reaction kinetics of $E'$ with H$_2$ is not purely
diffusion-limited.

\chapter{Temperature dependence of the generation and decay of $E'$ center}\label{PRL}
This chapter completes our experimental investigation by the study
of the temperature dependence of the generation and decay of $E'$
center in natural silica by 4.7\,eV laser radiation.
\section{Introduction}
\hspace{0.8cm}The experimental results have shown that natural
\sil{} is an interesting system in that it allows to investigate the
generation of $E'$ by 4.7\,eV pulsed laser radiation, the decay of
$E'$ due to reaction with H$_2$, and some conversion processes of
Ge-related defects. Although the experiments presented up to now
have provided many important information on these phenomena, a
thorough investigation must necessarily include also the temperature
dependence of these processes.

For what concerns $E'$ ($\equiv$Si$^{\bullet}$), temperature may be
expected \emph{a priori} to influence both its generation and its
stability through several effects. In particular, the investigation
of the temperature dependence of the post-irradiation decay of the
defect due to reaction with H$_2$:
\begin{equation} \label{reacteh2repreprep}
\equiv\text{Si}^{\bullet}\text{ + H$_2$ } \longrightarrow \text{
}\equiv\text{Si\,--\,H + H}\end{equation} is motivated by a variety
of reasons. Indeed, process (\ref{reacteh2repreprep}) basically
represents the passivation of a silicon dangling bond ($E'$) in the
"archetypal" \sil{} amorphous solid by the ubiquitous diffusing
species H$_2$; from this point of view, it is to be regarded as one
of the basic interaction processes between point defects and mobile
species in amorphous solids. Besides, from the technological point
of view, this process takes an important place in the general topic
of passivation of many defects detrimental for applications, such as
the NBOHC, the other fundamental paramagnetic defect in silica glass
\cite{GriscomJNCS84,KajiharaPRL02,KajiharaPRB06}, and the P$_b$
center, (Si$_3$$\equiv$Si$^{\bullet}$), common at the Si/SiO$_2$
interface in metal-oxide-semiconductor (MOS) devices
\cite{BrowerPRB88,Poindexterreview,StesmansAPL96}. The reaction of
$E'$ with H$_2$ has been discussed in literature by several
experimental and theoretical works, aiming to estimate the kinetic
parameters of the process (subsection \ref{reactewithh2section}).
One of the most debated issues has been whether the reaction
kinetics is diffusion- or reaction-limited: in fact, contrary to the
common assumption that the reaction kinetics of defects with
migrating species in solids are diffusion-limited (due to the
slowness of the diffusion process in solids), some results in
literature and in the previous chapters have indicated the reaction
of $E'$ with H$_2$ to proceed more slowly than expected for a purely
diffusion-limited process; consistently, computational
investigations proposed the interaction between $E'$ and H$_2$ to be
characterized by an activation barrier
\cite{EdwardsJNCS94,EdwardsJNCS95,Erice,VitielloJPCA00}. The
possibility that the diffusion-limited model may be unapplicable to
such a "basic" reaction deserves to be thoroughly investigated. In
this sense, besides the specific interest in elucidating the
properties of a very common point defect in \sil{}, the interest of
this problem extends well beyond the "silica community".

Nevertheless, until now detailed kinetic studies of reaction
(\ref{reacteh2repreprep}) have been carried out only for surface
$E'$ centers\footnote{The \emph{surface} $E'$ is a
$\equiv$Si$^{\bullet}$ isolated dangling bond existing on the
surface of \sil{}. It features slightly different spectroscopic
properties with respect to bulk
$E'$.\cite{SkujaReview98,SkujaSPIE01} } \cite{RadtzigJPC95}, or for
$E'$ centers in thin silica films \cite{LiJNCS90}. The advantage of
these systems is that the reaction can be directly investigated by
exposure of the surfaces to gaseous H$_2$. These two works report
disagreeing results: the measured rates of the reaction at
$T$=300\,K differ by two orders of magnitude, while the activation
energies are 0.4\,eV and 0.3\,eV respectively
\cite{RadtzigJPC95,LiJNCS90}. The process has been also discussed by
computational studies, where it was shown to require an activation
energy of a few tenth of eV
\cite{Erice,EdwardsJNCS94,EdwardsJNCS95,VitielloJPCA00,PacchioniPRB98c}.
In contrast, for $E'$ centers in bulk silica, the task of studying
experimentally the reaction becomes more difficult, since here the
concentration of available H$_2$ is not an external controllable
parameter, and diffusion of H$_2$ in the bulk \sil{} matrix becomes
a necessary step to bring the reagents in contact. Furthermore, it
is needed to measure \emph{in situ} the kinetics of the transient
defects in temperature-controlled experiments. For these reasons,
even if process (\ref{reacteh2repreprep}) has been repeatedly
observed to spontaneously passivate radiation-induced $E'$ centers
\emph{in bulk}\cite{ImaiPRB91,CannasJNCS04,SmithAO00,KuzuuPRB91},
its thermal activation properties have still to be thoroughly
investigated. Studying these problems by \emph{in situ} OA
spectroscopy is the purpose of this chapter.

\section{Experiments and Results}
\hspace{0.8cm}The experiments were performed on I301 natural dry
specimens, placed in the He flow cryostat and irradiated with
4.7\,eV laser radiation of 40\,mJcm$^{-2}$ energy density per pulse
and 1\,Hz repetition rate. We performed isothermal irradiations at
temperatures 10\,K$<$T$<$475\,K on different as-grown samples. Each
sample was exposed to laser radiation for a time
$\tau$=2$\times$10$^3$\,s at a given temperature, and its OA
spectrum was monitored \emph{in situ} during exposure. Then, the
measurements continued for a few 10$^3$\,s after the end of exposure
while keeping the specimen at the same temperature at which it had
been irradiated.

\begin{figure}[htb!]
\begin{center}
\includegraphics[width=.9\textwidth]{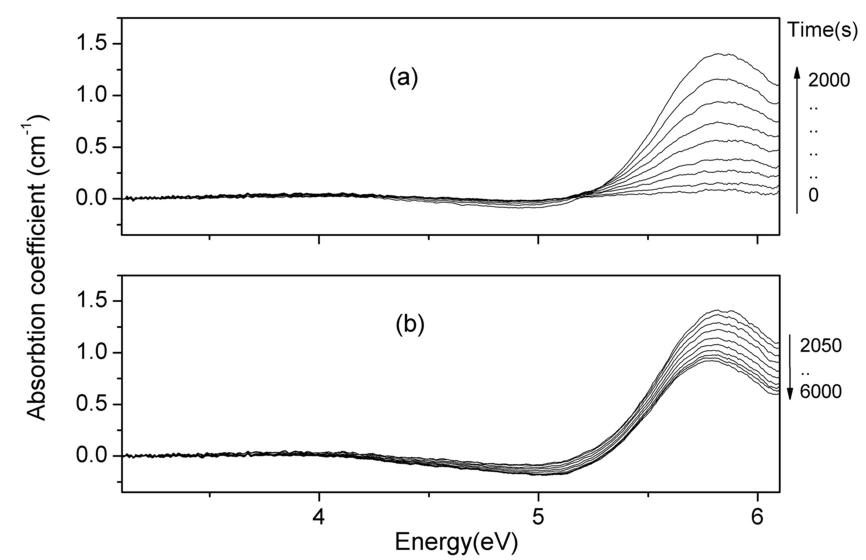}
\end{center}
\caption{Difference absorption induced on natural dry silica by
laser irradiation with 2000 pulses of 40\,mJcm$^{-2}$ energy density
per pulse and 1\,Hz repetition rate at 250\,K, as detected \emph{in
situ} at several time instants during (a) and after the end (b) of
the irradiation session.} \label{insitua250K}\end{figure}

In \figurename~\ref{insitua250K} the typical absorption profile is
reported as observed at different times during (a) and after the end
(b) of the irradiation session at 250\,K.
 Similarly
to what we observe at room temperature (see chapter \ref{overview})
on the same materials, the main detected signal is the 5.8\,eV band
of the $E'$ centers, accompanied by the negative component near
5.1\,eV due to bleaching of the B$_{2\beta}$ band and by a very weak
and broad component\footnote{This signal, detected only in the
natural dry I301 samples, has been ascribed to Ge-related
impurities\cite{Nalwa}. However, it is of no concern for our present
purposes.} peaked at $\sim$4\,eV. The profiles in
\figurename~\ref{insitua250K} are representative of all the
investigated temperatures.
\begin{figure}[htb!]
\begin{center}
\includegraphics[width=.7\textwidth]{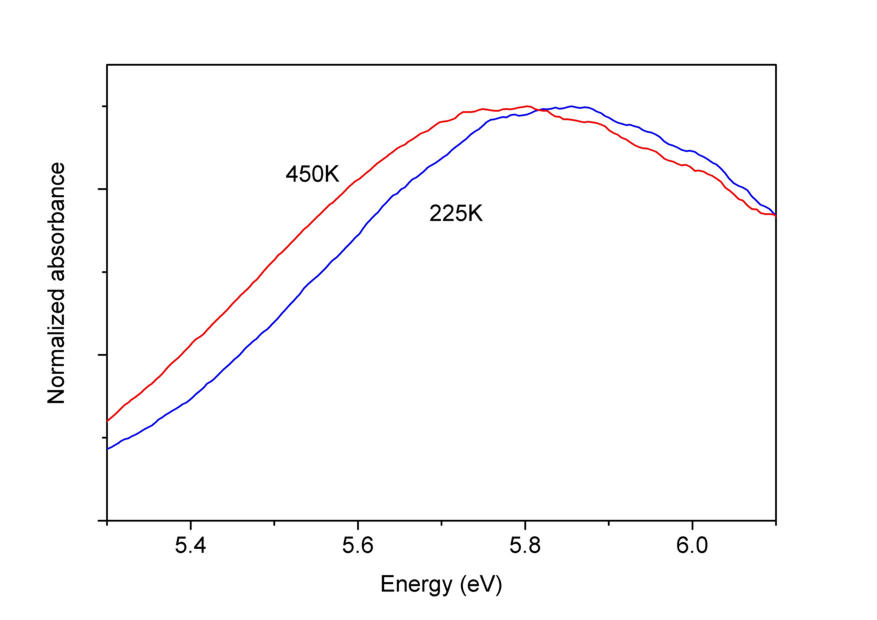}
\end{center}
\caption{Shape of the 5.8\,eV band induced by laser irradiation in
natural dry silica at two representative temperatures.}
\label{shapecomp}\end{figure}
\begin{figure}[htb!]
\begin{center}
\includegraphics[width=.7\textwidth]{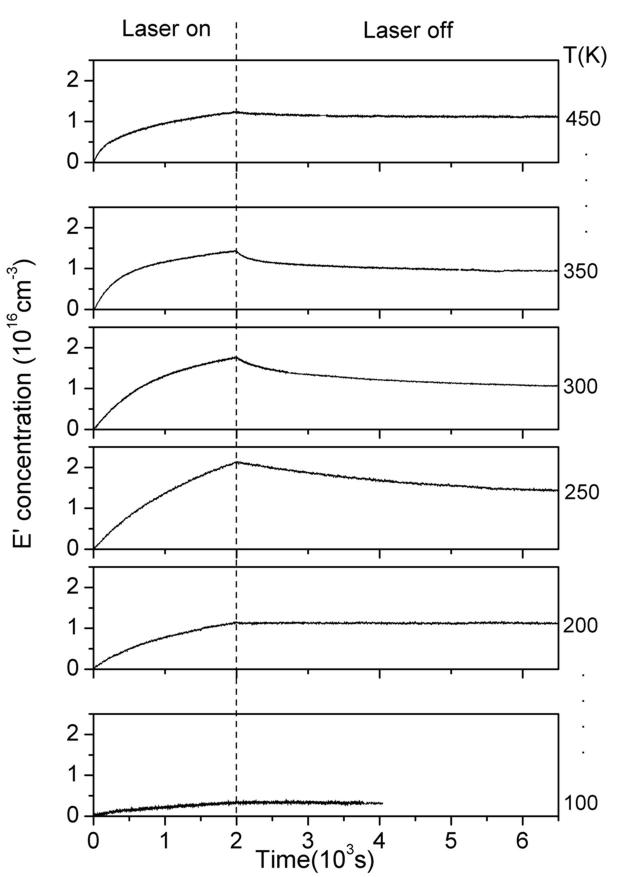}
\end{center}
\caption{Kinetics of $E'$ concentration induced by laser irradiation
at several representative temperatures.}
\label{kinvstemptutte}\end{figure}The shape of the laser-induced
5.8\,eV band weakly depends on temperature\footnote{In contrast, the
oscillator strength of the 5.8\,eV band (controlling the ratio
between [$E'$] and OA intensity), does not depend on temperature
\cite{upd}. }. For example, as shown in \figurename~\ref{shapecomp},
in the experiment at 225\,K the band is peaked at
(5.85$\pm$0.02)\,eV with a (0.68$\pm$0.02)\,eV FWHM, while at 450\,K
the peak is at (5.78$\pm$0.02)\,eV with (0.71\,eV$\pm$0.02)\,eV
FWHM. Nevertheless, such variations are sufficiently small to be
neglected for what concerns the procedure used to estimate the $E'$
concentration. Hence, also here [$E'$] can be calculated in a good
approximation by multiplying the peak amplitude of the band by the
known cross section, similarly to what we have seen in chapter
\ref{overview}. These calculated kinetics [$E'$](t,$T$) at different
representative temperatures are reported in
\figurename~\ref{kinvstemptutte}. The concentration of $E'$ grows
during the irradiation session of duration $\tau$ to a final value
[$E'$]($\tau$,$T$). Inspection of the curves shows that
[$E'$]($\tau$,$T$) has a maximum at $T$=250\,K. As for the
post-irradiation stage, only for the experiments at $T$$>$200\,K we
observe a measurable decay of $E'$, supposedly due to reaction
(\ref{reacteh2repreprep}). Both the entity of the decay and its
typical time scale are influenced by temperature, in a way that is
going to be analyzed in the following.

Let us focus now our attention on the 200\,K--475\,K temperature
range, i.e. we temporarily exclude from our considerations the
kinetics at temperatures so low that the diffusion and reaction of
H$_2$ with $E'$ are quenched. Consistently with our results at room
temperature, we are going to assume in the following that reaction
(\ref{reacteh2repreprep}) is the only cause of $E'$ decay in all
this temperature range\footnote{As usual, the parallel reaction of
$E'$ with \emph{atomic} hydrogen is automatically accounted for when
treating reaction (\ref{reacteh2repreprep}) within the stationary
state approximation.}. In addition we assume that hydrogen is
consumed only by $E'$: indeed, the reaction (\ref{formHII}) with
GLPC of H produced at the right side of reaction
(\ref{reacteh2repreprep}) involves only a minor portion of the total
hydrogen population, and NBOHC centers, which are known to react
with H$_2$ \cite{GriscomJNCS84,KajiharaPRL02,KajiharaPRB06}, are
virtually absent, as inferred from the lack of their 4.8\,eV OA and
1.9\,eV PL optical activities. Hence, we write again the rate
equation accounting for the time variations of [$E'$] in the
post-irradiation stage due to reaction (\ref{reacteh2repreprep}):
\begin{equation}
\frac{d[\text{$E'$}]}{dt}=2\frac{d[\text{H$_2$}]}{dt}=-2k_0(T)[\text{$E'$}][\text{H$_2$}]\label{rateeqrep}
\end{equation}
A consequence of this equation is that the total decay
$\Delta$[$E'$]($T$) equals twice the amount of radiolytic hydrogen
[H$_2$]($\tau$,$T$) available at the end of irradiation:
\begin{equation}
\Delta[E']\text{($T$)}=\text{[$E'$]($\tau$,$T$)}-[\text{$E'$}]_\infty\text{($T$)}=2\text{[H$_2$]($\tau$,$T$)}\label{deltarelatedtoh2}
\end{equation}where [$E'$]$_\infty$($T$) is the stationary asymptotic concentration of $E'$
at the end of the decay. On the one hand, this equation permits to
indirectly estimate the concentration of available hydrogen if both
the maximum and the stationary concentrations of $E'$ are known. On
the other hand, eq.~(\ref{deltarelatedtoh2}) uniquely determines
[$E'$]$_\infty$($T$) at each temperature given [$E'$]($\tau$,$T$)
and [H$_2$]($\tau$,$T$); in fact, the decrease of $E'$ only depends
on the amount of hydrogen made available by irradiation, since the
decay of the defect ends as soon as H$_2$ is completely exhausted.
This simple observation has an important consequence: if a sample is
irradiated at a given $T$ and then heated up (cooled down) to room
temperature some time after the end of irradiation, this does not
alter [$E'$]$_\infty$($T$), which depends only on the concentrations
of $E'$ and hydrogen created by irradiation, [$E'$]($\tau$,$T$) and
[H$_2$]($\tau$,$T$) respectively. The only effect of returning to
room temperature is an acceleration (slowing down) of the decay. On
this basis, for each irradiation, after observing \emph{in situ} for
a few 10$^3$\,s the kinetics of $E'$, we brought back the sample at
room temperature, where we continued to monitor the decay of [$E'$]
for a few days after irradiation, so as to determine
[$E'$]$_\infty$($T$) by \emph{ex situ} ESR measurements.

\begin{figure}[htb!]
\begin{center}
\includegraphics[width=.8\textwidth]{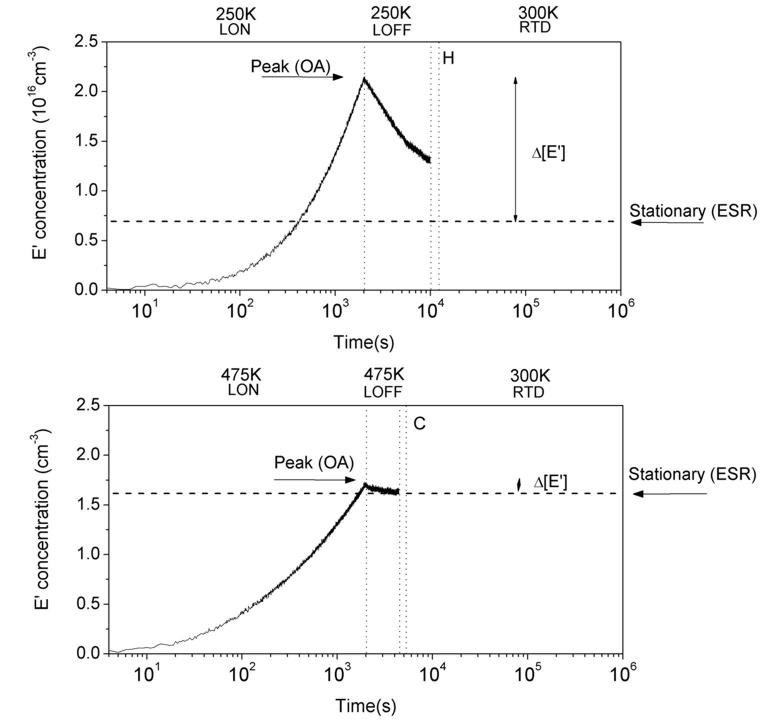}
\end{center}
\caption{Representation of the experimental procedure used to
extract some relevant parameters from the kinetics [$E'$](t) induced
by laser irradiation at two representative temperatures, 250\,K
(upper panel) and 475\,K (lower panel). After remaining a few
10$^3$\,s at the irradiation temperature, the specimen returns to
room temperature, where the decay of $E'$ continues till the
stationary concentration is measured by ESR. Consistent results are
obtained by measuring the stationary concentration by \emph{ex situ}
OA, but ESR allows for a higher precision when the concentration is
a few 10$^{15}$cm$^{-3}$. LON stands for "laser on", LOFF stands for
"laser off", H stands for "heating up at 300\,K", C for "cooling at
300\,K", RTD stands for Room Temperature Decay.}
\label{illustrtech}\end{figure}

Measuring the stationary concentration \emph{after} that the sample
has returned to room temperature is particularly convenient for low
temperature experiments, where an \emph{in situ} measurement of
[$E'$]$_\infty$($T$) would require to follow the slow decay kinetics
for a very long time. Two examples of this approach are displayed in
\figurename~\ref{illustrtech}.
\begin{figure}[htb!]
\begin{center}
\includegraphics[width=.9\textwidth]{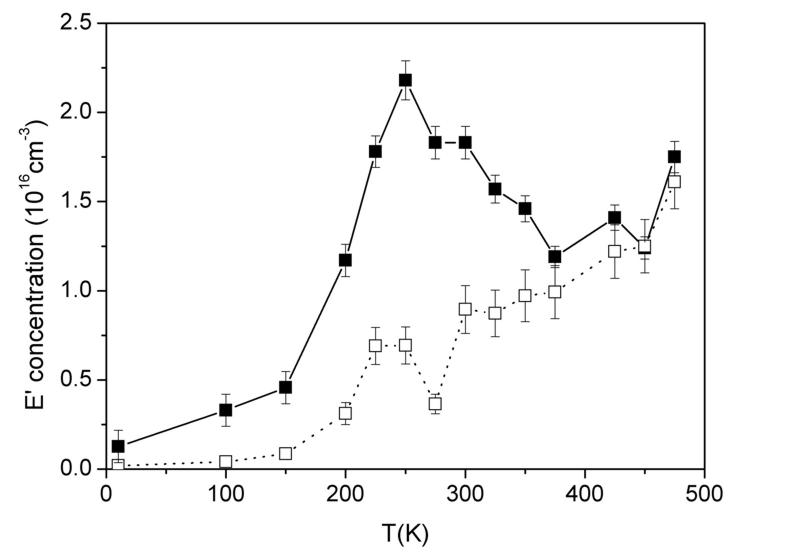}
\end{center}
\caption{Concentration of $E'$ measured \emph{in situ} at the end of
laser irradiation (full symbols) and stationary concentrations
measured by ESR after the end of the decay process (empty symbols),
as a function of the irradiation temperature.}
\label{maxminvsT}\end{figure}
\begin{figure}[htb!]
\begin{center}
\includegraphics[width=.8\textwidth]{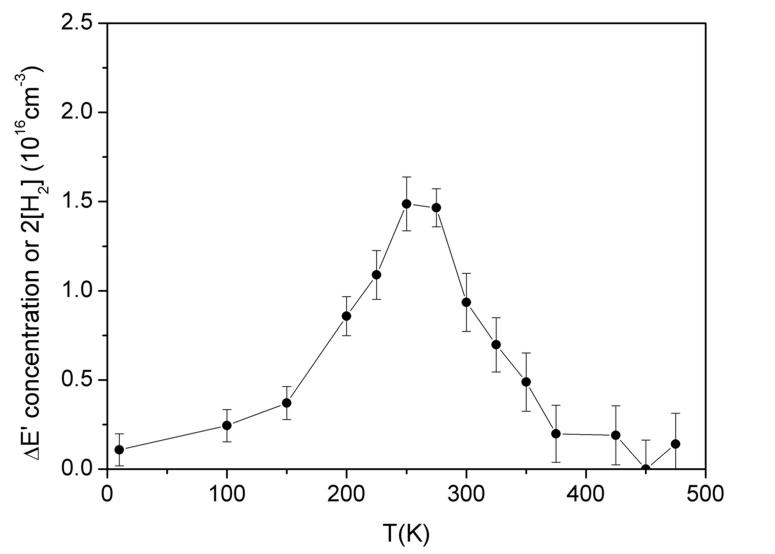}
\end{center}
\caption{Concentration decrease of $E'$ in the post-irradiation
stage (which equals two times the concentration of radiolytic
hydrogen 2[H$_2$]) after irradiation with 2000 laser pulses, as a
function of the irradiation temperature.}
\label{deltavsT}\end{figure}The upper panel shows the \emph{in situ}
kinetics at 250\,K. After a few hours from the end of exposure, when
the decay is still in progress, the specimen was heated at room
temperature. Then, a few days after the experiment we measured by
ESR the stationary concentration represented in the \figurename. The
same approach was used at all temperatures, even if for $T$
significantly higher than $>$300\,K (e.g. lower panel), the
permanence of the sample at $T$ for a few 10$^3$\,s is enough to
complete the kinetics, so that the last \emph{in situ} measurement
agrees with the stationary ESR measurement performed after cooling
back at room temperature.

Finally, at the beginning of this discussion we had momentarily left
aside the low temperature ($T$$<$200\,K) kinetics, i.e. the regime
in which H$_2$ diffusion and reaction are quenched. In this case,
the stationary concentration [$E'$]$_\infty$($T$) (measured after
returning to room temperature) assumes a slightly different meaning.
Indeed, even if we are not going to discuss in detail the
post-irradiation stability of $E'$ at low temperatures, we expect
that when process (\ref{reacteh2repreprep}) is quenched,
[$E'$]$_\infty$($T$) measured at room temperature does not coincide
with the asymptotic concentration theoretically reached at the end
of the isothermal decay. In fact, in this case the post-irradiation
stage is controlled by \emph{atomic} hydrogen diffusion only, as
H$_2$ is immobile; hence, after conclusion of the post-irradiation
stage, the $E'$ centers can still coexist with a concentration of
stationary (immobile) H$_2$, after that H has been exhausted by
dimerization and reaction with $E'$. In particular, in the extreme
case of the 10\,K kinetics also H is immobile, so that \emph{all}
$E'$ are expected to be stationary before heating. For comparison,
we stress that there is no stationary H$_2$ when process
(\ref{reacteh2repreprep}) is active (above 200\,K). However,
notwithstanding the fact that the values [$E'$]$_\infty$($T$) at the
three lowest investigated temperatures (150\,K, 100\,K and 10\,K)
cannot be literally interpreted as asymptotic isothermal
concentrations, the differences $\Delta[E']\text{($T$)}$ continue to
represent the amount of hydrogen made available by
irradiation\footnote{which at the irradiation temperature is
expected to be stored partly in atomic and partly in molecular
form.}. In fact, this property depends only on the circumstances
that reaction(s) with H and H$_2$ are the only cause of $E'$ decay,
and each hydrogen atom (or half a hydrogen molecule) passivates a
single $E'$.

With these considerations in mind, the values of [$E'$]($\tau$,$T$)
and [$E'$]$_\infty$($T$) measured with the above procedures are
plotted in \figurename~\ref{maxminvsT}. Apart from a certain
scattering of the data, mainly due to limited repeatability of the
irradiation conditions, the basic qualitative features of the
process are \textbf{(i)}: the stationary concentration is very low
below $\sim$150\,K, and above this threshold it monotonically
increases with temperature; \textbf{(ii)}: also the maximum \emph{in
situ} concentration increases starting from $\sim$150\,K; above this
temperature, it grows to a maximum at $\sim$250\,K, and then it
decreases and joins the stationary concentration curve above
$\sim$400\,K. The variation $\Delta$[$E'$]($T$) in the
post-irradiation stage, calculated as the difference between the two
curves in \figurename~\ref{maxminvsT}, is reported in
\figurename~\ref{deltavsT}; its temperature dependence is roughly
bell-shaped and features a maximum near 250\,K.

From a phenomenological point of view, we can summarize the results
as follows. First, the generation of $E'$ is a thermally activated
process, quite inefficient below $\sim$150\,K. Above this threshold,
a general feature of the generated $E'$ is their post-irradiation
instability, which leads to a difference $\Delta$[$E'$]($T$) between
the maximum \emph{in situ} concentrations and the stationary
concentrations. As a consequence, we stress that up to $\sim$400\,K,
a thorough study of laser-induced $E'$ cannot rely only on
measurements of the stationary concentration; on the contrary, the
\emph{in situ} approach is mandatory to have a full picture of the
laser-induced process based on the observation of the kinetics of
the defects. Above $\sim$400\,K, $\Delta$[$E'$]($T$) tends to zero,
meaning that the generated $E'$ are basically stable in the
post-irradiation stage, and the stationary concentration almost
coincides with the maximum \emph{in situ} concentrations. The
conceptual distinction between the two measurements is lost and in
these conditions the \emph{in situ} and \emph{ex situ} techniques
yield the same information.

Before going on to discuss the results, we also note that the
behaviors in \figurename~\ref{maxminvsT} and
\figurename~\ref{deltavsT} have been observed for a given number
(2000) of laser shots; hence, although the qualitative features of
the process are expected to be basically independent of irradiation
dose, detailed features such as the position of the peak in
\figurename~\ref{deltavsT} may vary with the number of pulses.
However, this consideration does not apply to the value of the
reaction constant between $E'$ and H$_2$,\footnote{Even if the
maximum or asymptotic concentration of $E'$ (as well as the
concentration of radiolytic hydrogen) depend on the number of laser
shots received by the specimen, the reaction constant clearly does
not, being an intrinsic property of the reaction process determined
only by temperature.} which is calculated and analyzed as a function
of temperature in the following section.

\section{Discussion I: Character of the reaction \\ between $E'$ center and \texorpdfstring{H$_2$}{hydrogen}}
\subsection{Rate constant of the reaction}
\hspace{0.8cm}We are going now to focus on the post-irradiation
stage of the laser-induced processes, and estimate from experimental
data the reaction constant between $E'$ and H$_2$. By inverting
eq.~(\ref{rateeqrep}) calculated at the end of irradiation
(t=$\tau$), and using (\ref{deltarelatedtoh2}), the rate constant
$k_0$(T) of the reaction can be expressed as follows:

\begin{eqnarray}
k_0(T)&=&-\frac{d\text{[$E'$]}}{dt}(\tau,T)
\frac{1}{2\text{[$E'$]($\tau$,$T$)}\text{[H$_2$]($\tau$,$T$)}}
\nonumber
\\
&=&-\frac{d\text{[$E'$]}}{dt}(\tau,T)\frac{1}{\text{[$E'$]($\tau$,$T$)}(\text{[$E'$]($\tau$,$T$)}-\text{[$E'$]$_\infty$($T$)})}\label{calcolarek}
\end{eqnarray}
This equation generalizes to dry natural silica, and at all
temperatures, the line of reasoning that permitted in chapter
\ref{JPCMs} to estimate the reaction constant at room
temperature\linebreak $k_0$($T$=300\,K)=
8.3$\times$10$^{-20}$cm$^3$s$^{-1}$. Also in this case, the initial
decay slopes d[$E'$]/dt($\tau$,$T$) appearing in (\ref{calcolarek})
can be estimated by a linear fit in the first $\sim$50\,s of the
post-irradiation stage of the kinetics of
\figurename~\ref{kinvstemptutte}.
\begin{figure}[h!]
\begin{center}
\includegraphics[width=\textwidth]{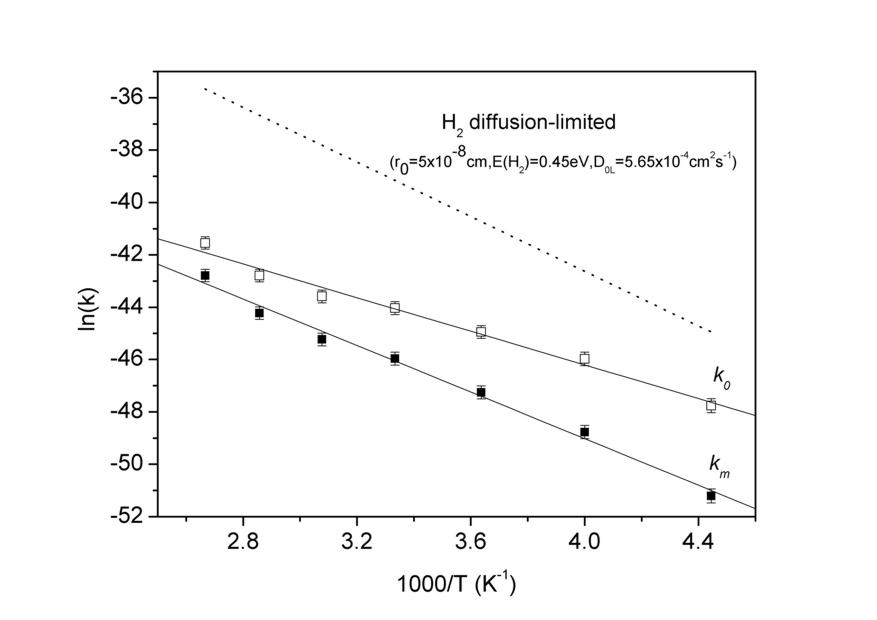}
\end{center}
\caption{Reaction constant between $E'$ and H$_2$, taking ($k_m$,
full symbols) or not taking ($k_0$, empty symbols) into account the
distribution of activation energy. Dotted line: prediction based
upon a purely diffusion-limited reaction model. Full lines: fits
with Arrhenius equations.} \label{costantereazionevsT}\end{figure}
Hence, inserting in (\ref{calcolarek}) these decay slopes and the
values in \figurename~\ref{maxminvsT}, we calculate $k_0$($T$),
which is plotted against temperature in
\figurename~\ref{costantereazionevsT} as empty symbols. The
uncertainties were estimated by repeating many times the experiment
at $T$=300\,K. The useful temperature range where it was possible to
estimate $k_0$ extends from 225\,K to 375\,K, outside which interval
the initial decay slope is respectively too slow or too fast to be
accurately measured. Data are consistent with an Arrhenius
dependence:
\begin{equation}
k_0=A_i×\exp{\left(-\frac{E_i}{k_BT}\right)};
\end{equation} where the index "i" in $E_i$ and $A_i$ reminds us that $k_0$ was
determined from the initial decay slope. From a linear fit, we
obtain the following values for the activation energy and for the
pre-exponential factor: $E_i$=(0.28$\pm$0.01)\,eV and
$A_i$=3$\times$10$^{-15}$cm$^3$s$^{-1}$.

We must now take into account the effects of the disorder in the
glass matrix, causing an inhomogeneous distribution of the
activation energy, and consequently also of the reaction constant
$k$. Indeed, according to what is known from literature
\cite{GriscomJNCS84,KajiharaPRL02,KajiharaPRB06}, we have
demonstrated in the previous chapter that the post-irradiation
kinetics of $E'$ in natural silica at room temperature can be fitted
only by linear combinations of solutions of the chemical rate
equations obtained with different reaction constants
$k=A\exp(-\epsilon/k_BT)$ weighted by a Gaussian distribution of
$\epsilon$ centered at a mean value $E_m$. The pre-exponential
factor, as usual in literature, was taken as undistributed. In
particular, we estimated the FWHM of the distribution to be
(0.12$\pm$0.02)\,eV, corresponding to a standard deviation of
$\sigma_0$=(0.05$\pm$0.01)\,eV.

This bears important consequences on the interpretation of the
constant $k_0$ evaluated from the initial decay slope. In fact, due
to such inhomogeneity effects, $k_0$ actually corresponds to the
\emph{mean} $<$$k$$>$ calculated on the distribution of $\epsilon$:
\begin{eqnarray}
  k_0=<k> & = & \int A\exp{\left(-\frac{\epsilon}{k_B
T}\right)}\times\frac1{\sigma\sqrt{2\pi}}exp\left[-{\frac{(E_m-\epsilon)^2}{2\sigma^2}
}\right]d\epsilon \nonumber \\
   & = & A\exp{\left[\frac{\sigma^2}{2k_B^2T^2}\right]}\exp{\left(-\frac{E_m}{k_B
T}\right)} \label{valmedkdistribuzione} \\
 & = & k_m\exp{\left[\frac{\sigma^2}{2k_B^2T^2}\right]} \nonumber
\end{eqnarray}
where
\begin{equation}
k_m=A\exp{\left(-\frac{E_m}{k_BT}\right)}<k_0
\label{relazionetrakmek0}
\end{equation}
 Hence, $k_0$ is \emph{larger} than the value $k_m$
corresponding to the mean activation energy $E_m$: for this reason,
the above derived value $E_i$=0.28\,eV must be interpreted as a
phenomenological \emph{effective} energy, describing the fastest
stage of the decay. The analysis based on the initial decay slope
actually samples only the most reactive subset within the whole
population of species participating in the reaction. Strictly
speaking, only $k_m$ is expected to exhibit an Arrhenius behavior,
whereas $k_0$ should deviate from it due to the term in 1/$T^2$
introduced by eq.~(\ref{valmedkdistribuzione}). Likely, the
investigated temperature range is not wide enough to directly
observe these deviations. We also observe that our estimate in
chapter \ref{JPCMs} of the reaction constant at room temperature on
wet natural silica,
$k_0$($T$=300\,K)=8.3$\times$10$^{-20}$cm$^3$s$^{-1}$ is to be
interpreted as well as a \emph{mean} value on the distribution, and
is in a good agreement with the value at $k_0$(300\,K) of
\figurename~\ref{costantereazionevsT} estimated here on dry natural
silica\footnote{In dry silica $E'$ center are supposedly generated
by two different channels, only one of which is active in wet \sil{}
(see chapter \ref{JPCMs}). The good agreement between $k_0$ in dry
and wet silica is consistent with a simple picture in which the
reaction properties of $E'$ centers with H$_2$ are the same
regardless the origin of the defect, despite the fact that different
generation mechanisms may result in structural variations. This idea
is strongly confirmed by the comparison (see below) with
\emph{surface} $E'$.}$^,$\footnote{We apologize for a slight
difference in the notation used here with respect to the previous
chapter. Here, $k_0$ represents (as in chapter \ref{JPCMs}) the mean
value of the distributed reaction constant $k$. In contrast, in
section \ref{modelingkin} we implicitly used $k_0$ to indicate what
here is referred to as $k$.}.

From the physical point of view, the parameter $k_m$ appears most
suitable to be compared with theoretical predictions. For this
reason, we calculate $k_m$ from $k_0$ by inverting
eq.~(\ref{valmedkdistribuzione}). In the simplest assumption, in
this calculation we fix $\sigma$=$\sigma_0$ at all temperatures.
$k_m$ is reported as full symbols in
\figurename~\ref{costantereazionevsT}, and is consistent with an
Arrhenius dependence from $T$, the best fit values for activation
energy and pre-exponential factor being E$_m$=(0.38$\pm$0.04)\,eV
and $A$=3$\times$10$^{-14}$cm$^3$s$^{-1}$. The uncertainty in E$_m$
mainly derives from the error on $\sigma_0$.

Before going on with the discussion it is worth comparing the mean
activation energy $E_m$ with the value $<$$\epsilon$$>$=0.59\,eV
found in the previous chapter from room temperature data. As already
discussed, measurements at a single temperature do not allow to
distinguish the contribution to $k_0$ of the pre-exponential factor
from that of the activation energy; hence, the 0.59\,eV value does
not have an absolute meaning, because it is based upon a specific
choice $A_W$ of the pre-exponential factor coming from the Waite
theory of diffusion-limited reactions. Recalling
eq.~(\ref{expressionA}) we see that:
\begin{equation}
A_W=4\pi r_0D_{0L}\sim 3.55\times10^{-10}cm^3s^{-1}
\label{Waiterepreprep}
\end{equation}using r$_0$=5$\times$10$^{-8}$cm. Now, data presented here show
that the \emph{real} value $A$=3$\times$10$^{-14}$cm$^3$s$^{-1}$,
estimated from the temperature-dependence of $k_0$, is 4 orders of
magnitude lower than $A_W$. Therefore, one can redefine \emph{a
posteriori} the pre-exponential factor used in the previous chapter,
this leading to a \emph{shift} of the statistical distribution of
the activation energy that makes $<$$\epsilon$$>$ coincide with
$E_m$, as verified by the following equation
\begin{equation}
A_W\exp{\left(\frac{-0.59\,\text{eV}}{k_B\times300\,\text{K}}\right)}\sim
A\exp{\left(\frac{-0.38\,\text{eV}}{k_B\times300\,\text{K}}\right)}
\end{equation} valid within the experimental errors of the two
activation energies. In conclusion, present results are actually
consistent with the treatment proposed in the previous chapter,
although the observation of the temperature dependence of the decay
permits now to attribute to $E_m$, $A$ (and to $A_i$, $E_i$) an
\emph{absolute} meaning.

\subsection{Comparison with theoretical models and previous literature}
\hspace{0.8cm}Reactions of bulk defects in solids with mobile
species are generally assumed to be diffusion-limited, meaning that
their rate is determined by the mobility of the diffuser, whose
migration is usually the bottleneck of the process. In the Waite
treatment of diffusion-limited reactions, the rate constant of
eq.~(\ref{reacteh2repreprep}) is predicted to be (from
eq.~\ref{waiterateconstant}):
\begin{equation}
k_{WD}=4 \pi r_0D_{0L} \exp{\left(
\frac{-E(\text{H$_2$})}{k_BT}\right)}
\end{equation}$k_{WD}$, calculated with a value $r_0$=5$\times$10$^{-8}$ for the capture
radius of $E'$ and $E$(H$_2$), D$_{0L}$ set to the literature
parameters for H$_2$ diffusion in
\sil{},\cite{LeeJCP63,GriscomJNCS84} is reported (dashed line) in
\figurename~\ref{costantereazionevsT} for comparison with present
results. We observe that $k_{WD}$ is everywhere much larger than the
experimental data.\footnote{This remains true with any reasonable
choice of $r_0$, or using any of the slightly different values of
$E$(H$_2$) (0.38\,eV--0.45\,eV) which have been reported in
literature\cite{LeeJCP62,LeeJCP63,PerkinsJPC71,Shelby77}. } This
result is consistent with our repeated observations that the
reaction proceeds more slowly than expected for a purely
diffusion-limited reaction, and with the unrealistically small
values of the capture radius $r_0$, which were found in previous
works trying to reproduce the kinetics of reaction
(\ref{reacteh2repreprep}) at $T$=300\,K within the Waite theory
\cite{ImaiPRB91,CannasJNCS04}.

For the measured macroscopic reaction rate to be so small, it must
be limited also by the rate of the reaction itself, i.e. the
interaction between H$_2$ and $E'$ below the pair separation $r_0$,
rather than by the diffusion of H$_2$ only. These effects can be
included in the expression for the rate constant as discussed in
chapter \ref{backII}\cite{Collins,Waite57,Kuchinsky}: it is assumed
that below $r_0$ the reaction is no longer diffusion-controlled, but
proceeds by first order chemical kinetics, characterized by a rate
constant w[s$^{-1}$]. Hence, the generalized expression for the
overall rate constant is found to be eq.~(\ref{kwaitereaction}),
which we rewrite here in a slightly simplified form:
\begin{equation}
k_{WR}=4\pi r_0 D\frac{w}{w+Dr_{0}^{-2}} \label{kwaitereactionrep}
\end{equation}where $D=D_{0L}\exp{\left(\frac{-E(\text{H$_2$})}{k_BT}\right)}$. When $w\gg Dr_0^{-2}$,
$k_{WR}\sim k_{WD}$ and the reaction is diffusion-limited, but if
$w\ll Dr_0^{-2}$, $k_{WR}\sim 4 \pi r_0^3 w$; in this case, the
overall rate $k_{WR}$ is much smaller than $k_{WD}$ and, most
important, it is not related to the diffusion constant D. From
\figurename~\ref{costantereazionevsT} it is now clear that reaction
(\ref{reacteh2repreprep}) falls in the latter case: the theory of
diffusion-limited reactions is inapplicable to the reaction between
$E'$ and H$_2$, whose rate is mainly determined by the reaction
process in itself rather than by the migration of H$_2$. We stress
that this conclusion is independent of the simplifications used to
treat the effects of the statistical distribution of the activation
energy (such as having supposed a $\sigma$ independent of T),
because it is founded only on the result: $k_m<k_0\ll k_{WD}$. For
the same reason, it would be valid also assuming a non-Gaussian
distribution of $\epsilon$.

\begin{figure*}[t!]
\begin{center}
\fcolorbox{postitA}{postitA}{\begin{minipage}[c]{.95\textwidth}\linespread{1.2}\hrule\vspace{0.2cm}
\textbf{\textsc{Summary I. }}\textsl{\textcolor{postitC}{The
reaction constant $k_0$ between $E'$ and H$_2$ can be extracted by
eq.~(\ref{calcolarek}) from the post-irradiation kinetics of $E'$
measured at several temperatures (\figurename~\ref{kinvstemptutte}).
In this procedure, one must take into account the fact that the
reaction constant is statistical distributed, as shown in the
previous chapter. As a consequence, the mean value $k_0$ directly
extracted from data must be conveniently manipulated by
eq.~(\ref{valmedkdistribuzione}), in order to obtain a parameter
k$_m$ suitable to evaluate the mean activation energy of the
reaction. A detailed study of the temperature dependence of the
reaction constant (\figurename~\ref{costantereazionevsT}) and the
comparison with theoretical models reveal an important result: the
kinetics of the passivation of $E'$ by H$_2$ is reaction-limited
rather than diffusion-limited. Aside from clarifying the reaction
properties of the $E'$ center, this finding is important in that it
is remarkably different from the properties of other defects in
\sil{} reported by previous works.
\vspace{0.2cm}}\hrule}\end{minipage}}
\end{center}
\end{figure*}

Our result on E' strikingly differs from what was found on the other
basic defect in silica, the NBOHC, whose reaction with H$_2$ is
diffusion-limited \cite{GriscomJNCS84,KajiharaPRL02,KajiharaPRB06}.
This remarkable difference indicates that, differently from NBOHC,
the recombination of $E'$ with H$_2$ is not spontaneous, but
requires overcoming an energy barrier, to such an extent that the
character of the reaction is overwhelmingly determined by the local
features of the defect rather than by the matrix in which it is
embedded. The different reactivities of NBOHC and $E'$ with H$_2$
had been anticipated by some computational works
\cite{Erice,VitielloJPCA00}.

Since the reaction rate is unrelated to $D$, our best estimates for
the mean activation energy $E_m$=(0.38$\pm$0.04)\,eV and for the
pre-exponential factor A=3$\times$10$^{-14}$cm$^3$s$^{-1}$, must be
interpreted as features of the local reaction, not related to the
diffusion of H$_2$. In this sense, it is worth noting that the
similarity between $E_m$ and the activation energy 0.45\,eV for
hydrogen diffusion in \sil{} is purely accidental
\cite{LeeJCP63,Doremus}. What's more, as anticipated in the previous
chapter, also the existence \emph{per se} of a distribution in the
activation energy, and its width $\sigma_0$, are not interpretable
here as features of H$_2$ diffusion in the amorphous solid. In this
sense, one must distinguish between two physically different effects
of disorder: inhomogeneity affecting the \emph{diffusion process} in
\sil{}, and inhomogeneity affecting the \emph{reaction properties}
of a specific point defect. The former shows up in the randomization
of the diffusion constant $D$, and in turn of any diffusion-limited
reaction constant. In contrast, the latter effect is probed by a
defect whose distributed reaction constant is \emph{not}
diffusion-limited. The physics of \sil{} provides good examples of
these two conceptually different phenomena in the two basic defects,
the NBOHC and the $E'$ respectively.

Present experimental results may be used as a benchmark to be
compared with computational works, in which several estimates of the
activation energy for process (\ref{reacteh2repreprep}) ranging from
0.2\,eV to 0.7\,eV were proposed, depending on the calculation
method
\cite{Erice,VitielloJPCA00,EdwardsJNCS94,EdwardsJNCS95,PacchioniPRB98c}.
In this sense, the closest agreement is found with the $\sim$0.5\,eV
value reported by Vitiello \emph{et al}\cite{VitielloJPCA00}.

For what concerns the reaction of surface $E'$ with H$_2$, where
diffusion is clearly not a rate limiting factor, the mean activation
energy was reported to be (0.43$\pm$0.04)\,eV, consistent with our
value $E_m$ \cite{RadtzigJPC95}. Also the mean rate constant $k$ at
room temperature
$k$($T$=300\,K)=8.2$\times$10$^{-20}$\,cm$^3$s$^{-1}$, closely
agrees with our estimate
$k_0$($T$=300\,K)=8.3$\times$10$^{-20}$cm$^3$s$^{-1}$. The two
observations lead also to consistent pre-exponential factors. Since
the reaction properties of surface and bulk $E'$ with H$_2$ are
found to be very similar, we conclude that the surroundings of the
defect do not influence its interaction with the H$_2$ molecule, and
only the O$_3$$\equiv$Si$^{\bullet}$ moiety determines the reaction
dynamics\footnote{Put in other words, with respect to the reaction
with H$_2$ the $E'$ centers (at least the type(s) of $E'$ found in
our laser-irradiated samples) basically behave as isolated dangling
bonds}. Such a simple conclusion is far from being obvious \emph{a
priori}, since surface and bulk varieties of $E'$ are known to
differ in basic spectroscopic features, such as the OA peak (which
falls at energies higher than 6\,eV for surface centers) and ESR
signal\cite{Erice}. For comparison, we also observe that the P$_b$
centers (Si$_3$$\equiv$Si$^{\bullet}$) on Si/SiO$_2$ interfaces have
been found to react with H$_2$ with a much higher ($\sim$1.7\,eV)
activation energy \cite{BrowerPRB88,EdwardsPRB92,StesmansAPL96},
this meaning that the nature of the three silicon bonds is a factor
that strongly influences the activation energy.

The reaction kinetics of $E'$ with H$_2$ was also investigated in a
pioneering work by Li \emph{et al.} dealing with $E'$ embedded in
thermal SiO$_2$ films exposed to H$_2$ in a vessel\cite{LiJNCS90},
(see \figurename~\ref{lee}), where it was estimated an activation
energy of 0.3\,eV and a rate constant at room temperature of
$\sim$5$\times$10$^{-22}$cm$^3$s$^{-1}$, which is two orders of
magnitude lower than our $k_0$(300K). Consequently, the authors had
qualitatively drawn the same conclusion that is rigorously confirmed
here, i.e. reaction (\ref{reacteh2repreprep}) is not
diffusion-limited. Although the comparison with our findings could
lead to conclude that reaction parameters of $E'$ in thin films are
significantly different from bulk $E'$, these results cannot be
directly compared to ours. In fact, within the experimental approach
by Li \emph{et al.}, one must take into account also the entry and
permeation processes of H$_2$ in SiO$_2$ as additional steps in the
process which may condition the measure of the reaction parameters,
as clearly discussed by the same authors.

\section{Discussion II: Temperature dependence of the generation process}
\begin{figure}[htb!]
\begin{center}
\includegraphics[width=.7\textwidth]{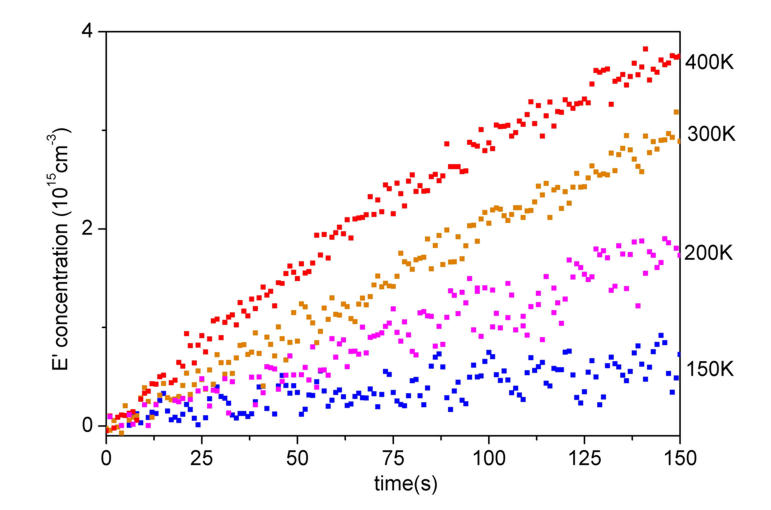}
\end{center}
\caption{Zoom of the first stage of the growth kinetics of
\figurename~\ref{kinvstemptutte} at 4 representative temperatures.
It is apparent the progressive increase of the generation efficieny
with temperature. } \label{Rvstemp}\end{figure}

\begin{figure}[htb!]
\begin{center}
\includegraphics[width=.8\textwidth]{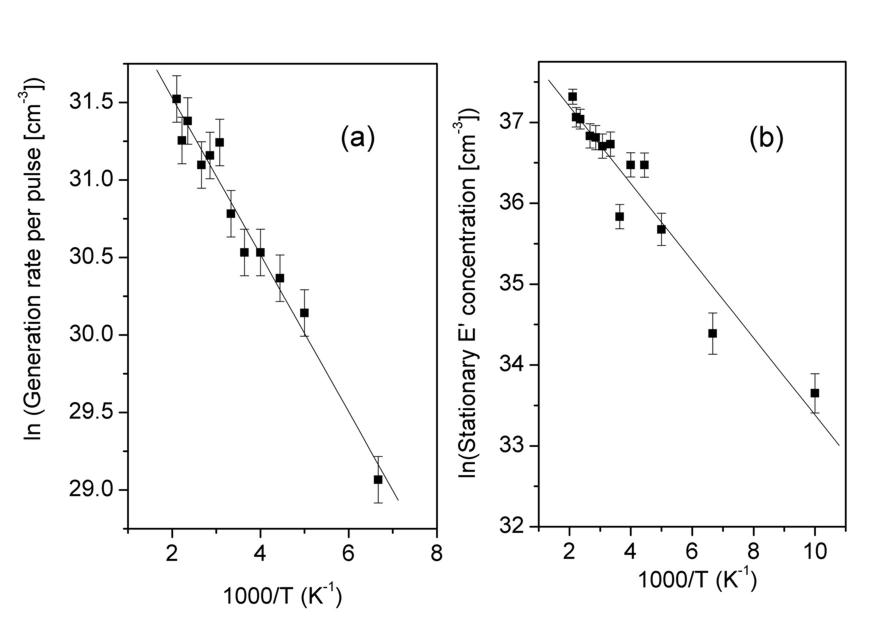}
\end{center}
\caption{Panel (a): Arrhenius plot of the initial generation rate
per pulse of $E'$ in the 150\,K--475\,K interval, extracted from the
kinetics in \figurename~\ref{kinvstemptutte}. Panel (b): Arrhenius
plot of the stationary $E'$ concentration (empty symbols in
\figurename~\ref{maxminvsT}) in the 100\,K--475\,K interval.}
\label{attivazione}\end{figure}

\hspace{0.8cm}Apart from reaction (\ref{reacteh2repreprep}), also
the \emph{generation} of $E'$ appears to be (see
\figurename~\ref{maxminvsT}) a thermally activated process, quenched
below $\sim$150\,K. The activation and the progressive increase of
the generation efficiency with temperature is even more apparent in
\figurename~\ref{Rvstemp}, where we report a zoom of the first
150\,s of the kinetics in \figurename~\ref{kinvstemptutte} at some
representative temperatures. Further information comes from the
analysis of the initial generation rate per pulse $R$ [cm$^{-3}$] at
the beginning of irradiation, defined in chapter \ref{JPCMs}, which
can be estimated by a linear fit in the first $\sim$100\,s of the
exposure session\footnote{We recall that $R$ is given by
$R$=d[$E'$]/dN(N=0)=$\nu_r^{-1}$d[$E'$]/dt(t=0), where $\nu_r$ is
the repetition rate of laser pulses. As $R$ is derived from the
initial linear range of the \emph{in situ} growth curve of $E'$, it
represents the "pure" generation efficiency, not yet affected by the
competition with any concurrent annealing processes of the defect.
}. $R$ is plotted against temperature in the 150\,K--475\,K range in
\figurename~\ref{attivazione}-(a). The values below 150\,K were too
small to be accurately estimated. We see that $R$ is consistent with
an Arrhenius dependence from $T$, with activation energy of 44\,meV
as estimated from a least-square linear fit.

In literature, the temperature-dependence of \emph{stationary}
laser-generated $E'$ was investigated in a few
works\cite{DevineMRSSP86,DevineNIMB94,RothschildAPL89}, which
reported a result not dissimilar to the present one, namely the
formation of $E'$ is quenched below $\sim$150\,K, above which its
efficiency monotonically increases with $T$ (see for example
\figurename~\ref{devinetemp}). This finding was explained as the
consequence of $E'$ being generated by non-radiative decay of
self-trapped excitons (STE), based on the anti-correlated
temperature dependence of the luminescence decay time of STE.
However, to understand if also in the present case STE are involved
in $E'$ generation, more experiments are necessary in order to
directly observe STE by detecting \emph{in situ} their luminescence
signal (reported\cite{Devinebook,Trukhinreviewexc} between
2.0\,eV\,--\,2.3\,eV) during low temperature irradiations, and to
verify if also here their luminescence lifetime is anti-correlated
to the $E'$ generation efficiency.

As proposed in chapter \ref{JPCMs}, in dry natural silica under
laser irradiation the $E'$ centers arise partly from Si\,--\,H
precursors, whose laser-induced breaking generates the defect
together with H:
\begin{equation}
\text{Si\,--\,H
}\stackrel{h\nu}{\longrightarrow}\equiv\text{Si$^{\bullet}$ +
H}\label{generationreprep} \end{equation} and partly from a second
unknown precursor $X$. After irradiation, the $E'$ centers partially
decay, all of them participating in the reaction with H$_2$
independently of their origin. Within this scheme, the parameter $R$
in \figurename~\ref{attivazione}-(a) accounts for both generation
channels. In contrast, the contribution to the $E'$ population
arising from the precursor $X$ can be isolated as it equals the
stationary concentration [$E'$]$_\infty$ of
\figurename~\ref{maxminvsT} (on this point see subsection
\ref{dns}). As shown in \figurename~\ref{attivazione}-(b), also
[$E'$]$_\infty$ approximately follows an Arrhenius
temperature-dependence from 100\,K to 475\,K with 41\,meV activation
energy, although the agreement with the Arrhenius curve is less
satisfactory than for the parameter $R$.\footnote{The value at 10\,K
(not included in the plot) is not consistent with the Arrhenius
dependence; however, the availability of a single data point below
100\,K does not permit for the moment to discuss possible deviations
from the Arrhenius behavior in the low temperature range.} Within
our interpretation scheme, on the whole, the data in
\figurename~\ref{attivazione} suggest that both generation
mechanisms of $E'$ are thermally activated and feature comparable
activation energies. Although this behavior is very clear from a
purely phenomenological point of view, it is difficult at the moment
to provide a satisfactory interpretation without further
experiments\footnote{Clearly, the circumstance that in the present
experimental conditions two distinct processes supposedly contribute
to the formation of $E'$ complicates the interpretation. This
difficulty could be overcome by performing a similar investigation
on \emph{wet} natural silica, where $E'$ was proposed to arise only
from Si\,--\,H breaking. Actually, the experiments presented in this
chapter were carried out on dry specimens because they feature a
higher $E'$ generation efficiency, thereby allowing for more
accurate measurements of the reaction constant $k_0$, this being the
primary aim of this investigation.}. We just remark that the
similarity of the two activation energies seems to suggest a
correlation between the two generation mechanisms. The simplest
explanation of this finding would be that the two processes are both
controlled by non-radiative decay of STE. Nevertheless, other
interpretations are possible, and more experiments are needed to
clarify this point.

As a last point, we discuss qualitatively the data in
\figurename~\ref{deltavsT}. The variation $\Delta$[$E'$]($T$) in the
post-irradiation stage, is an indirect measurement of the amount of
radiolytic hydrogen [H$_2$]($\tau$,$T$) made available by
irradiation\footnote{Although at the lowest temperatures hydrogen is
likely to be stored in atomic form before heating causes its
dimerization.}. Its temperature-dependence is roughly bell-shaped
with a maximum near 250\,K. This finding may be understood as
follows. In chapter \ref{JPCMs} it was proposed a rate equation able
to model the growth of $E'$ by process (\ref{generationreprep})
during irradiation of wet natural silica. The basic idea was the
competition between laser-induced rupture of the precursor, and the
concurrent annealing of $E'$ by H$_2$ due to reaction
(\ref{reacteh2repreprep}). This mathematical model cannot be
directly applied here because of the second channel contributing in
dry specimens to the generation of $E'$. Even so, a similar line of
reasoning allows to understand the temperature-dependence of
$\Delta$[$E'$]. In detail, also here radiolytic hydrogen and a
portion of $E'$ are generated by Si\,--\,H breaking, which is a
thermally activated process (\figurename~\ref{maxminvsT}); hence,
$\Delta$[$E'$]($T$)=2[H$_2$]($\tau$,$T$) remains very low under
$\sim$150\,K. Above this threshold $\Delta$[$E'$]($T$) grows till
the onset ($\sim$200\,K) of H$_2$ diffusion activates the
$E'$--H$_2$ reaction: when this occurs, the accumulation of H$_2$
coming from process (\ref{generationreprep}) during the 2000 pulses
irradiation is limited by the interplay with the concurrent reaction
with $E'$, which at the same time consumes the produced H$_2$. The
reaction constant $k_0$ grows with temperature, and an "optimal"
condition is reached around 250\,K, where generation and annealing
of radiolytic H$_2$ balance at the highest concentration. Then, at
even higher temperatures, the reaction becomes so fast that it
prevents H$_2$ to grow during irradiation, because it is too rapidly
consumed by the simultaneously growing $E'$. For this reason,
$\Delta$[$E'$]($T$) decreases back to zero. The main difference from
the simpler case of wet natural silica, where process
(\ref{generationreprep}) leads to [$E'$]=[H$_2$]/2, is that here,
due to the presence of a second generation channel for $E'$, we have
[$E'$]($\tau$,$T$)$>$[H$_2$]($\tau$,$T$)/2, so that the
concentration of $E'$ can be nonzero even if H$_2$ tends to zero.
Consistently, at $T$$>$400\,K, the generation of $E'$ is still
efficient, even if H$_2$=$\Delta$[$E'$]($T$)/2 fails to grow during
irradiation so that the post-irradiation decay of $E'$ almost
disappears. It is expected that in a similar experiment in wet
natural silica at $T$$>$400\,K, one should observe \emph{both}
[$E'$]($\tau$,$T$) and [H$_2$]($\tau$,$T$) decreasing to zero. Once
understood qualitatively the bell-shaped temperature dependence of
$\Delta$[$E'$]($T$) and the Arrhenius behavior of
[$E'$]$_\infty$($T$) (\figurename~\ref{attivazione}-(b)), the
peculiar temperature dependence of the maximum \emph{in situ}
concentration
[$E'$]($\tau$,$T$)=[$E'$]$_\infty$($T$)+$\Delta$[$E'$]($T$) in
\figurename~\ref{maxminvsT} can be simply regarded as the sum of the
two effects, thus completing our qualitative understanding of data
in \figurename~\ref{maxminvsT} and \figurename~\ref{deltavsT}. Work
is in progress to translate these considerations in a system of rate
equations fitting experimental data.

\section{Conclusions}
\hspace{0.8cm}We investigated by \emph{in situ} optical absorption
spectroscopy the generation and post-irradiation decay of $E'$
centers induced at several temperatures on dry natural silica by
laser-irradiation. We provided a measurement of the temperature
dependence of the reaction constant between bulk $E'$ in \sil{} and
mobile H$_2$. The values of the activation energy and
pre-exponential factor of the process lead to the conclusion that
the reaction kinetics is not diffusion-limited. This result
remarkably differs from what is known for the other basic point
defect in \sil{}, the NBOHC, whose reaction with H$_2$ is purely
diffusion-limited. The comparison with properties of surface $E'$
centers investigated in previous works suggests that the reaction
properties of $E'$ are independent of its surroundings. Besides the
reaction with H$_2$, also the generation of $E'$ appears to be
thermally activated. The competition between the laser-induced
breaking of Si\,--\,H that produces radiolytic hydrogen, and the
concurrent reaction of H$_2$ with $E'$, leads to a characteristic
temperature-dependence of the concentration of the defect at the end
of irradiation, and to the factual deactivation of the
post-irradiation decay process above $\sim$400\,K.

\chapter{Conclusions}\label{chaconc}
\hspace{0.8cm}In this Thesis we have reported an experimental study
of the generation and decay kinetics of point defects induced by
4.7\,eV pulsed laser radiation on amorphous silicon dioxide. Our
investigation was founded on the combined use of several
experimental techniques, mainly \emph{in situ} optical absorption
(OA), electron spin resonance (ESR) and photoluminescence (PL). The
results show that natural \sil{} samples, produced by fusion of
quartz, feature an articulated landscape of processes triggered by
exposure to laser light.

One of them is the generation of the silicon dangling bond defect
($\equiv$Si$^{\bullet}$), a paramagnetic center known as $E'$ in the
specialized literature, which is one of the most important point
defects in silica. A characteristic feature of the laser-induced
$E'$ centers, is that after the end of exposure they are unstable
and undergo a spontaneous decay with a typical time scale of a few
10$^3$\,s. The growth kinetics of $E'$ during irradiation and the
first stages of the post-irradiation decay have been studied by
\emph{in situ} optical absorption spectroscopy. Starting from a few
minutes from the end of exposure, the decay can also be monitored by
ESR, with consistent results. Although laser-induced $E'$ has been
investigated by plenty of works in the specialized literature, the
current knowledge on its laser- or radiation-induced generation is
mainly based on stationary measurements unable to monitor the
kinetics of the defects. Only a few works have studied \emph{in
situ} the generation and decay of the center; furthermore, most
previous \emph{in situ} studies  mainly reported single-wavelength
measurements, while the observation of the complete absorption
profile provides important information that goes beyond the estimate
of the concentration of $E'$.

Simultaneously to the post-irradiation decay of $E'$, we have
observed by ESR the growth of a defect related to the presence of Ge
impurities, known as H(II) (=Ge$^{\bullet}$\,--\,H). Based on the
structure of H(II), which may be regarded as a probe of the presence
in the solid of mobile hydrogen, and on the typical time scale and
temperature dependence of the post-irradiation processes, we have
interpreted both the decay of $E'$ and the formation of H(II) as
effects of molecular hydrogen diffusion in the silica matrix. In our
interpretation, the decay of $E'$ is due to its reaction with mobile
H$_2$, which converts the paramagnetic defect to a Si\,--\,H group
and makes available an hydrogen atom H, while the formation of H(II)
is a consequence of subsequent trapping of H on already present
twofold coordinated Ge impurities (=Ge$^{\bullet\bullet}$), known as
GLPC. This model has been proved by showing that the
post-irradiation kinetics of GLPC and H(II) are anti-correlated, and
that the time dependence of the concentrations of $E'$, GLPC and
H(II) after the end of laser exposure can be reproduced by a
chemical rate equation model describing the diffusion and reactions
of mobile H$_2$ and H. In addition, we have shown that the features
of the kinetics are a fingerprint of the disordered structure of the
amorphous solid, which manifests itself in a statistical
distribution of the activation energy controlling the reaction
constant of $E'$ and H$_2$. Till now, this kind of property had been
pointed out only for the other "basic" defect in amorphous silica,
the oxygen dangling bond, known as Non Bridging Oxygen Hole Center
(NBOHC).

The detailed study \emph{in situ }of the kinetics of $E'$ during and
after irradiation, has led us to conclude that this defect is
generated by laser radiation by breakage of pre-existing Si\,--\,H
bond. This process is also the origin of the radiolytic hydrogen
responsible for the post-irradiation processes; in this sense, the
instability of the generated $E'$ centers is, in the present
context, an intrinsic property of their very generation process.
Although generation of $E'$ from Si\,--\,H had already been proposed
in literature, we have shown that natural (wet) silica allows to
isolate this process from other formation mechanisms of $E'$.
Exploiting this favorable and rather new circumstance, we have been
able to carry out a quantitative study of the generation kinetics of
$E'$: the growth of the defect has been found to be conditioned by
the interplay between laser-induced generation and concurrent
annealing of the induced defects by H$_2$. Based upon this idea, we
proposed a rate equation model able to describe the kinetics and
successfully tested it against experimental data. Moreover, for the
first time to our knowledge, we have provided a clear experimental
proof that Si\,--\,H bond breakage occurs by a two photon absorption
process, thus giving support to previous theoretical calculations.

As a further step, we have studied the temperature dependence of the
kinetics of $E'$ induced by laser irradiation. For what concerns the
post-irradiation reaction of $E'$ with H$_2$, previous literature
data existed only for surface $E'$ and for $E'$ in thin films, and
they were contradictory. In contrast, our study has focused on the
$E'$ in bulk silica, and has permitted to clarify the character of
the reaction between $E'$ and H$_2$, whose kinetics is
reaction-limited rather than diffusion-limited. This is a remarkable
new result that contrasts with the common assumption that kinetics
of reactions in solids driven by diffusion of mobile species are
diffusion-limited due to the slowness of the diffusion process.
Moreover, our result is remarkably different from what is known for
the NBOHC center in silica, whose reaction with H$_2$ is
diffusion-limited. It is worth noting that our results lead to
alternative interpretations of the randomization of the activation
energy for chemical reactions typical of processes in an amorphous
solid.

Another interesting aspect of natural silica is that it allows to
study selectively the laser-induced conversion processes of a
specific and important Ge-related defect, the GLPC. Our findings
strongly suggest that a significant portion of GLPC is converted by
radiation to a diamagnetic defect. This is at variance with most
literature investigations, carried out on Ge-doped silica samples,
in which the GLPC is generally supposed to be converted to
paramagnetic defects. Also, our results underline the importance of
monitoring the kinetics of Ge-related conversion processes, choice
which provides information not available by common studies focusing
only on the stationary concentration of defects. Then, we have
studied the effects of repeated irradiations, demonstrating that
after each high-dose exposure the samples lose memory of their
previous history, namely the post-irradiation kinetics repeat
identically after each new irradiation. A necessary step for the
achievement of such an effect, is the destruction of already-formed
H(II) centers by laser light, which are back-converted to GLPC by
breaking of the Ge-H bond. We have measured the cross section for
this process and have proposed that it occurs by direct single
photon absorption at the defect site.

We argue that these results constitute significant advancements in
the understanding of laser-induced processes in silica. From the
experimental and technological points of view, they prove the
usefulness of \emph{in situ} time-resolved detection of absorption
spectra to perform comprehensive studies on transient point defect
conversion processes and their effect on the transparency of optical
materials during UV exposure.

Several interesting questions remain open and could be investigated
by further experiments. To mention some, more details on the
generation process of $E'$ could be inferred by studying the
dependence of the process from the laser wavelength or by observing
the process with pump-and-probe techniques on the femtosecond scale.
Also, an increase of the available time resolution, possible with
slight modifications of our experimental apparatus, would permit to
study the fast ($<$1\,s) stages of the decay of $E'$ supposedly
involving atomic hydrogen. Loading with O$_2$ or H$_2$O and an
extension of the temperature range of the experiments would allow to
comprehensively investigate the reaction properties of $E'$ with
other mobile species. Finally, our \emph{in situ} approach could be
applied also to other systems, such as pure silica under ionizing
($\gamma$ or $\beta$) radiation, or Ge-doped silica under laser
exposure.

\backmatter
\chapter*{List of related papers\markboth{List of related papers}{}}
\hypertarget{LRP}{}%
\addcontentsline{toc}{chapter}{List of related papers}

\renewcommand{\labelenumi}{$\bullet$}
\small
\begin{enumerate}
\item M. Cannas, S. Agnello, R. Boscaino, S. Costa, F. M. Gelardi, and F. Messina, \emph{Growth of H(II) centers in natural silica after UV laser exposure}, J. Non-Cryst. Solids \textbf{322}, 90 (2003).
\item M. Cannas and F. Messina, \emph{Bleaching of optical activity induced by UV laser exposure in natural silica}, J. Non-Cryst. Solids \textbf{345}, 433 (2004).
\item F. Messina, M. Cannas, and R. Boscaino, \emph{Influence of hydrogen on paramagnetic defects induced by UV laser exposure in natural silica}, Phys. Stat . Sol. (C) \textbf{2}, 616 (2005).
\item M. Cannas and F. Messina, \emph{Nd:YAG laser induced E' centers probed by in situ absorption measurements}, J. Non-Cryst. Solids \textbf{351}, 1780 (2005).
\item F. Messina, M. Cannas, and R. Boscaino, \emph{H(II) centers in natural silica under repeated UV laser irradiations}, J. Non-Cryst. Solids \textbf{351}, 1770 (2005).
\item F. Messina and M. Cannas, \emph{In situ observation of the generation and annealing kinetics of E' centers induced in amorphous SiO$_2$ by 4.7\,eV laser irradiation}, J. Phys.: Condens. Matter \textbf{17}, 3837 (2005).
\item F. Messina, M. Cannas, K. M\'{e}djahdi, A. Boukenter, and Y. Ouerdane, \emph{UV-Photoinduced defects in Ge-doped optical fibers}, Proceedings of 2005 IEEE/LEOS Workshop on Fibres and Optical Passive Components, ISBN 0-7803-8949-2, S. Riva-Sanseverino, M. Artiglia (Eds.), pag. 313 (2005).
\item F. Messina and M. Cannas, \emph{Hydrogen-related conversion processes of Ge-related point defects in silica triggered by UV laser irradiation}, Phys. Rev. B \textbf{72}, 195212 (2005).
\item K. M\'{e}djahdi, A. Boukenter, Y. Ouerdane, F. Messina, and M. Cannas, \emph{Ultraviolet-induced paramagnetic centers and absorption changes in singlemode Ge-doped optical fibers}, Optics Express \textbf{14}, 5885 (2006).
\item F. Messina and M. Cannas, \emph{Photochemical generation of $E'$ centres from Si\,--\,H in amorphous SiO$_2$ under pulsed ultraviolet laser irradiation}, J. Phys.: Condens. Matter \textbf{18}, 9967 (2006).
\item F. Messina, S. Agnello, R. Boscaino, M. Cannas, S. Grandi, and E. Quartarone, \emph{Optical properties of Ge-oxygen defect center embedded in silica films}, in press on J. Non-Cryst. Solids (2007).
\item K. M\'{e}djahdi, A. Boukenter, Y. Ouerdane, F. Messina, and M. Cannas, \emph{Role of diffusing molecular hydrogen on relaxation processes in Ge-doped glass}, in press on J. Non-Cryst. Solids (2007).
\item F. Messina and M. Cannas, \emph{Stability of $E'$ centers induced by 4.7\,eV laser irradiation in SiO$_2$}, in press on J. Non-Cryst. Solids (2007).
\item F. Messina, M. Cannas, R. Boscaino, S. Grandi, and P. Mustarelli, \emph{Optical absorption induced by UV laser radiation in Ge-doped amorphous silica probed by in situ spectroscopy}, in press on Physica Status Solidi (2007).
\item S. Agnello, G. Buscarino, M. Cannas, F. Messina, S. Grandi, and A. Magistris, \emph{Structural inhomogeneity of Ge-doped amorphous SiO$_2$ probed by photoluminescence lifetime measurements under synchroton radiation}, in press on Physica Status Solidi (2007).
 \end{enumerate}
\clearpage{\pagestyle{empty}\cleardoublepage}

\renewcommand{\bibname}{References}
\footnotesize
\linespread{0}

\includepdf[pages={1,2}, scale=1, offset=0 -1cm]{Figure/2pagvuote.pdf}
\end{document}